\documentclass[11pt,a4paper]{article}

\usepackage{jheppub}
\usepackage{chessfss}
\pdfoutput=1
\pdfpageattr {/Group << /S /Transparency /I true /CS /DeviceRGB>>}
\hypersetup{colorlinks=true,linkcolor=blue,citecolor=blue}
\usepackage{amsmath,amssymb,amsfonts}
\usepackage{caption,subcaption}
\usepackage[usenames,dvipsnames]{xcolor}
\usepackage{tikz}
\usepackage{multirow}

\newcommand{\SU}{\mathrm{SU}}
\newcommand{\UU}{\mathrm{U}}

\definecolor{darkergreen}{rgb}{0,0.667,0}
\definecolor{gold}{rgb}{1.,0.843137,0.}
\definecolor{deepskyblue}{rgb}{0.,0.74902,1.}
\definecolor{mediumorchid}{rgb}{0.729412,0.333333,0.827451}

\usetikzlibrary{arrows}
\usetikzlibrary{positioning}
\usetikzlibrary{decorations.pathmorphing}
\usetikzlibrary{decorations.markings}
\def\arrowhead{angle 90}
\tikzset{>=\arrowhead}
\pgfarrowsdeclaredouble{<<}{>>}{\arrowhead}{\arrowhead}
\pgfarrowsdeclaretriple{<<<}{>>>}{\arrowhead}{\arrowhead}
\pgfarrowsdeclaredouble{<<<<}{>>>>}{<<}{>>}
\tikzstyle{W}=[draw, circle, minimum size=1em, scale=1, inner sep=2pt]
\tikzstyle{B}=[draw,circle,fill=black,scale=1]
\tikzstyle{D}= [circle, minimum size=1em]
\tikzstyle{H}=[draw,circle,fill=gray,scale=1]
\tikzstyle{every picture}=[scale=1,baseline=(current bounding box.south)]

\newcommand{\beq}{\begin{equation}}
\newcommand{\eeq}{\end{equation}}
\newcommand{\bea}{\begin{eqnarray}}
\newcommand{\eea}{\end{eqnarray}}

\newcommand{\CC}{{\mathcal C}}
\newcommand{\CD}{{\mathcal D}}

\newcommand{\CN}{{\mathcal N}}
\newcommand{\CO}{{\mathcal O}}

\newcommand{\CS}{{\mathcal S}}

\newcommand\U{{\rm U}}
\newcommand\SO{{\rm SO}}

\newcommand{\intsec}[2]{{\langle{#1},{#2}\rangle}}

\setboardfontsize{8}

\title{BPS spectrum of Argyres-Douglas theory via spectral network}

\preprint{CALT-68-2948}

\author{Kazunobu Maruyoshi,}
\affiliation{California Institute of Technology \\ Pasadena, CA 91125, USA}
\emailAdd{maruyosh@caltech.edu}

\author{Chan Y.~Park, and}
\emailAdd{splendid@caltech.edu}

\author{Wenbin Yan}
\emailAdd{wbyan@theory.caltech.edu}

\abstract{%
  We study the BPS spectrum of four-dimensional $\CN=2$ superconformal field theory of Argyres-Douglas type, obtained via twisted compactification of six-dimensional $A_{N-1}$ $(2,0)$ theory on a sphere with an irregular puncture, by using spectral networks.
  We give strong evidence of the equivalence of $\CN=2$ superconformal field theories from six-dimensional theories of different ranks by systematically comparing the chamber structure and wall-crossing phenomena.
}


\begin{document}
\setcounter{tocdepth}{2}
\maketitle


\section{Introduction and Summary of Results}
\label{sec:intro}
  The BPS spectrum of supersymmetric field theory is a remarkable object
  which is protected by supersymmetry and includes a lot of information of the theory.
  With $\CN=2$ supersymmetry in four dimensions, we have much more control over the BPS spectrum
  on the Coulomb branch moduli space via the Seiberg-Witten theory \cite{Seiberg:1994rs, Seiberg:1994aj}.
  Not only massless BPS states can be found at the loci where the Seiberg-Witten curve degenerates,
  but the full spectrum of massive BPS states is also tractable. 
  
  There are special points on the Coulomb branch of an $\CN=2$ theory
  where mutually non-local BPS states become massless, first found by Argyres and Douglas \cite{Argyres:1995jj}.
  Theories at these points have a superconformal symmetry, and are strongly coupled:
  there is no known Lagrangian description of such theories.
  Recently, a lot of superconformal field theories (SCFTs) of this Argyres-Douglas type have been found and studied
  by making use of the six-dimensional (2,0) theory viewpoint of four-dimensional $\CN=2$ theory
  \cite{Gaiotto:2009we,Gaiotto:2009hg}:
  the topologically twisted compactification on the Riemann sphere with an irregular puncture\footnote{
    When a Seiberg-Witten differential $\lambda = x\, dt$ has a singularity at a point on a Riemann surface
    of degree more than one we call the point an irregular puncture.
    When the degree is equal to or less than one we call it a regular puncture.
    }
  leads to a four-dimensional SCFT of Argyres-Douglas type
  \cite{Gaiotto:2009hg, Cecotti:2010fi, Cecotti:2011rv, Alim:2011ae, Bonelli:2011aa, Xie:2012hs,
  Cecotti:2012jx, Kanno:2013vi, Cecotti:2013lda}.
  We will focus on the SCFTs obtained from the $A_{N-1}$ $(2,0)$ theory,
  whose Seiberg-Witten curve is the $N$-sheeted cover of the sphere.

  The aim of this paper is to find the BPS spectra of the theories on the Coulomb branch of these SCFTs.
  The main tool we will use is spectral network \cite{Gaiotto:2012rg,Gaiotto:2012db}, which was recently used to study an interesting behavior of BPS spectrum of $\CN=2$ $\SU(3)$ super Yang-Mills theory in \cite{Galakhov:2013oja}.
  A spectral network consists of a set of real one-dimensional lines on the Riemann surface (in our present case the sphere).
  These are refered to as $\CS$-walls and each $\CS_{jk}$-wall is obtained as a solution of
    \bea
    \frac{\partial \lambda_{jk}}{\partial \tau}
     =     e^{i \theta},
    \eea
  where $\lambda_{jk} := \lambda_{j}-\lambda_{k}$, $\lambda_{j}$ is the Seiberg-Witten differential
  on the $j$-th sheet, and $\tau$ is a real parameter along the $\CS_{jk}$-wall.
  By sweeping $\theta$ from $0$ to $2\pi$, there could be critical phases $\theta_{c}$
  where $\CS$-walls corresponding to BPS states of the four-dimensional theory 
  with central charge of phase $\theta_{c}$ appear.
  By this method we find all the BPS states
  at a general point of Coulomb branch moduli space, including strongly-coupled regimes.
  
  One merit of the method using spectral network is that one can examine the BPS spectra over the whole Coulomb moduli space of a theory of class $\CS$ and find various chambers depending on the moduli. These chambers are separated by co\-dimension-one lines, called walls of marginal stability, across which the BPS spectrum is changed. In the chamber where the number of the BPS states are minimal, the spectrum of the SCFTs which we will study here is specified by the $A$- and the $D$-type Dynkin diagrams, agreeing with the results in \cite{Shapere:1999xr, Cecotti:2010fi, Cecotti:2011rv, Alim:2011ae}.

%

  We study BPS spectra of various SCFTs obtained from the $A_{1}$ theory
  compactified on a punctured sphere. 
  There are infinitely many theories depending on the singularity of the punctures. 
  When the sphere has one irregular puncture, the theory corresponds to the maximal conformal point 
  of the pure $\SU(n)$ SYM theory, 
  and when the sphere has two punctures with one being irregular and the other being regular, 
  the theory corresponds to the maximal conformal point of the $\SU(n)$ gauge theory with two flavors.

  As pointed out in \cite{Cecotti:2010fi}, the SCFTs from the $A_{1}$ theory discussed above have another realizations from the $A_{N-1}$ theory, which can be expected from the fact that the ``original'' pure $\SU(N)$ SYM theory and $\SU(N)$ theory with two flavors can be constructed in the $A_{N-1}$ theory framework.
  Spectral network enables us to deal with the theory from the higher rank $A_{N-1}$ theory ($N>2$), whose spectral networks differ from those from the $A_1$ theory by the existence of a joint of multiple $\CS$-walls. Using spectral network, we elucidate the BPS spectra of the SCFTs obtained from the $A_{N-1}$ theory compactified on a punctured sphere.

  We get strong evidence that each SCFT from the $A_{1}$ theory considered above is equivalent
  to an SCFT from the $A_{N-1}$ theory on a sphere with one puncture of a particular singularity
  by showing that the equivalence of the chamber structure, minimal and maximal BPS spectra of the two when they have the maximal flavor symmetry, i.e.\ when all mass parameters vanish.
  We also see the enhancement of the flavor symmetry discussed in \cite{Argyres:2012fu},
  by checking that the BPS states indeed form representations of the flavor symmetry
  when we set the associated mass parameters to vanish.

This paper is organized as follows. Section \ref{sec:4d N=2 SCFTs and spectral networks} describes four-dimensional SCFTs that we will consider in this paper, and reviews how to construct a spectral network and find 4d BPS states from it. In Sections \ref{sec:SCFTs_in_A_n_class} and \ref{sec:SCFTs_in_D_n_class} we study spectral networks of SCFTs whose minimal BPS spectra can be represented with $A_n$ - and $D_n$-quiver, respectively, and propose the equivalence between the theories in each class. In Section \ref{sec:Discussion and outlook} we discuss interesting questions that can be addressed with the approach presented in this paper. In Appendix \ref{sec:AD points of SU(3) SQCD} we describe in detail how to obtain various SCFTs at Argyres-Douglas fixed points starting from high-energy 4d $\CN=2$ theories. In Appendix \ref{sec:IntNumberSWall} we explain how to read out low-energy gauge charges of BPS states from spectral network.


\section{$\CN=2$ SCFTs and Spectral Networks}
\label{sec:4d N=2 SCFTs and spectral networks}
  In this section we introduce the $\CN=2$ SCFTs analyzed in this paper
  and its construction from the six-dimensional $A_{N-1}$ $(2,0)$ theory.
  We then give a quick review of spectral network that will be used to analyze the BPS spectra of the SCFTs in the subsequent sections.

\subsection{$\CN=2$ SCFTs of Class $\CS$}
\label{sec:A class of N=2 SCFTs}
  By partially twisted compactification of the six-dimensional $A_{N-1}$ $(2,0)$ theory
  on $\mathbb{R}^{1,3} \times \CC$ where $\CC$ is a Riemann surface,
  we have a class of four-dimensional $\CN=2$ SCFTs on $\mathbb{R}^{1,3}$,
  specified by the rank of $A_{N-1}$ and by the Riemann surface $\CC$ \cite{Gaiotto:2009we,Gaiotto:2009hg}.
  The Seiberg-Witten curve $\Sigma$ which determines the low energy effective theory on the Coulomb branch
  is given by a curve in $(x, t) \in T^{*} \CC$
  where $x$ and $t$ are the coordinates of the fiber and the base respectively:
    \bea
    x^{N} + \sum_{k=2}^{N} \phi_{k}(t) x^{N-k}
     =     0.
           \label{SW}
    \eea
  The moduli of the $k$-th differentials $\phi_{k}$ on $\CC$ are identified with the Coulomb moduli.
  In terms of the Seiberg-Witten differential which is given in the coordinates as $\lambda = x dt$,
  the central charge of a BPS state is calculated as
    \bea
    Z
     =     \oint_{\gamma} \lambda,
    \eea
  where $\gamma$ is a two-cycle in the curve \eqref{SW}.
  We will see a way to determine $\gamma$ in the next subsection.

  We allow the Riemann surface to have punctures, at which we place codimension-two defects in the $(2,0)$ theory.
  A puncture is associated with a flavor symmetry of four-dimensional theory,
  and is classified into the following two types: 
  regular where $\lambda$ has a simple pole; 
  irregular where $\lambda$ has a higher order pole.
  Equipped with punctures, we denote the class of theories as $\CS[A_{N-1}, \CC; \CD_{1}, \CD_{2}, \ldots]$
  where $\CD$ represents a puncture or defect. 

\paragraph{$A_{1}$ case}
  Let us first focus on the rank one case, namely $\CS[A_{1}, \CC; \CD]$.
  In this case the puncture is simply specified by the degree of the pole of the quadratic differential.
  When the degree is two, $\lambda$ has a simple pole, thus it is regular.
  This is related with an $\SU(2)$ flavor symmetry and denoted as $\CD_{\rm reg}$.
  When the degree is more than two, it is irregular.
  If all the punctures are regular, the four-dimensional theory is a class of SCFTs.
  There are various weak coupling descriptions of this class associated with possible degeneration limits of $\CC$,
  where all the $\SU(2)$ gauge groups have vanishing one-loop beta functions.
  Thus, the only building block of this theory is three-punctured sphere
  associated with four free hypermultiplets.
  Conversely, connecting two punctures corresponds to the gauging of the diagonal $\SU(2)$ symmetry
  by an $\CN=2$ vector multiplet.

  Allowing irregular punctures gives us two more building blocks:
  one is one-punctured sphere with an irregular puncture; the other is two-punctured sphere 
  where one of them is regular and the other is irregular.
  The former is ``isolated'' in the sense that this cannot be used to construct a bigger theory.
  This is classified by the degree $n+5$ ($n>1$) of the pole of the quadratic differential at the puncture.
  Thus let us denote this puncture as $\CD_{n+5}$.
  The four-dimensional theory resulting from the compactification of the six-dimensional theory on this sphere is 
  indeed the nontrivial SCFT of Argyres-Douglas type:
  the maximal superconformal point of $\SU(n+1)$ pure SYM theory \cite{Xie:2012hs,Argyres:2012fu}.
  (The central charges, $a$ and $c$, have been computed in \cite{Shapere:2008zf,Xie:2012hs,Xie:2013jc, Cecotti:2013lda}.)
  We will denote this SCFT as $\CS[A_{1}; \CD_{n+5}]$.
  We omit $\CC$ here and below, since we always consider the case where $\CC$ is a sphere.

  The latter also corresponds to SCFTs of Argyres-Douglas type.
  We denote it as $\CS[A_{1}; \CD_{\rm reg}, \CD_{n+2}]$.
  Because of the existence of the regular puncture, this is not isolated in the sense mentioned previously.
  When $n=1$, the theory is trivial.
  By gauging the diagonal $\SU(2)$ flavor symmetry coming from the two regular punctures
  of two $\CS[A_{1}; \CD_{\rm reg}, \CD_{3}]$'s, we obtain the pure $\SU(2)$ SYM theory.
  When $n=2$, the theory is just two free hypermultiplets.
  When $n>2$, we have the nontrivial SCFT which is the maximal superconformal point of
  $\SU(n-1)$ gauge theory with two flavors \cite{Bonelli:2011aa}.
  Indeed, for $n>2$ the quadratic differential has moduli, indicating that the theory is nontrivial.
  Note that these SCFTs have at least an $\SU(2)$ flavor symmetry associated with the regular puncture.

\begin{figure}[h]
	\centering
	\begin{subfigure}[b]{.3\textwidth}
		\centering
		\begin{tikzpicture}
			\draw (0,0) circle (1);
			\node at (0,0) {$A_1$};
			\node at (-1.35,.7) {$\CD_{n+5}$};
			\node at (-.6,.6) {$\star$};
		\end{tikzpicture}
		\caption{$\CS[A_{1}; \CD_{n+5}]$}
	\end{subfigure}
	\hspace{1em}
	\begin{subfigure}[b]{.3\textwidth}
		\centering
		\begin{tikzpicture}
			\draw (0,0) circle (1);
			\node at (0,0) {$A_1$};
			\node at (-.6,.6) {$\star$};
			\node at (-1.35,.7) {$\CD_{n+2}$};
			\node at (.6,-.6) {$\bullet$};
			\node at (1.2,-.7) {$\CD_{\rm{reg}}$};
		\end{tikzpicture}
		\caption{$\CS[A_{1}; \CD_{\rm{reg}}, \CD_{n+2}]$}
	\end{subfigure}
	\caption{4d SCFTs from 6d $A_1$ theory.} 
\end{figure}
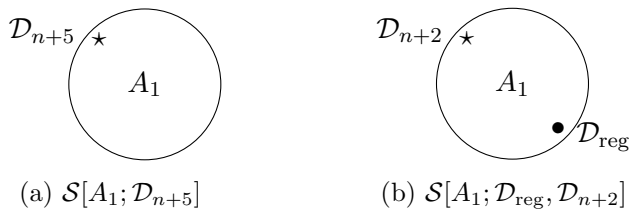

  These two classes are the only possibilities which can be constructed in the $A_{1}$ theory
  on the sphere with the irregular puncture.
  We will elucidate how to obtain the BPS spectra of these two classes of SCFTs 
  in Sections \ref{sec:S[A_1,C;D_n+5]} and \ref{subsec:A1dn}.

\paragraph{$A_{N-1}$ case}
  We then consider the higher rank theory $\CS[A_{N-1}; \CD]$.
  The four-dimensional SCFTs of Argyres-Douglas type are again constructed by the compactification
  on the one-punctured sphere and on two-punctured sphere with one of them being regular.
  We will focus on the case with the regular puncture having an $\SU(N)$ flavor symmetry.
  Let us denote this puncture as $\CD_{\rm reg}$.

  There are various choices of the irregular puncture.
  Among them we will study the following three types of SCFTs in the subsequent sections.
  The one is associated with the one-punctured sphere where the degree $d_{k}$ of the pole
  of the differential $\phi_{k}$ is
    \bea
    (d_{2}, d_{3}, \ldots, d_{N-1}, d_{N})
     =     (4, 6, \ldots, 2N-2, 2N+2).
           \label{Ipole}
    \eea
  We will refer to this as $\CD_{{\rm I}}$.
  This type of SCFTs $\CS[A_{N-1};\CD_{{\rm I}}]$ can be obtained
  as the maximal conformal point of an $\SU(N)$ pure SYM theory, as shown in Appendix \ref{sec:ISYM}. 
  Note that we obtained $\CS[A_{1}; \CD_{N+4}]$ from the same four-dimensional $\CN=2$ theory, 
  so we propose that the two SCFTs are equivalent, though the constructions of these SCFTs are quite different.
  This equivalence was also proposed in \cite{Cecotti:2010fi}.

  The second type is associated with the one-punctured sphere with the following singularity
    \bea
    (d_{2}, d_{3}, \ldots, d_{N-2}, d_{N-1}, d_{N})
     =     (4, 6, \ldots, 2N-4, 2N, 2N+2),
           \label{IIpole}
    \eea
  which we will refer to as $\CD_{{\rm II}}$.
  This type of SCFTs, $\CS[A_{N-1}; \CD_{{\rm II}}]$, can be obtained
  as the maximal conformal point of $\SU(N)$ gauge theory with two flavors,
  as shown in Appendix \ref{sec:II2flavors}.
  Because we get $\CS[A_{1}; \CD_{\rm reg}, \CD_{N+3}]$ from the maximal conformal point 
  of the same four-dimensional theory, we propose that the two SCFTs are equivalent.
  Note that when $N=3$, the singularity is $(d_{2},d_{3}) = (6,8)$.

  The third type is the one associated with the sphere with one regular puncture and one irregular puncture.
  As an illustration of the inclusion of a regular puncture,
  we consider in this paper only one example of SCFT from 6d $A_2$ theory with a regular puncture of the following singularity
    \bea
    (d_{2}, d_{3})
     =     (3,5),
    \eea
  which we will denote as $\CD_{{\rm III}}$.
  We show in appendix \ref{sec:III3flavors} that this is obtained from $\SU(3)$ gauge theory with three flavors
  as the maximal superconformal point \cite{Kanno:2013vi}.

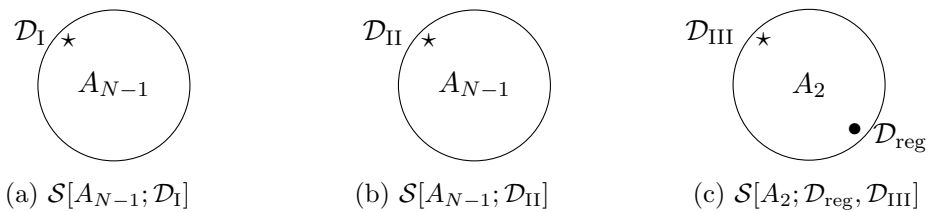
\begin{figure}[h]
	\centering
	\begin{subfigure}[b]{.3\textwidth}
		\centering
		\begin{tikzpicture}
			\draw (0,0) circle (1);
			\node at (0,0) {$A_{N-1}$};
			\node at (-1.1,.7) {$\CD_{\rm I}$};
			\node at (-.6,.6) {$\star$};
		\end{tikzpicture}
		\caption{$\CS[A_{N-1};\CD_{{\rm I}}]$}
	\end{subfigure}
	\begin{subfigure}[b]{.3\textwidth}
		\centering
		\begin{tikzpicture}
			\draw (0,0) circle (1);
			\node at (0,0) {$A_{N-1}$};
			\node at (-1.2,.7) {$\CD_{\rm II}$};
			\node at (-.6,.6) {$\star$};
		\end{tikzpicture}
		\caption{$\CS[A_{N-1};\CD_{{\rm II}}]$}
	\end{subfigure}
	\begin{subfigure}[b]{.3\textwidth}
		\centering
		\begin{tikzpicture}
			\draw (0,0) circle (1);
			\node at (0,0) {$A_{2}$};
			\node at (-.6,.6) {$\star$};
			\node at (-1.3,.7) {$\CD_{\rm III}$};
			\node at (.6,-.6) {$\bullet$};
			\node at (1.2,-.7) {$\CD_{{\rm reg}}$};
		\end{tikzpicture}
		\caption{$\CS[A_{2}; \CD_{\rm reg}, \CD_{\rm III}]$}
	\end{subfigure}
	\caption{4d SCFTs from 6d $A_{N-1}$ theory.}
\end{figure}

  The central charges and some properties of these SCFTs have been considered
  in \cite{Cecotti:2010fi,Xie:2013jc, Cecotti:2013lda}, 
  and the matching of the central charges of the SCFTs supports the proposed equivalences. 
  Matching their BPS spectra provides more powerful evidence for the claims, 
  and we will find the BPS spectra of these SCFTs 
  in Sections \ref{sec:S[A_N-1,C;D_I]}, \ref{sec:S[A_N-1,C;D_II]}, and \ref{sec:S[A_2,C;D_reg,D_III]}.

\paragraph{Equivalence classes of SCFTs}

  One way to summarize the proposed equivalences of the SCFTs is to introduce a notion of equivalence classes of them. Because all the theories above mentioned that are proposed to flow to the same IR fixed point have the same minimal BPS spectrum, which in turn can be conveniently represented by a quiver based on a Dynkin diagram $\Gamma$, we will denote such an equivalence class of SCFTs as a $\Gamma$-class. 
  
  In Table \ref{tab:various_SCFTs_and_BPS_quivers} we summarized the SCFTs in the way that each row corresponds to SCFTs that are in the same class and therefore have the same BPS spectrum \& the same chamber structure. We also described in the table from which 4d $\CN=2$ gauge theories we can obtain the SCFT of each row.

\begin{table}[h]
	\begin{center}
	\begin{tabular}{c|c|c|c}
			$\Gamma$ & 6d $A_1$ & 6d $A_{N-1}$ & UV 4d gauge theory \\ \hline
			$A_{n=N-1}$, $N \geq 3$ & $\CS[A_{1}; \CD_{n+5}]$ & $\CS[A_{N-1};\CD_{{\rm I}}]$  & pure $\SU(N)$ \\ \hline
			\multirow{2}{*}{$D_3 = A_3$} & $\CS[A_{1}; \CD_{8}]$ & \multirow{2}{*}{$\CS[A_3;\CD_{{\rm I}}]$} & $\SU(2)$, $N_{\rm f}=2$ (pure $\SO(6)$) \\
			 & $\CS[A_{1}; \CD_{\rm reg},\CD_{5}]$ & & pure $\SU(4)$ \\ \hline
			\multirow{2}{*}{$D_{4}$} & \multirow{2}{*}{$\CS[A_{1}; \CD_{\rm reg},\CD_{6}]$} & $\CS[A_{2}; \CD_{{\rm II}}]$ & $\SU(3)$, $N_{\rm f}=2$ (pure $\SO(8)$) \\
			& & $\CS[A_{2}; \CD_{\rm reg},\CD_{{\rm III}}]$ & $\SU(3)$, $N_{\rm f}=3$ \\ \hline
			\multirow{2}{*}{$D_{n=N+1}$, $N \geq 4$} & \multirow{2}{*}{$\CS[A_{1}; \CD_{\rm reg},\CD_{n+2}]$} & \multirow{2}{*}{$\CS[A_{N-1}; \CD_{{\rm II}}]$} & $\SU(N)$, $N_{\rm f}=2$ \\
			& & & (pure $\SO(2N+2)$)
		\end{tabular}
	\end{center}
	\caption{Various SCFTs from the 6d $(2,0)$ $A_1$ and $A_{N-1}$ theories in the same $\Gamma$-class.}
	\label{tab:various_SCFTs_and_BPS_quivers}
\end{table}

\subsection{Spectral Network}
\label{sec:SpectralNetwork}
Spectral network is introduced in \cite{Gaiotto:2012rg} as an extension of the analysis done in \cite{Klemm:1996bj,Shapere:1999xr}, building on the previous related work of \cite{Gaiotto:2009hg,Gaiotto:2010be,Gaiotto:2011tf}. Here we will briefly review the topic of spectral network and introduce a couple of additional ingredients in its construction. 

\subsubsection{$\CS$-walls}
\label{sec:AroundABranchPointOfRamificationIndexN}
A spectral network consists of $\CS$-walls, and each $\CS$-wall carries two indices. One convenient picture to have in mind is that, when we consider the low energy effective theory of a four-dimensional $\CN=2$ gauge theory on the Coulomb branch as coming from an M5-brane that wraps a punctured Riemann surface as an $N$-sheeted cover over it \eqref{SW}, these $\CS$-walls correspond to the projections of the boundaries of M2-branes stretched between two sheets of the M5-brane onto the Riemann sphere, and the indices indicate which two sheets the boundaries are. This is not a precise statement, though, as pointed in \cite{Mikhailov:1997jv}, but in some limit of the metric that the M-branes live the correspondence works. More precise statement is understanding an $\CS$-wall as a self-dual string on the Riemann sphere, as explained in \cite{Klemm:1996bj}. However we expect both will give the same answer for the existence of a BPS state and the value of its central charge thanks to supersymmetry. 

Each $\CS$-wall follows the path described by the Seiberg-Witten curve and differential of the four-dimensional theory. When we have a Seiberg-Witten curve $f(t,x)=0$ as a multi-sheeted cover over the $t$-plane and the corresponding Seiberg-Witten differential $\lambda = x dt$, an $\CS_{jk}$-wall of a spectral network satisfies
\begin{align}
	\frac{\partial \lambda_{jk}}{\partial \tau} = \left( \lambda_j (t,x) - \lambda_k (t,x) \right) \frac{dt}{d \tau} = e^{i \theta}, \label{eq:S-wall}
\end{align}
where $\lambda_j$ is the value of $\lambda$ on the $j$-th sheet of $x$ and $\tau$ is a real parameter along the $\CS_{jk}$-wall.

An $\CS$-wall starts either from a branch point or from a supersymmetric joint of $\CS$-walls and flows in general into a puncture. In the following we will provide local descriptions of such cases. By patching the local pictures with the flow that (\ref{eq:S-wall}) describes we construct a spectral network at a value of $\theta$.  

\paragraph{Around a Branch Point of Ramification Index $N$}
\begin{figure}[t]
	\centering
	\begin{subfigure}{.3\textwidth}
		\centering
		\includegraphics[width=\textwidth]{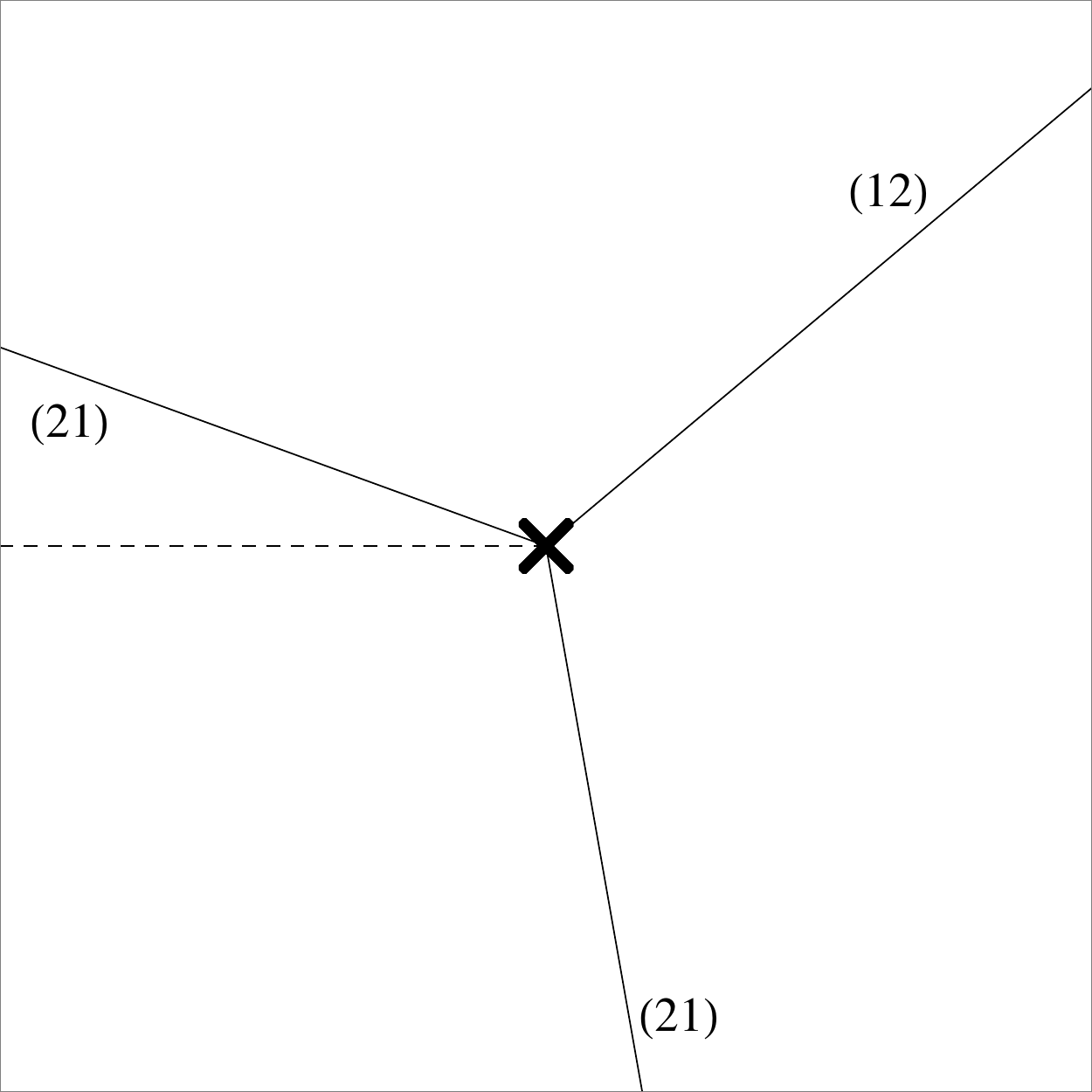}
		\caption{index 2}
		\label{fig:bn2SN}
	\end{subfigure}
	\begin{subfigure}{.3\textwidth}
		\centering
		\includegraphics[width=\textwidth]{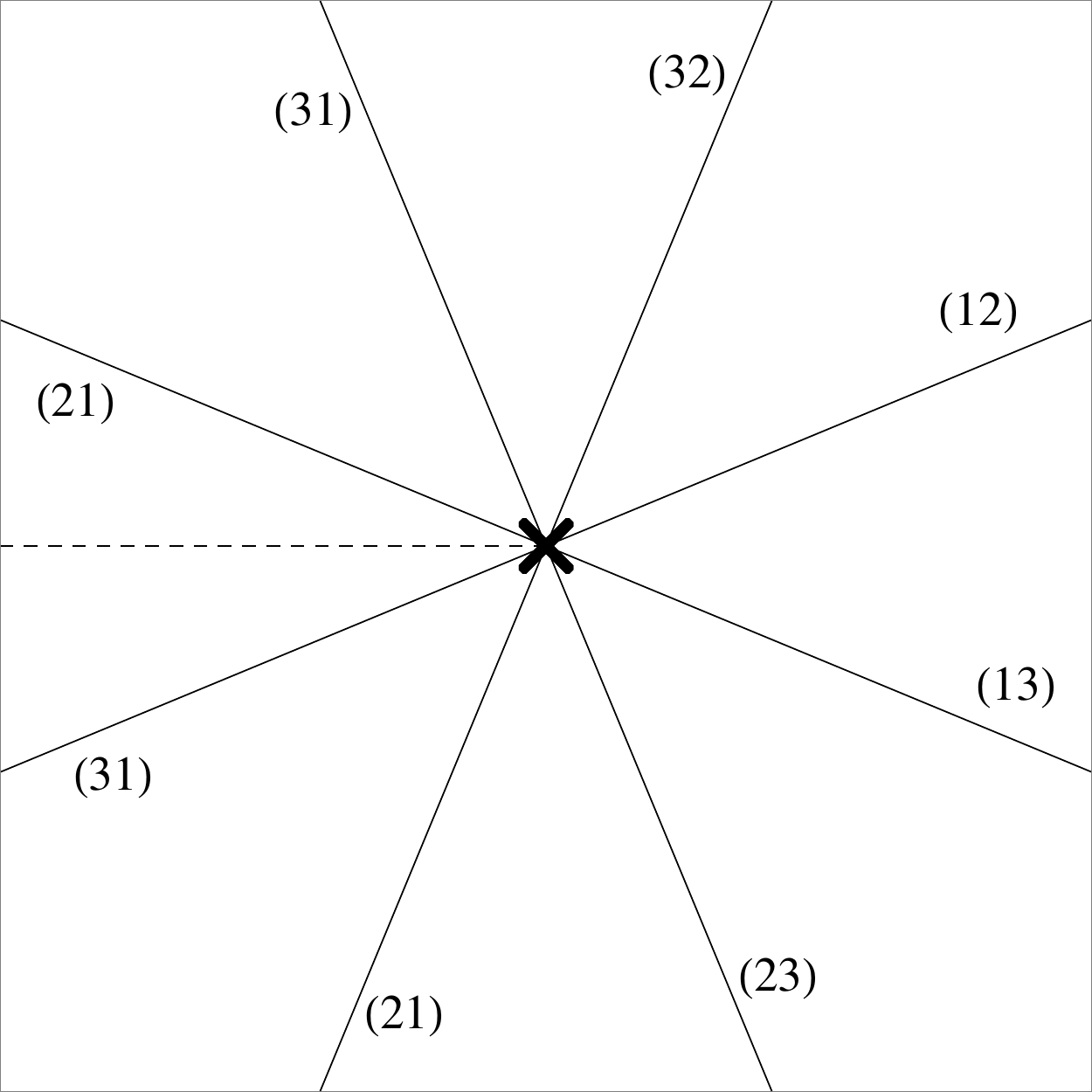}
		\caption{index 3}
		\label{fig:bn3SN}
	\end{subfigure}
	\begin{subfigure}{.3\textwidth}
		\centering
		\includegraphics[width=\textwidth]{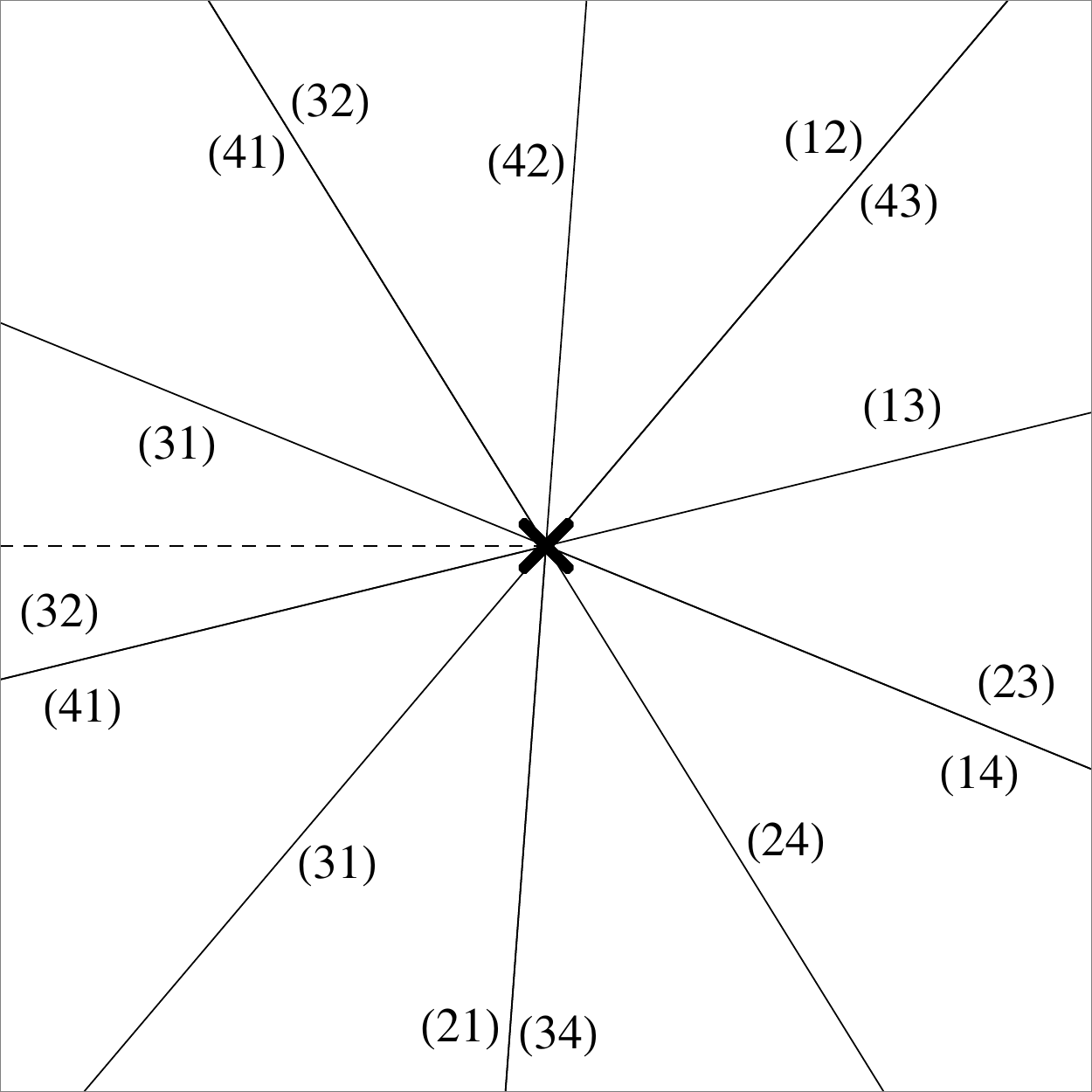}
		\caption{index 4}
		\label{fig:bn4SN}
	\end{subfigure}
	\caption{$\CS$-walls around a branch point.
	         $\CS_{jk}$-walls are denoted by solid lines with $(jk)$. 
	         The broken line denotes the branch cut.}
	\label{fig:S-walls around a branch point}
\end{figure}
First consider $\CS$-walls on the curve $t=x^2$ around the branch point $t=0$, with $\lambda = x\,dt$. On the $t$-plane, each $\CS$-wall travels from the branch point along a real one-dimensional path defined by (\ref{eq:S-wall}).
For a fixed $\theta$, each $\CS$-wall starts at the branch point and goes to infinity 
as shown in Figure \ref{fig:bn2SN}. 
Let us find out the equation that describes each $\mathcal{S}$-wall. We get two branches from the curve,
\begin{align}
	x_1 = \sqrt{t},\ x_2 = -\sqrt{t},
\end{align}
which give us two differential equations for each $\theta$,
\begin{align}
	\lambda_{12}(t) \frac{dt}{d\tau} = 2 \sqrt{t} \frac{dt}{d\tau} = \exp(i\theta),\\
	\lambda_{21}(t) \frac{dt}{d\tau} = -2 \sqrt{t} \frac{dt}{d\tau} = \exp(i\theta).
\end{align}
We will call the $\mathcal{S}$-walls obtained from the first equation $\mathcal{S}_{12}$, and the $\mathcal{S}$-walls from the second equation $\mathcal{S}_{21}$. By changing $\theta \to \theta + \pi$ we can also absorb the sign difference of $\lambda_{12}$ and $\lambda_{21}$, which implies in practice we only need to solve the equation for $\theta \in [0,\pi)$ because the spectral network for $\theta + \pi$ can be obtained by flipping the indices of every $\mathcal{S}$-walls.

\begin{figure}[t]
	\centering
	\includegraphics[width=.3\textwidth]{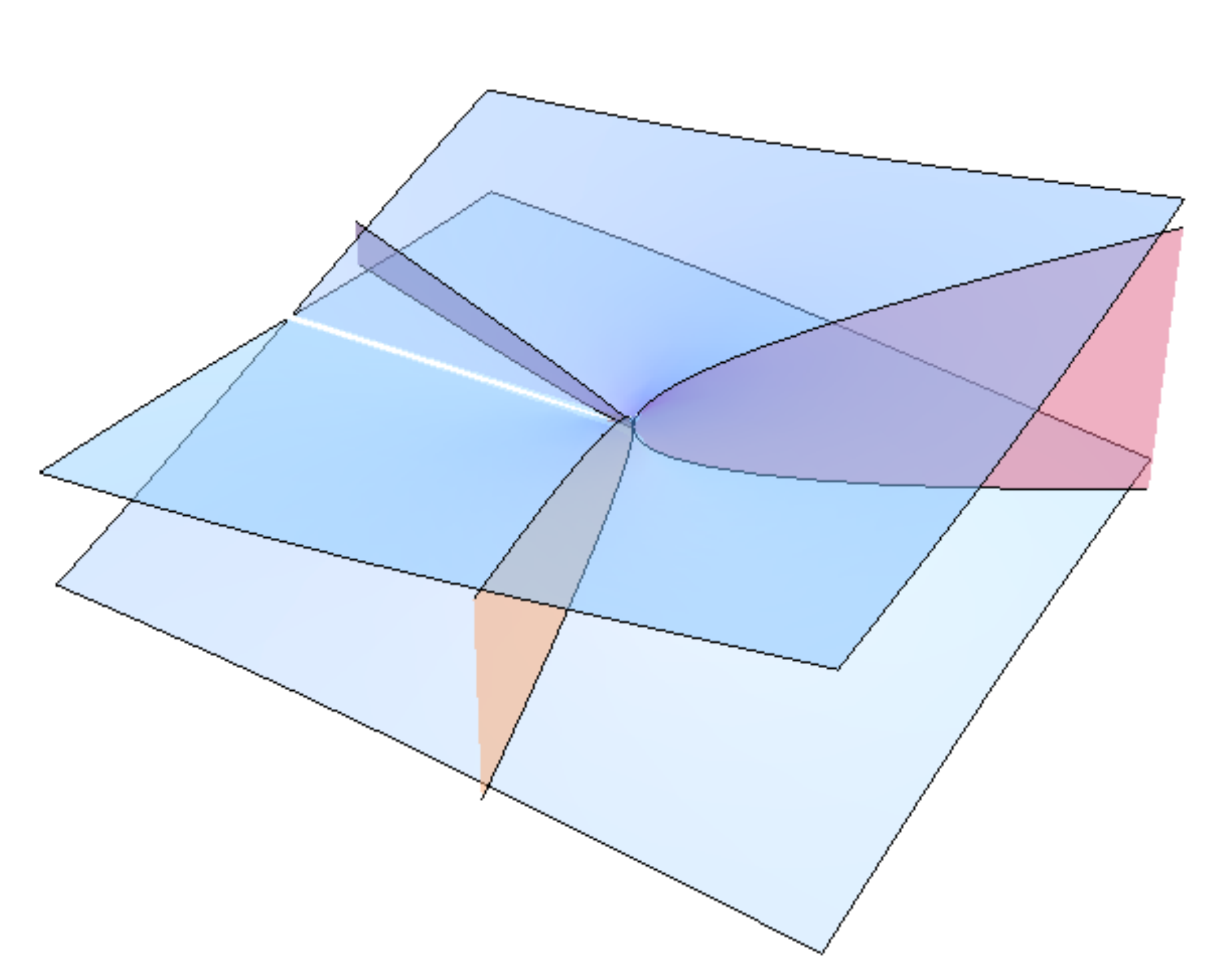}
	\caption{A Seiberg-Witten curve and $\CS$-walls around a branch point of index 2}
	\label{fig:index_2_curve_and_S_wall}
\end{figure}

It is easy to solve the differential equations. We get, for $\mathcal{S}_{12}$,
\begin{align}
	t(\tau) =  \exp\left(\frac{2}{3} i\theta\right) \tau^{2/3}, \label{eq:SWall near a branch point of index 2}
\end{align}
after an appropriate redefinition of $\tau$. Now consider the cases when $\theta$ is changed by a multiple of $2\pi$. When $\theta \to \theta + 2\pi$, the solution gets rotated around the branch point by $4\pi/3$, and it should be another $\mathcal{S}$-wall. However there is a branch cut on the $x$-plane, and if the rotation by $4\pi/3$ makes the $\mathcal{S}_{12}$ go through the branch cut, then the $\mathcal{S}$-wall becomes $\mathcal{S}_{21}$, otherwise it is another $\mathcal{S}_{12}$. 
When we change $\theta$ by $\pi$, the overall spectral network rotates by $\frac{2\pi}{3}$, which can be easily understood from (\ref{eq:SWall near a branch point of index 2}), modulo the flip of the indices of the $\mathcal{S}$-walls as mentioned above. A spectral network around a branch point should be consistent under these monodromies, 
therefore we have three $\mathcal{S}$-walls as shown in Figure \ref{fig:bn2SN}. Figure \ref{fig:index_2_curve_and_S_wall} illustrates the Seiberg-Witten curve and real two-dimensional surfaces that ends on the curve along the $\CS$-walls. Here the Seiberg-Witten curve is represented by plotting the real part of $x$ of the curve over the $t$-plane around a branch point of index 2.

Now let us generalize this analysis to the spectral network from a branch point of index $N$ \cite{Hori:2013ewa}. When we have a branch point of ramification index $N$ at $t = 0$, the corresponding curve is $t=x^N$, and the differential equation that governs the behavior of each $\mathcal{S}_{ij}$ on the $t$-plane is
\begin{align}
	\omega_{ij} t^{1/N} \frac{\partial t}{\partial \tau} = \exp(i \theta),
\end{align}
where $\omega_{ij} = \omega_i - \omega_j$ and
\begin{align}
	\omega_k = \exp \left( \frac{2\pi i}{N} k \right),\ k=0,1,\ \ldots,\ N-1.
\end{align}
Then the solution for an $\mathcal{S}_{ij}$ is
\begin{align}
	t_{ij}(\tau) =\left(\frac{\tau}{\omega_{ij}} \right)^{N/N+1} \exp \left( \frac{N}{N+1} i \theta \right)
\end{align}
after rescaling $\tau$ to absorb a real numerical coefficient. 
From the factor $1/\omega_{ij}$ we find $N(N-1)$ walls, and the exponent $\frac{N}{N+1}$ makes the angles between the walls to be multiplied by the factor $\frac{N}{N+1}$ from the differences of $\arg(1/\omega_{ij})$'s. 
As in the $N=2$ case, the whole spectral network rotates by $\frac{2Nk\pi}{N+1}$ when we change $\theta$ from $0$ to $2k\pi$.
Consistency of a spectral network under this rotation requires $N-1$ additional walls and we have
$N^{2}-1$ $\CS$-walls around the branch point.
The indices of $\CS$-walls are determined by choosing the branch cut.
Figures \ref{fig:bn3SN}, \ref{fig:bn4SN} shows spectral networks around a branch point of index 3 and 4, respectively.

\paragraph{Around a Regular Puncture of Ramification Index $N$}
\label{sec:Around a puncture of ramification index N}
Let us first consider a regular puncture that carries an $\SU(2)$ flavor symmetry in the $A_{1}$ theory. 
The residue of the Seiberg-Witten differential at the puncture is the Cartan of the flavor symmetry, in this case a mass parameter $m$. Consider such a regular puncture at $t=0$, having $m \neq 0$. Then the corresponding (local) Seiberg-Witten curve is
\begin{align}
	t = (v-m)(v+m) = v^2 - m^2
\end{align}
and the Seiberg-Witten differential is
$\lambda = \frac{v}{t}dt$.
When we project the curve on the $t$-plane, we have one branch point of index 2 at $t = -m^2$ and one puncture at $t=0$.

\begin{figure}[ht]
	\centering
	\begin{subfigure}{.4\textwidth}
		\centering
		\includegraphics[width=\textwidth]{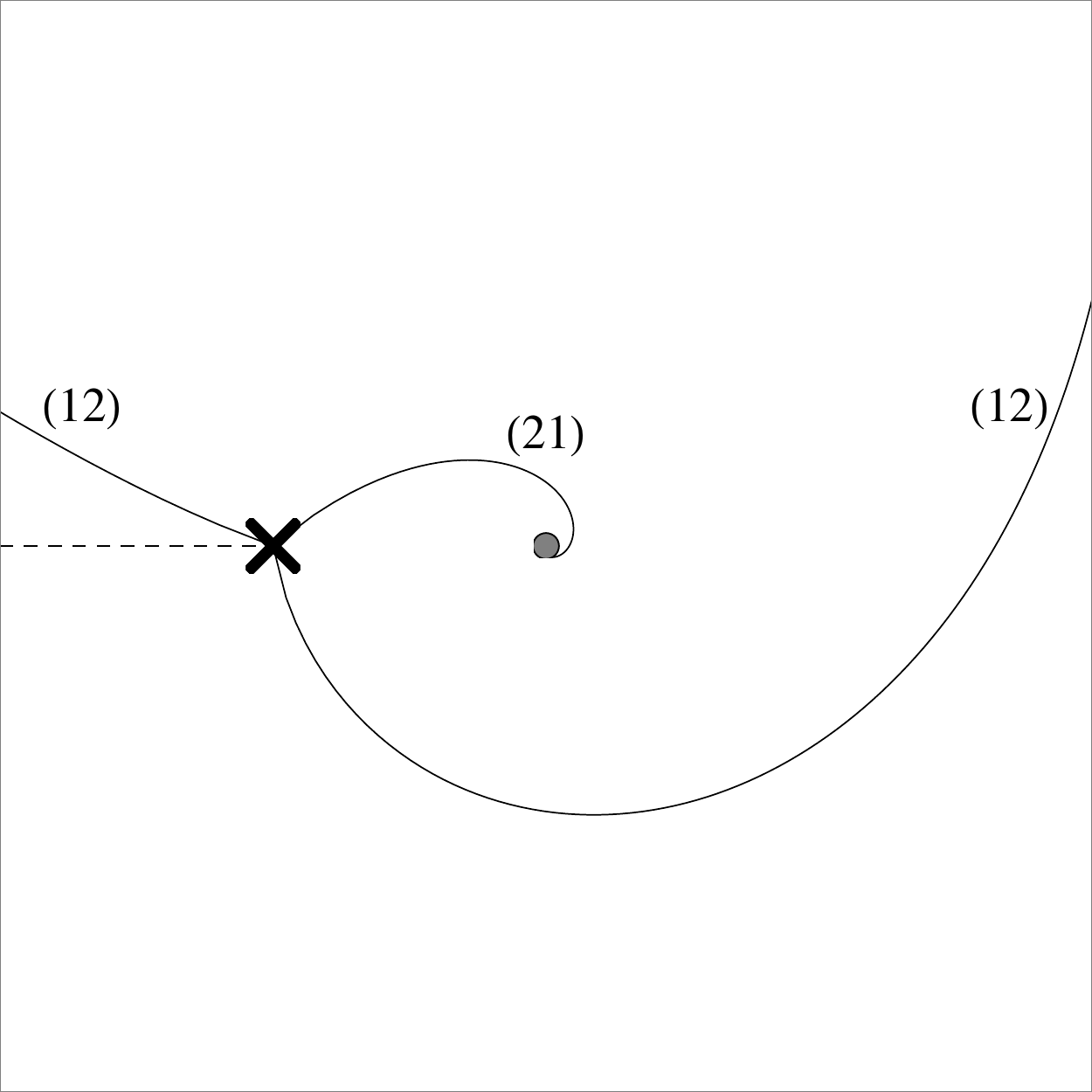}
		\caption{$\theta < \theta_\text{c}$}
		\label{fig:pn2SN_1}
	\end{subfigure}
	\begin{subfigure}{.4\textwidth}
		\centering
		\includegraphics[width=\textwidth]{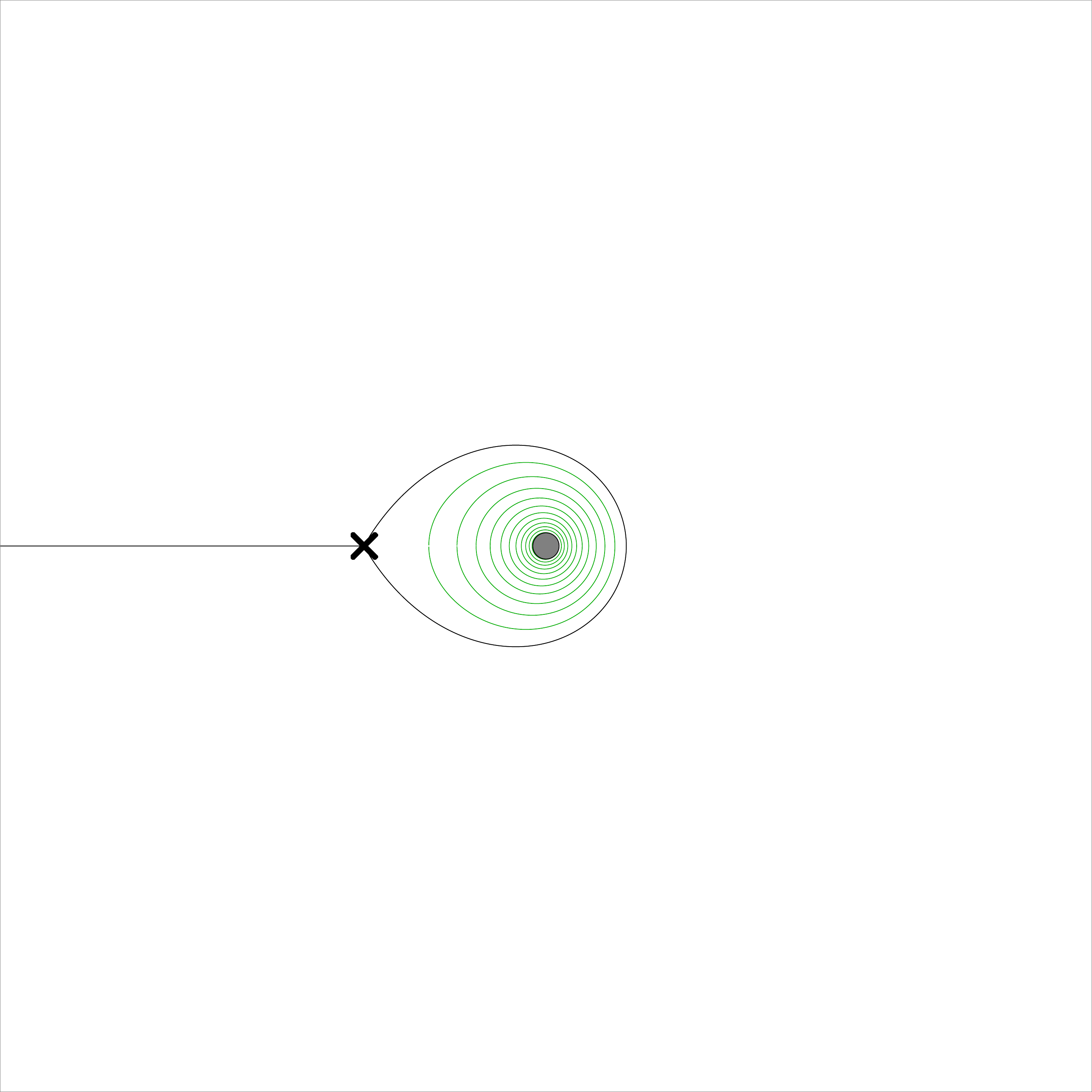}
		\caption{$\theta = \theta_\text{c}$}
		\label{fig:pn2SN_2}
	\end{subfigure}
	
	\begin{subfigure}{.4\textwidth}
		\centering
		\includegraphics[width=\textwidth]{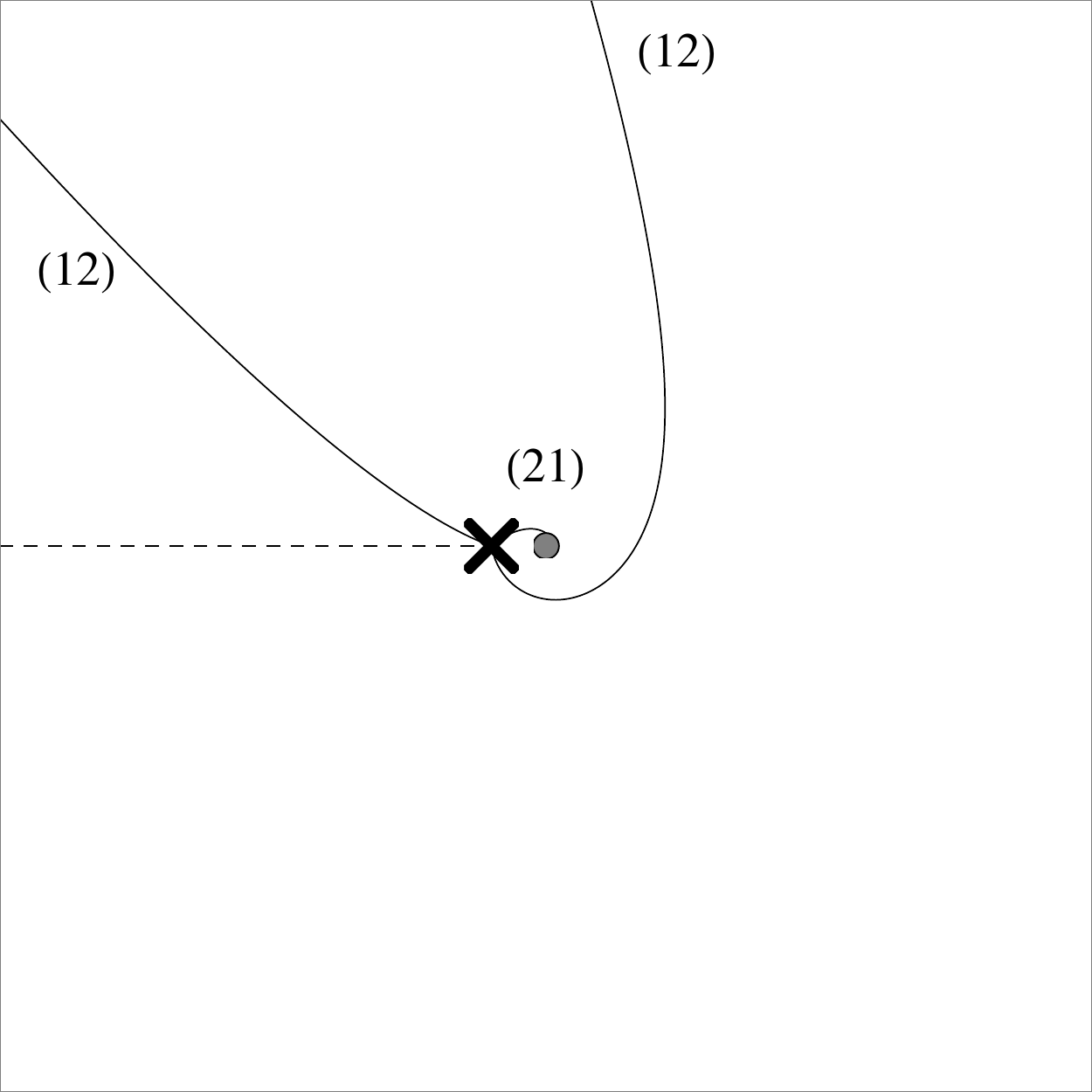}
		\caption{$m \to 0$}
		\label{fig:pn2SN_3}
	\end{subfigure}
	\begin{subfigure}{.4\textwidth}
		\centering
		\includegraphics[width=\textwidth]{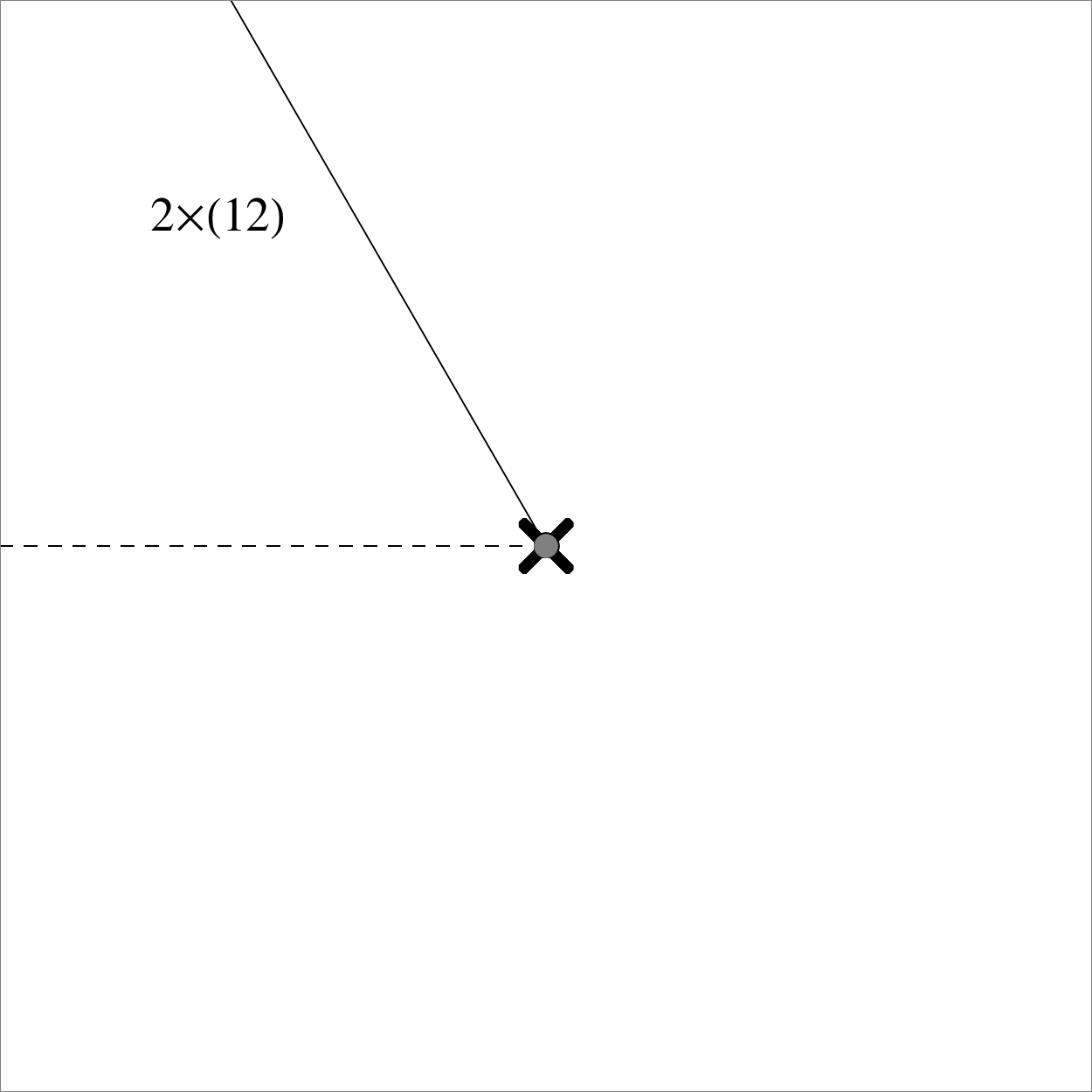}
		\caption{$m=0$}
		\label{fig:pn2SN_4}
	\end{subfigure}
	\caption{$\CS$-walls around an $\SU(2)$ puncture}
	\label{fig:SU2_SWall}
\end{figure}

When $m \neq 0$, we can start with a spectral network from a branch point of index 2, as shown in Figure \ref{fig:pn2SN_1}. Note that one $\CS$-wall flows into the puncture, while the other two escapes to infinity
\cite{Gaiotto:2009hg}. 
When $\theta = \theta_\text{c}$, where $\theta_\text{c} = \arg(m_1 - m_2)+\pi/2 = \arg(2m)+\pi/2$, closed $\CS$-walls can form around the puncture. This $\CS$-wall has a topology of a cylinder, with its boundaries lying along the $\CS$-walls on the two sheets. Therefore it corresponds to a BPS state carrying an $\SU(2)$ flavor charge. This is consistent with the fact that an $\CN=2$ vector multiplet corresponds to an M2-brane with a topology of a cylinder, and when we gauge the flavor symmetry the $\CS$-wall corresponds to a vector multiplet. 
Now consider the limit of $m \to 0$. Then the branch point moves toward the puncture as shown in Figure \ref{fig:pn2SN_3}, and when the two collide, we have a doublet of $\CS$-walls emanating from the puncture. 

Let us then consider the puncture with an $\SU(N)$ flavor symmetry in the $A_{N-1}$ theory.
The curve around the puncture is described by
  \bea
  t
   =     \prod_{i=1}^{N}(v - m_{i}),
  \eea
where $\sum_{i} m_{i} =0$ and the Seiberg-Witten differential is $\lambda = \frac{v}{t} dt$.
Let us focus on the massless limit where $t=0$ becomes the branch point of index $N$, in addition to being the puncture.
The asymptotic behavior of the $\CS$-walls is obtained by solving
\begin{align}
	\int_0^t \omega_{ij} \frac{t'^{1/N}}{t'} dt' = e^{i \theta} \tau,
\end{align}
where we get $t(\tau) = \left( e^{iN\theta} / \omega_{ij}^N \right) \tau$ after rescaling real parameter $\tau$. 
There are $N-1$ sets of asymptotic directions for a value of $\theta$ due to the factor $1/\omega_{ij}^{N}$, 
and along each direction $N$ $\CS$-walls of same indices flow from the puncture. 
In total there are $N(N-1)$ $\CS$-walls from the massless puncture.

\paragraph{BPS Joint of $\CS$-Walls}

When we consider the spectral network in (the compactification of) the $A_{N-1}$ theory, $N > 2$, 
then there are more than two types of $\CS$-walls. When there is a set of $n$ $\CS$-walls $\CS_{i_1 i_2}, \CS_{i_2 i_3}, \ldots, \CS_{i_n i_1}$, there can be a joint of the $\CS$-walls. This is because $\lambda_{i_1 i_2} + \lambda_{i_2 i_3} + 
\cdots + \lambda_{i_n i_1} = 0$ is satisfied at the joint such that it preserves supersymmetry.

\begin{figure}[t]
	\centering
	\begin{subfigure}[b]{.3\textwidth}
		\centering
		\includegraphics[width=\textwidth]{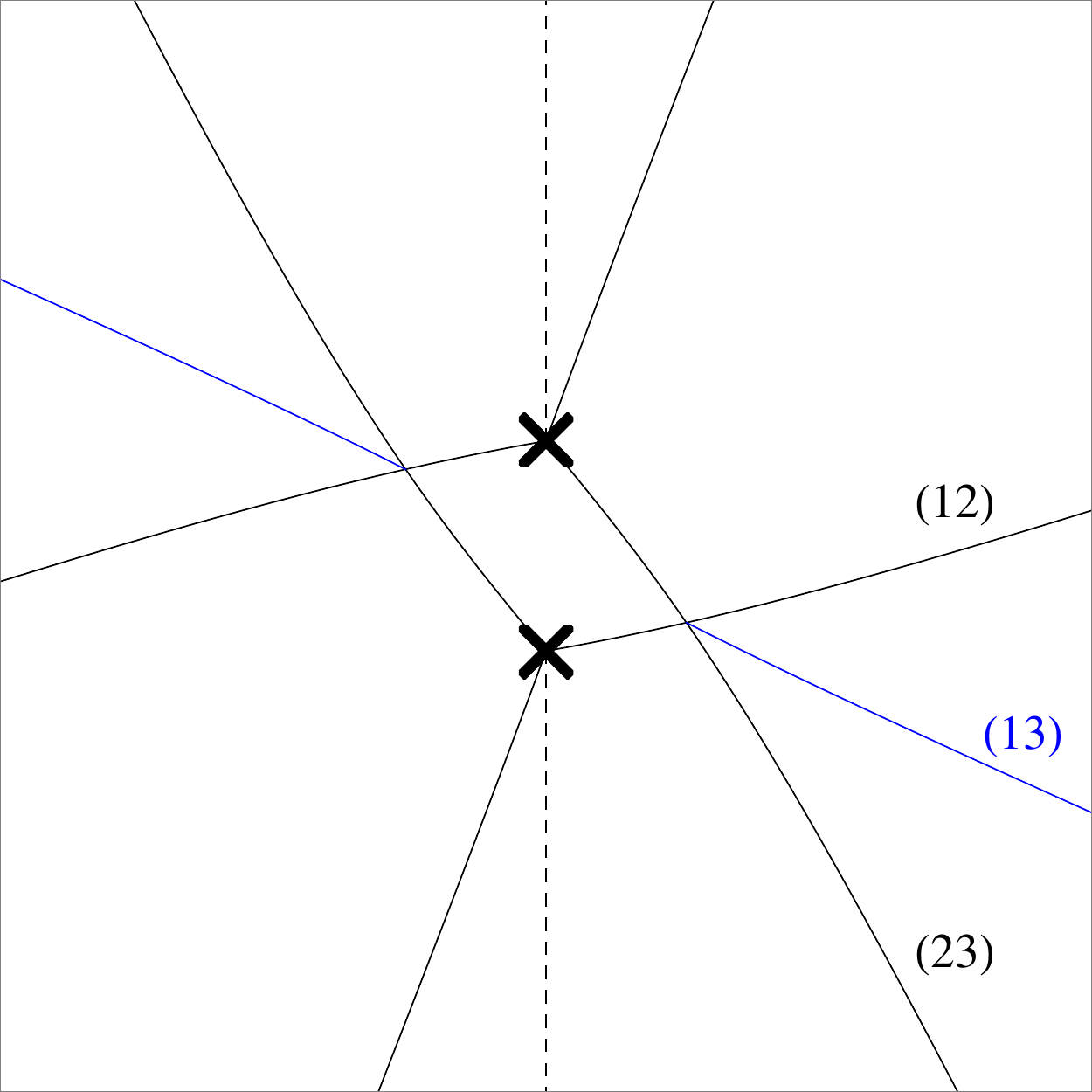}
		\caption{spectral network}
		\label{fig:bn2A2SN}
	\end{subfigure}
	\begin{subfigure}[b]{.4\textwidth}
		\centering
		\includegraphics[width=\textwidth]{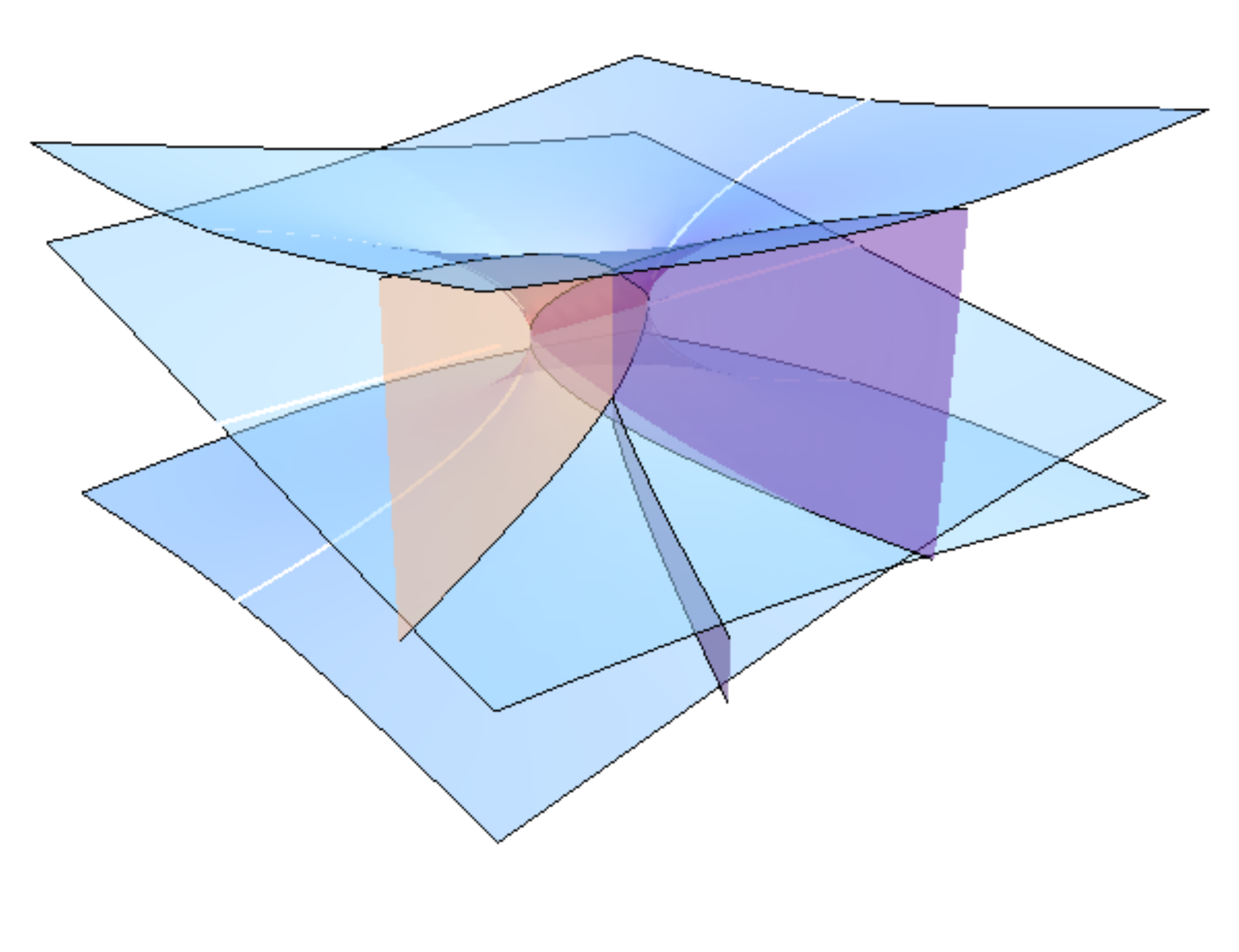}
		\caption{Seiberg-Witten curve and $\CS$-walls}
		\label{fig:bn2A2}
	\end{subfigure}
	\caption{$\CS$-walls forming a joint}
	\label{fig:b2A2_joint}
\end{figure}

Figure \ref{fig:bn2A2SN} shows the spectral network of the $A_2$ theory with two branch points of index 2, where we have $\CS_{13}$ coming from the joint of $\CS_{12}$ and $\CS_{23}$. Figure \ref{fig:bn2A2} illustrates the Seiberg-Witten curve and the three $\CS$-walls that form a joint.

\subsubsection{BPS States from Spectral Network}
\label{sec:4dBPSFromSwall}
\begin{figure}[ht]
	\centering
	\begin{subfigure}{.3\textwidth}
		\centering
		\includegraphics[width=\textwidth]{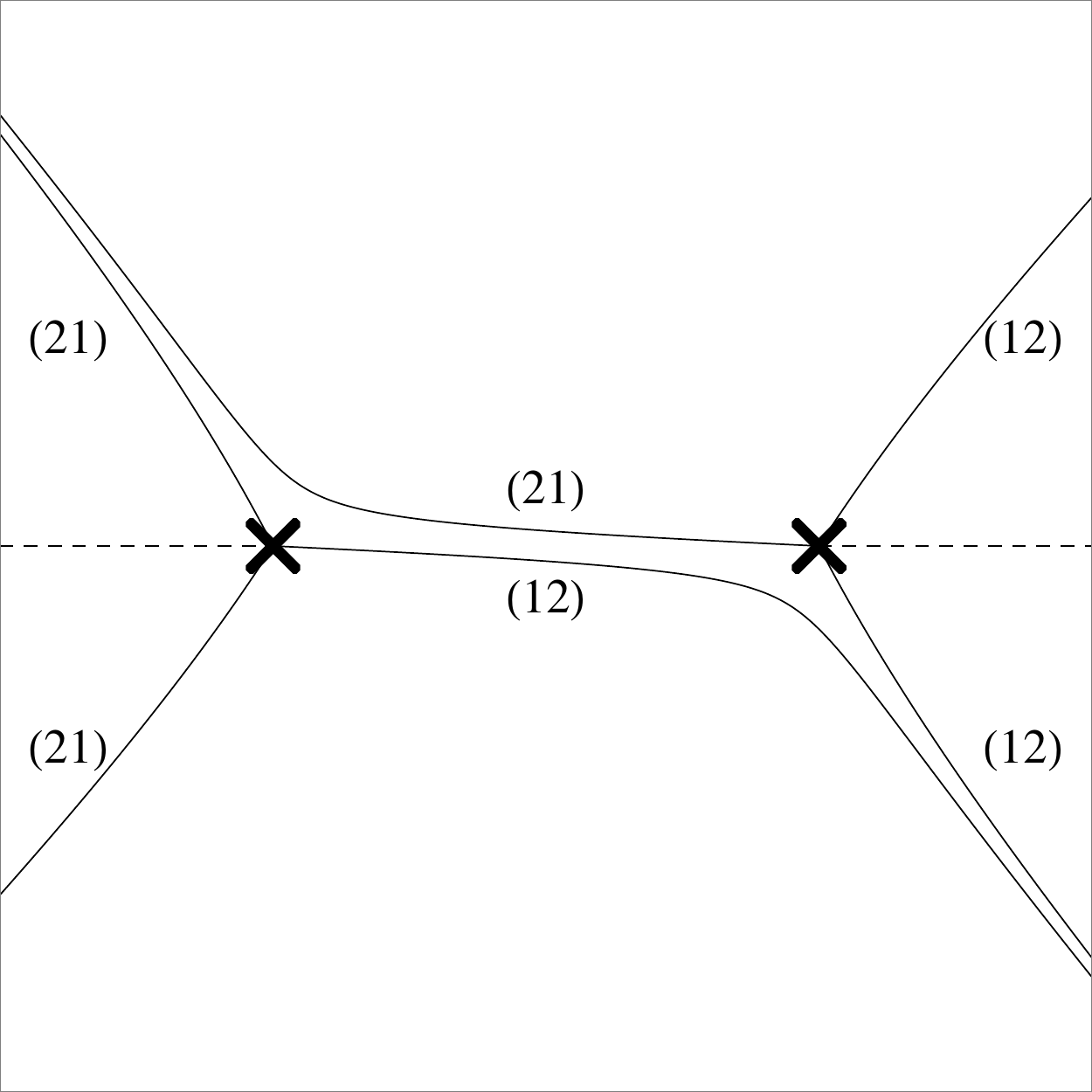}
		\caption{$\theta < \theta_\text{c}$}
		\label{fig:a1SN1}
	\end{subfigure}
	\begin{subfigure}{.3\textwidth}
		\centering
		\includegraphics[width=\textwidth]{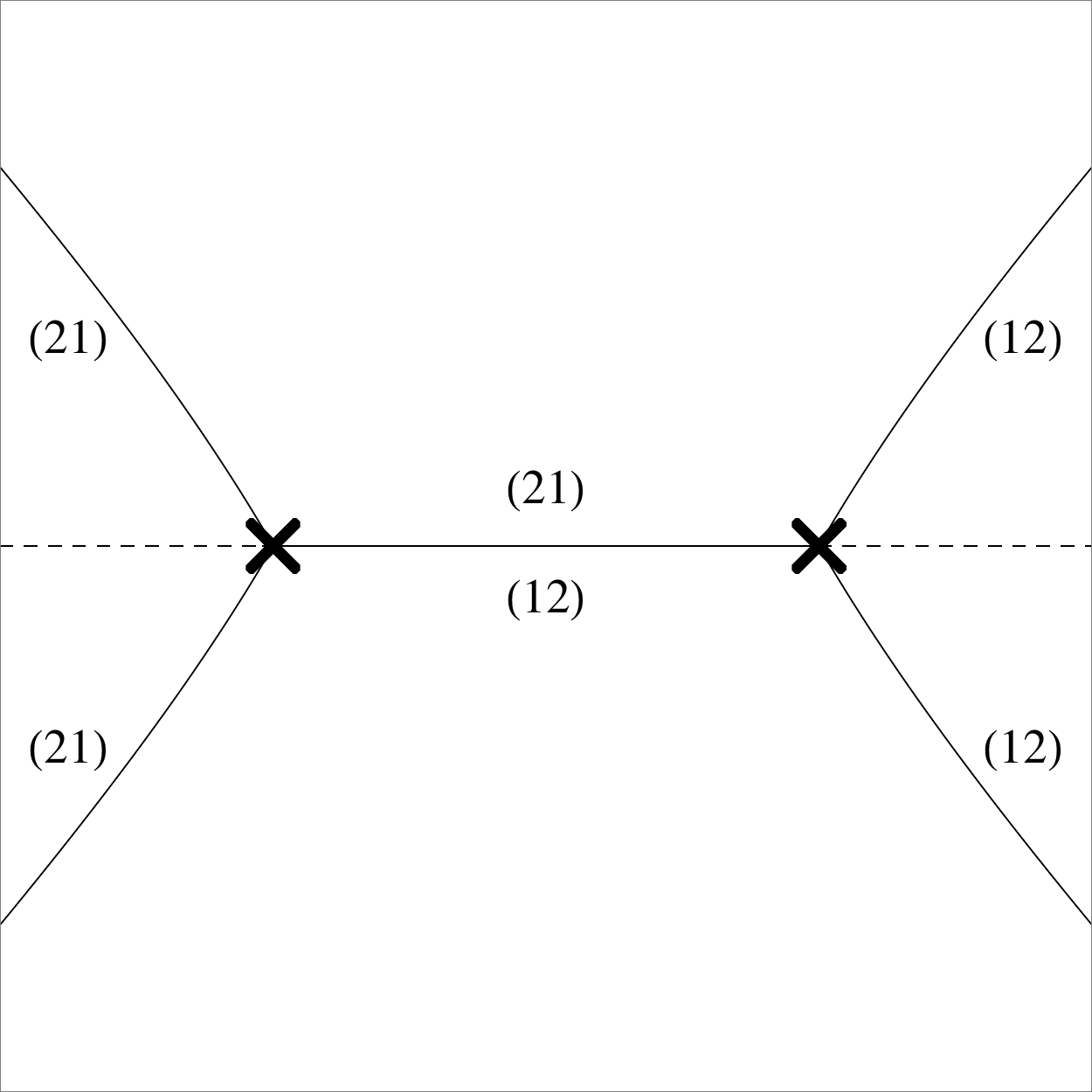}
		\caption{$\theta = \theta_\text{c}$}
		\label{fig:a1SN2}
	\end{subfigure}
	\begin{subfigure}{.3\textwidth}
		\centering
		\includegraphics[width=\textwidth]{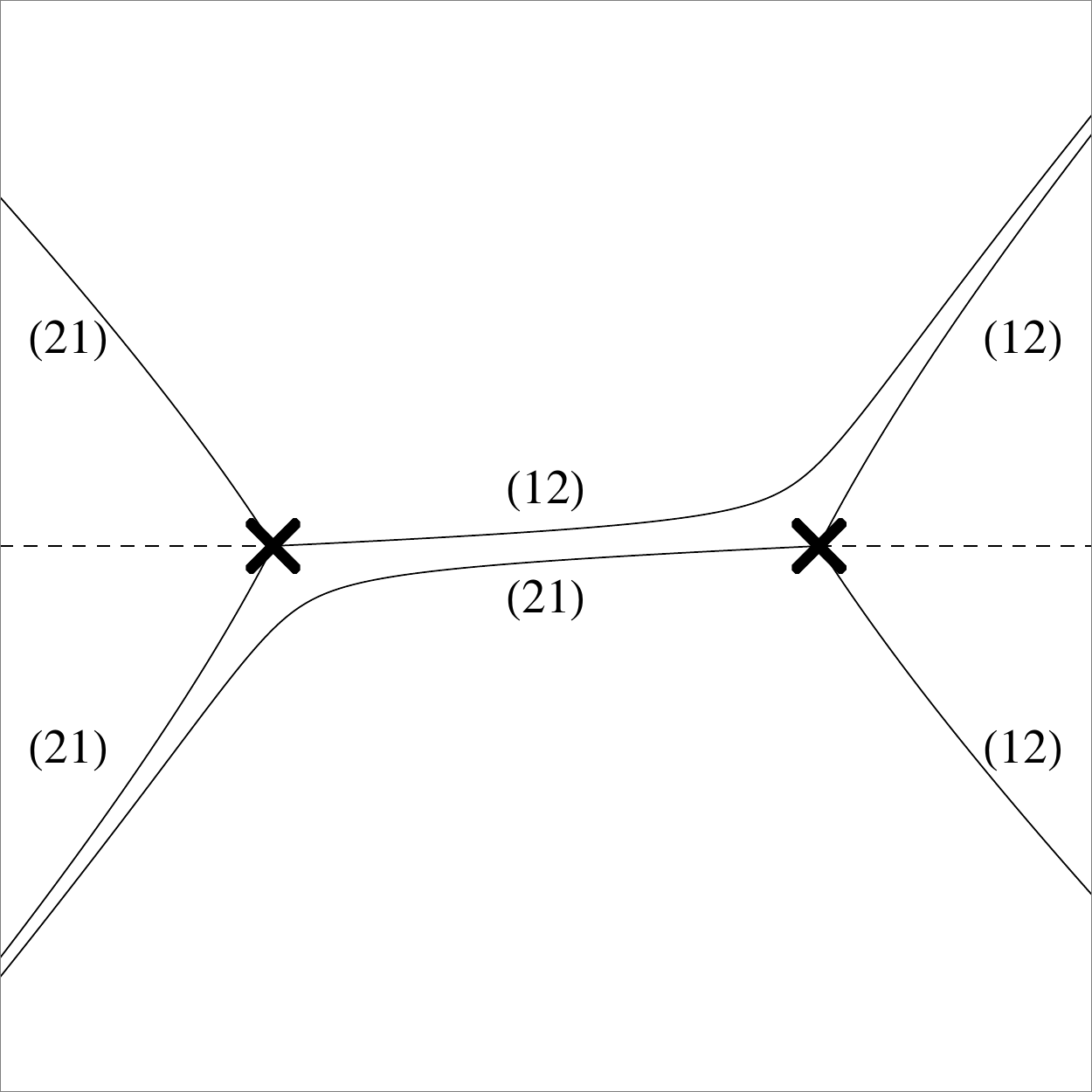}
		\caption{$\theta > \theta_\text{c}$}
		\label{fig:a1SN3}
	\end{subfigure}
	\caption{Finite $\CS$-wall}
	\label{fig:a1SN}
\end{figure}

Using the spectral network, we can identify a BPS state of the low energy effective theory, 
which corresponds to a cycle of the Seiberg-Witten curve, with a finite $\CS$-walls \cite{Gaiotto:2009hg,Gaiotto:2012rg}. 
This is an $\CS$-wall that has a finite value of
\begin{align}
	\int_{\CS_{jk}} \lambda_{jk} = \int_{\tau_\text{i}}^{\tau_\text{f}} \lambda_{jk} (t) \frac{\partial t}{\partial \tau} d\tau = \int_{\tau_\text{i}}^{\tau_\text{f}} e^{i \theta_\text{c}} d\tau = Z,
\end{align}
where $\theta_\text{c}$ is the value of $\theta$ when such a finite $\CS$-wall appears, as shown in Figure \ref{fig:a1SN2}, and $Z$ is the central charge of the corresponding BPS state.
To find out the whole set of BPS states, we evolve a spectral network from $\theta = 0$ to $\theta = 2\pi$ and identify finite $\CS$-walls.

\begin{figure}[ht]
	\centering
	\includegraphics[width=.3 \textwidth]{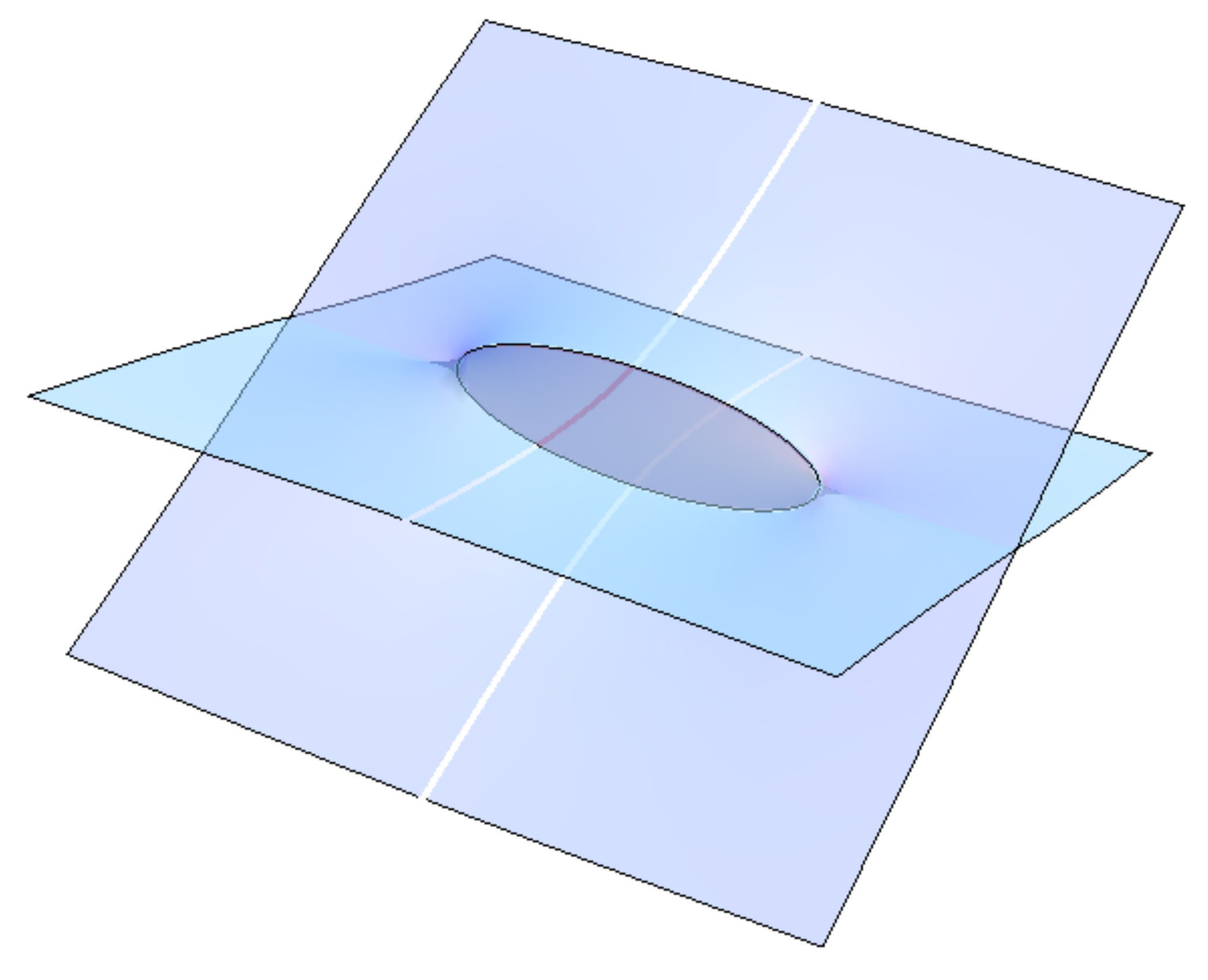}
	\caption{Seiberg-Witten curve from $A_1$ and a finite $\CS$-wall}
	\label{fig:a1_curve_and_finite_S_wall}
\end{figure}

Figure \ref{fig:a1SN} shows how a finite $\CS$-wall forms at $\theta = \theta_\text{c}$. When $\theta < \theta_\text{c}$, the corresponding spectral network, shown in Figure \ref{fig:a1SN1}, has two $\CS$-walls of opposite indices that approach each other. When $\theta = \theta_\text{c}$, the two collide and this indicates that there is a finite $\CS$-wall connecting the two branch points, forming a 1-cycle of the Seiberg-Witten curve. Figure \ref{fig:a1_curve_and_finite_S_wall} illustrates the Seiberg-Witten curve and the finite $\CS$-wall.

A spectral network provides more information than just the existence and the central charge of each BPS state: it also allows us to calculate the electric and the magnetic charges of the BPS state with respect to the IR gauge group. This is done by considering the intersections of the cycles corresponding to finite $\CS$-walls, and in order to do that we need 
to put down the indices of every $\CS$-wall of a spectral network.

We have already studied how to put indices to the $\CS$-walls from a branch point and those from a joint, so the only question remaining is how to patch the $\CS$-walls in the right way. The indices are changed only when an $\CS$-wall crosses a relevant branch cut, which is from the trivialization of the covering map from the Seiberg-Witten curve to the base space. Suppose we have an $\CS_{ij}$-wall crossing a $(jk)$-cut. Then after crossing the cut it becomes an $\CS_{ik}$-wall \cite{Gaiotto:2012rg}, as shown in Figure \ref{fig:S-wall_crossing_a_branch_cut}.
\begin{figure}[ht]
	\centering
	\begin{tikzpicture}
		\draw[
			decoration={markings, mark=at position 0.25 with {\arrow{>}}},
			decoration={markings, mark=at position 0.75 with {\arrow{>}}},
        			postaction={decorate}
        		] (-1,2) .. controls (-1,0) and (0,-1) .. (1,-2);
		\draw[decorate,decoration=snake] (-2,-1) -- (2,1); 
		\node at (-1.25,1) {$ij$};
		\node at (-.2,-1.2) {$ik$};
		\node at (-2.25,-1.25) {$jk$};
	\end{tikzpicture}
	\caption{$\CS$-wall crossing a branch cut}
	\label{fig:S-wall_crossing_a_branch_cut}
\end{figure} 

We can give indices for all the $\CS$-walls of a spectral network in a globally consistent manner, however local information around each intersection of finite $\CS$-walls is enough for us to find out the IR charges of the corresponding BPS states. Appendix \ref{sec:IntNumberSWall} explains how to calculate the intersections of the cycles from finite $\CS$-walls of a spectral network.

\subsubsection{Study of Argyres-Douglas Fixed Points via Spectral Networks}
In Section \ref{sec:AroundABranchPointOfRamificationIndexN} we have discussed which SCFTs from Argyres-Douglas fixed points we are interested in. Here we provide a schematic description of how to study the SCFTs using spectral network.

The distances between branch points depend on the Coulomb branch parameters of the theory, and therefore the mass of the BPS states from finite $\CS$-walls connecting those branch points vanish as we make the parameters to vanish. This could give the mutually nonlocal massless states required for an Argyres-Douglas fixed point. Therefore the limit of Coulomb branch parameters that results in the collisions of branch points are indications of interesting physics, especially for Seiberg-Witten theories from the six-dimensional $A_{N-1}$ $(2,0)$ theory \cite{Park:2011is}.
In general we have many complex parameters controlling the locations of the branch points, which in turn determine the shape of the Seiberg-Witten curve that is a multi-sheeted cover over a punctured Riemann surface. Studying BPS spectra when those parameters have general values is an extensive, often practically impossible task. Instead, here we will pick a few choices of the parameters that are both physically interesting and practically less demanding.

First we maintain as much flavor symmetry as necessary. This makes mass parameters vanish and at the same time simplifies the analysis of spectral networks. When we maintain the maximal flavor symmetry it is easier to discuss an upper bound of the number of BPS states of the theory.

Next, to identify interesting points in the moduli space we consider the discriminant of the equation that describes the locations of the branch points on the $t$-plane. For a Seiberg-Witten curve $f(t,x;u_i)$, where $u_i$ are Coulomb branch parameters, by eliminating $x$ from the following equations
\begin{align}
	f(t,x) = 0,\ \partial_t f(t,x) = 0,
\end{align}
we get a polynomial equation $g(t;u_i)=0$, whose solutions are the locations of the branch points. Therefore the solutions of the discriminant of $g(t;u_i)$, $\Delta_t g (u_i)$, denotes the loci on the Coulomb branch where branch points collide. When we describe the Seiberg-Witten curve as a multi-sheeted cover, not all the choices correspond to singularities of the curve, however some of the choices can result in the collisions of the branch points that connect the same sheets, then it is exactly where we have the singularity of the curve and therefore it may result in a massless hypermultiplet. Even if a solution of the discriminant does not correspond to a singularity, it correspond to having a branch point of higher ramification index, and as we have seen above it results in a more symmetric configuration of spectral network which is easier to analyze.


\section{4d SCFTs in $A_n$-class}
\label{sec:SCFTs_in_A_n_class}

\subsection{$\CS[A_{1}; \CD_{n+5}]$ Theories}
\label{sec:S[A_1,C;D_n+5]}
\label{subsec:A1an}
Let us start studying the spectral network of the $\CS[A_{1}; \CD_{n+5}]$ theories, which are in $A_{n}$-class.
The Seiberg-Witten curve of (the deformation of) this theory is
\begin{align}
	x^2 = t^{n+1}+u_2\, t^{n-1}+ \cdots + u_{n+1},
\end{align}
and the Seiberg-Witten differential is $\lambda = x\, dt$ 
which has one irregular puncture of degree $n+5$ at $t = \infty$ and no regular puncture.

\subsubsection{$\CS[A_{1}; \CD_{7}]$ in $A_2$-class}
\label{sec:S[A_1,C;D_7]}
The simplest theory is $\CS[A_{1}; \CD_{7}]$.
This corresponds to the original example found in \cite{Argyres:1995jj}, and its spectral network, which is studied in \cite{Gaiotto:2009hg}, is the building block of that of $n>2$. The Seiberg-Witten curve is
\begin{align}
	x^2 = t^3 + c_2 t + v_2.
\end{align}
The parameters $v_{2}$ and $c_{2}$ are respectively the vev of the relevant deformation operator 
and its coupling constant with scaling dimensions $6/5$ and $4/5$.
Thus $v_{2}$ is considered as the moduli of the theory.
For a fixed value of $c_2$, there is a single, closed BPS wall on the $v_2$-plane encircling $v_2=0$ \cite{Gaiotto:2009hg}.

\paragraph{Minimal BPS spectrum}

\begin{figure}[h]
	\centering
	\begin{subfigure}[b]{.25\textwidth}	
		\includegraphics[width=\textwidth]{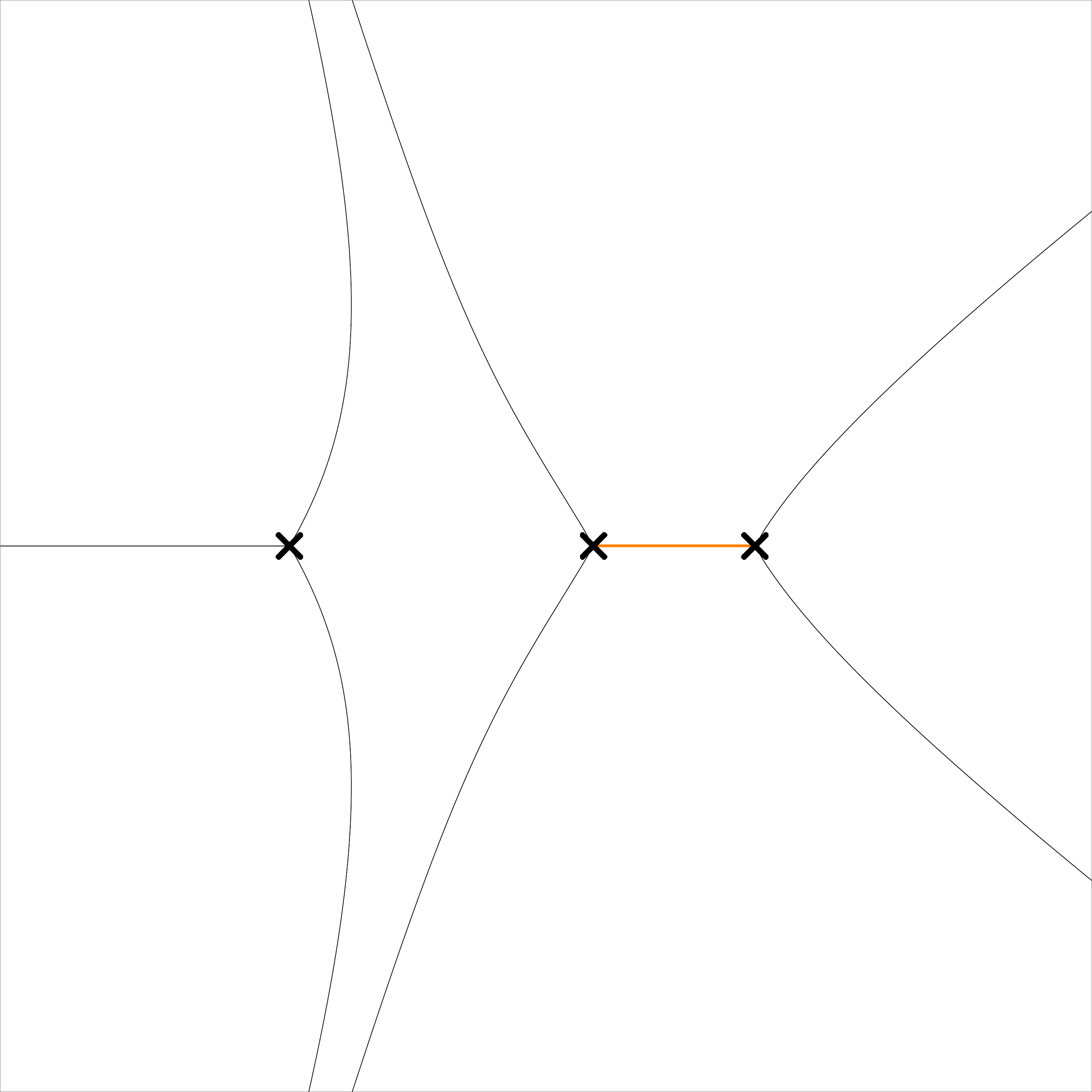}
		\caption{$\theta = \arg(Z_1)$}
		\label{fig:a2SN_minimal_1}
	\end{subfigure}
	\begin{subfigure}[b]{.25\textwidth}	
		\centering
		\includegraphics[width=\textwidth]{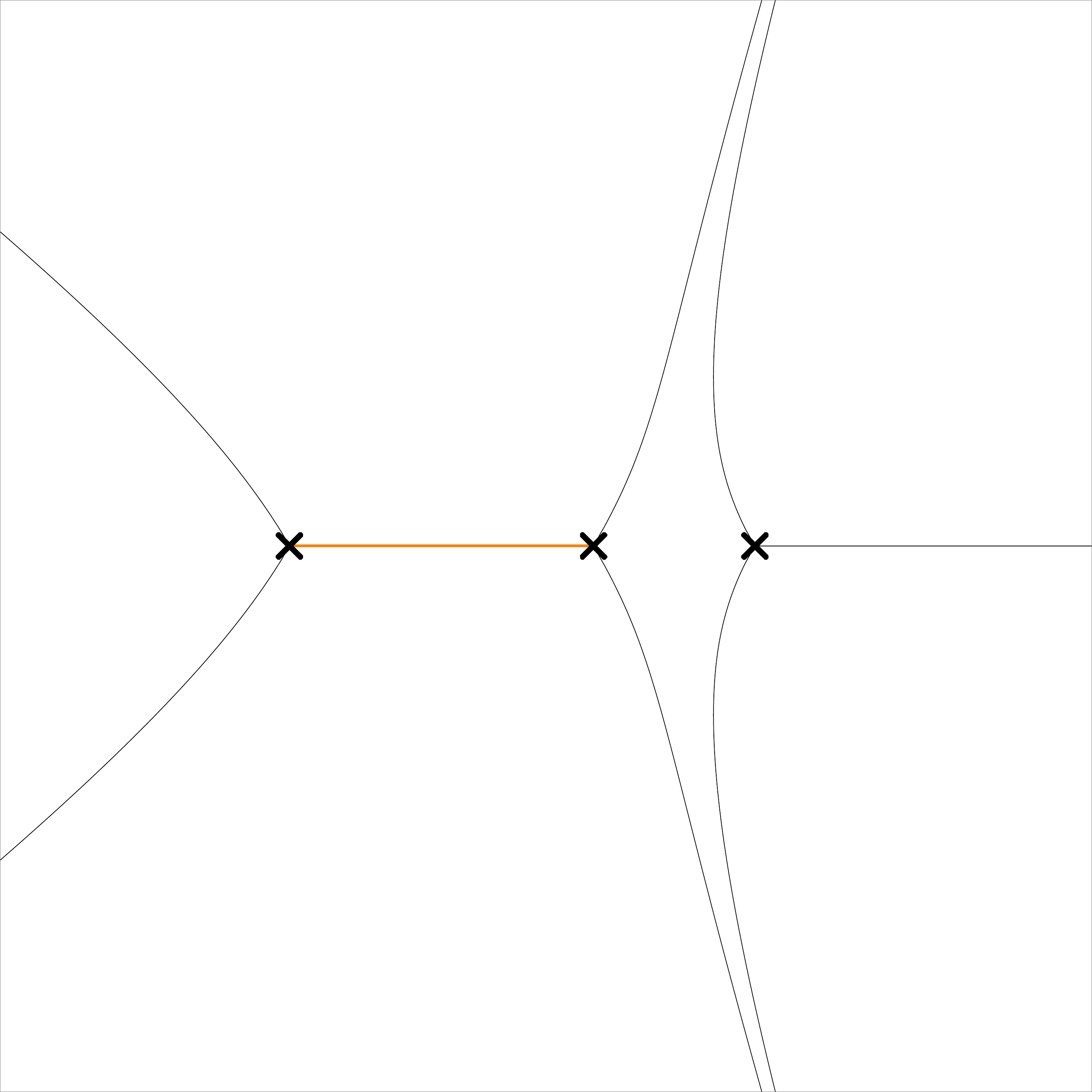}
		\caption{$\theta = \arg(Z_2)$}
		\label{fig:a2SN_minimal_2}
	\end{subfigure}
	\begin{subfigure}[b]{.25\textwidth}	
		\centering
		\includegraphics[width=\textwidth]{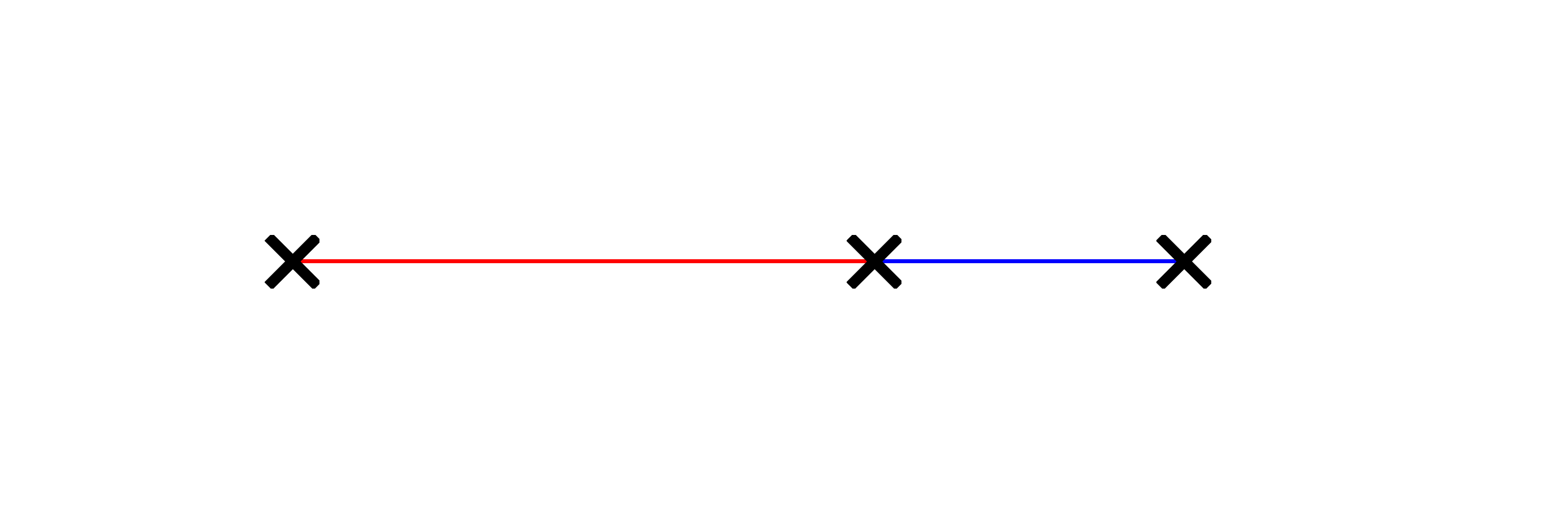}
		\vspace{2.5em}
		\caption{Finite $\CS$-walls}
		\label{fig:a2SN_minimal_finite}
	\end{subfigure}

	\caption{Spectral network of $\CS[A_{1}; D_{7}]$ when the BPS spectrum is minimal.}
	\label{fig:a2SN_minimal}
\end{figure}

We first start inside the BPS wall, where the BPS spectrum is minimal. The spectral networks of $\CS[A_{1}; \CD_{7}]$ when $c_2, v_2$ are inside the BPS wall and when $\theta = \arg(Z_i)$ such that there is a finite $\CS$-wall corresponding to a BPS state are shown in Figure \ref{fig:a2SN_minimal}. The animated version of Figure \ref{fig:a2SN_minimal} can be found at \href{http://www.its.caltech.edu/~splendid/spectral_network/a2SN_minimal.gif}{this website}. 

\begin{figure}[h]
	\centering
	\begin{subfigure}[b]{.3\textwidth}	
		\centering
		\includegraphics[width=\textwidth]{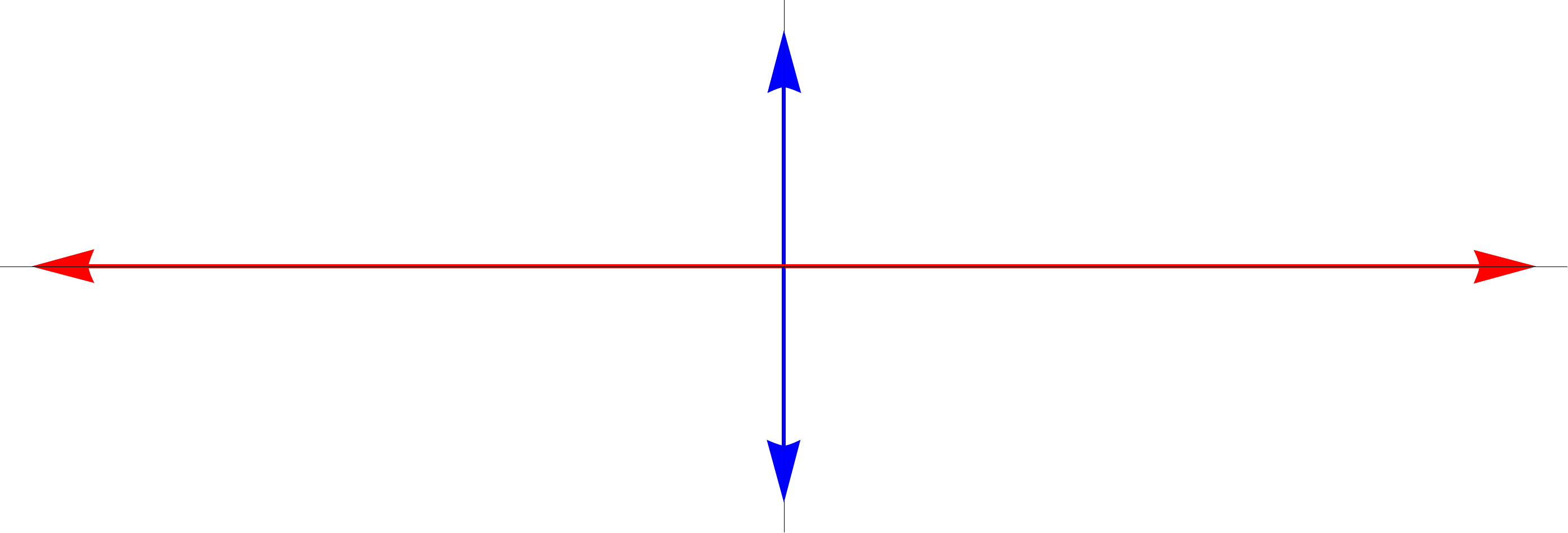}
		\caption{central charges}
		\label{fig:a2SN_minimal_Z}
	\end{subfigure}
	\begin{subfigure}[b]{.3\textwidth}	
		\centering
		\begin{tabular}{c|c}
		   state  & $(e,m)$ \\ \hline
		  1 & $(1,0)$ \\ \hline
		  2 & $(0,1)$
		\end{tabular}
		\caption{IR charges}
		\label{tbl:a2SN_minimal_IR}
	\end{subfigure}	
	\begin{subfigure}[b]{.2\textwidth}	
		\centering
		\begin{tikzpicture}
			\node[W,red] (1) at (0,0){1};
		 	\node[W,blue] (2) at (2,0) {2};
			\draw[->] (1)--(2);
		\end{tikzpicture}
		\vspace{1em}
		\caption{BPS quiver}
		\label{fig:a2SN_minimal_quiver}
	\end{subfigure}	
	\renewcommand{\figurename}{Figure \& Table}
	\caption{Minimal BPS spectrum of $\CS[A_{1}; D_{7}]$.}
	\label{figntbl:a2SN_minimal_BPS}
\end{figure}

To summarize the results from the spectral network, let us collect the finite $\CS$-walls and calculate the central charge of each BPS state from the corresponding finite $\CS$-walls, which is shown in Figure \ref{fig:a2SN_minimal_Z}. One of the two $\CS$-walls is an A-cycle of the Seiberg-Witten curve of $\CS[A_{1}; \CD_{7}]$, and the other is a B-cycle. Therefore the BPS spectrum of the theory has two BPS states, which is the minimal BPS spectrum of $\CS[A_{1}; \CD_{7}]$. 

We can determine the low-energy $\UU(1)$ charges of the BPS states from Figure \ref{fig:a2SN_minimal} after picking up a suitable basis. If one chooses a branch cut along one of the finite $\CS$-walls, say $\CS_{1}$, of Figure \ref{fig:a2SN_minimal_finite}, it is natural to define the cycle corresponding to that finite $\CS$-wall as an A-cycle and the one corresponding to the other finite $\CS$-wall, $\CS_2$, as a B-cycle. Their intersection number is $\intsec{\CS_{1}}{\CS_{2}}=1$ with a proper choice of orientations of the cycles,  see Appendix \ref{sec:IntNumberSWall} for the details. The IR charges of the BPS states are summarized in Table \ref{tbl:a2SN_minimal_IR} and their anti-states have charges of opposite sign. 
Using the $\UU(1)$ charges, this BPS spectrum can be represented with an $A_2$ quiver shown in Figure \ref{fig:a2SN_minimal_quiver}, where the direction and the number of the heads of an arrow correspond 
to the inner product of the charges connected by the arrow. 

\begin{figure}[h]
	\centering
	\begin{subfigure}[b]{.22\textwidth}	
		\includegraphics[width=\textwidth]{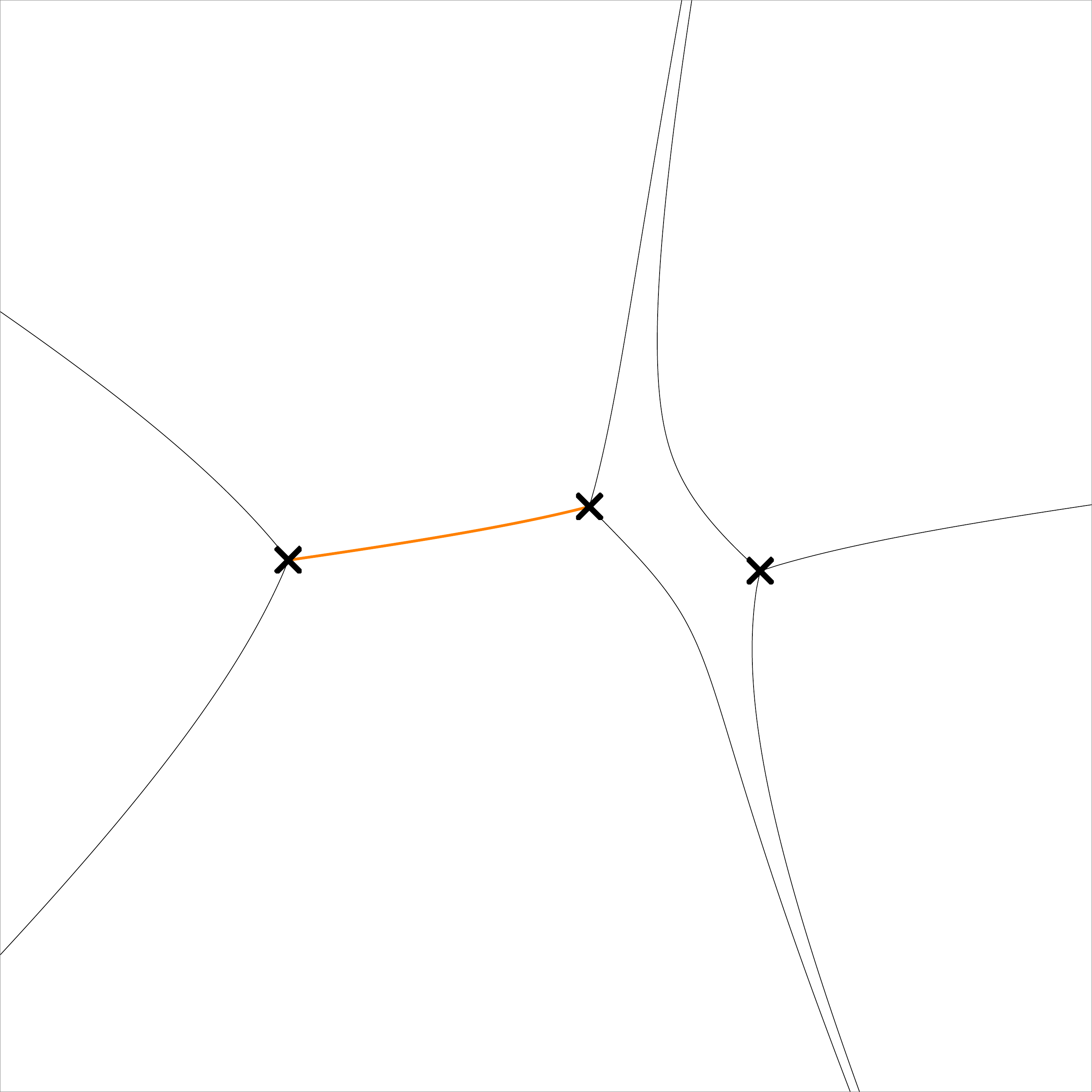}
		\caption{$\theta = \arg(Z_1)$}
		\label{fig:a2SN_deformed_minimal_1}
	\end{subfigure}
	\begin{subfigure}[b]{.22\textwidth}	
		\centering
		\includegraphics[width=\textwidth]{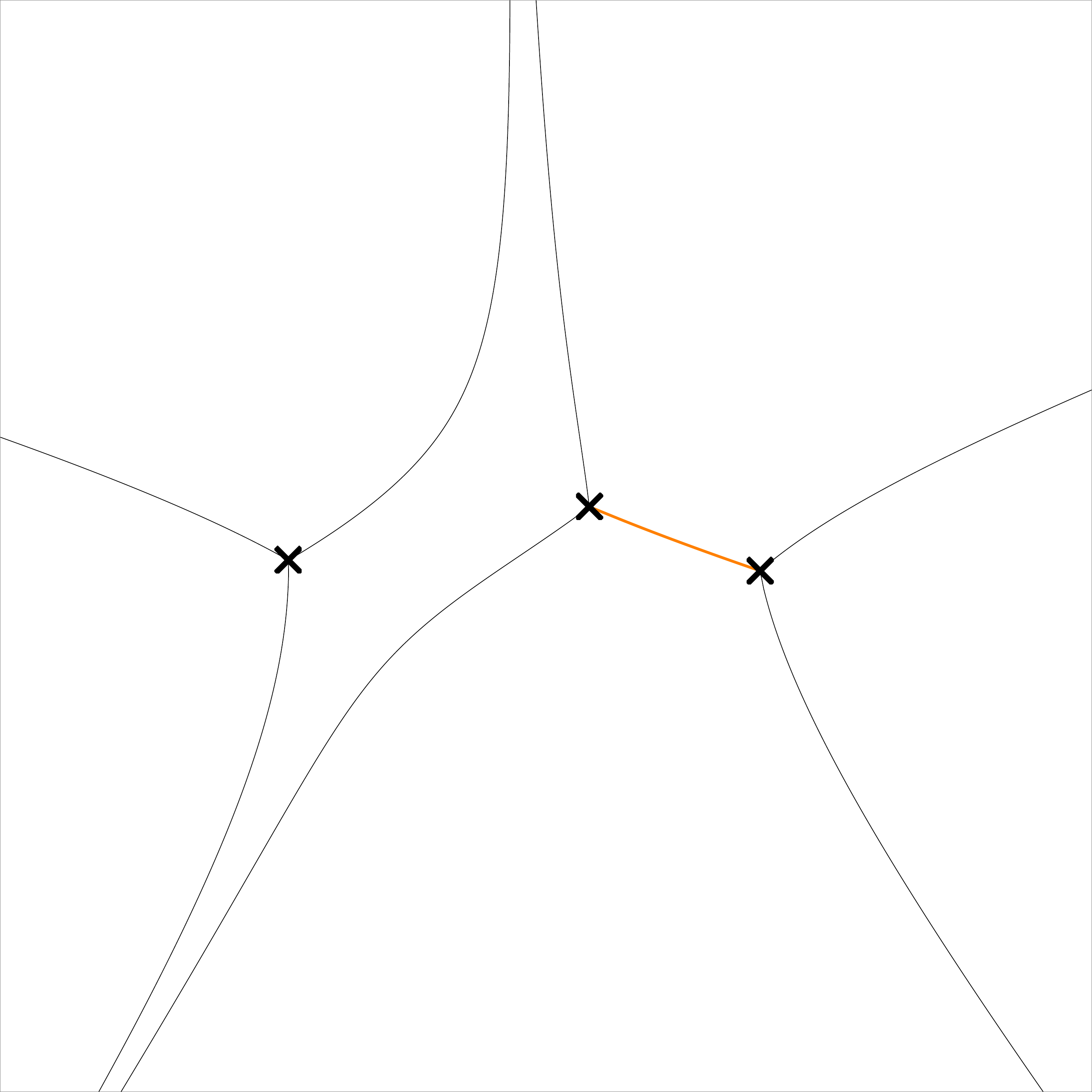}
		\caption{$\theta = \arg(Z_2)$}
		\label{fig:a2SN_deformed_minimal_2}
	\end{subfigure}
	\begin{subfigure}[b]{.22\textwidth}	
		\includegraphics[width=\textwidth]{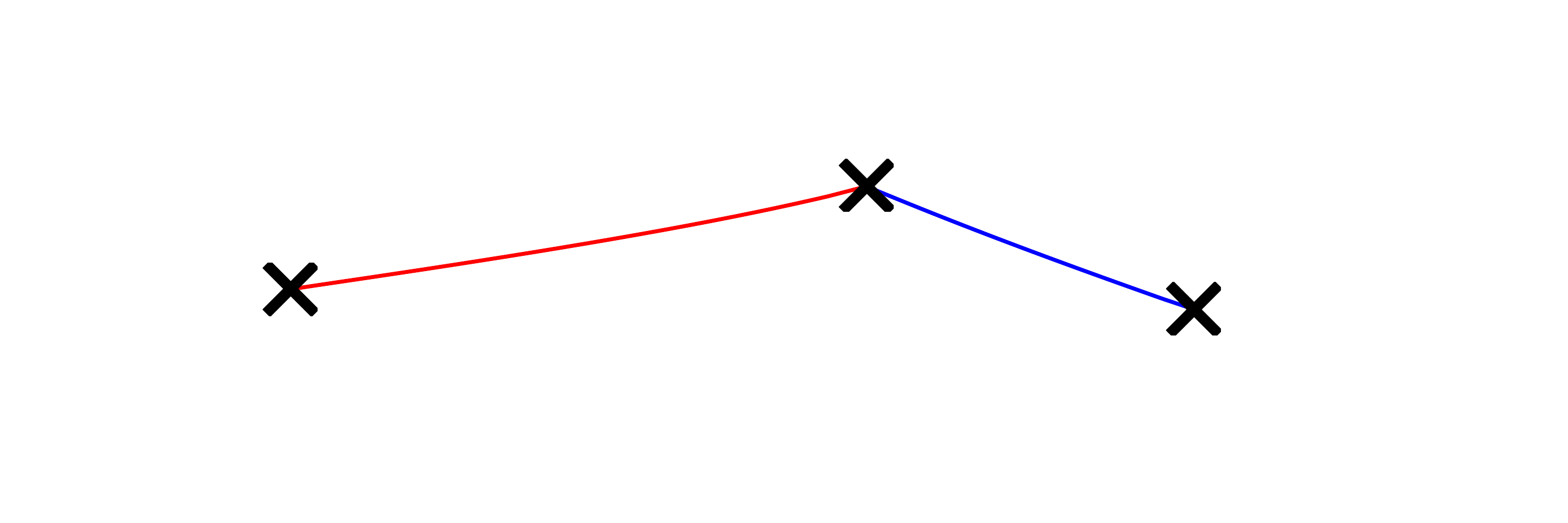}
		\vspace{2.5em}
		\caption{finite $\CS$-walls}		
		\label{fig:a2SN_deformed_minimal_finite}
	\end{subfigure}
	\begin{subfigure}[b]{.22\textwidth}	
		\includegraphics[width=\textwidth]{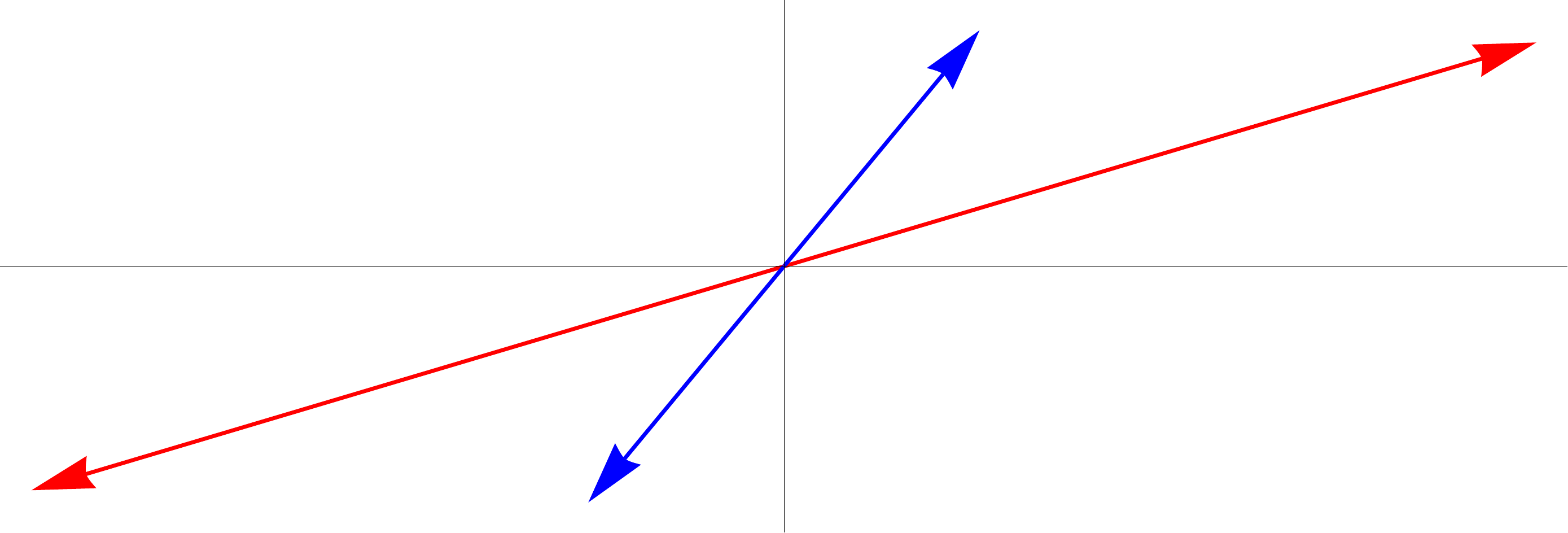}
		\vspace{2.5em}
		\caption{central charges}		
		\label{fig:a2SN_deformed_minimal_Z}
	\end{subfigure}

	\caption{Spectral network of $\CS[A_{1}; D_{7}]$ from different choices of parameters 
	         with minimal BPS spectrum.
	         (c) and (d): finite $\CS$-walls corresponding to two BPS states and their central charges.}
	\label{fig:a2SN_deformed_minimal}
\end{figure}

When we deform $v_2$ a bit, as long as the parameters are inside the BPS wall, the number of BPS states does not change, which is shown in Figure \ref{fig:a2SN_deformed_minimal}. The animated version of Figure \ref{fig:a2SN_deformed_minimal} can be found at \href{http://www.its.caltech.edu/~splendid/spectral_network/a2SN_deformed_minimal.gif}{this website}. 
Figure \ref{fig:a2SN_deformed_minimal_Z} describes the central charges of the BPS states. Now each BPS state has a different value of central charge from those in Figure \ref{fig:a2SN_minimal_Z}, however the $\UU(1)$ charges are the same as shown in Table \ref{tbl:a2SN_minimal_IR} and therefore the BPS spectrum is also represented by an $A_2$ quiver.

\begin{figure}[h]
	\centering
	\begin{subfigure}[b]{.225\textwidth}	
		\includegraphics[width=\textwidth]{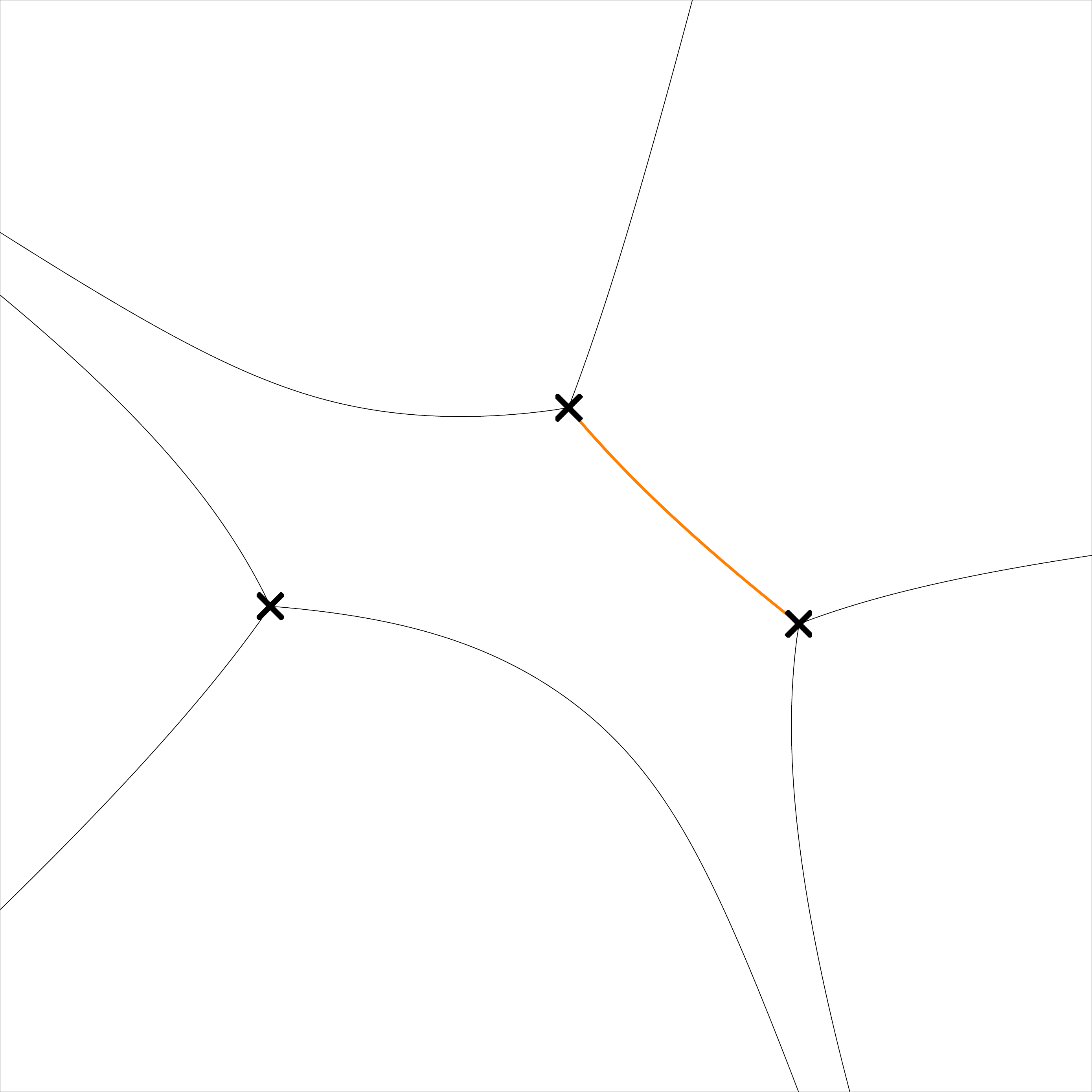}
		\caption{$\theta = \arg(Z_2)$}
		\label{fig:a2SN_maximal_1}
	\end{subfigure}
	\begin{subfigure}[b]{.225\textwidth}	
		\centering
		\includegraphics[width=\textwidth]{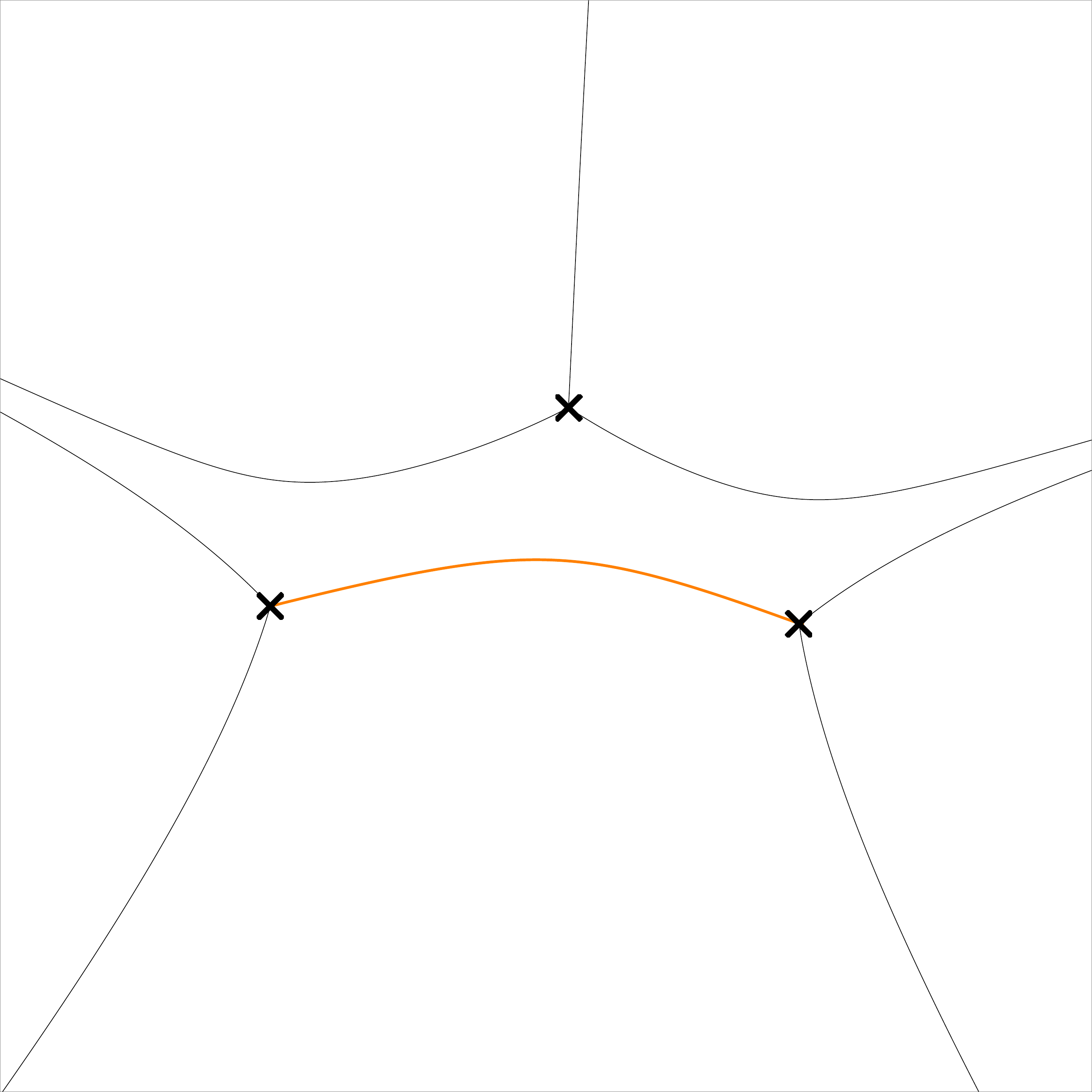}
		\caption{$\theta = \arg(Z_3)$}
		\label{fig:a2SN_maximal_2}
	\end{subfigure}
	\begin{subfigure}[b]{.225\textwidth}	
		\centering
		\includegraphics[width=\textwidth]{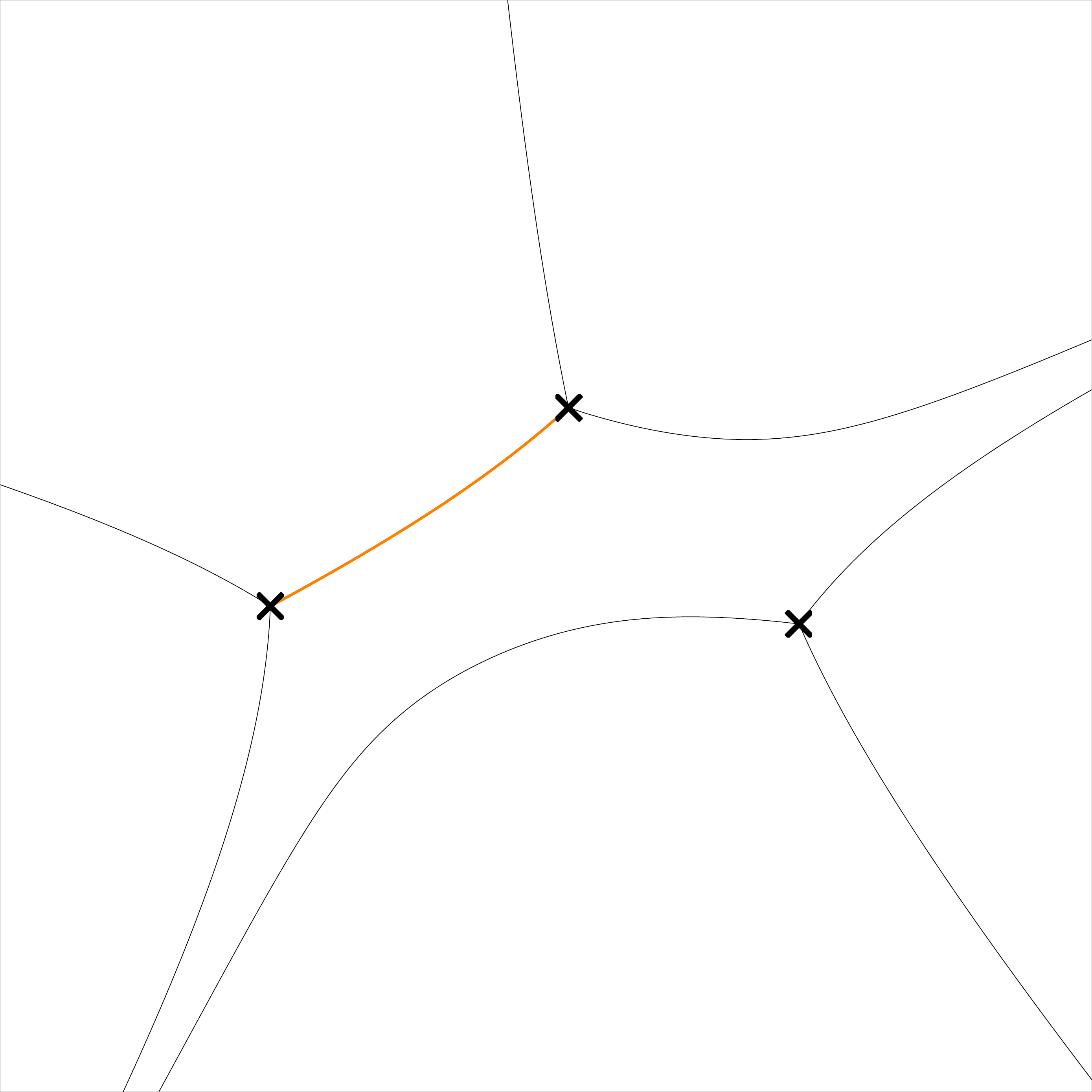}
		\caption{$\theta = \arg(Z_1)$}
		\label{fig:a2SN_maximal_3}
	\end{subfigure}	
	\begin{subfigure}[b]{.3\textwidth}	
		\includegraphics[width=\textwidth]{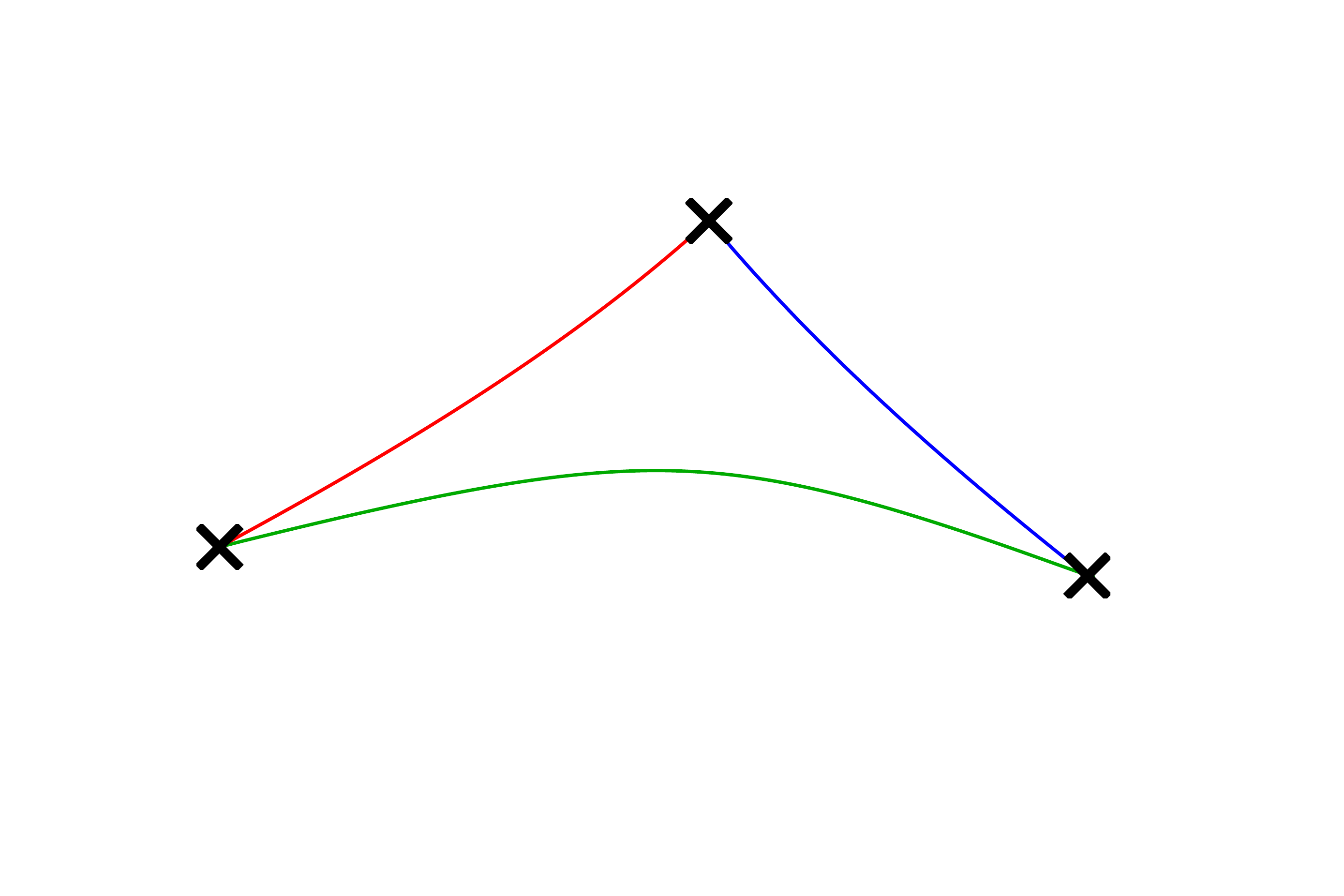}
		\caption{finite $\CS$-walls}
		\label{fig:a2SN_maximal_finite}
	\end{subfigure}

	\caption{Spectral network of $\CS[A_{1}; D_{7}]$ with a maximal BPS spectrum}
	\label{fig:a2SN_maximal}
\end{figure}

\paragraph{Maximal BPS spectrum}
When $v_2$ is further changed so that now the parameters are on the other side of the BPS wall, we observe  a wall-crossing phenomenon of spectral network \cite{Gaiotto:2009hg} as shown in Figure \ref{fig:a2SN_maximal}, where we have an additional finite $\CS$-wall. An animated version of the spectral network can be found at  
\href{http://www.its.caltech.edu/~splendid/spectral_network/a2SN_maximal.gif}{this website}.

\begin{figure}[h]
	\centering
	\begin{subfigure}[b]{.3\textwidth}	
		\centering
		\includegraphics[width=\textwidth]{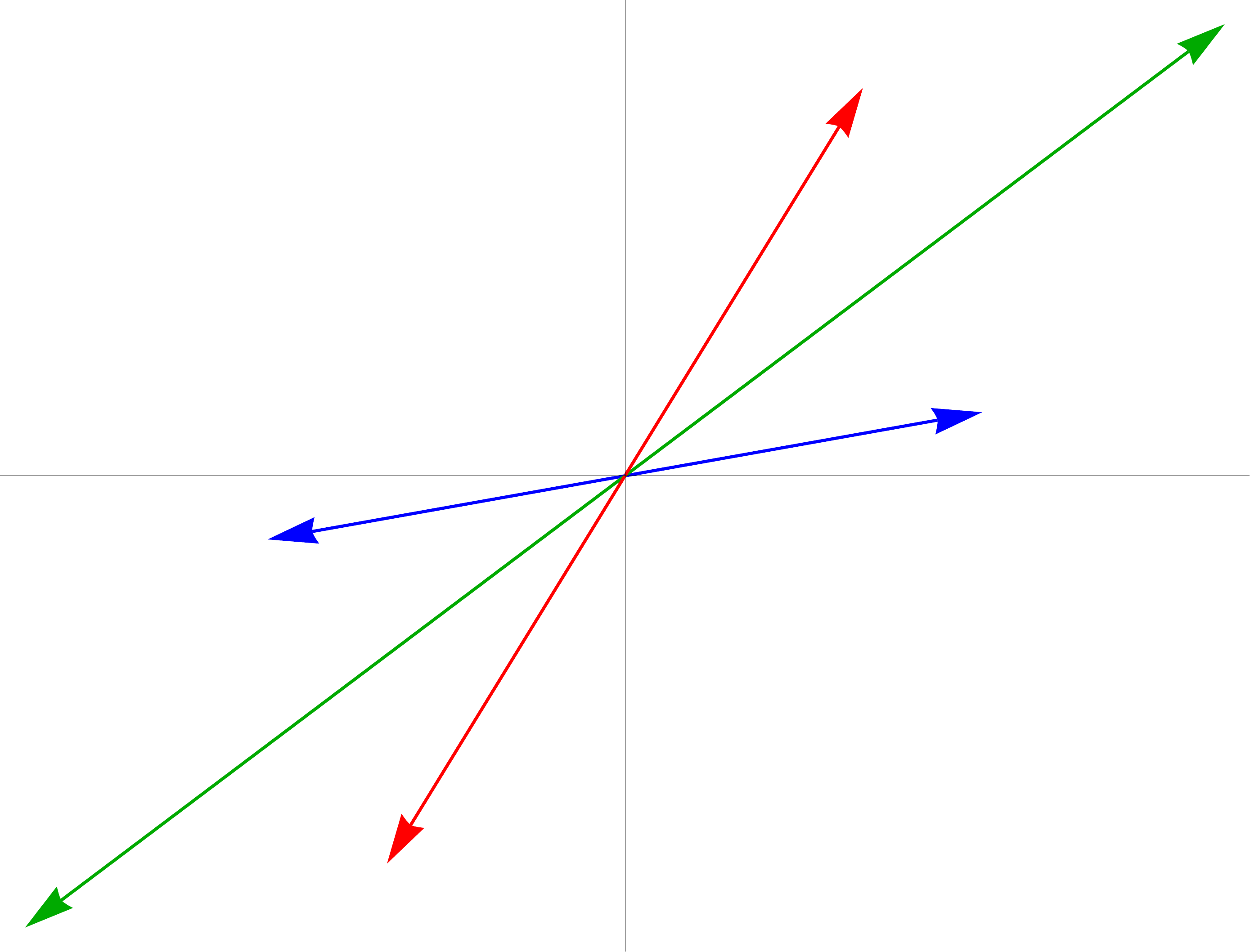}
		\caption{central charges}
		\label{fig:a2SN_maximal_Z}
	\end{subfigure}
	\begin{subfigure}[b]{.3\textwidth}	
		\centering
		\begin{tabular}{c|c}
		   state  & $(e,m)$ \\ \hline
		  1 & $(1,0)$ \\ \hline
		  2 & $(0,1)$ \\ \hline
		  3 & $(1,1)$
		\end{tabular}
		\vspace{1em}
		\caption{IR charges}
		\label{tbl:a2SN_maximal_IR}
	\end{subfigure}	
	\renewcommand{\figurename}{Figure \& Table}
	\caption{Maximal BPS spectrum of $\CS[A_{1}; \CD_{7}]$.}
	\label{figntbl:a2SN_maximal_BPS}
\end{figure}

The BPS spectrum of the theory after the wall-crossing is described in Figure \ref{fig:a2SN_maximal_Z} and Table \ref{tbl:a2SN_maximal_IR}, where we have an additional BPS state from the third finite $\CS$-wall. This is the maximal BPS spectrum of $\CS[A_{1}; \CD_{7}]$.

\subsubsection{$\CS[A_{1}; \CD_{n+5}]$ in $A_{n}$-class, $n>2$}
Now we generalize the previous analysis to general $n$. 
The Seiberg-Witten curve of the $\CS[A_{1}; \CD_{n+5}]$ theory when $n=2k$ is 
\begin{align}
	x^2 = t^{2k+1}+c_2 t^{2k-1}+ \cdots + c_{k+1} t^k + v_{k+1} t^{k-1} + \cdots + v_2
\end{align}
with dimensions $\Delta(v_{i}) = 2-\frac{2i}{2k+3}$ and $\Delta(c_{i})=\frac{2i}{2k+3}$,
and when $n=2k-1$ the Seiberg-Witten curve is
\begin{align}
	x^2 = t^{2k}+c_2 t^{2k-2}+ \cdots + c_{k} t^k + c_{k+1} t^{k-1} + v_{k} t^{k-2} + \cdots + v_2
\end{align}
with dimensions $\Delta(v_{i}) = 2-\frac{i}{k+1}$, $\Delta(c_{i})=\frac{i}{k+1}$, where $i=2,\ldots,k$,
and $\Delta(c_{k+1}) = 1$.
$v_{i}$ and $c_{i}$ ($i=2, \ldots, [n/2]+1$) are the vevs of the relevant deformation operators and their couplings respectively.
We can see that the Coulomb moduli space is of $[n/2]$ dimensional.
When $n = 2k-1$, $\lambda$ has a nonzero residue at $t = \infty$, which is $\frac{1}{2} c_{k+1} + f(c_i)$, where $f(c_i)$ is a polynomial with $i \leq k$ and homogeneous of scaling dimension one \cite{Argyres:2012fu}.

\paragraph{Wall-crossing of $\CS[A_{1}; \CD_{n+5}]$}

\begin{figure}[t]
	\centering
	\includegraphics[width=.35\textwidth]{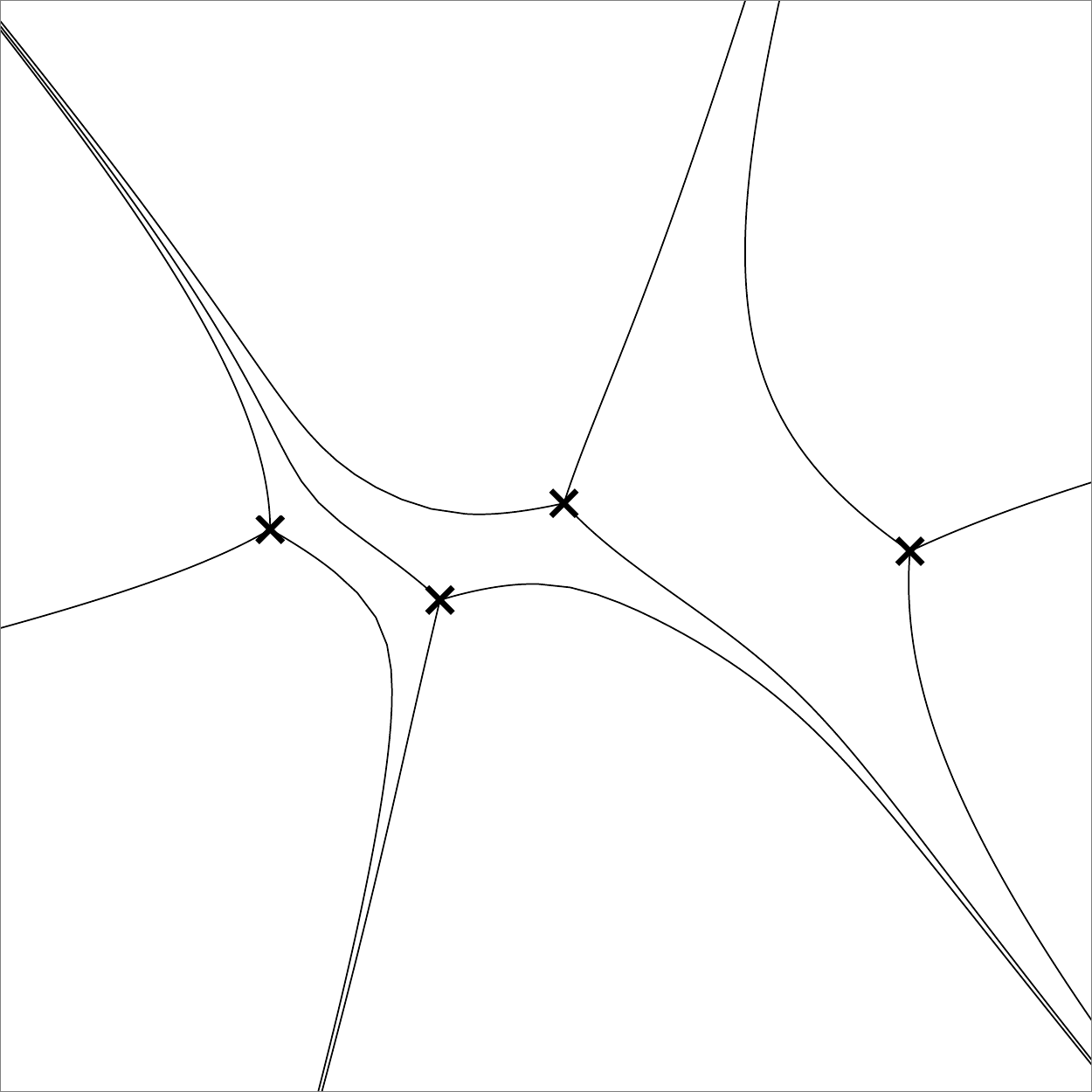}
	\caption{Spectral network of $\CS[A_{1}; \CD_{8}]$.}
	\label{fig:a3SN_general}
\end{figure}

\begin{figure}[t]
	\centering
	\begin{subfigure}{.32\textwidth}	
		\includegraphics[width=\textwidth]{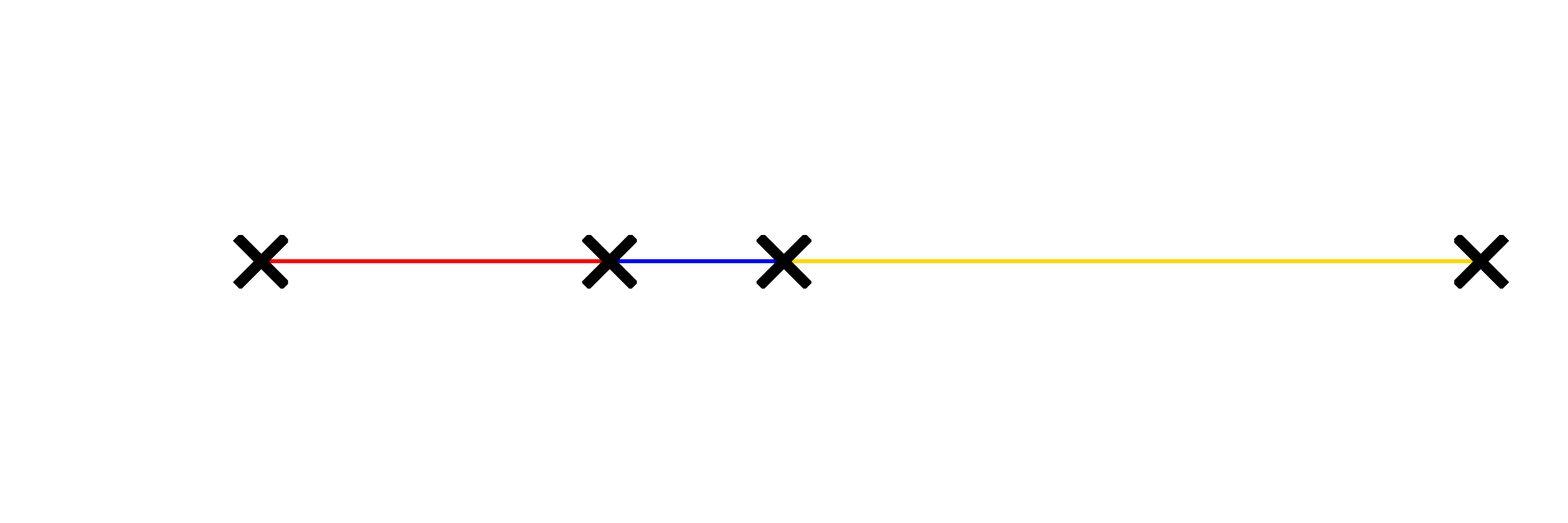}
	\end{subfigure}
	\begin{subfigure}{.22\textwidth}	
		\centering
		\includegraphics[width=\textwidth]{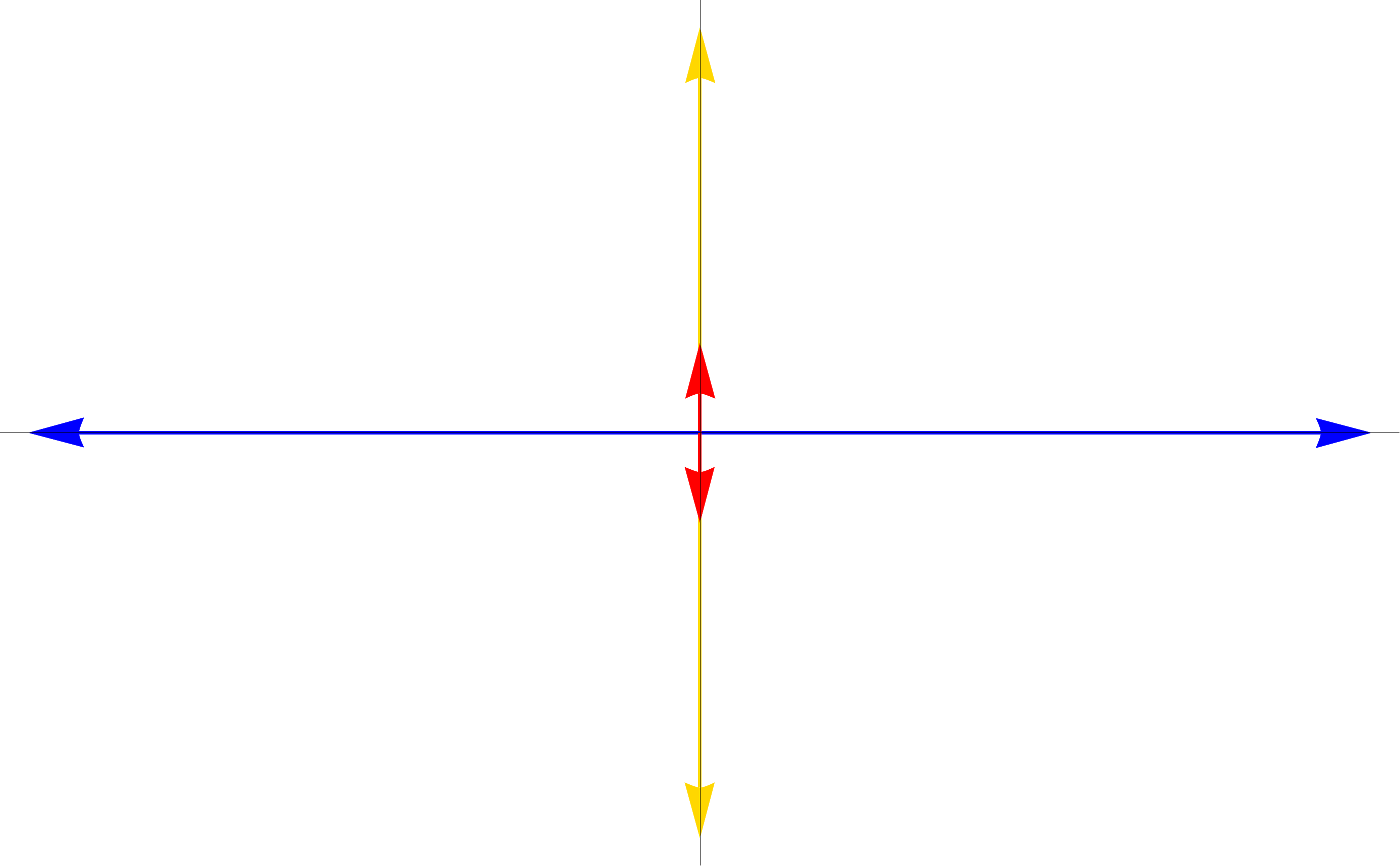}
	\end{subfigure}

	\begin{subfigure}{.32\textwidth}	
		\includegraphics[width=\textwidth]{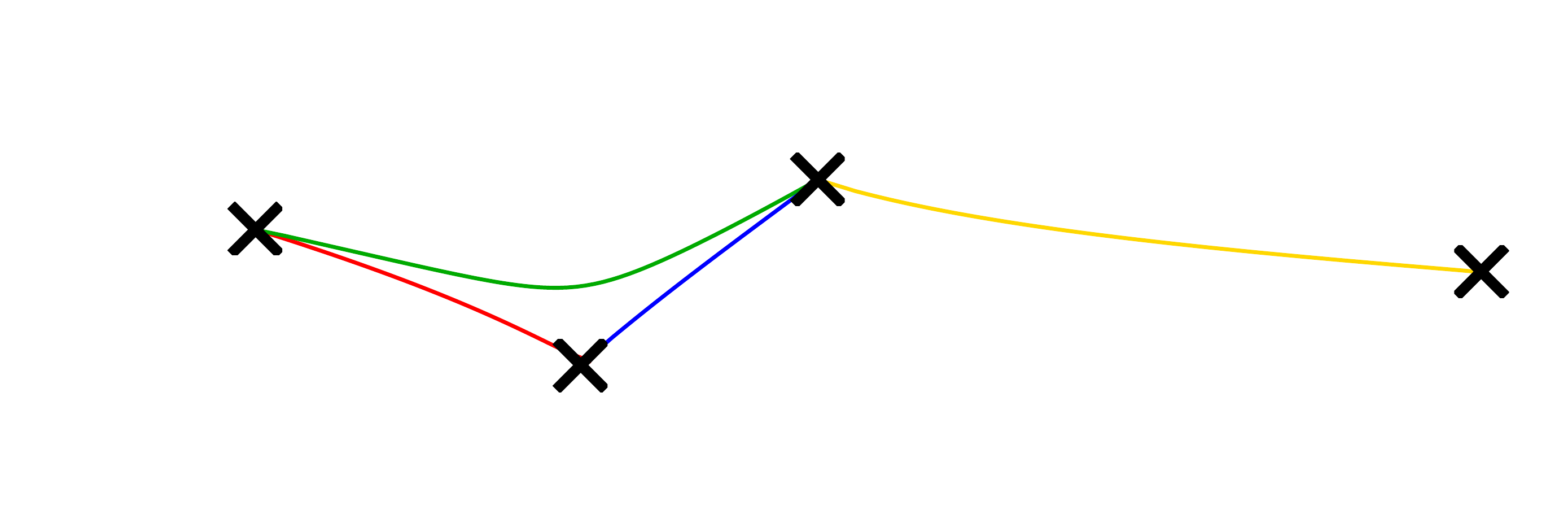}
	\end{subfigure}
	\begin{subfigure}{.22\textwidth}	
		\centering
		\includegraphics[width=\textwidth]{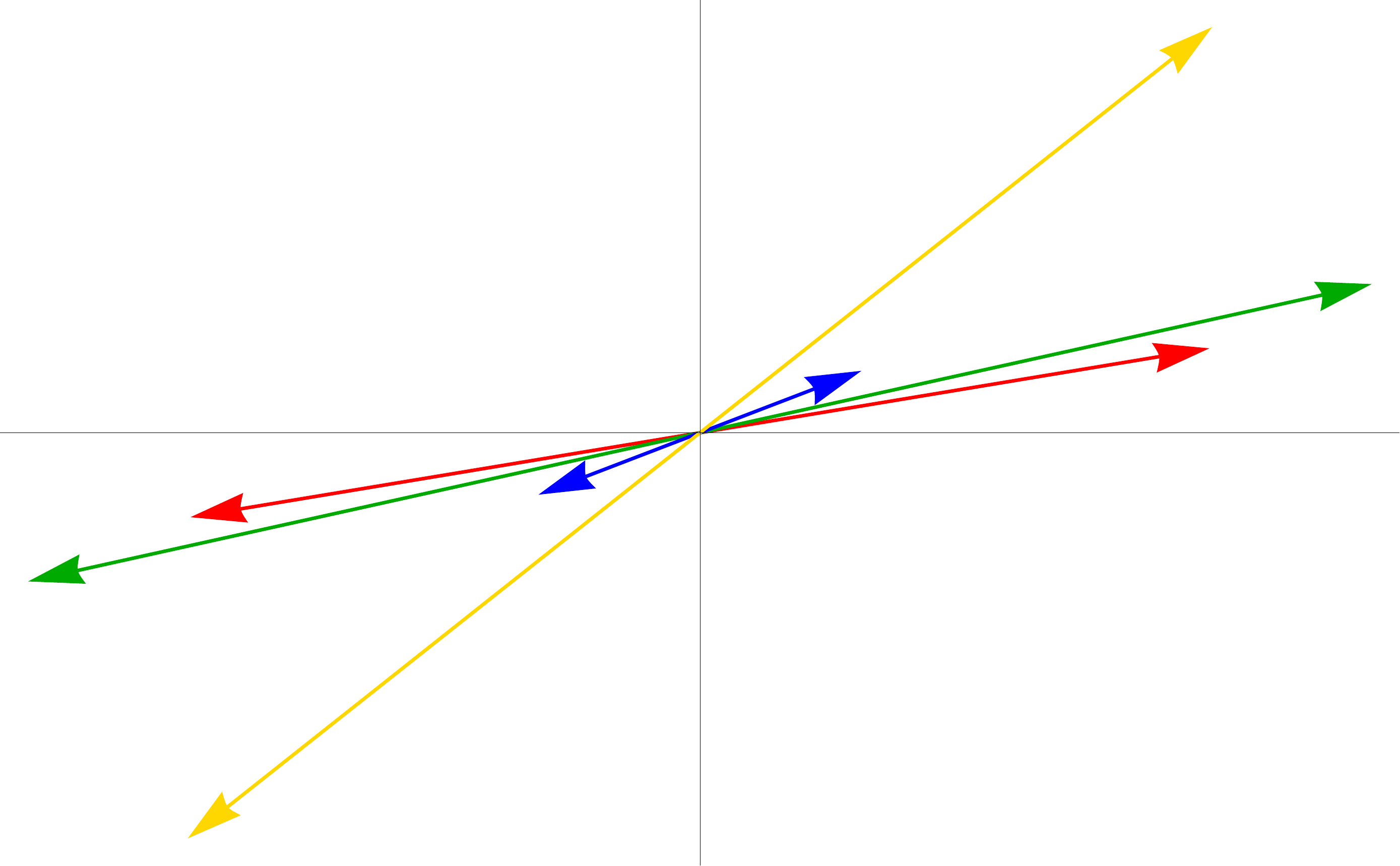}
	\end{subfigure}
	
	\begin{subfigure}{.32\textwidth}	
		\includegraphics[width=\textwidth]{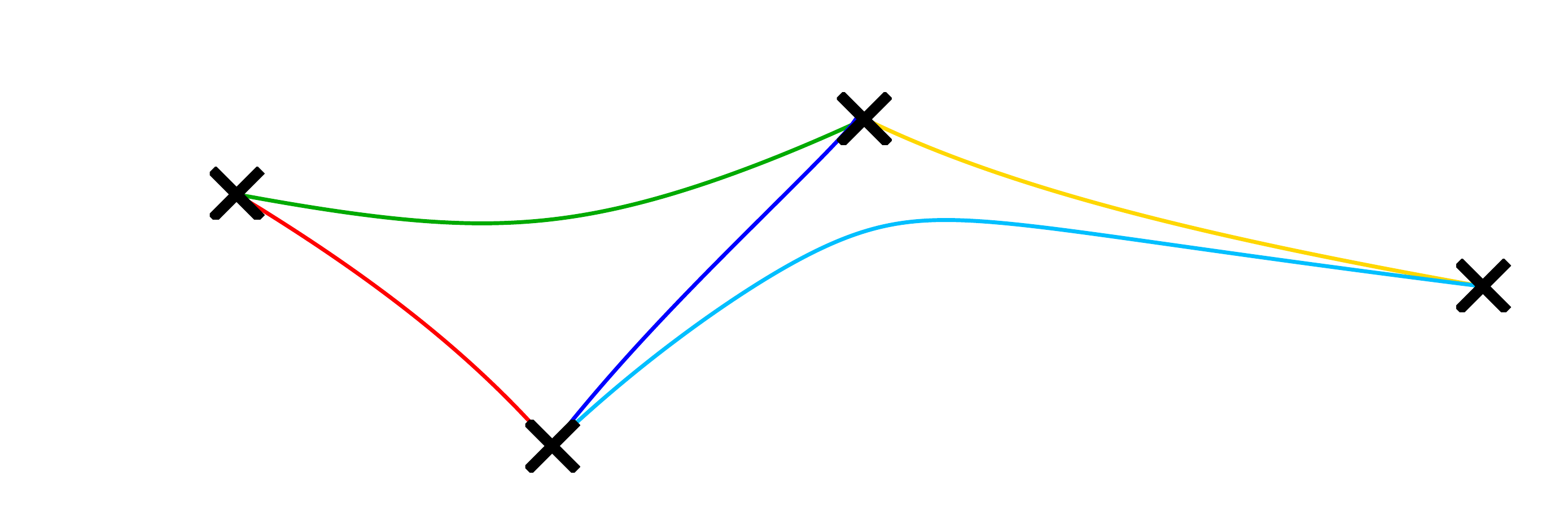}
	\end{subfigure}
	\begin{subfigure}{.22\textwidth}	
		\centering
		\includegraphics[width=\textwidth]{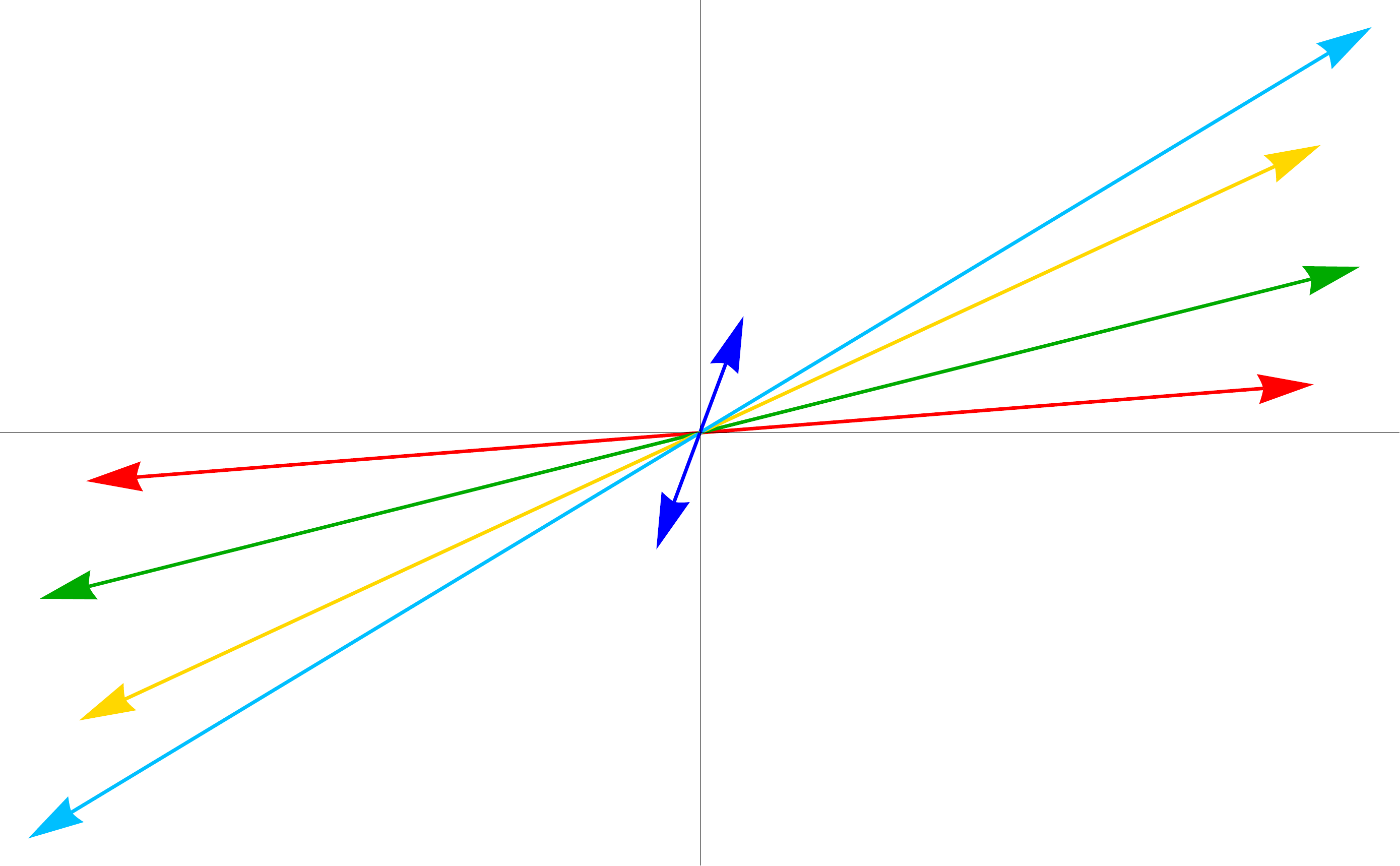}
	\end{subfigure}
	
\vspace{1em}
	
	\begin{subfigure}{.32\textwidth}	
		\includegraphics[width=\textwidth]{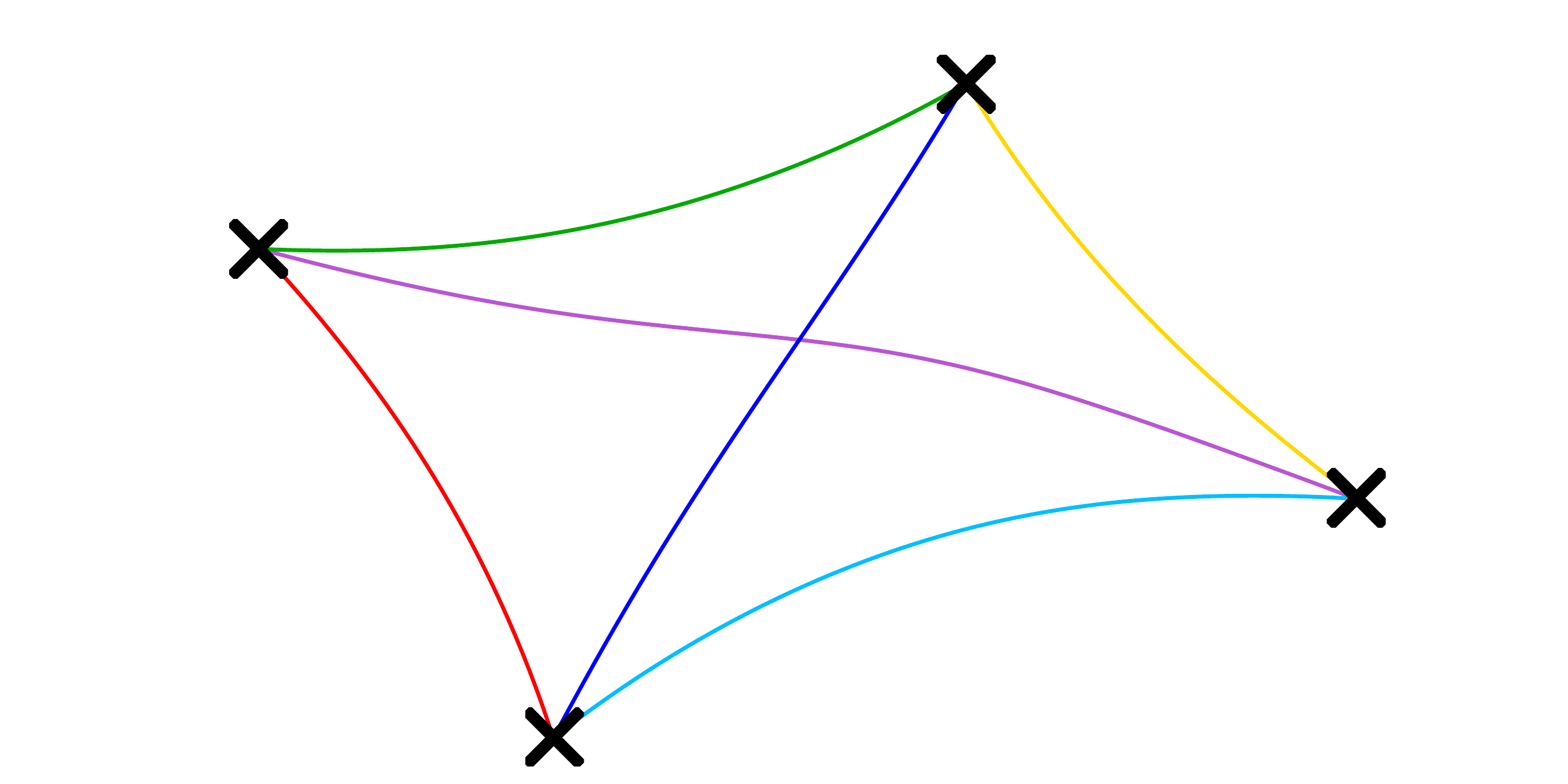}
	\end{subfigure}
	\begin{subfigure}{.22\textwidth}	
		\centering
		\includegraphics[width=\textwidth]{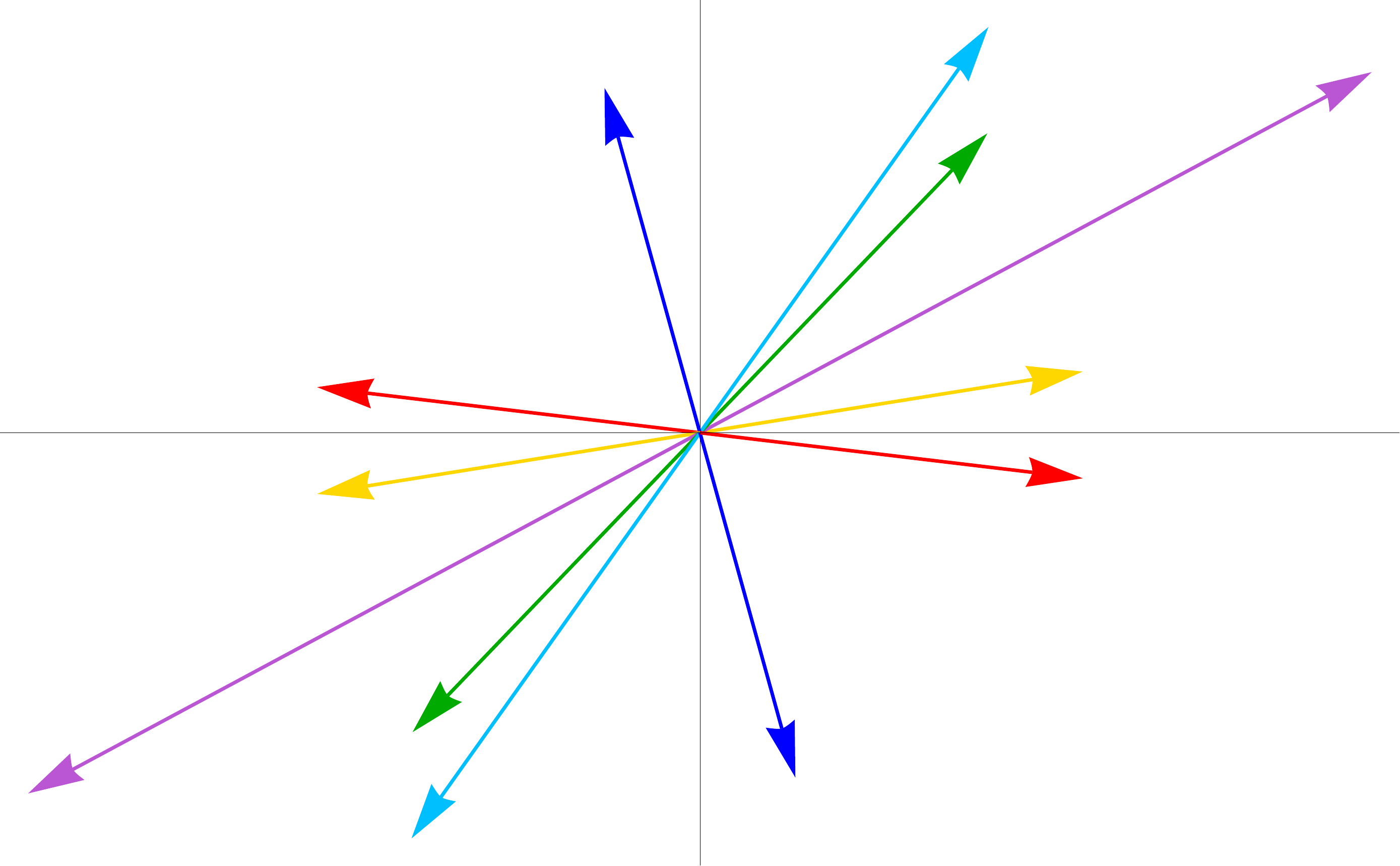}
	\end{subfigure}
	
	\caption{BPS spectra of $\CS[A_{1}; \CD_{8}]$ and their wall-crossings.}
	\label{fig:a3SN_finite}
\end{figure}

Studying the spectral network of $\CS[A_{1}; \CD_{8}]$ gives us a good idea of the generalization to general $\CS[A_{1}; \CD_{n+5}]$. The Seiberg-Witten curve of $\CS[A_{1}; \CD_{8}]$ is
\begin{align}
	x^2 = t^4 + c_2 t^2 + c_3 t + v_2.
\end{align}
An example of its spectral network is shown in Figure \ref{fig:a3SN_general}.
After analyzing the spectral networks for various values of the parameters, which is done in \cite{Gaiotto:2009hg}, we obtain the finite $\CS$-walls and corresponding BPS spectra as shown in Figure \ref{fig:a3SN_finite}.

The first row of Figure \ref{fig:a3SN_finite} shows finite $\CS$-walls corresponding to three states in the BPS spectrum and the central charges of the states when the residue of $\lambda$ at $x=\infty$ is not zero, i.e.\ $\frac{1}{2} c_3 + f(c_2) = \frac{1}{2} c_3 \neq 0$. ($f$ is zero in this case.) This is a minimal BPS spectrum of $\CS[A_{1}; \CD_{8}]$, but this does not have the $\SU(2)$ flavor symmetry and we have three BPS states of different central charges. This spectrum can be represented with a quiver diagram shown in Figure \ref{fig:a3SN_BPS_quiver}, which is based on an $A_3$ Dynkin diagram.

\begin{figure}[h]
	\centering
	\begin{tikzpicture}[scale=.75]
		\node[W,red] (1) at (0,0) {1};
		\node[W,blue] (2) at (2,0) {2};
		\node[W,gold] (3) at (4,0) {3};
		\path (1) edge[->] (2);
		\path (2) edge[<-] (3);
	\end{tikzpicture}
	\caption{BPS quiver of the minimal BPS spectrum of $\CS[A_{1}; \CD_{8}]$.}
	\label{fig:a3SN_BPS_quiver}
\end{figure}

As we change the value of $v_2$ while fixing $c_i$, we can observe three wall-crossings as shown in the bottom three rows of Figure \ref{fig:a3SN_finite}. Each wall-crossing is similar to that of $\CS[A_{1}; \CD_{7}]$: each wall-crossing adds an additional BPS state to the spectrum, and at the end of the series of wall-crossings, we get the maximal BPS spectrum as shown in the last row of Figure \ref{fig:a3SN_finite}, where we have a BPS state from an $\CS$-wall connecting every pair of branch points as found in \cite{Shapere:1999xr}. The IR gauge charges of the states in the maximal BPS spectrum is described in Table \ref{tbl:a3SN_IR}.

\begin{table}[h]
	\centering
	\begin{tabular}{c|c}
	   state  & $(e,m)$ \\ \hline
	  1 & $(1,0)$ \\ \hline
	  2 & $(0,1)$ \\ \hline
	  3 & $(1,0)$ \\ \hline
	  4 & $(1,1)$\\ \hline
	  5 & $(1,1)$\\ \hline
	  6 & $(2,1)$
	\end{tabular}
	\caption{IR charges of the BPS states in the BPS spectrum of $\CS[A_{1}; \CD_{8}]$.}
	\label{tbl:a3SN_IR}		
\end{table}

We can generalize this to $n > 3$ using the fact that the spectral network of $\CS[A_{1}; \CD_{7}]$ serves as the building block of that of $\CS[A_{1}; \CD_{n+5}]$. When $\CS[A_{1}; \CD_{n+5}]$ has a minimal BPS spectrum, its spectral network can be considered as a combination of $(n-1)$ number of $\CS[A_{1}; \CD_{7}]$ spectral networks, where $(k-1)$-th, $k$-th, and $(k+1)$-th branch points of $\CS[A_{1}; \CD_{n+5}]$ are in the $k$-th building block. Starting from that configuration, the spectral network of $\CS[A_{1}; \CD_{n+5}]$ undergoes a series of the wall-crossing of $\CS[A_{1}; \CD_{7}]$ when the theory moves away from the minimal chamber in the Coulomb branch moduli space, at the end of which we get the maximal BPS spectrum having $\binom{n}{2}$ states.

\paragraph{Wall-crossing of $\CS[A_{1}; \CD_{8}]$ with an $\SU(2)$ flavor symmetry}

\begin{figure}[t]
	\centering
	\begin{subfigure}{.32\textwidth}	
		\includegraphics[width=\textwidth]{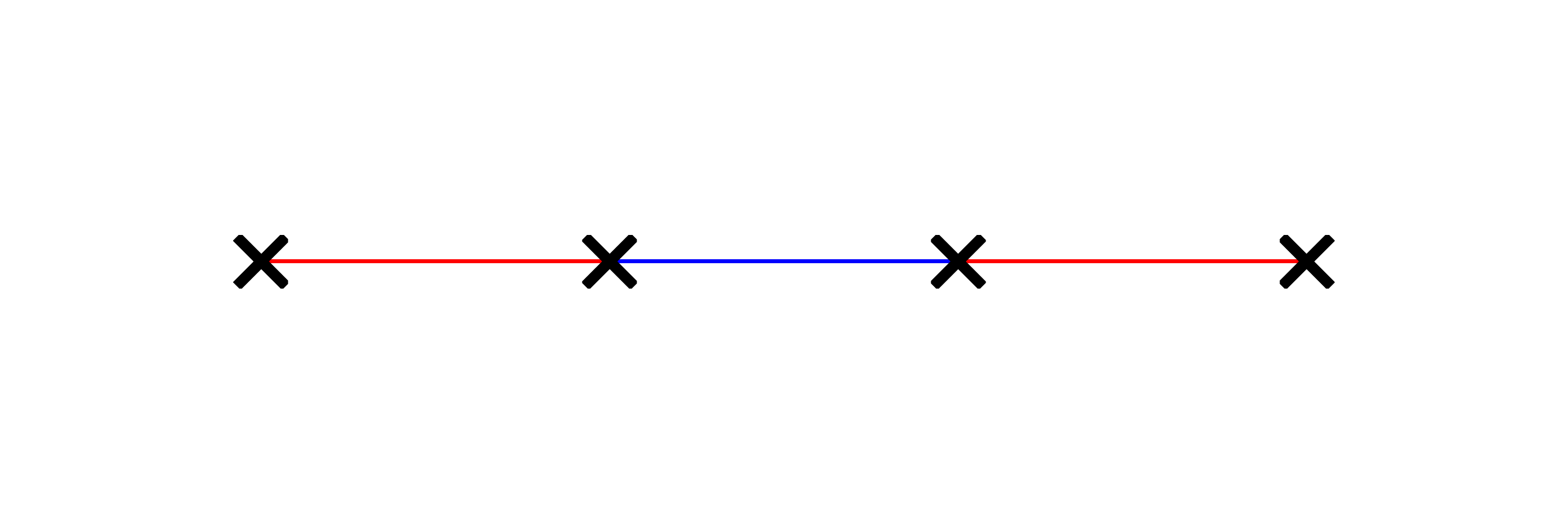}
	\end{subfigure}
	\begin{subfigure}{.22\textwidth}	
		\centering
		\includegraphics[width=\textwidth]{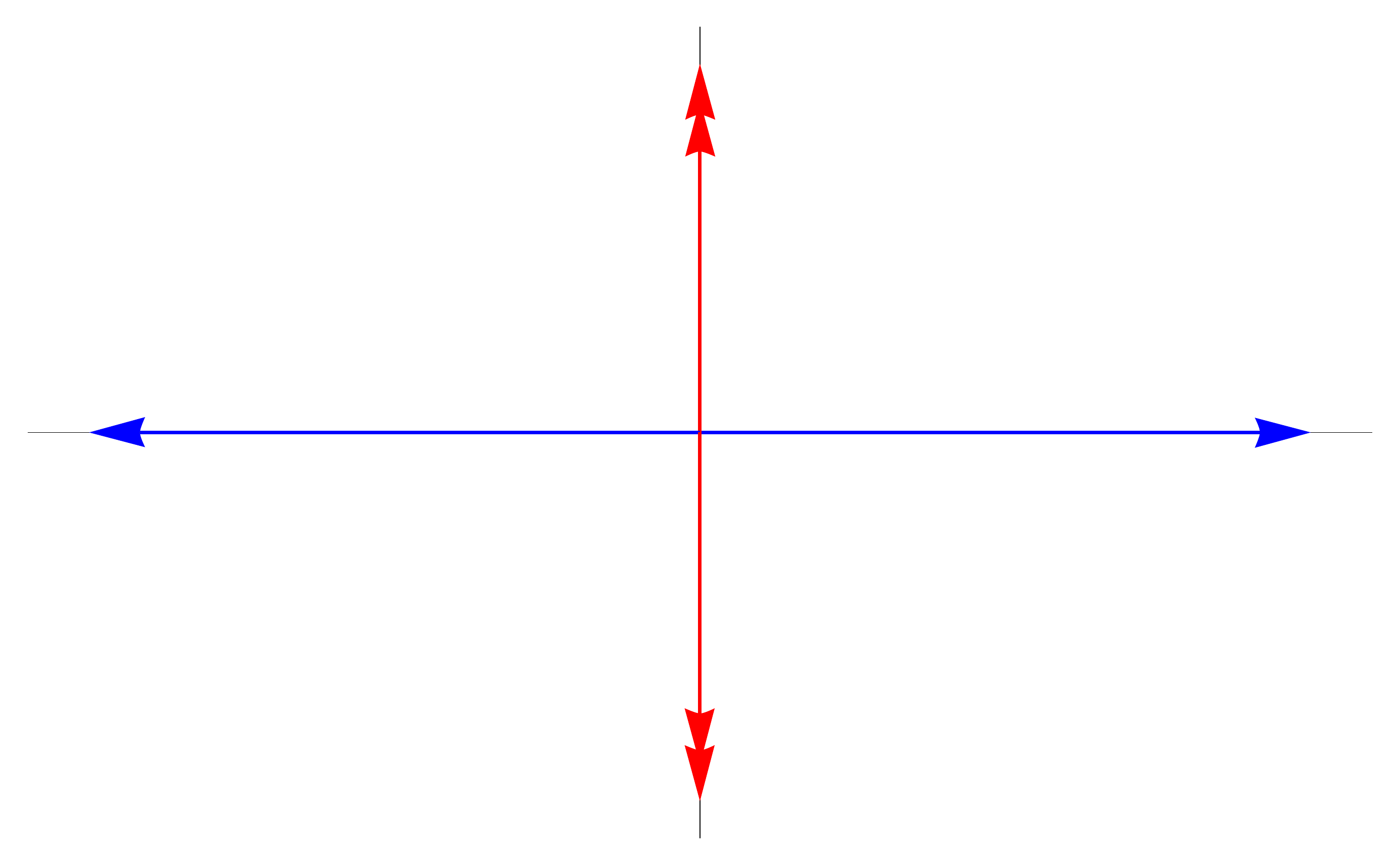}
	\end{subfigure}

	\begin{subfigure}{.32\textwidth}	
		\includegraphics[width=\textwidth]{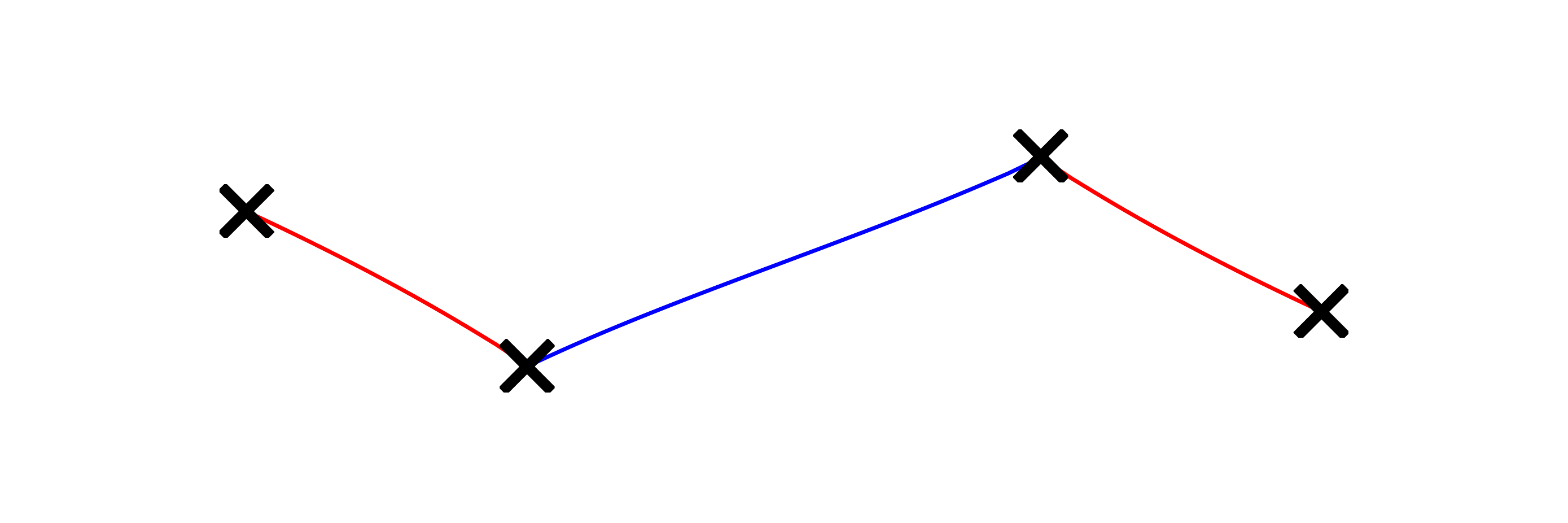}
	\end{subfigure}
	\begin{subfigure}{.22\textwidth}	
		\centering
		\includegraphics[width=\textwidth]{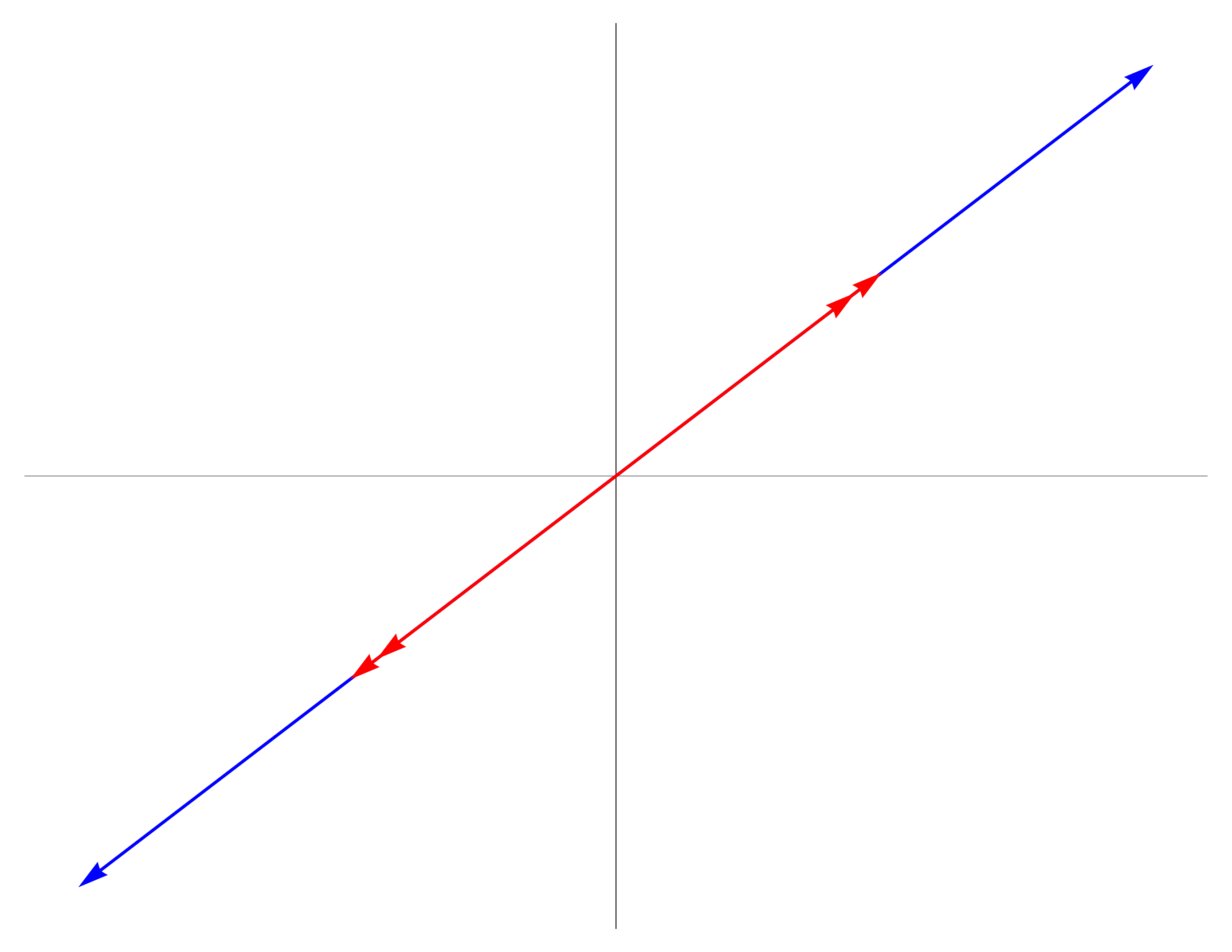}
	\end{subfigure}
	
	\begin{subfigure}{.32\textwidth}	
		\includegraphics[width=\textwidth]{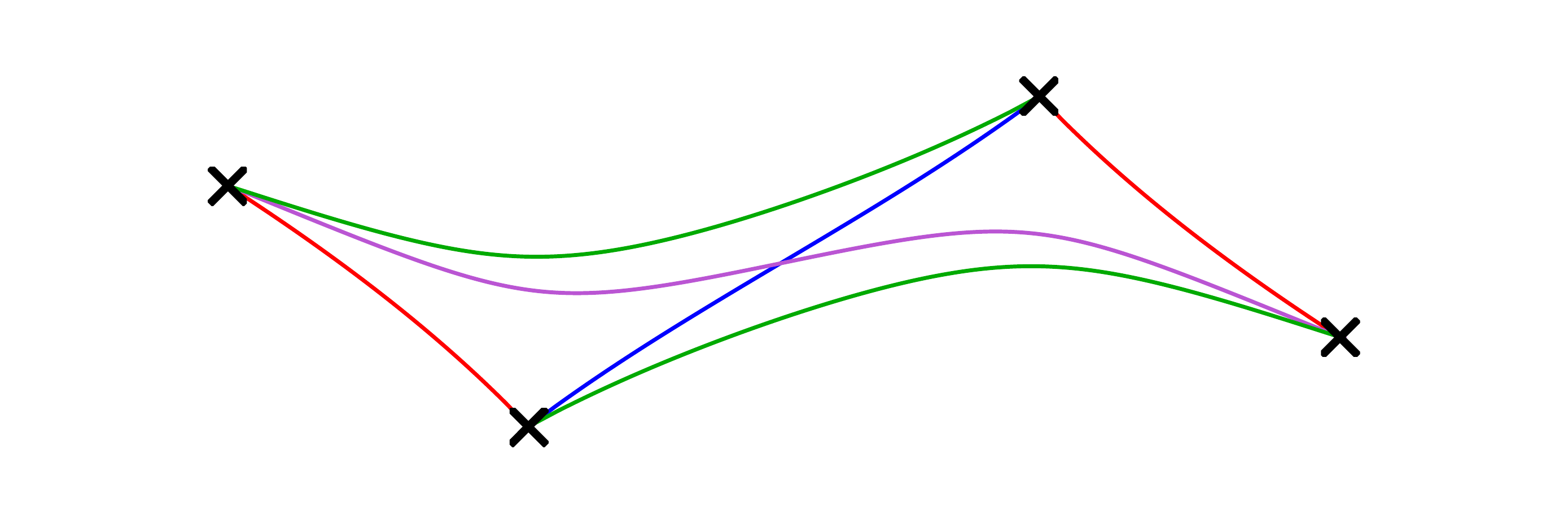}
	\end{subfigure}
	\begin{subfigure}{.22\textwidth}	
		\centering
		\includegraphics[width=\textwidth]{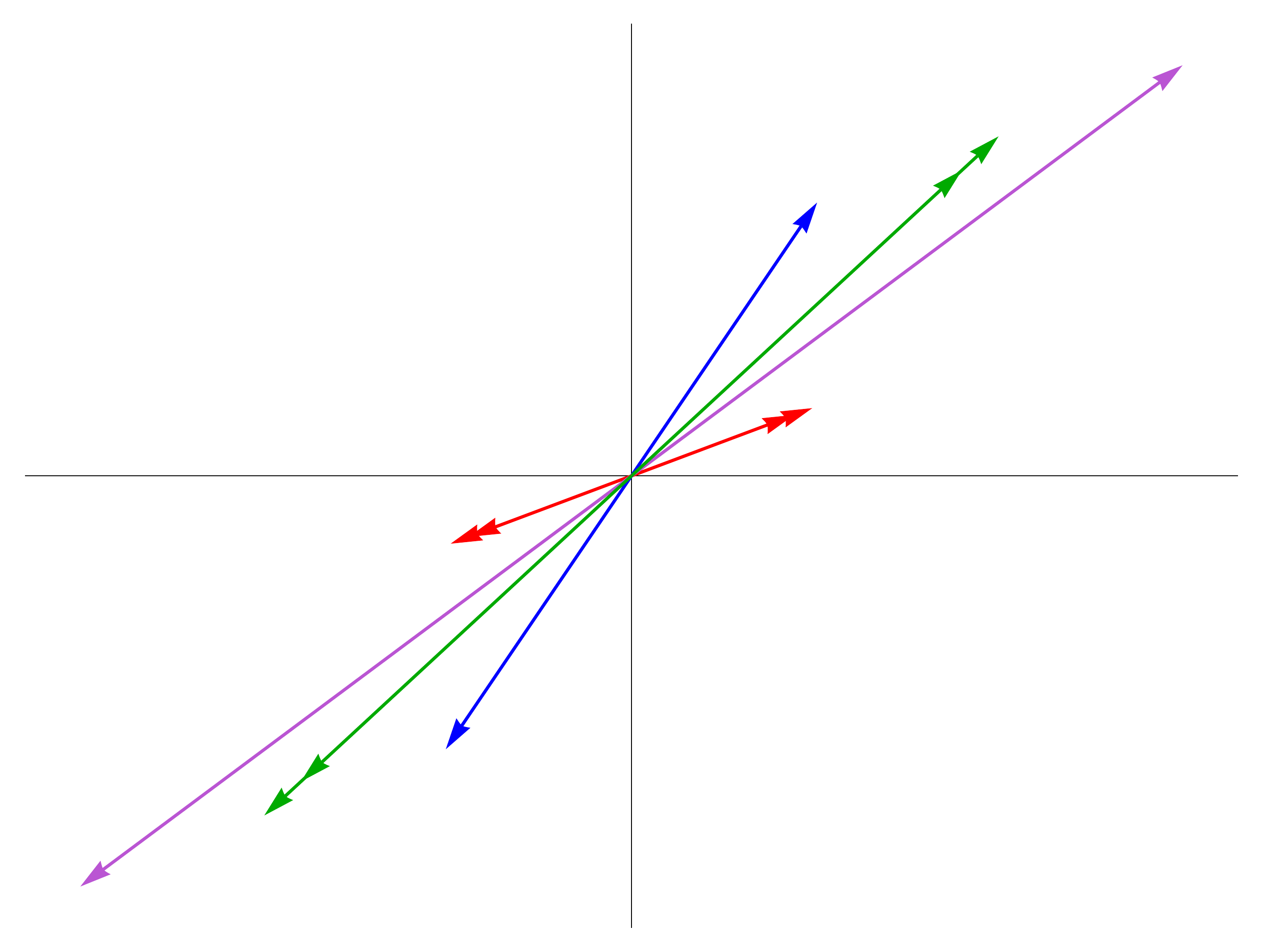}
	\end{subfigure}	
	
	\caption{BPS spectra of $\CS[A_{1}; \CD_{8}]$ and their wall-crossings with an $\SU(2)$ flavor symmetry.}
	\label{fig:a3SN_SU_2_finite}
\end{figure}

Now let us consider the wall-crossing from the minimal BPS spectrum to the maximal one for $\CS[A_{1}; \CD_{8}]$ 
with the $\SU(2)$ flavor symmetry, i.e. the residue of $\lambda$ at $t=\infty$ vanishes: $c_3 = 0$. 
The first row of Figure \ref{fig:a3SN_SU_2_finite} describes its BPS spectrum. 
Because of the vanishing residue, two of the three BPS states form an $\SU(2)$ doublet \cite{Argyres:2012fu}: the two cycles of the Seiberg-Witten curve corresponding to the two red $\CS$-walls are the same cycle when the residue vanishes. This is a minimal BPS spectrum of $\CS[A_{1}; \CD_{8}]$ with the $\SU(2)$ flavor symmetry, which can be represented with a quiver of Figure \ref{fig:a3SN_SU_2_f_BPS_quiver}.

\begin{figure}[h]
	\centering
	\begin{tikzpicture}[scale=.75]
		\node[W,red] (1) at (0,0) {1};
		\node[W,blue] (2) at (2,0) {2};
		\node[W,red] (3) at (4,0) {1};
		\path (1) edge[->] (2);
		\path (2) edge[<-] (3);
	\end{tikzpicture}
	\caption{BPS quiver of the minimal BPS spectrum of $\CS[A_{1}; \CD_{8}]$ with an $\SU(2)$ flavor symmetry.
	         The left and the right nodes correspond to the BPS states forming the doublet of $\SU(2)$.}
	\label{fig:a3SN_SU_2_f_BPS_quiver}
\end{figure}

\begin{figure}[t]
	\centering
	\includegraphics[width=.35\textwidth]{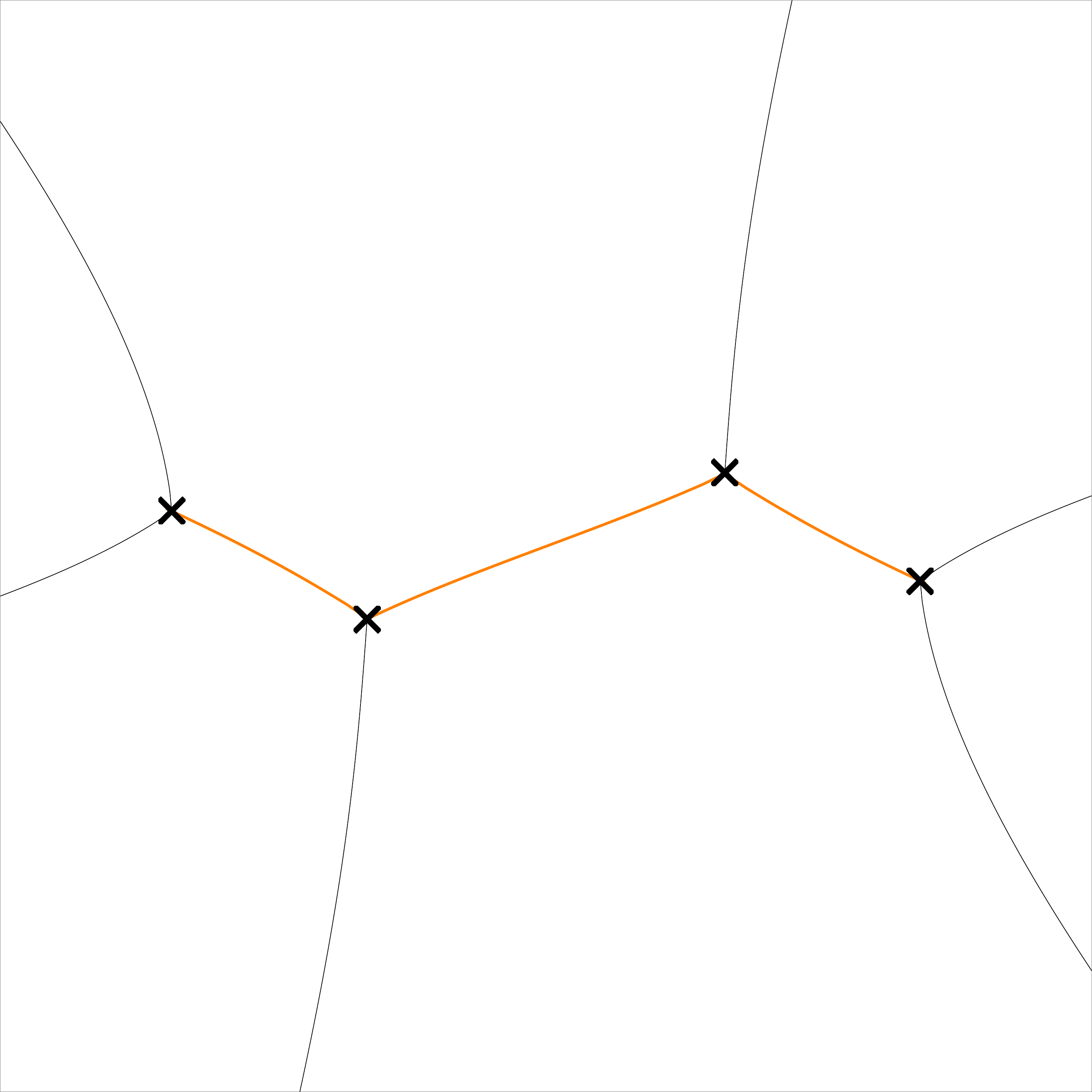}
	\caption{Spectral network of $\CS[A_{1}; \CD_{8}]$ with an $\SU(2)$ flavor symmetry on the BPS wall.}
	\label{fig:a3SN_on_BPS_wall_21}
\end{figure}

Note that when $c_3$ vanishes, the Seiberg-Witten curve is $x^2 = t^4 + c_2 t^2 + v_2$, so for general values of $c_2$ and $v_2$, we have four branch points where each pair is located symmetrically across $t = 0$. Because of the symmetry and the vanishing residue at $t = \infty$, we can see in Figure \ref{fig:a3SN_SU_2_finite} that the three wall-crossings in Figure \ref{fig:a3SN_finite} happens now at the same time, after which we have three additional BPS states, two of them forming another doublet of the $\SU(2)$ flavor symmetry. Note that maintaining the maximal flavor symmetry simplifies the analysis of wall-crossings of spectral network: we have less number of free parameters, which constrains the motion of branch points and in the end we have one wall-crossing rather than three between the minimal and the maximal BPS spectrum. Figure \ref{fig:a3SN_on_BPS_wall_21} shows the spectral network when the theory is on the BPS wall and $\theta$ has a value such that the finite $\CS$-walls appear at the same time.

\subsection{$\CS[A_{N-1}; \CD_{{\rm I}}]$ Theories}
\label{sec:S[A_N-1,C;D_I]}
  As we stated in section \ref{sec:A class of N=2 SCFTs},
  this SCFT is obtained as the maximal conformal point of $\SU(N)$ pure SYM theory.
  See appendix \ref{sec:ISYM} for the detailed derivation.
  We claim this SCFT is the same as $\CS[A_1; \CD_{N+4}]$.

  The Seiberg-Witten curve of the SCFT is of canonical form
  $x^{N} + \sum_{i=2}^{N} \phi_{i}(t) x^{N-i} = 0$ where
    \bea
    \phi_{i}
    &=&    c_{i},~~(i=2,\ldots,[\frac{N+2}{2}]),
           \nonumber \\
    \phi_{i}
    &=&    v_{N-i+2},~~(i=[\frac{N+2}{2}]+1, \ldots, N-1),~~{\rm and}~~
    \phi_{N}
     =     t^{2} + v_{2},
    \eea
  and the Seiberg-Witten differential is $\lambda = x\, dt$.
  The scaling dimensions are given by
  $\Delta(v_{i}) = 2 - \frac{2i}{N+2}$ and $\Delta(c_{i}) = \frac{2i}{N+2}$ for $i=2,\ldots,[\frac{N+1}{2}]$.
  When $N=2k$, there is a mass parameter $c_{k+1}$ with dimension $1$.
  We can see that the dimensions of the operators are the same as those of $\CS[A_{1}; \CD_{N+4}]$.

  We see that there is a singularity only at $t=\infty$
  where the differentials $\phi_{k}$ have poles as described in \eqref{Ipole}.
  Note that from this curve we can obtain the curve of $\CS[A_1; \CD_{N+4}]$ by changing $t \to i x$ \cite{Gaiotto:2012rg}, which illustrates that this description boils down to how to project the complex one-dimensional curve living in a complex two-dimensional space onto a complex plane as either a 2-to-1 or an $N$-to-1 mapping. However the Seiberg-Witten differentials for the two theories differ by an exact 1-form.

\subsubsection{$\CS[A_{2}; \CD_{{\rm I}}]$ in $A_2$-class}
The building block of the spectral network of $\CS[A_{N-1}; \CD_{{\rm I}}]$ is that of $\CS[A_{2}; \CD_{{\rm I}}]$, which is studied in \cite{Gaiotto:2012rg}. We will reproduce their result and provide a configuration of the spectral network that is useful in studying general $N$ case. The Seiberg-Witten curve and the differential are
\begin{align}
	x^3 + c_2 x + v_2 + t^2=0,\ \lambda = x\, dt,
\end{align}
where $\Delta(c_2) = \frac{4}{5}$ and $\Delta(v_2) = \frac{6}{5}$. 

\begin{figure}[h]
	\centering
	\begin{subfigure}[b]{.3\textwidth}	
		\includegraphics[width=\textwidth]{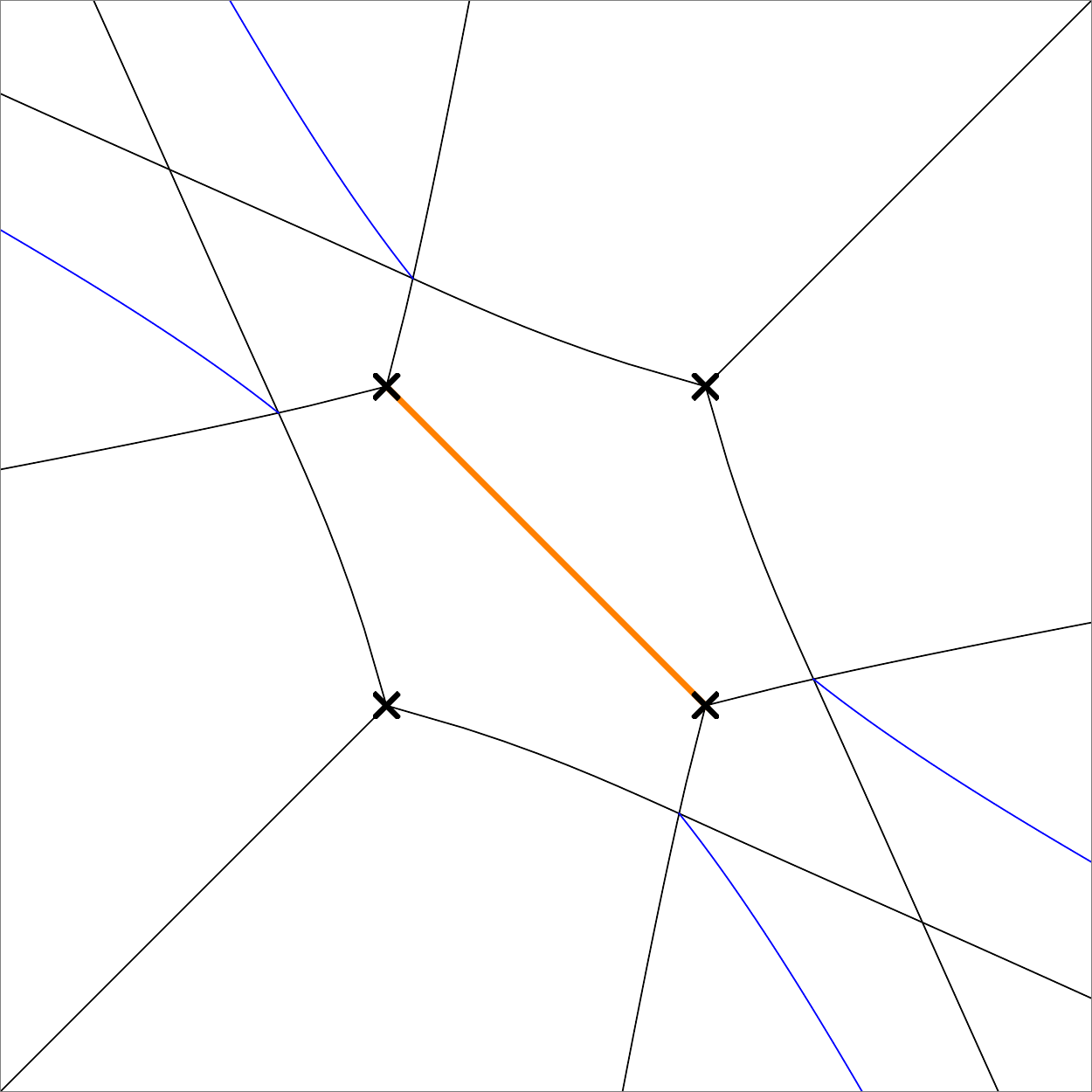}
		\caption{$\theta = \arg(Z_{1})$}
		\label{fig:a2SNFromA2_rho_inf_25}
	\end{subfigure}
	\begin{subfigure}[b]{.3\textwidth}	
		\includegraphics[width=\textwidth]{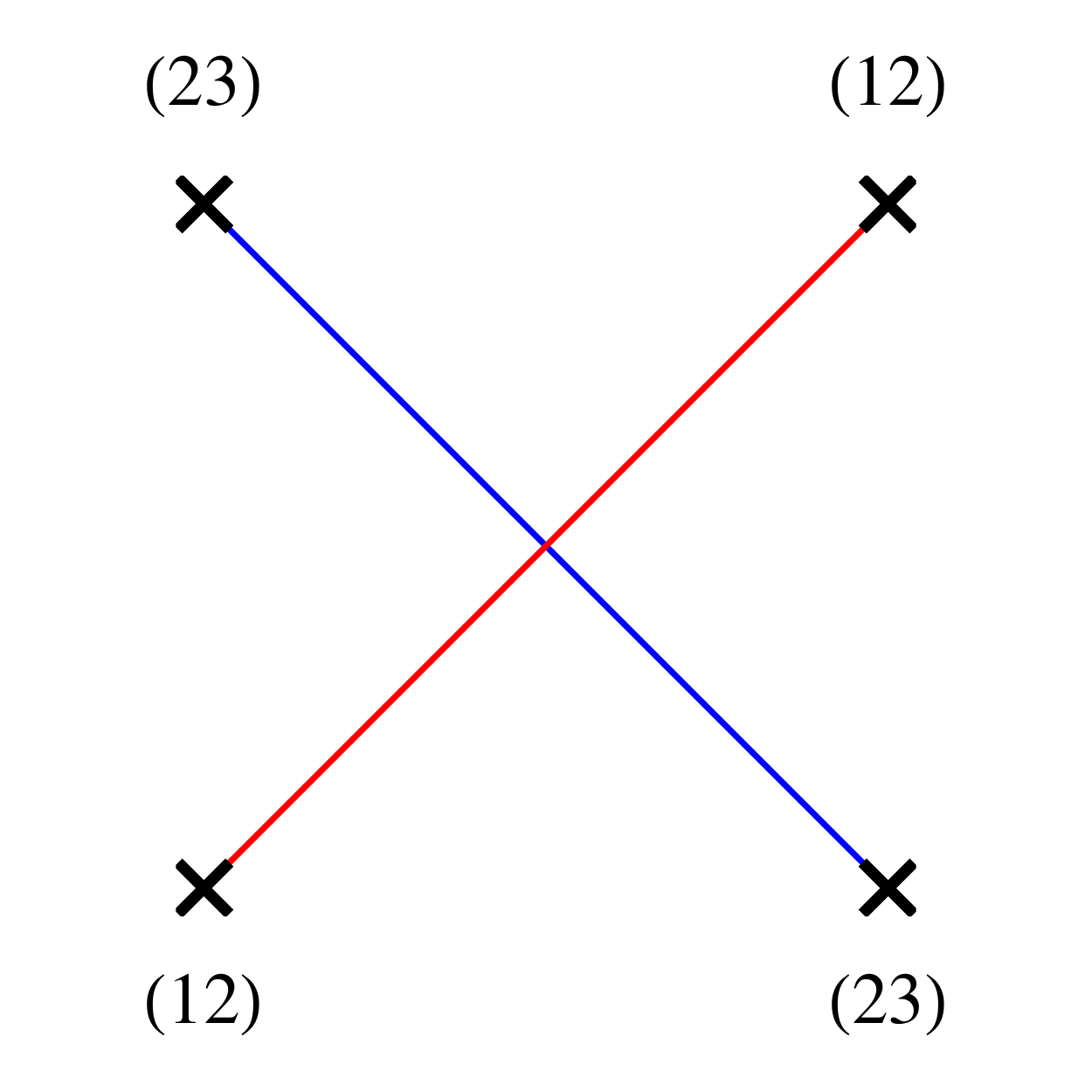}
		\caption{finite $\CS$-walls}
		\label{fig:a2SNFromA2_rho_inf_finite_t}
	\end{subfigure}
	\caption{Spectral network of $\CS[A_{2}; \CD_{{\rm I}}]$ with minimal BPS spectrum.}
	\label{fig:a2SNFromA2_rho_inf_SN}
\end{figure}

\paragraph{Minimal BPS spectrum}
We first consider the case of $c_2 \neq 0$ and $v_2 = 0$, when the theory has a minimal BPS spectrum. 
This choice of parameters results in four branch points of index 2 at finite $t$ and one branch point of index 3 at $t=\infty$, which is the location of the irregular puncture. A spectral network at a value of $\theta$ that contains a finite $\CS$-wall is shown in Figure \ref{fig:a2SNFromA2_rho_inf_25}. The animated version of the spectral network can be found at \href{http://www.its.caltech.edu/~splendid/spectral_network/a2SNFromA2_rho_inf.gif}{this website}. There will be another spectral network at $\pi- \theta$ that has the second finite $\CS$-wall, and both of the $\CS$-walls are shown in Figures \ref{fig:a2SNFromA2_rho_inf_finite_t}. 

\begin{figure}[h]
	\centering	
	\begin{subfigure}[b]{.23\textwidth}	
		\centering
		\includegraphics[width=\textwidth]{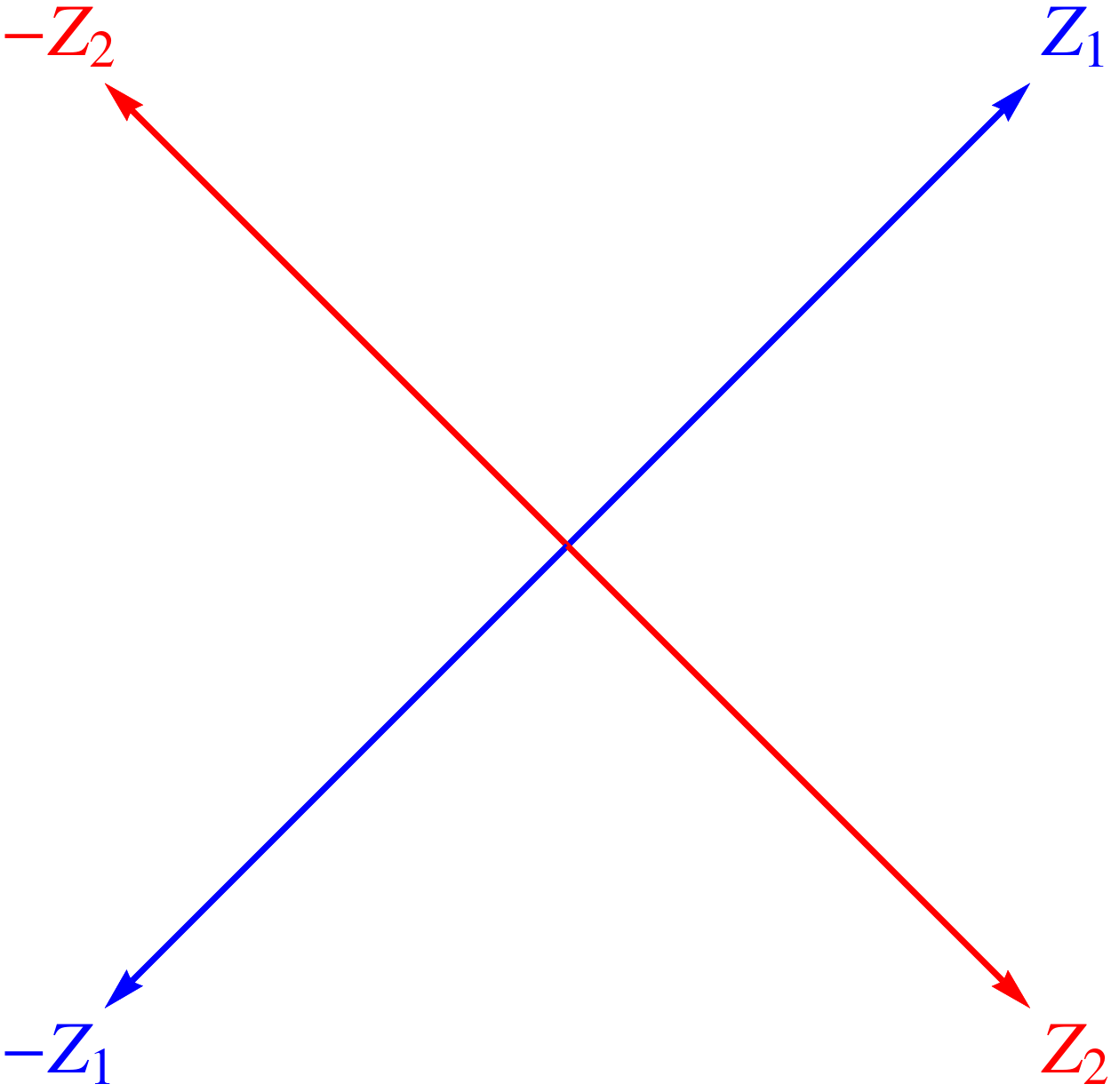}
		\caption{central charges}
		\label{fig:a2SNFromA2_rho_inf_Z}
	\end{subfigure}	
	\begin{subfigure}[b]{.23\textwidth}	
		\centering
		\begin{tabular}{c|c}
		   state  & $(e,m)$ \\ \hline
		  1 & $(1,0)$ \\ \hline
		  2 & $(0,1)$ 
		\end{tabular}
		\vspace{3.5em}
		\caption{IR charges}
		\label{tbl:a2SNFromA2_rho_inf_IR}
	\end{subfigure}	
	\begin{subfigure}[b]{.23\textwidth}	
		\centering
		\begin{tikzpicture}
			\node[W,blue] (1) at (0,0){1};
		 	\node[W,red] (2) at (2,0) {2};
			\draw[->] (1)--(2);
		\end{tikzpicture}
		\vspace{5em}
		\caption{BPS quiver}
		\label{fig:a2SNFromA2_rho_inf_quiver}
	\end{subfigure}	
	\renewcommand{\figurename}{Figure \& Table}
	\caption{Minimal BPS spectrum of $\CS[A_{2}; \CD_{{\rm I}}]$.}
	\label{figntbl:a2SNFromA2_rho_inf_BPS}

\end{figure}

Figure \ref{fig:a2SNFromA2_rho_inf_Z} shows the central charges of the BPS states, and their low-energy $U(1)$-charges are in Table \ref{tbl:a2SNFromA2_rho_inf_IR}, which can be read out from the intersections of the cycles that are homologically equivalent to the $\CS$-walls. This BPS spectrum can be represented by the BPS quiver shown in Figure \ref{fig:a2SNFromA2_rho_inf_quiver}, which is an $A_2$ quiver. Note that this is exactly the same BPS spectrum as the minimal spectrum of $\CS[A_1;\CD_7]$, see Figure \& Table \ref{figntbl:a2SN_minimal_BPS}.

\paragraph{Wall-crossing to a maximal BPS spectrum}
Now we fix $c_2$ and set $v_2$ to be nonzero. When $v_2$ is small, this deforms the previous configuration of spectral networks but does not cause a wall-crossing. After the value of $v_2$ is over a certain threshold, now the theory is on the other side of a BPS wall in the Coulomb branch moduli space. 

\begin{figure}[t]
	\centering
	\begin{subfigure}{.30\textwidth}	
		\includegraphics[width=\textwidth]{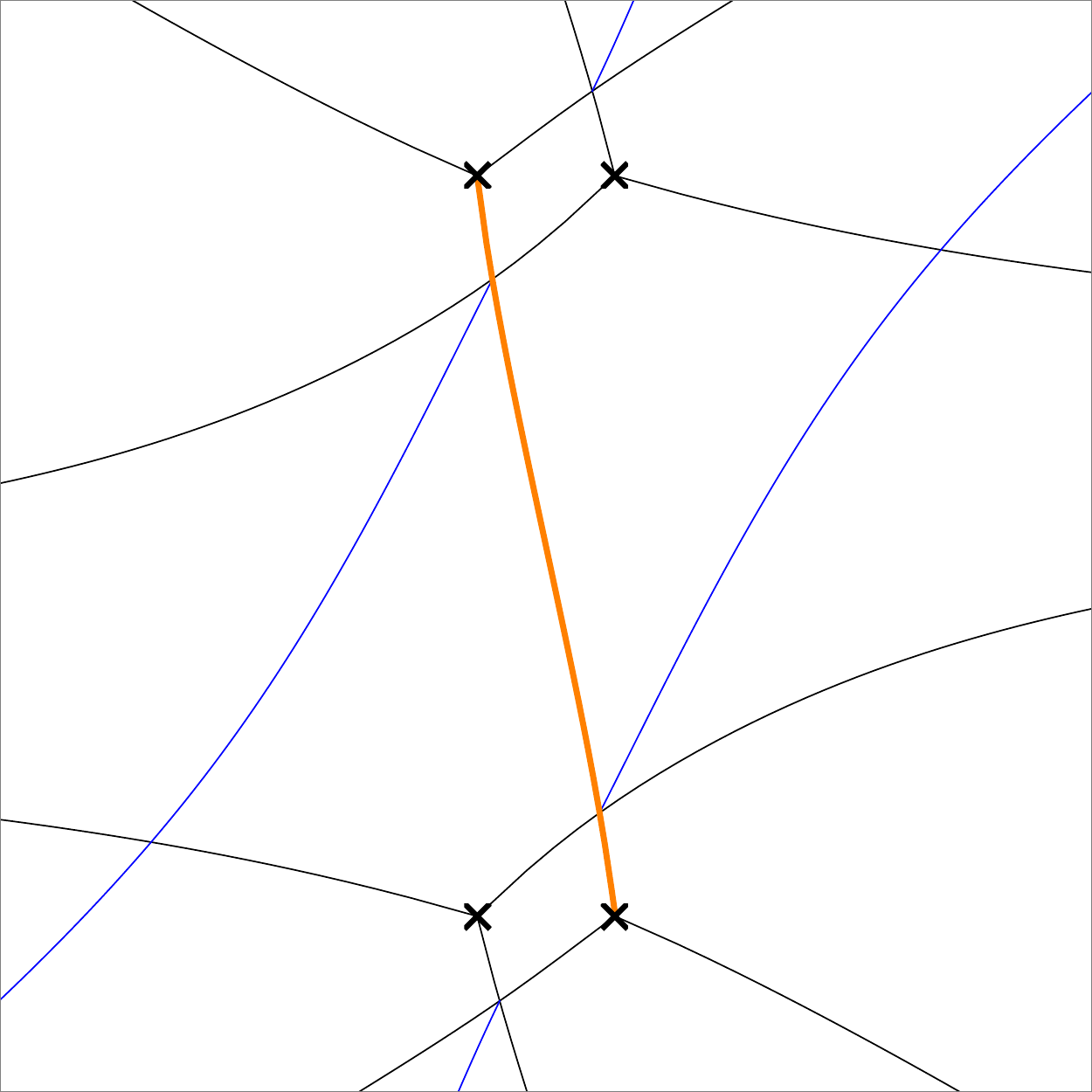}
		\caption{$\theta = \arg(Z_{1})$}
		\label{fig:a2SNFromA2_rho_1_83}
	\end{subfigure}
	\begin{subfigure}{.30\textwidth}	
		\centering
		\includegraphics[width=\textwidth]{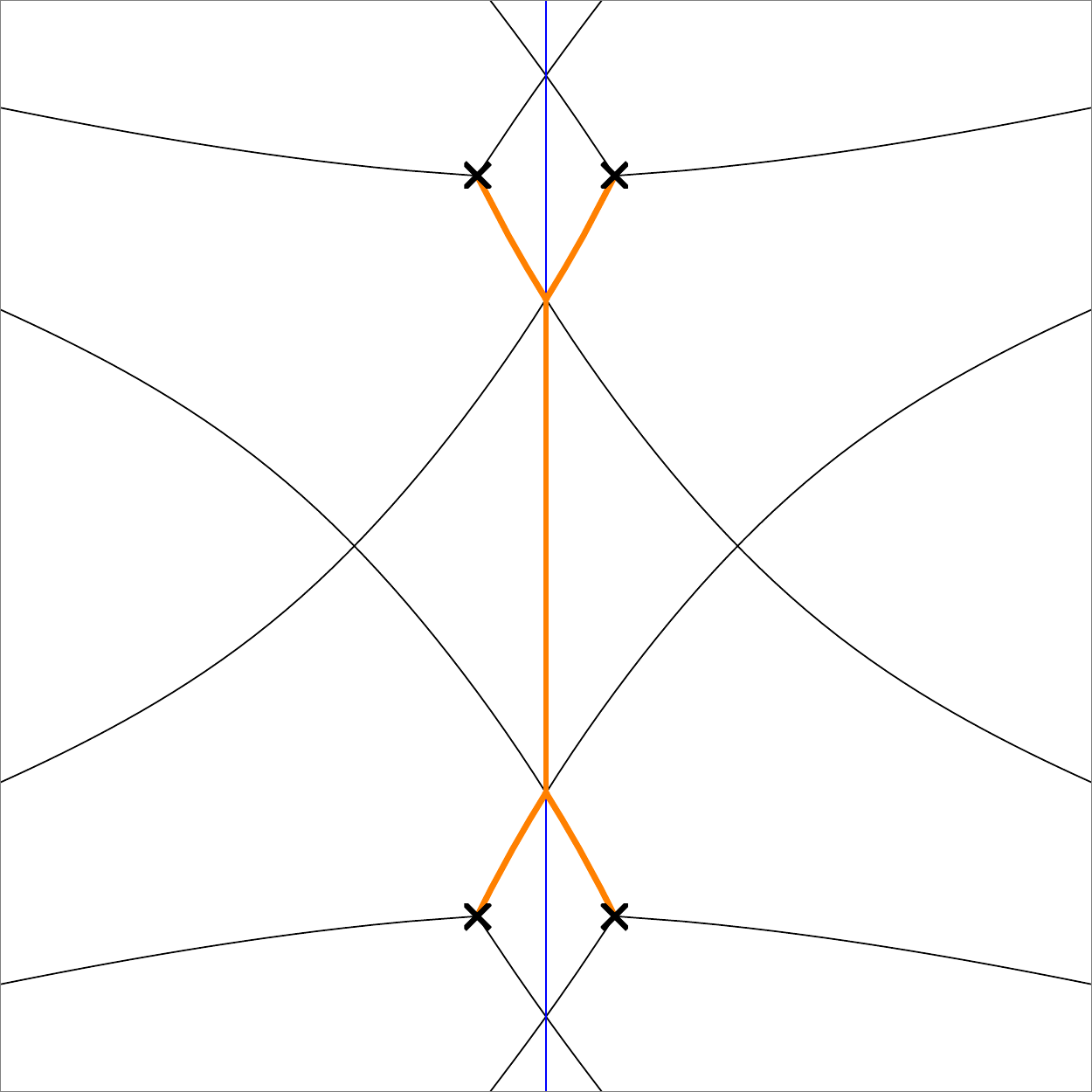}
		\caption{$\theta = \arg(Z_{3})$}
		\label{fig:a2SNFromA2_rho_1_00}
	\end{subfigure}
	\begin{subfigure}{.18\textwidth}
		\centering	
		\includegraphics[scale=.33]{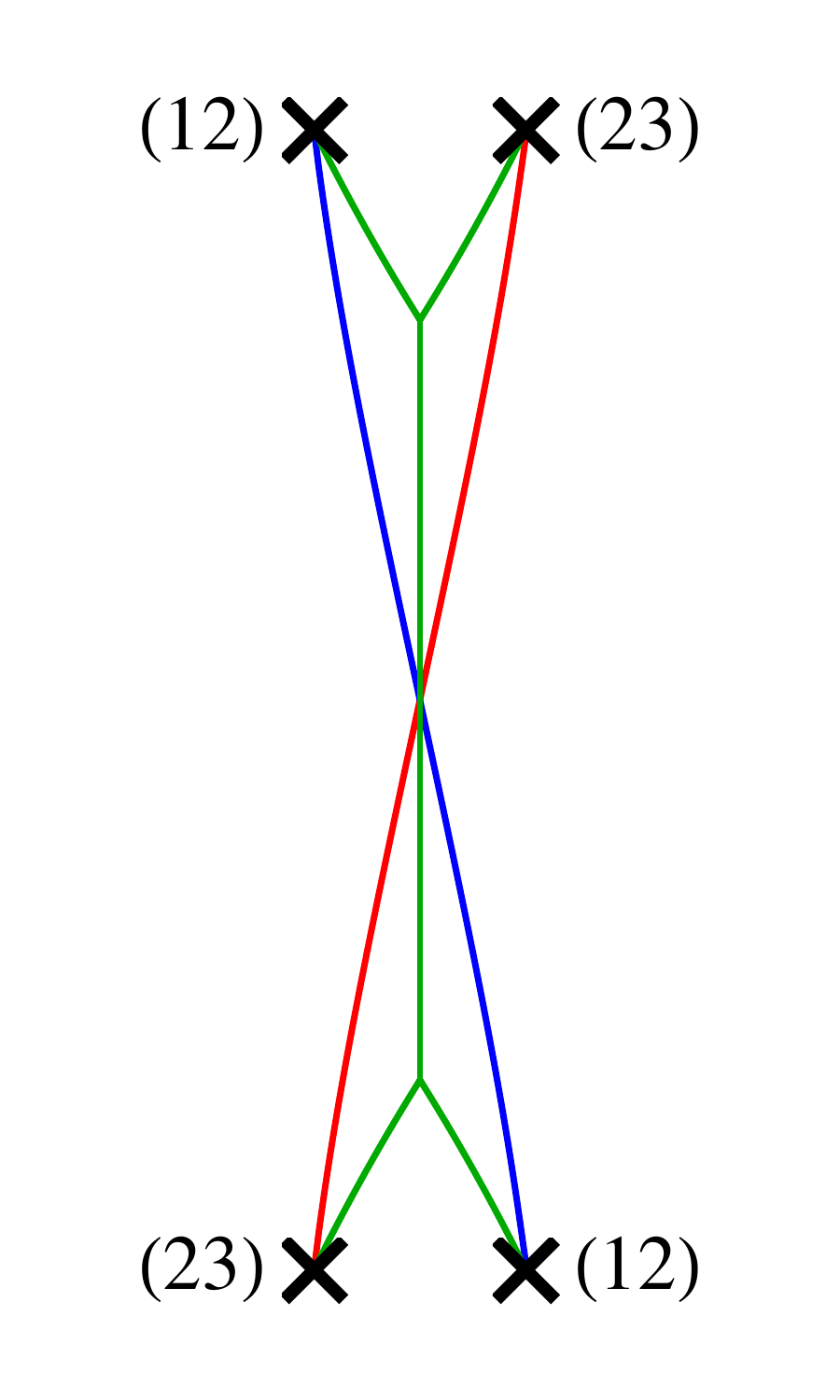}
		\caption{finite $\CS$-walls}
		\label{fig:a2SNFromA2_rho_1_finite_t}
	\end{subfigure}
	
	\caption{Spectral network of $\CS[A_{2}; \CD_{{\rm I}}]$ with maximal BPS spectrum.}
	\label{fig:a2SNFromA2_rho_1}
\end{figure}

Figure \ref{fig:a2SNFromA2_rho_1_83} shows a spectral network with a finite $\CS$-wall that is similar to the one in Figure \ref{fig:a2SNFromA2_rho_inf_25}. But in Figure \ref{fig:a2SNFromA2_rho_1_00} we have a finite $\CS$-wall that appears only for $A_{N-1}$ with $N > 2$ \cite{Gaiotto:2012rg}. We can see there is a supersymmetric joint formed by three $\CS$-walls, where $\lambda_{12} + \lambda_{23} = \lambda_{13}$ is satisfied. The animated version of Figure \ref{fig:a2SNFromA2_rho_1} can be found \href{http://www.its.caltech.edu/~splendid/spectral_network/a2SNFromA2_rho_1.gif}{at this webpage}.

\begin{figure}[h]
	\centering
	\begin{subfigure}[b]{.45\textwidth}	
		\centering
		\includegraphics[width=\textwidth]{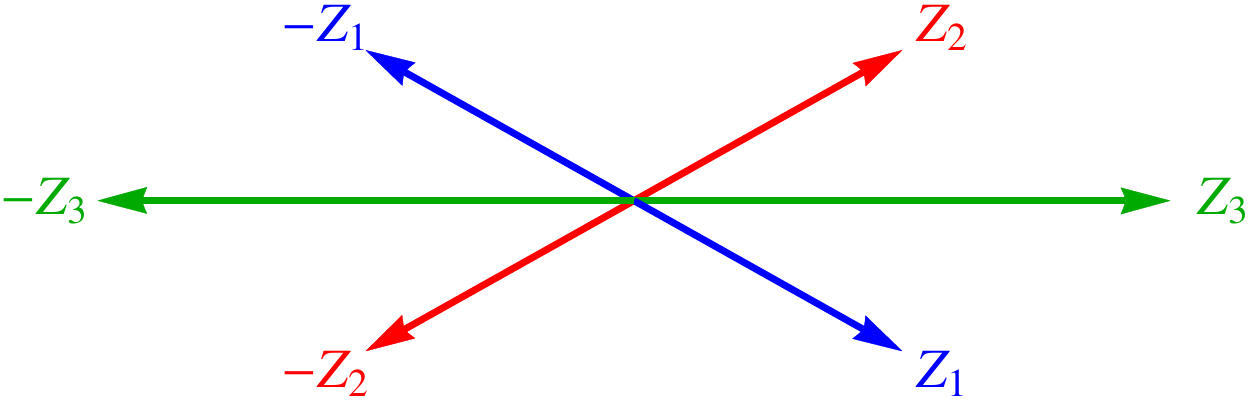}
		\caption{central charges}
		\label{fig:a2SNFromA2_rho_1_Z}
	\end{subfigure}
	\begin{subfigure}[b]{.3\textwidth}	
		\centering
		\begin{tabular}{c|c}
		   state  & $(e,m)$ \\ \hline
		  1 & $(1,0)$ \\ \hline
		  2 & $(0,1)$ \\ \hline
		  3 & $(1,1)$
		\end{tabular}
		\caption{IR charges}
		\label{tbl:a2SNFromA2_rho_1_IR}
	\end{subfigure}	
	\renewcommand{\figurename}{Figure \& Table}
	\caption{Maximal BPS spectrum of $\CS[A_{2}; \CD_{{\rm I}}]$.}
	\label{figntbl:a2SNFromA2_rho_1_BPS}
\end{figure}

Figure \ref{fig:a2SNFromA2_rho_1_finite_t} shows the three finite $\CS$-walls, and the central charges of the corresponding BPS states are shown in Figure \ref{fig:a2SNFromA2_rho_1_Z}. 
When we compare it with Figure \ref{fig:a2SNFromA2_rho_inf_Z}, we see that $Z_{1}$ and $Z_{2}$ approach and cross over each other when the wall-crossing happens, resulting in the creation of $Z_{3}$ \cite{Gaiotto:2012rg}. By considering the intersections of the cycles that are homologically equivalent to the finite $\CS$-walls, we can find out the IR charges of the BPS states as in Table \ref{tbl:a2SNFromA2_rho_1_IR}. This maximal BPS spectrum is the same as that of $\CS[A_1, \CD_7]$, see Figure \& Table \ref{figntbl:a2SN_maximal_BPS}.

\begin{figure}[h]
	\centering
	\begin{subfigure}[b]{.3\textwidth}
		\centering
		\includegraphics[width=\textwidth]{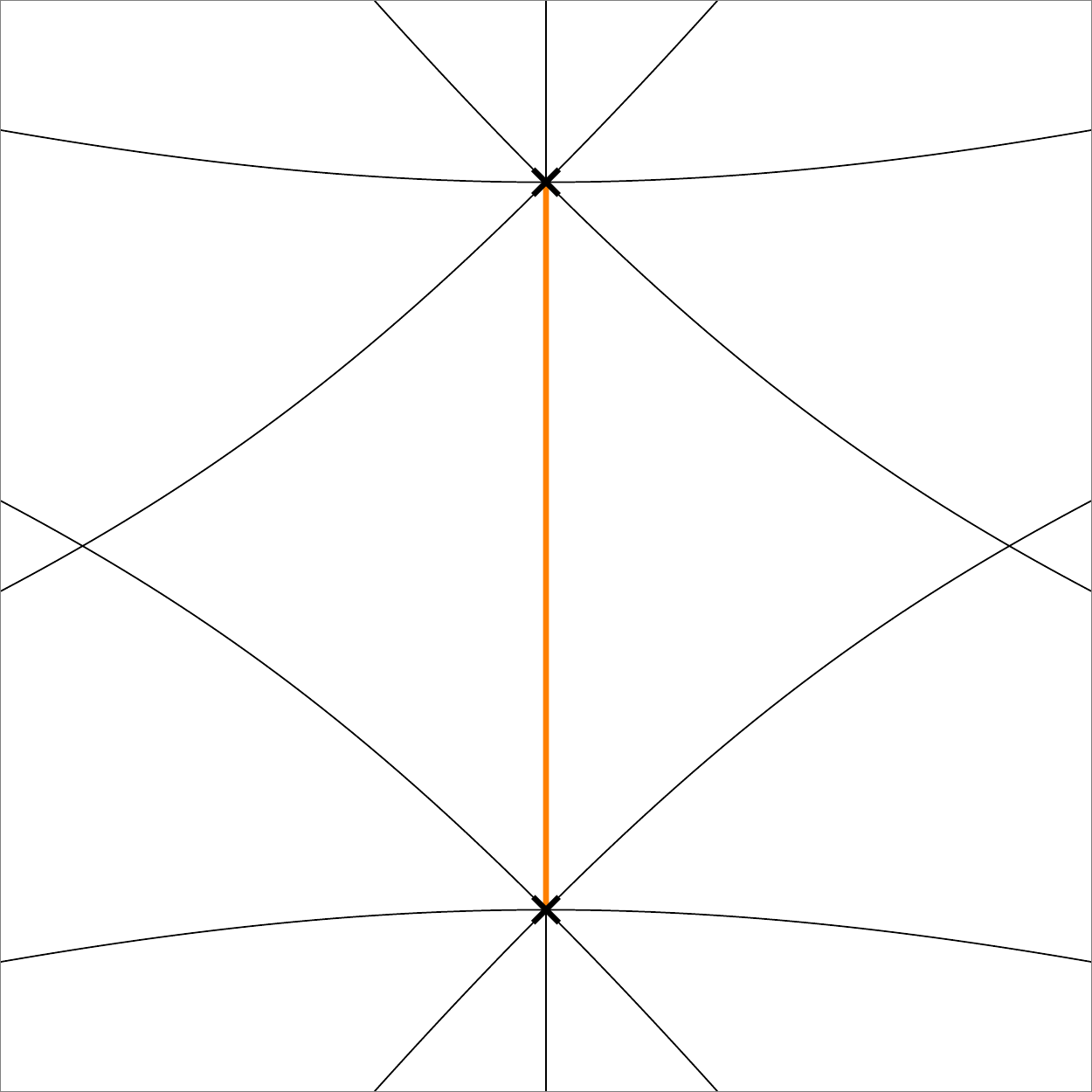}
		\caption{$\theta = \arg{Z_3}$}
		\label{fig:a2SNFromA2_rho_0_00}
	\end{subfigure}
	\begin{subfigure}[b]{.3\textwidth}
		\centering	
		\includegraphics[scale=.33]{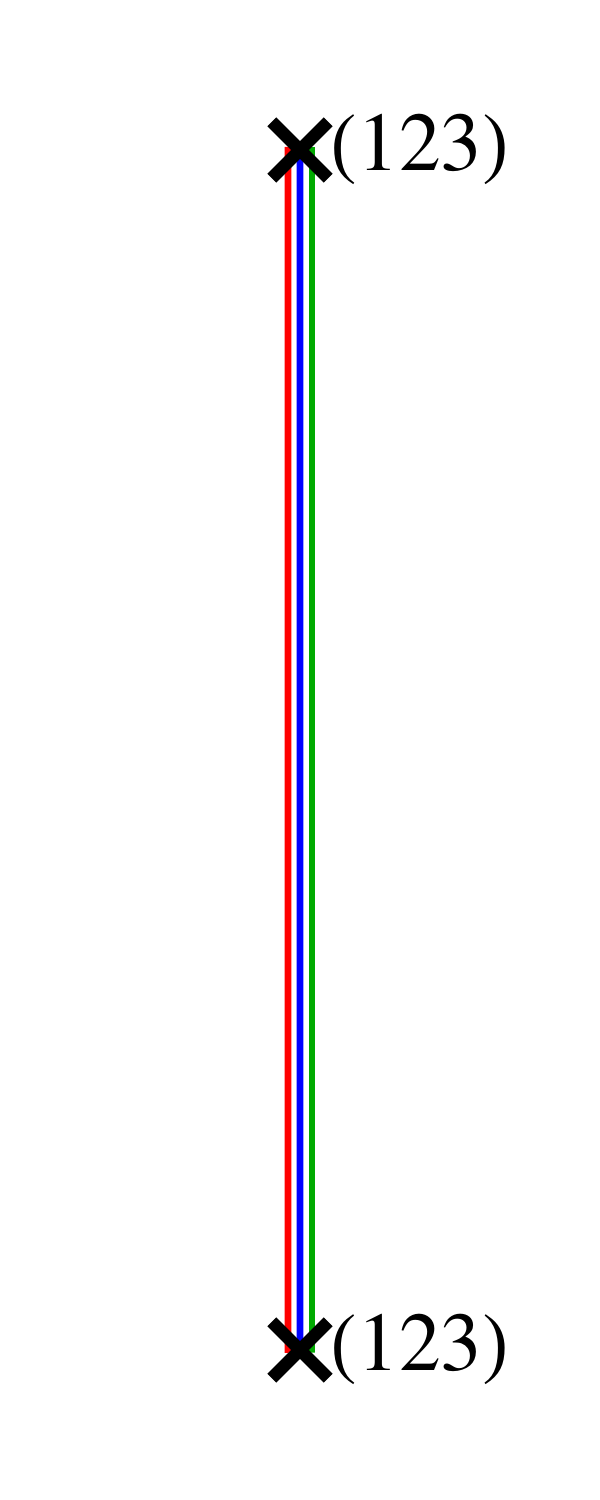}
		\caption{finite $\CS$-walls}
		\label{fig:a2SNFromA2_rho_0_finite_t}
	\end{subfigure}	
	\caption{Spectral network of $\CS[A_{2}; \CD_{{\rm I}}]$ with maximal, symmetric BPS spectrum.}
	\label{fig:a2SNFromA2_rho_0_SN}
\end{figure}

\paragraph{Maximal, symmetric BPS spectrum}
We can collide two pairs of branch points of index 2 at finite values of $t$ without causing a singularity, creating two branch points of index 3. This is obtained by setting $c_2 = 0$ and $v_3 \neq 0$, which gives us the maximal, symmetric BPS spectrum. Figure \ref{fig:a2SNFromA2_rho_0_00} shows a spectral network from the two branch points of index 3 when there is a finite $\CS$-wall.

\begin{figure}[ht]
	\centering
	\begin{subfigure}[b]{.3\textwidth}	
		\centering
		\includegraphics[width=\textwidth]{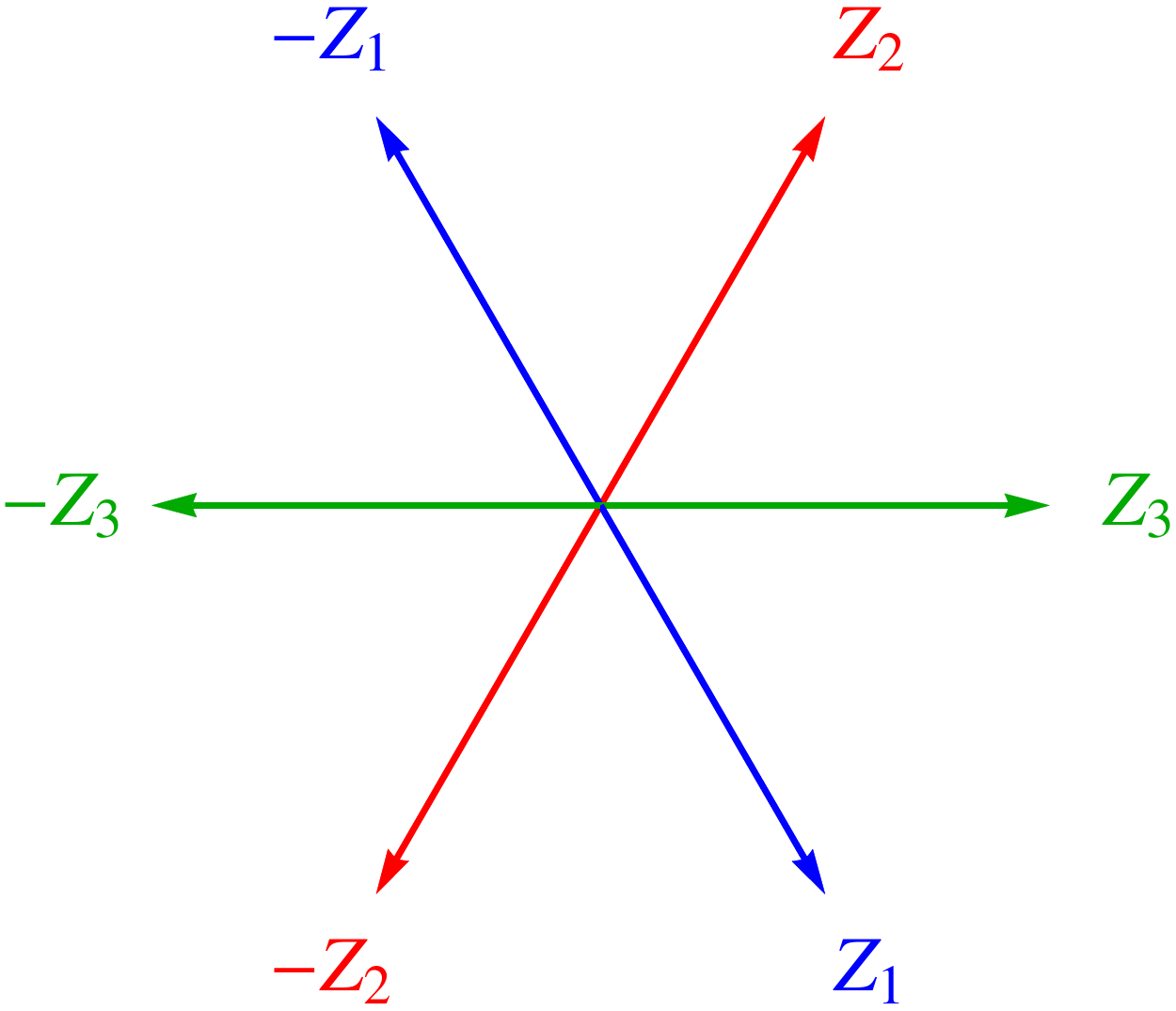}
		\caption{central charges}
		\label{fig:a2SNFromA2_rho_0_Z}
	\end{subfigure}	
	\begin{subfigure}[b]{.3\textwidth}	
		\centering
		\begin{tabular}{c|c}
		   state  & $(e,m)$ \\ \hline
		  1 & $(1,0)$ \\ \hline
		  2 & $(0,1)$ \\ \hline
		  3 & $(1,1)$
		\end{tabular}
		\vspace{2.5em}
		\caption{IR charges}
		\label{tbl:a2SNFromA2_rho_0_IR}
	\end{subfigure}	
	\renewcommand{\figurename}{Figure \& Table}	
	\caption{Maximal, symetric BPS spectrum of $\CS[A_{2}; \CD_{{\rm I}}]$.}
	\label{figntbl:a2SNFromA2_rho_0_BPS}
\end{figure}

There are three finite $\CS$-walls connecting the two $(123)$ branch points for $0 \leq \theta < \pi$. The animated version of Figure \ref{fig:a2SNFromA2_rho_0_00} can be found \href{http://www.its.caltech.edu/~splendid/spectral_network/a2SNFromA2_rho_0.gif}{at this webpage}. Figure \ref{fig:a2SNFromA2_rho_0_finite_t} shows the three finite $\CS$-walls connecting the two branch points. 
Figure \ref{fig:a2SNFromA2_rho_0_Z} and Table \ref{tbl:a2SNFromA2_rho_0_IR} describe the maximal symmetric BPS spectrum of $\CS[A_{2}; \CD_{{\rm I}}]$, which can be identified with the symmetric maximal BPS spectrum of $\CS[A_1;\CD_{7}]$. 

\subsubsection{$\CS[A_{N-1}; \CD_{{\rm I}}]$ in $A_{N-1}$-class, $N>3$}
Now we generalize the previous analysis to $\CS[A_{N-1}; \CD_{{\rm I}}]$. We will compare its spectral network with that of $\CS[A_1; \CD_{N+4}]$ to see that both SCFTs have the same minimal and maximal BPS spectra, and check for an example that the two theories have the same BPS chamber structure, which we expect to hold for general $N$ and is therefore good evidence that both theories are in $A_{N-1}$-class.

\paragraph{Minimal BPS spectrum}
For general values of parameters, the Seiberg-Witten curve of $\CS[A_{N-1}; \CD_{{\rm I}}]$ has $2(N-1)$ branch points of index 2 at finite $t$. In addition to that, when $N$ is odd, there is a branch point of index $N$ at $t = \infty$, and by applying the Riemann-Hurwitz formula we get the genus of the Seiberg-Witten curve $g = (N-1)/2$. 
When $N$ is even, there are two branch points of index $N/2$ at $t = \infty$, and the genus of the Seiberg-Witten curve is $g = N/2-1$.
We choose a branch cut between every branch point of finite $t$ and the irregular puncture at $t = \infty$.

With this, when the theory has a minimal BPS spectrum the configuration of spectral network has its branch points at finite $t$ being aligned along two perpendicular lines. There is a finite $\CS$-wall for every pair of branch points of the same kind, and none between the branch points of different kinds. This gives us the BPS spectrum that can be represented as an $A_{N-1}$ quiver diagram. 

\begin{figure}[t]
	\centering
	\begin{subfigure}[b]{.3\textwidth}
		\includegraphics[width=\textwidth]{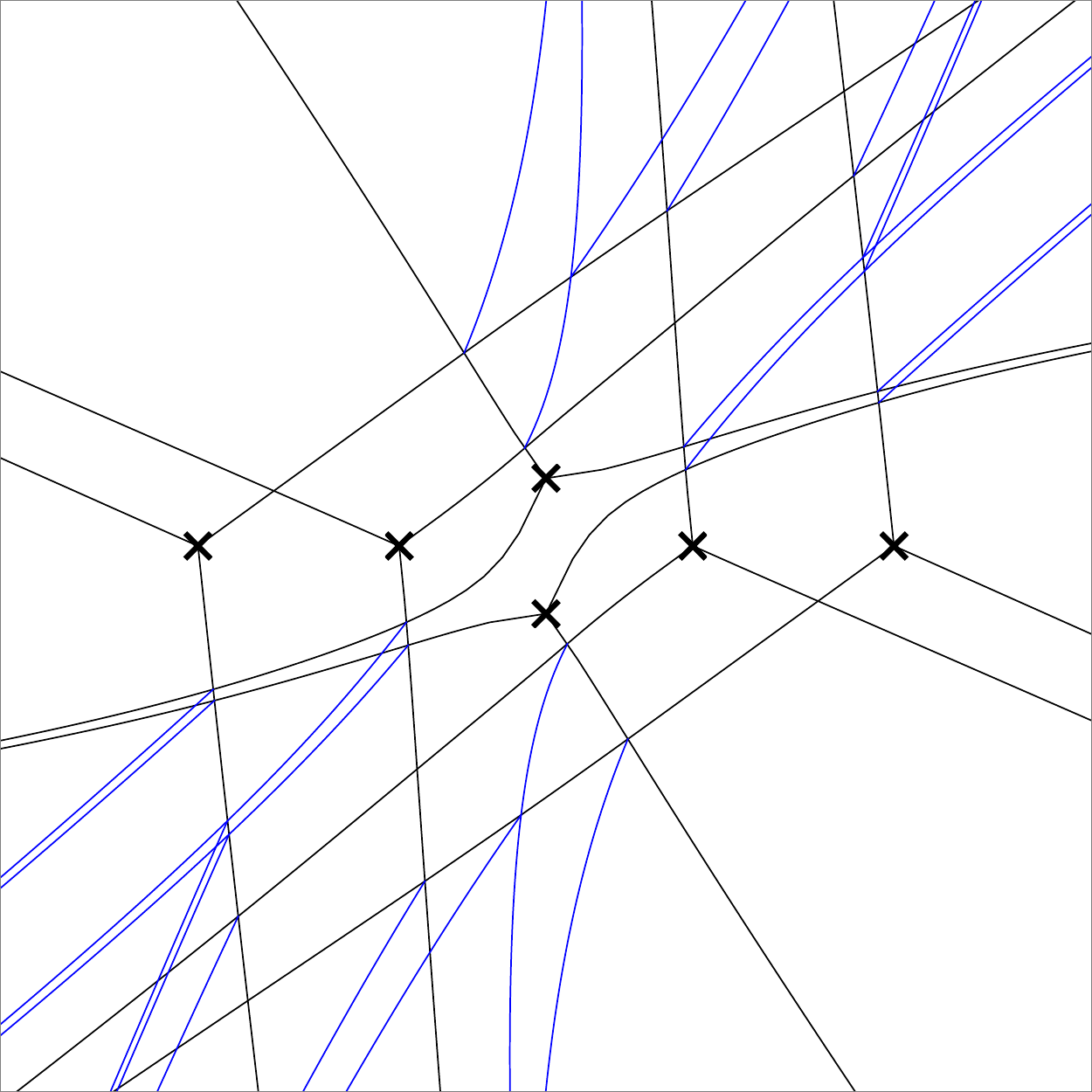}
		\caption{Spectral network}
		\label{fig:a3SNFromA3_05_30}
	\end{subfigure}
	\begin{subfigure}[b]{.37\textwidth}	
		\centering
		\includegraphics[width=\textwidth]{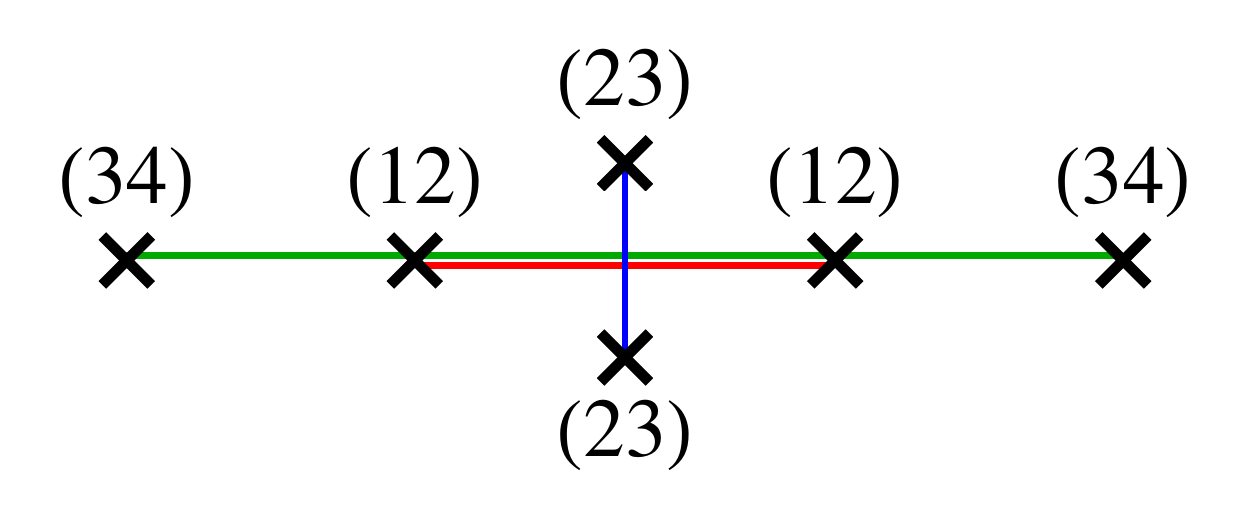}
		\vspace{1.5em}
		\caption{finite $\CS$-walls}
		\label{fig:a3SNFromA3_05_finite}
	\end{subfigure}		

	\caption{Spectral network of $\CS[A_{3}; \CD_{{\rm I}}]$ with minimal BPS spectrum.}
	\label{fig:a3SNFromA3_05_SN}
\end{figure}

The case of $N=4$ is a useful example to understand the generalization. Its Seiberg-Witten curve is
\begin{align}
	x^4 + c_2 x^2 + m x + v_2 + t^2 = 0,
\end{align}
where $\Delta(c_2) = \frac{2}{3}$, $\Delta(m) = 1$, $\Delta(v_2) = \frac{4}{3}$. $m$ is a mass parameter, and when it is zero we expect an $\SU(2)$ doublet to appear. Figure \ref{fig:a3SNFromA3_05_30} shows the spectral network at a general value of $\theta$ and Figure \ref{fig:a3SNFromA3_05_finite} shows finite $\CS$-walls of $\CS[A_{3}; D_{{\rm I}}]$ when it has a minimal BPS spectrum and when $m \neq 0$. When $m=0$, $(12)$- and $(34)$-branch points are at the same location, resulting in an $\SU(2)$ doublet 
of two finite $\CS$-walls.

\begin{figure}[ht]
	\centering
	\begin{subfigure}[b]{.4\textwidth}	
		\centering
		\includegraphics[width=\textwidth]{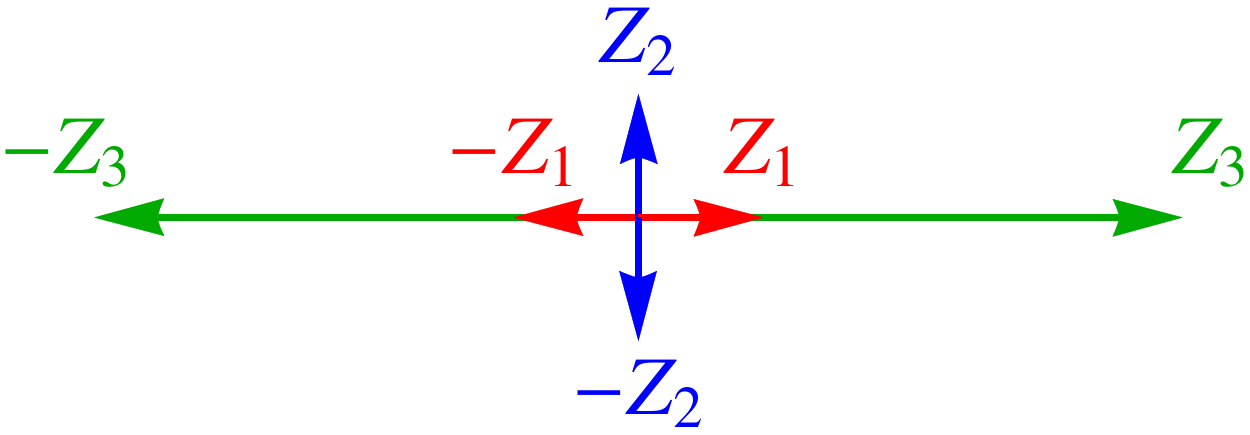}
		\caption{on the $Z$-plane}
		\label{fig:a3SNFromA3_05_Z}
	\end{subfigure}	
	\begin{subfigure}[b]{.3\textwidth}	
		\centering
		\begin{tabular}{c|c}
		   state  & $(e,m)$ \\ \hline
		  1 & $(1,0)$ \\ \hline
		  2 & $(0,1)$ \\ \hline
		  3 & $(1,0)$
		\end{tabular}
		\caption{IR charges}
		\label{tbl:a3SNFromA3_05_IR}
	\end{subfigure}	
	\begin{subfigure}[b]{.25\textwidth}
		\centering
		\begin{tikzpicture}[scale=.75]
			\node[W,red] (1) at (0,0) {1};
			\node[W,blue] (2) at (2,0) {2};
			\node[W,darkergreen] (3) at (4,0) {3};
			\path (1) edge[->] (2);
			\path (2) edge[<-] (3);
		\end{tikzpicture}
		\vspace{2em}
		\caption{BPS quiver}
		\label{fig:a3SNFromA3_05_quiver}
	\end{subfigure}
	\renewcommand{\figurename}{Figure \& Table}
	\caption{Minimal BPS spectrum of $\CS[A_{3}; \CD_{{\rm I}}]$.}
	\label{figntbl:a3SNFromA3_05_BPS}
\end{figure}

Figure \ref{fig:a3SNFromA3_05_Z} describes the central charges of the states in the minimal BPS spectrum of $\CS[A_{3}; \CD_{{\rm I}}]$ when $m \neq 0$. When $m = 0$, we have an $\SU(2)$ doublet  with $Z_1 = Z_3$. Table \ref{tbl:a3SNFromA3_05_IR} describes the IR $\UU(1)$-charges of the states, from which we can construct a BPS quiver as shown in Figure \ref{fig:a3SNFromA3_05_quiver}. This is the same BPS spectrum as the minimal BPS spectrum of $\CS[A_1; \CD_{{8}}]$, see Figures \ref{fig:a3SN_BPS_quiver} and \ref{fig:a3SN_finite}.

\paragraph{Wall-crossing to the maximal symmetric BPS spectrum}
From the minimal BPS spectrum of $\CS[A_{N-1}; \CD_{{\rm I}}]$, when its parameters have general nonzero values, it undergoes one wall-crossing of $\CS[A_{2}; \CD_{{\rm I}}]$ after another to reach the maximal BPS spectrum, where each wall-crossing adds a BPS state in the spectrum. Remember that each wall-crossing of $\CS[A_1; \CD_{N+4}]$ is that of $\CS[A_1; \CD_{7}]$, and from the minimal BPS spectrum after a series of such wall-crossings $\CS[A_1; \CD_{N+4}]$ reaches the chamber of the maximal BPS spectrum. This matching of both the BPS spectrum and its wall-crossings from the minimal chamber to the maximal one is good evidence for the equivalence of $\CS[A_{N-1}; \CD_{{\rm I}}]$ and $\CS[A_1; \CD_{N+4}]$.

\begin{figure}[ht]
	\centering
	\begin{subfigure}[b]{.3\textwidth}
		\includegraphics[width=\textwidth]{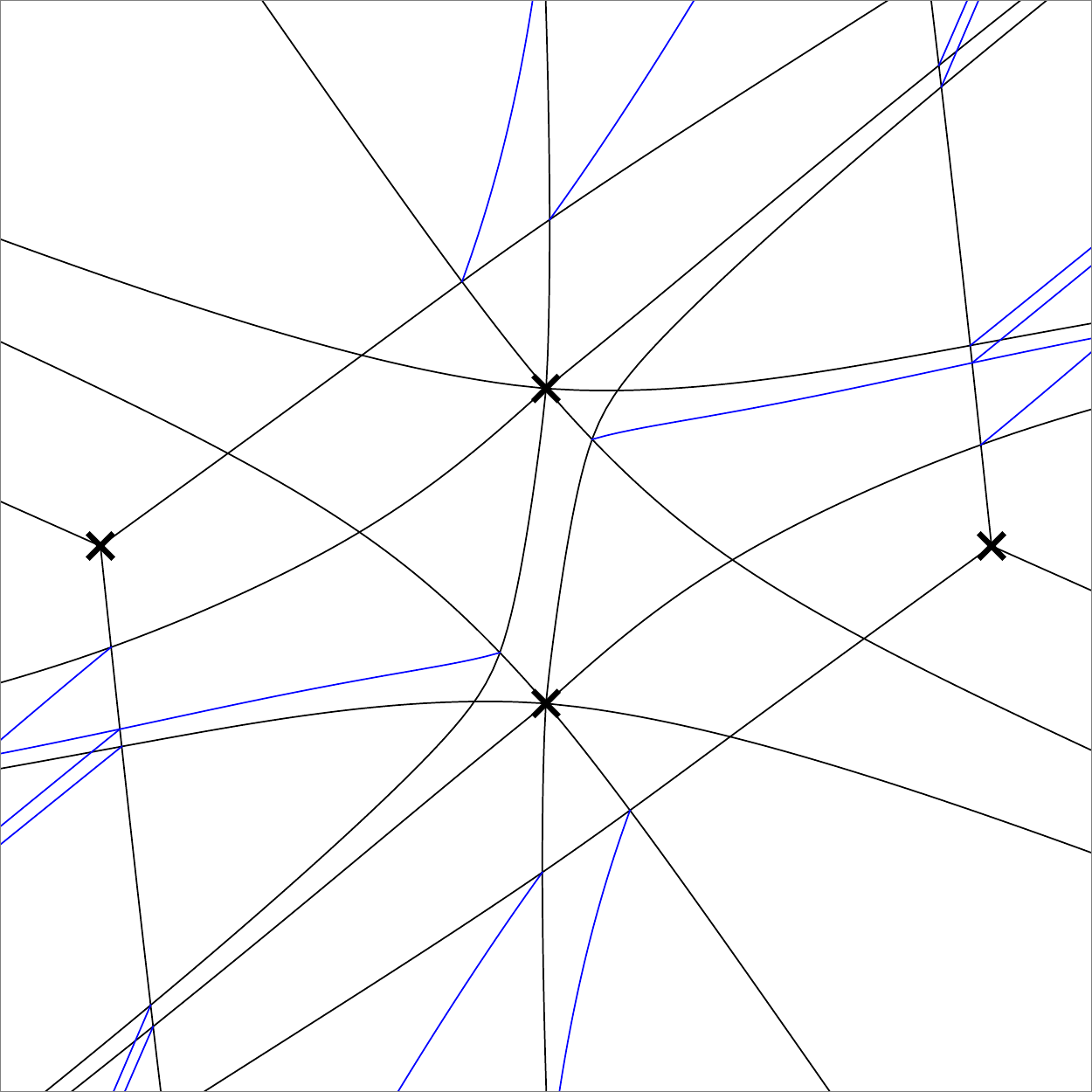}
		\caption{spectral network}
		\label{fig:a3SNFromA3_03_30_SN}
	\end{subfigure}
	\begin{subfigure}[b]{.3\textwidth}
		\centering	
		\includegraphics[width=\textwidth]{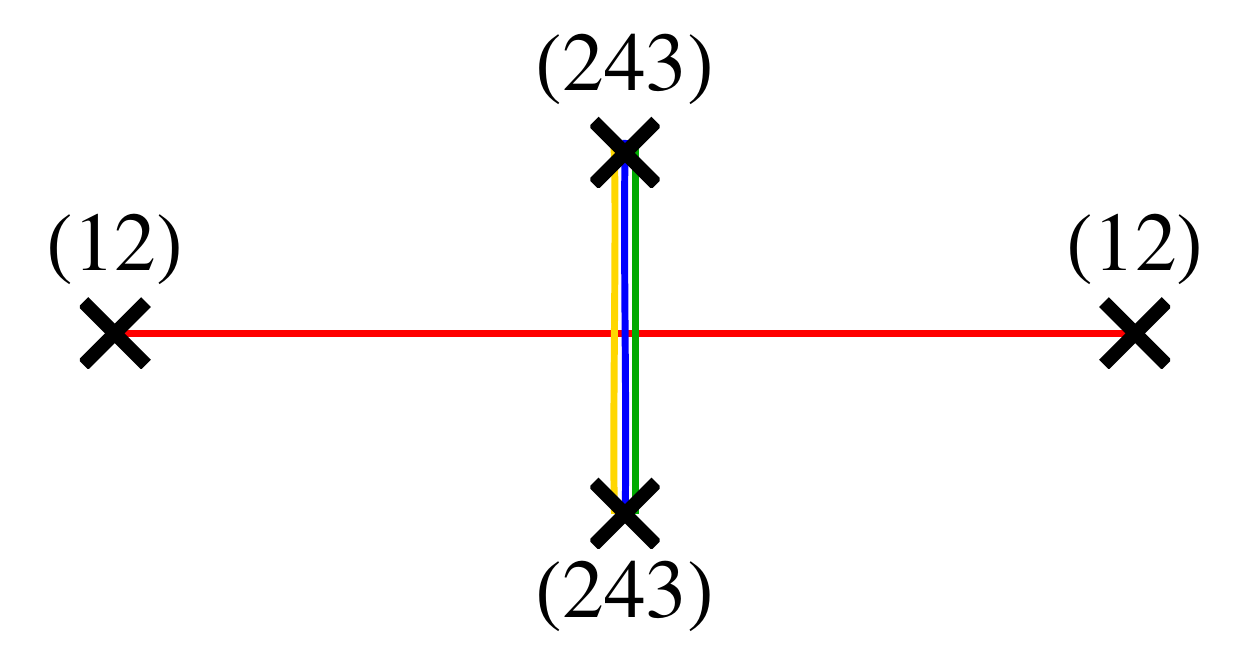}
		\vspace{1em}
		\caption{finite $\CS$-walls}
		\label{fig:a3SNFromA3_03_finite}
	\end{subfigure}
	\caption{Spectral network of $\CS[A_{3}; \CD_{{\rm I}}]$ after one wall-crossing from the minimal BPS spectrum.}
	\label{fig:a3SNFromA3_03_SN}
\end{figure}

We illustrate this procedure with the example of $\CS[A_{3}; \CD_{{\rm I}}]$. Starting from the configuration of Figure \ref{fig:a3SNFromA3_05_SN}, we change the parameters such that two pairs of branch points of index 2 collide with each other to form two branch points of index 3 as shown in Figure \ref{fig:a3SNFromA3_03_SN}. During the change we cross a BPS wall in the Coulomb branch moduli space, adding a BPS state in the spectrum.

\begin{figure}[h]
	\centering		
	\begin{subfigure}[b]{.4\textwidth}	
		\centering
		\includegraphics[width=\textwidth]{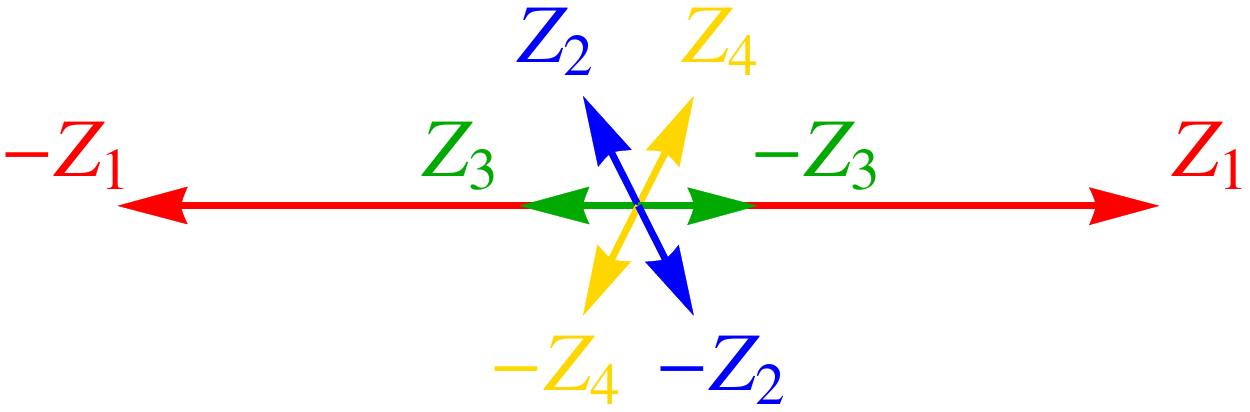}
		\caption{central charges}
		\label{fig:a3SNFromA3_03_Z}
	\end{subfigure}	
	\begin{subfigure}[b]{.4\textwidth}	
		\centering
		\begin{tabular}{c|c}
		   state  & $(e,m)$ \\ \hline
		  1 & $(1,0)$ \\ \hline
		  2 & $(0,1)$ \\ \hline
		  3 & $(1,0)$ \\ \hline
		  4 & $(1,1)$
		\end{tabular}
		\caption{IR charges}
		\label{tbl:a3SNFromA3_03_IR}
	\end{subfigure}
	\renewcommand{\figurename}{Figure \& Table}	
	\caption{Next-to-minimal BPS spectrum of $\CS[A_{3}; \CD_{{\rm I}}]$.}
	\label{figntbl:a3SNFromA3_03_BPS}
\end{figure}

When we compare Figure \ref{fig:a3SNFromA3_03_Z} with Figure \ref{fig:a3SNFromA3_05_Z}, we see that as we collide the two branch points, $Z_{3}$ move across $Z_{2}$ when the BPS spectrum gains a BPS state with charge $Z_{4}$. Analyzing the intersections of the cycles corresponding to the finite $\CS$-wals shown in Figure \ref{fig:a3SNFromA3_03_finite} gives the IR charges of the BPS states in Table \ref{tbl:a3SNFromA3_03_IR}. When we collide the two pairs of two branch points of different kinds, the BPS spectrum undergoes two additional wall-crossings, becoming a maximal symmetric BPS spectrum that we will see below. 

Note that if we keep $m = 0$ throughout the whole process, then $Z_1$ and $Z_3$ should move on the $Z$-plane together because they form an $\SU(2)$ doublet, and the wall-crossing from the minimal to the maximal BPS spectrum happens at once as the BPS spectrum of $\CS[A_1; \CD_{8}]$ with an $\SU(2)$ flavor symmetry does, see Figure \ref{fig:a3SN_SU_2_finite}.

\paragraph{Maximal, symmetric BPS spectrum}
When $v_{2} \neq 0$ and the other parameters are zero, the spectral network of $\CS[A_{N-1}; \CD_{{\rm I}}]$ has two branch points of index $N$. It results in the maximal symmetric BPS spectrum that contains $2 \times \binom{N}{2} = N(N-1)$ states, including anti-states having $\pi \leq \arg(Z) \leq 2\pi$. Thanks to the symmetric configuration, the central charges of the BPS states can be identified with the projections of root vectors connecting every pair of weights of the fundamental representation of $A_{N-1}$ in the weight space onto the $Z$-plane \cite{Hori:2013ewa}. The maximal symmetric BPS spectrum of $\CS[A_1; \CD_{N+4}]$ has the same structure: the corresponding spectral network comes from a symmetric configuration of its branch points that form vertices of an $N$-polygon, and there is a finite $\CS$-wall between every pair of the branch points. 

\begin{figure}[ht]
	\centering
	\begin{subfigure}[b]{.3\textwidth}	
		\includegraphics[width=\textwidth]{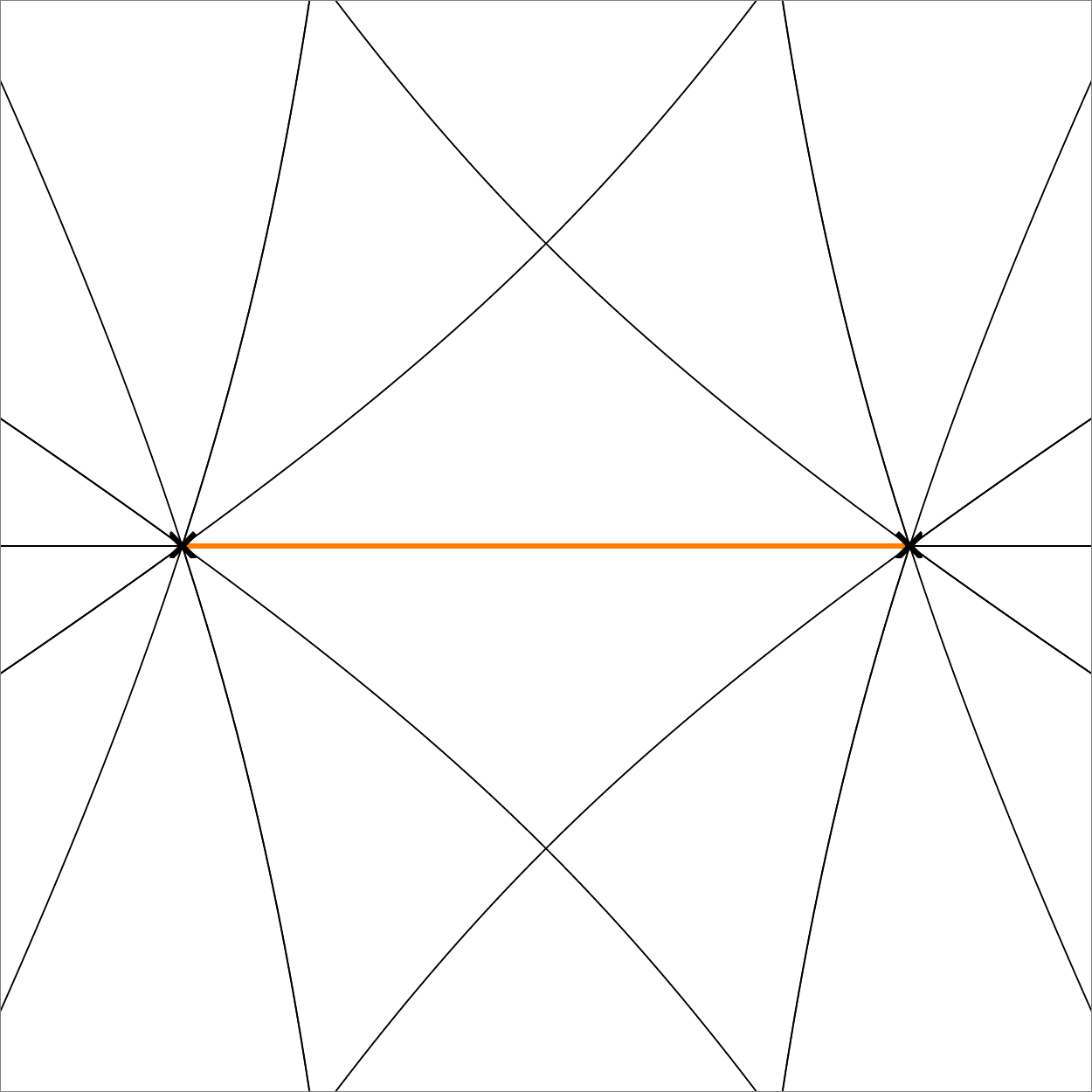}
		\caption{$\theta = \arg(Z_{1}^\mathbf{2})$}
		\label{fig:a3SNFromA3_01_25}
	\end{subfigure}
	\begin{subfigure}[b]{.3\textwidth}	
		\centering
		\includegraphics[width=\textwidth]{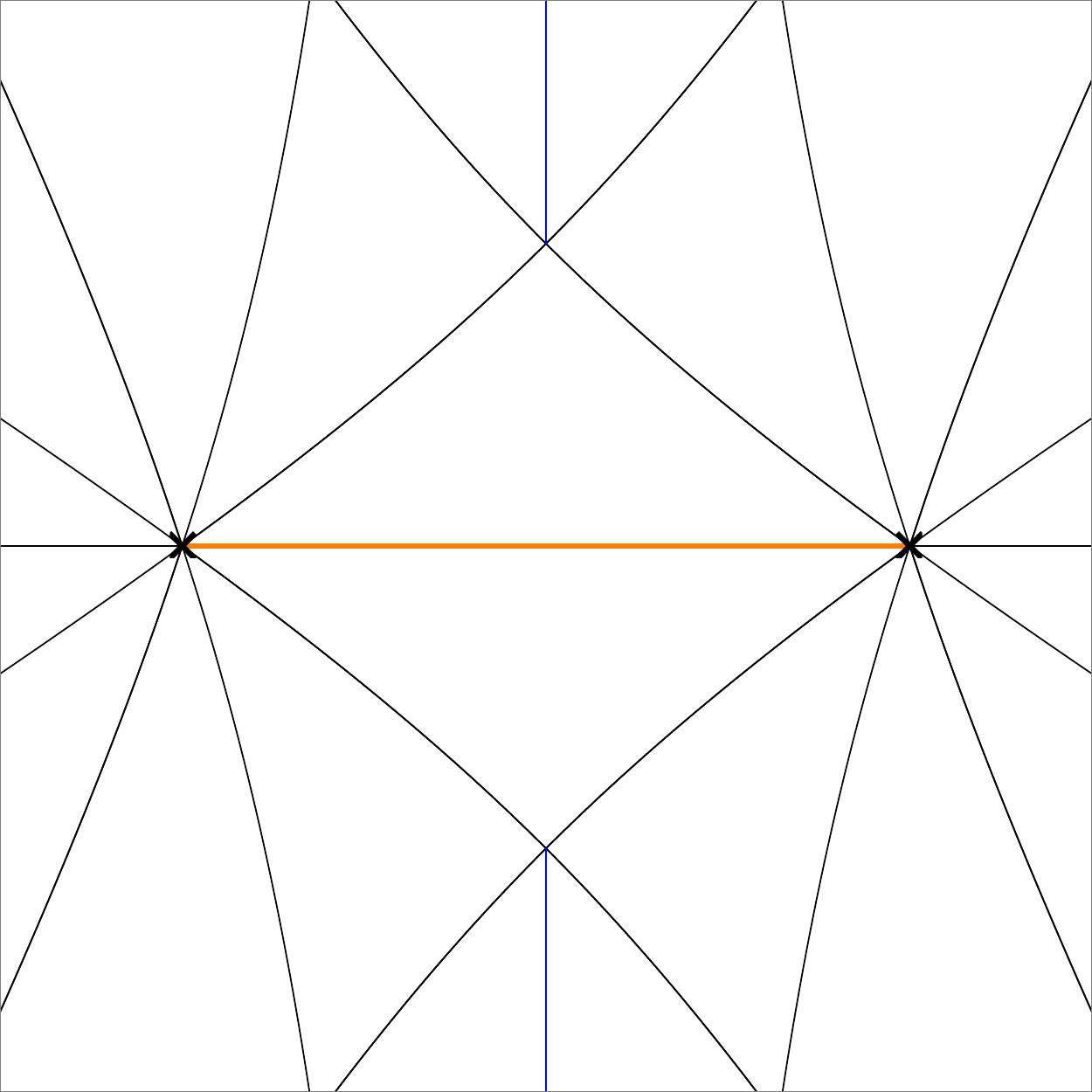}
		\caption{$\theta = \arg(Z_{2})$}
		\label{fig:a3SNFromA3_01_50}
	\end{subfigure}
	\caption{Spectral network of $\CS[A_3;\CD_{\rm I}]$ with maximal BPS spectrum.}
	\label{fig:a3SNFromA3_01}
\end{figure}

Figure \ref{fig:a3SNFromA3_01} shows the spectral network of $\CS[A_3,\CC;D_{\rm I}]$. Note that there is a value of $\theta$ that two $\CS$-walls appear at the same time, corresponding to the $\SU(2)$ doublet, 
as shown in Figure \ref{fig:a3SNFromA3_01_25}. Although the two $\CS$-walls are projected onto the same location on the $t$-plane, they are two distinct $\CS$-walls. There are two values of $\theta$ between $0$ and $\pi$ that a doublet appears, and at the other two values of $\theta$ only a single finite $\CS$-wall appears, thereby giving 6 BPS states and 6 anti-states.

\begin{figure}[ht]
	\centering	
	\begin{subfigure}[b]{.35\textwidth}	
		\centering
		\includegraphics[width=\textwidth]{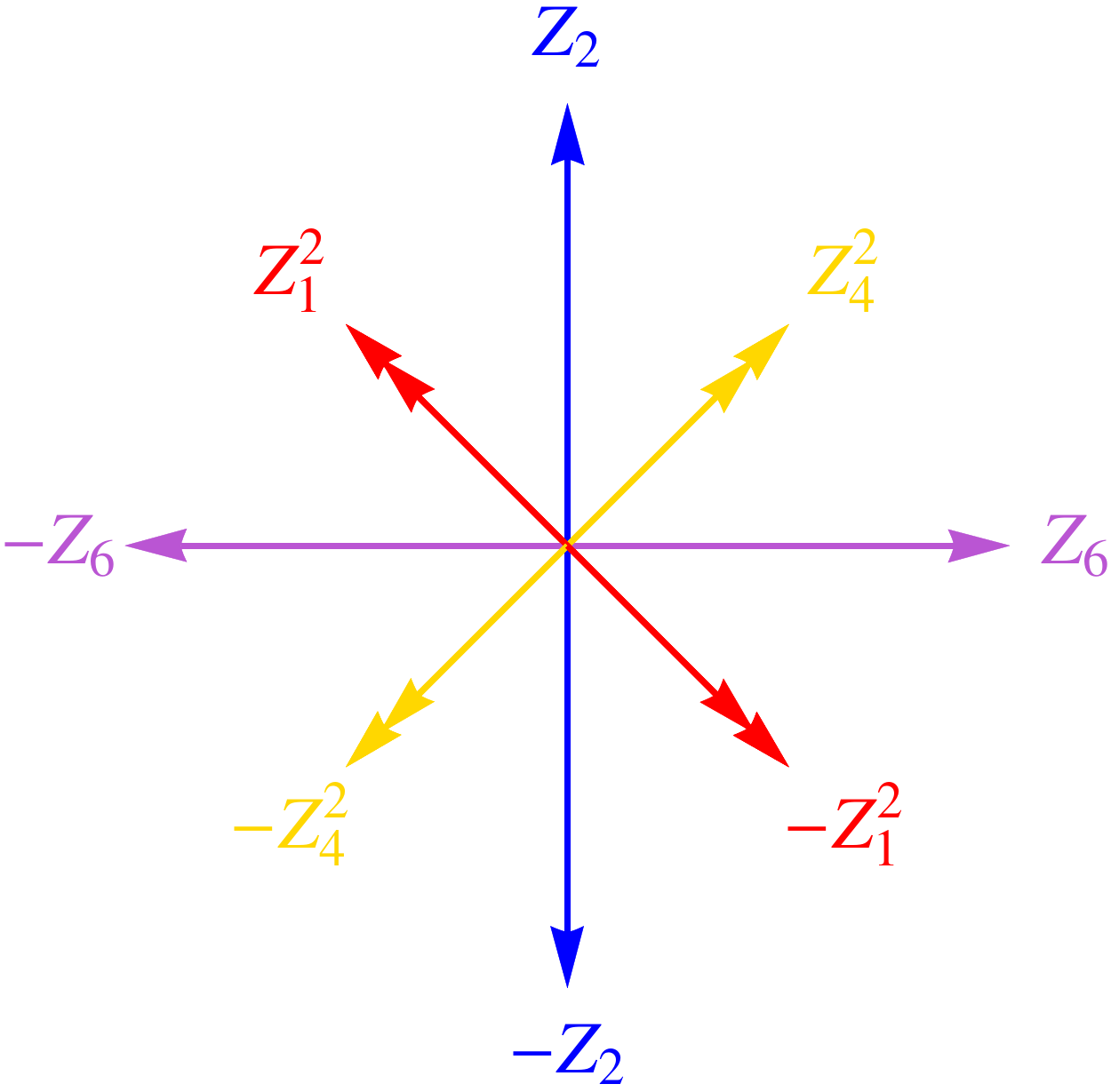}
		\caption{central charges}
		\label{fig:a3SNFromA3_01_Z}
	\end{subfigure}	
	\begin{subfigure}[b]{.4\textwidth}	
		\centering
		\begin{tabular}{c|c}
		   state  & $(e,m)$ \\ \hline
		  1 & $(1,0)$ \\ \hline
		  2 & $(0,1)$ \\ \hline
		  4 & $(1,1)$\\ \hline
		  6 & $(2,1)$
		\end{tabular}
		\vspace{3.5em}
		\caption{IR charges}
		\label{tbl:a3SNFromA3_03_IRs}
	\end{subfigure}	
	\renewcommand{\figurename}{Figure \& Table}
	\caption{Maximal, symmetric BPS spectrum of  $\CS[A_3,\CC;D_{\rm I}]$}
	\label{figntbl:a3SNFromA3_01_BPS}
\end{figure}

Figure \ref{fig:a3SNFromA3_01_Z} and Table \ref{tbl:a3SNFromA3_03_IRs} describe the maximal symmetric BPS spectrum of $\CS[A_3;\CD_{\rm I}]$, which can be identified with that of $\CS[A_{N-1}; \CD_{{8}}]$, see Figure \ref{fig:a3SN_SU_2_finite} and Table \ref{tbl:a3SN_IR}.

\section{4d SCFTs in $D_{n}$-class}
\label{sec:SCFTs_in_D_n_class}

\subsection{$\CS[A_{1}; \CD_\text{\rm reg}, \CD_{n+2}]$ Theories}
\label{subsec:A1dn}
Let us next consider the SCFTs obtained from the $A_{1}$ theory on a sphere with the irregular puncture $\CD_{n+2}$
and the regular puncture $\CD_{{\rm reg}}$.
This class of SCFTs $\CS[A_{1}; \CD_\text{\rm reg}, \CD_{n+2}]$ differs from $\CS[A_{1}; D_{n+5}]$ in that there is one regular puncture with a mass parameter. 
The Seiberg-Witten curve is given by
\begin{align}
	v^2 = t^{n} + \sum_{i=1}^{n-1} s_{2i} t^{n-i} + m^2,
\end{align}
and the Seiberg-Witten differential $\lambda = \frac{v}{t}\, dt$,
which has one regular puncture at $t=0$ and one irregular puncture at $t=\infty$.
The dimensions of the parameters are obtained as $\Delta(s_{2i})=\frac{2i}{n}$.
The parameter $m$ with dimension one is associated with the global $\SU(2)$ symmetry
which is a subgroup of the full flavor symmetry.

\subsubsection{$\CS[A_{1}; \CD_\text{\rm reg}, \CD_{5}]$ in $D_3$-class}
\label{sec:S[A_1,C;D_reg,D_5]}
Let us first study the simplest example $\CS[A_{1}; \CD_\text{\rm reg}, \CD_{5}]$, whose Seiberg-Witten curve is
\begin{align}
	v^2 = t^3 + c_1 t^2 + v_1 t + m^2,
\end{align}
where we denoted the parameters as $v_{1}$ and $c_{1}$ in order to emphasize 
that $v_{1}$ is the vev of the relevant operator and $c_{1}$ is its coupling. 

Here we describe how spectral networks can be used to show that $\CS[A_{1}; \CD_\text{\rm reg}, \CD_{5}]$ is in the same $D_3$-class ($=A_3$-class) as $\CS[A_{1}; \CD_{8}]$ and $\CS[A_3;\CD_{\rm I}]$ are. We will see here in particular that when the three theories have the $\SU(2)$ flavor symmetry they have the same BPS spectrum. However, the way that the $\SU(2)$ doublet of $\CS[A_{1}; \CD_\text{\rm reg}, \CD_{5}]$ appears is different from the other two theories due to the existence of a regular puncture.

\paragraph{Minimal BPS spectrum}

\begin{figure}[t]
	\centering
	\begin{subfigure}[b]{.27\textwidth}	
		\includegraphics[width=\textwidth]{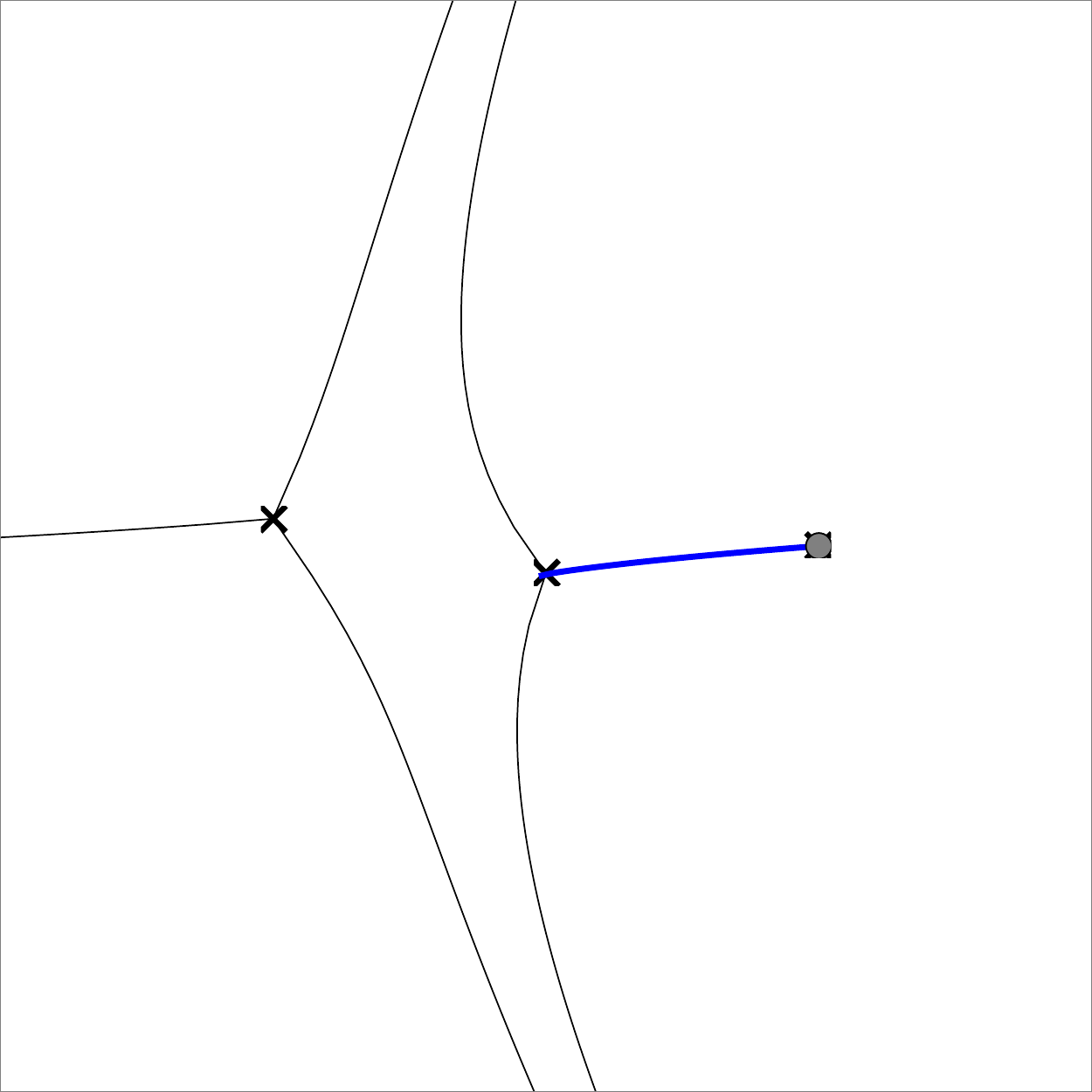}
		\caption{$\theta = \arg(Z_1^{\mathbf{2}})$}
		\label{fig:d3SN_m00_doublet}
	\end{subfigure}
	\begin{subfigure}[b]{.27\textwidth}	
		\includegraphics[width=\textwidth]{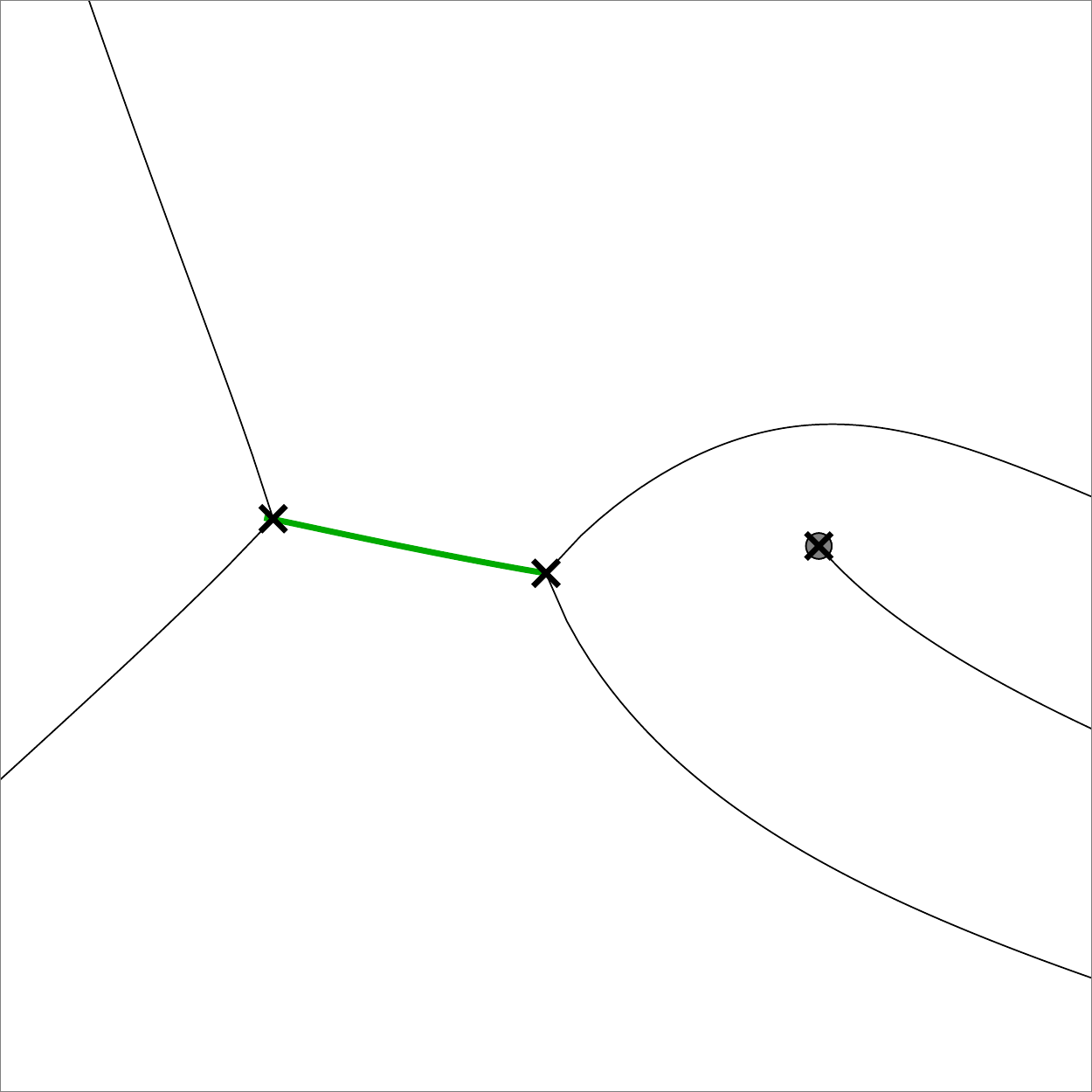}
		\caption{$\theta = \arg(Z_2)$}
		\label{fig:d3SN_m00_singlet}
	\end{subfigure}
	\begin{subfigure}[b]{.30\textwidth}	
		\includegraphics[width=\textwidth]{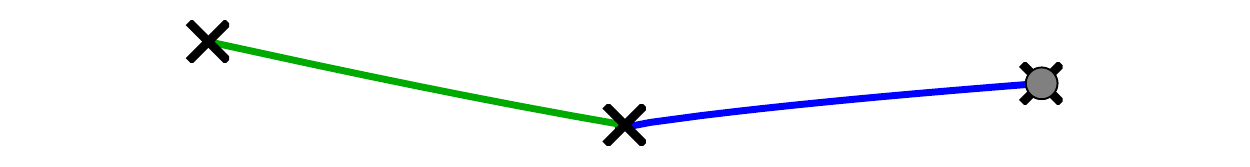}
		\vspace{3em}
		\caption{finite $\CS$-walls}
		\label{fig:d3SN_m00_finite}
	\end{subfigure}
	
	\caption{Spectral network of $\CS[A_{1}; \CD_\text{\rm reg}, \CD_{5}]$ with minimal BPS spectrum.}
	\label{fig:d3SN_doublet}
\end{figure}

When the parameters, $c_{1}$, $v_{1}$ and $m$, have general values there are three branch points on the $t$-plane. When we take $m \to 0$, one of the three branch points now collides with the regular puncture at $t=0$, as we discussed in Section \ref{sec:Around a puncture of ramification index N}, resulting in the spectral network shown in Figure \ref{fig:d3SN_doublet}, where we have a doublet of $\CS$-walls from the branch point on the puncture along the same direction reaching another branch point (Figure \ref{fig:d3SN_m00_doublet}). For a different value of $\theta$ there is another $\CS$-wall, now a singlet, connecting two branch points that are not on the puncture (Figure \ref{fig:d3SN_m00_singlet}). 

\begin{figure}[h]
	\centering
	\begin{subfigure}[b]{.3\textwidth}	
		\centering
		\includegraphics[width=\textwidth]{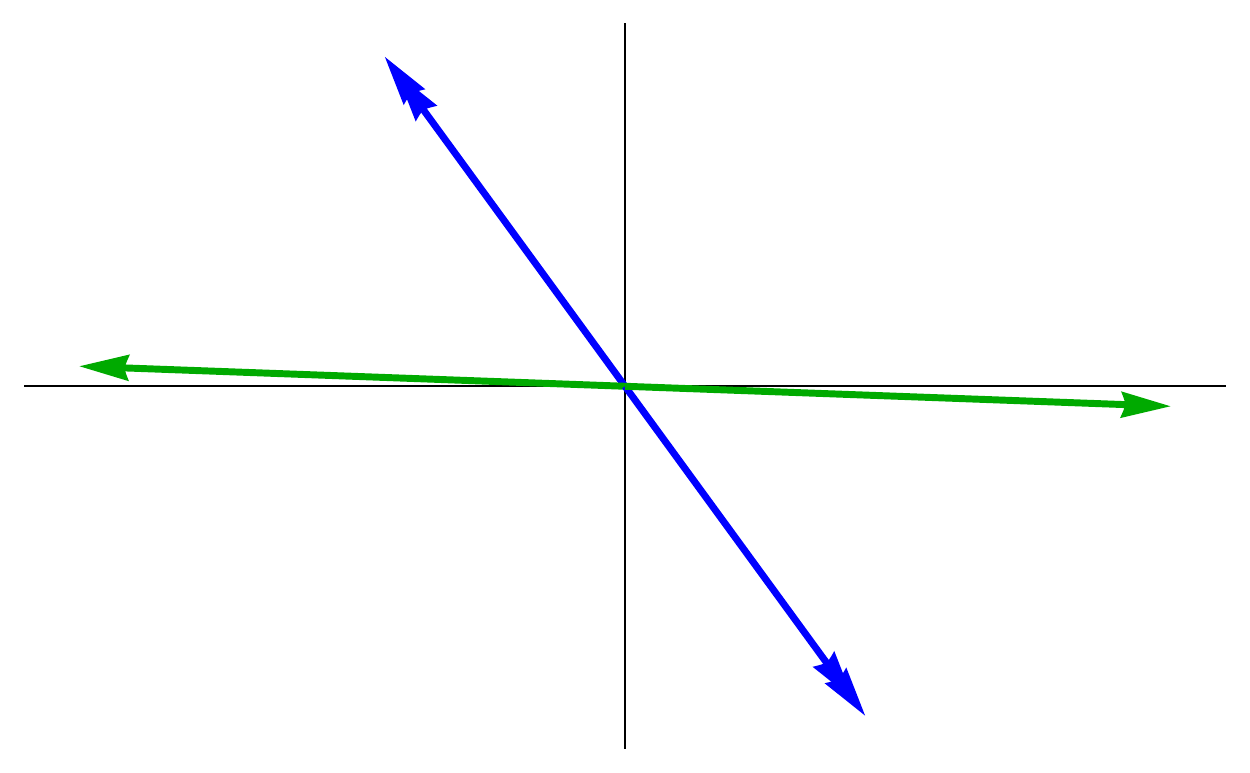}
		\caption{central charges}
		\label{fig:d3SN_minimal_Z}
	\end{subfigure}
	\begin{subfigure}[b]{.3\textwidth}	
		\centering
		\begin{tabular}{c|c}
		   state  & $(e,m)$ \\ \hline
		  1 & $(1,0)$ \\ \hline
		  2 & $(0,1)$
		\end{tabular}
		\vspace{2em}
		\caption{IR charges}
		\label{tbl:d3SN_minimal_IR}
	\end{subfigure}
	\begin{subfigure}[b]{.25\textwidth}	
		\centering
		\begin{tikzpicture}[scale=.75]
			\node[W,blue] (1) at (0,0) {1};
			\node[W,darkergreen] (2) at (2,0) {2};
			\node[W,blue] (3) at (4,0) {1};
			\path (1) edge[->] (2);
			\path (2) edge[<-] (3);
		\end{tikzpicture}
		\vspace{3em}
		\caption{BPS quiver}
		\label{fig:d3SN_minimal_quiver}
	\end{subfigure}
	\renewcommand{\figurename}{Figure \& Table}
	\caption{Minimal BPS spectrum of $\CS[A_{1}; \CD_\text{\rm reg}, \CD_{5}]$.}
	\label{fig:d3SN_minimal_BPS}
\end{figure}

From the finite $\CS$-walls we get the corresponding BPS states. Figure \ref{fig:d3SN_minimal_Z} describes their central charges, where we denote a doublet as a double-headed arrow. Table \ref{tbl:d3SN_minimal_IR} shows the IR $\UU(1)$-charges of the states. This is a minimal BPS spectrum of $\CS[A_{1}; \CD_\text{\rm reg}, \CD_{5}]$, which is represented by a $D_3 = A_3$ quiver shown in Figure \ref{fig:d3SN_minimal_quiver}. In fact it is the same as the minimal BPS spectrum of $A_3$-class theories with the $\SU(2)$ flavor symmetry. For example, see Figure \ref{fig:a3SN_SU_2_f_BPS_quiver} and the first row of Figure \ref{fig:a3SN_SU_2_finite} that illustrate the minimal BPS spectrum of $\CS[A_{1}; \CD_{8}]$.

\paragraph{Wall-crossing to the maximal BPS spectrum}

\begin{figure}[t]
	\centering
	
	\begin{subfigure}{.3\textwidth}	
		\includegraphics[width=\textwidth]{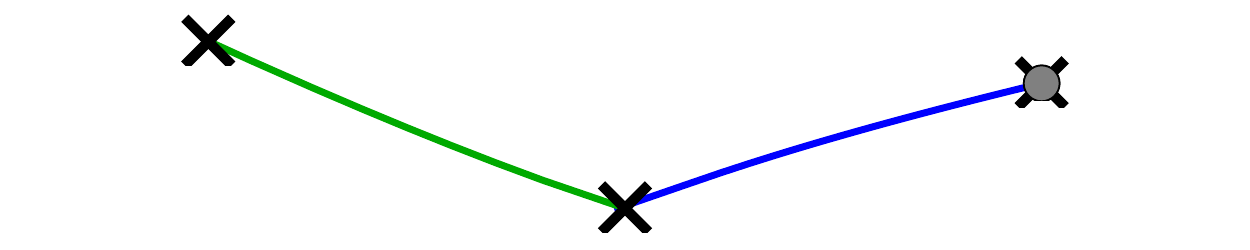}
	\end{subfigure}
	\begin{subfigure}{.22\textwidth}
		\centering
		\includegraphics[scale=.27]{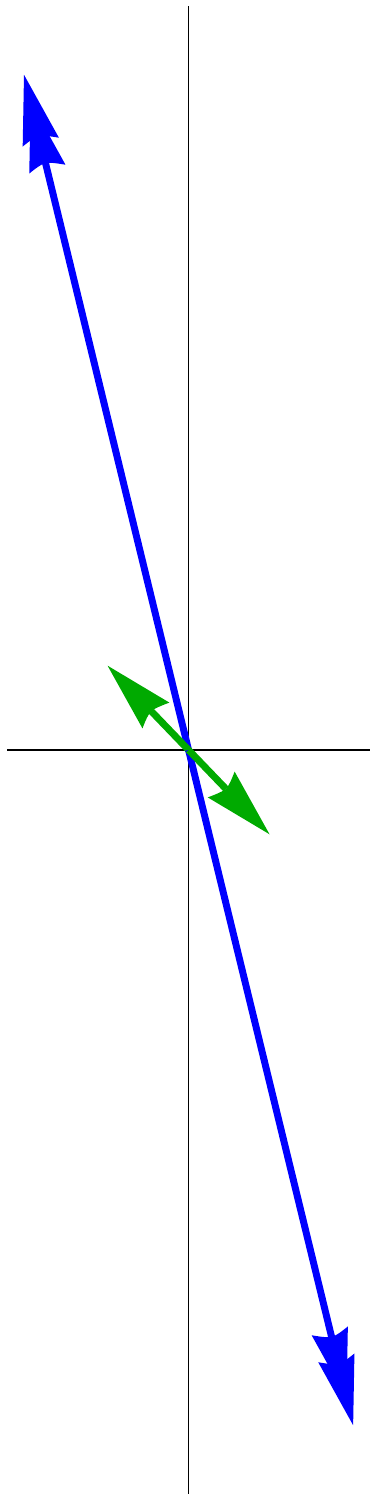}
	\end{subfigure}

	\begin{subfigure}{.3\textwidth}	
		\includegraphics[width=\textwidth]{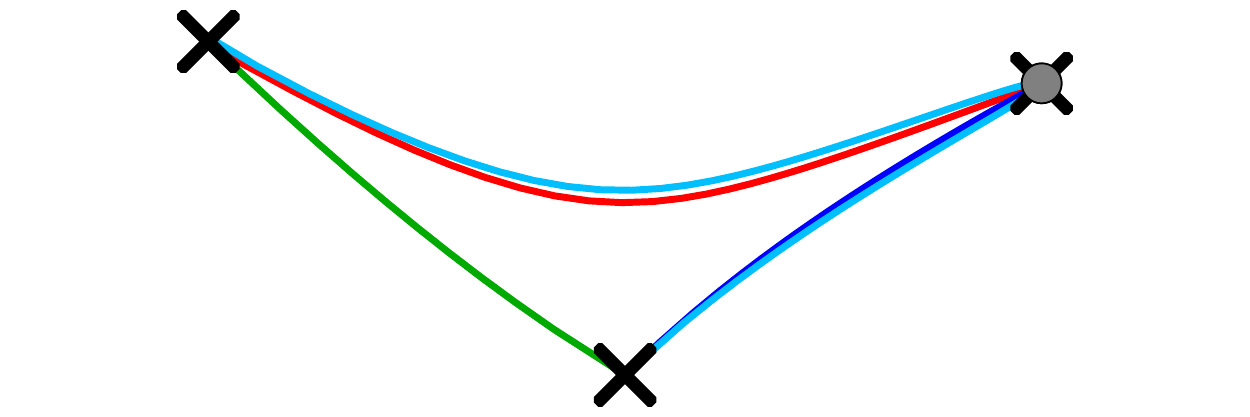}
	\end{subfigure}
	\begin{subfigure}{.22\textwidth}
		\centering
		\includegraphics[scale=.27]{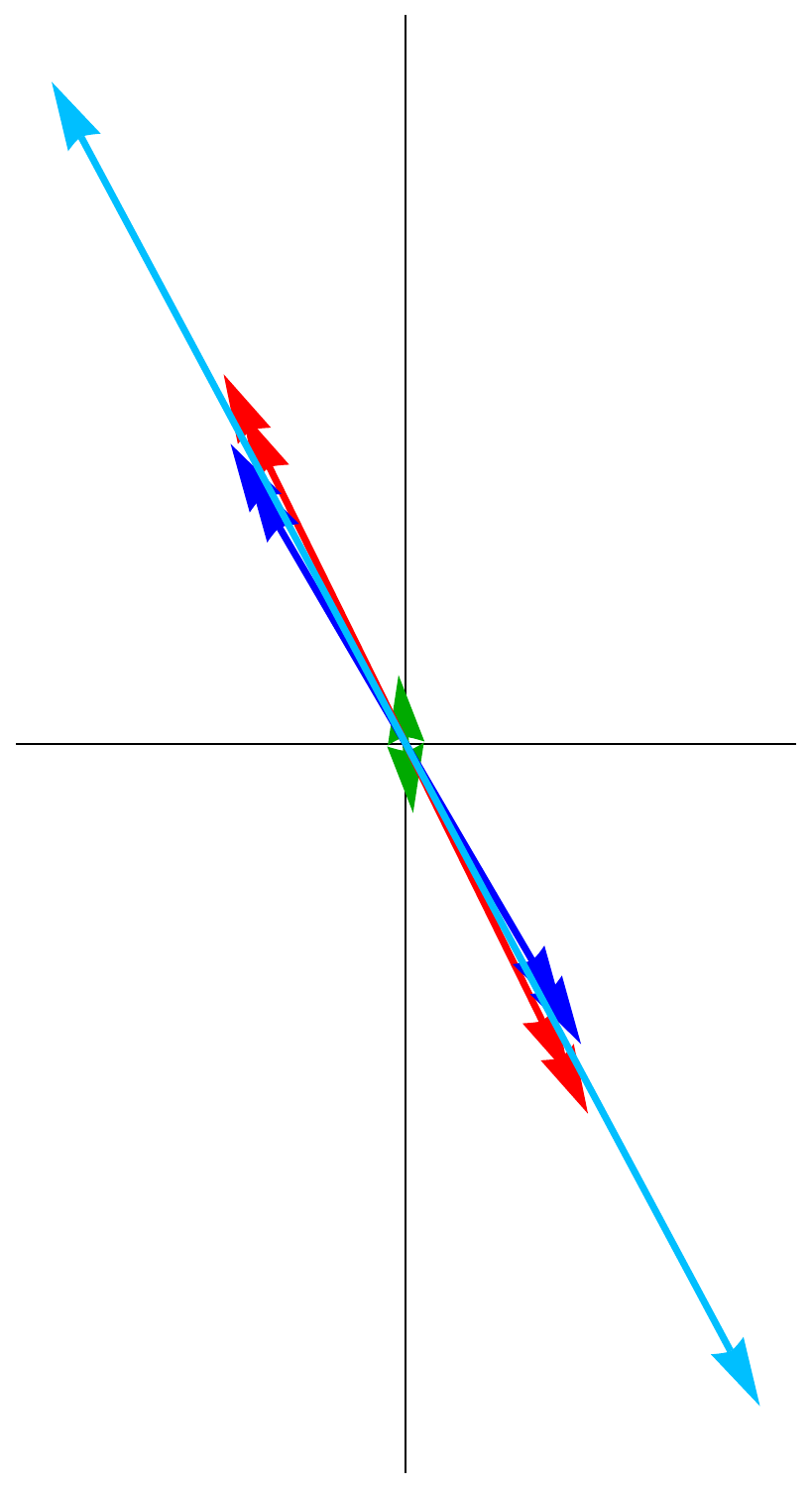}
	\end{subfigure}
		
	\begin{subfigure}{.3\textwidth}	
		\includegraphics[width=\textwidth]{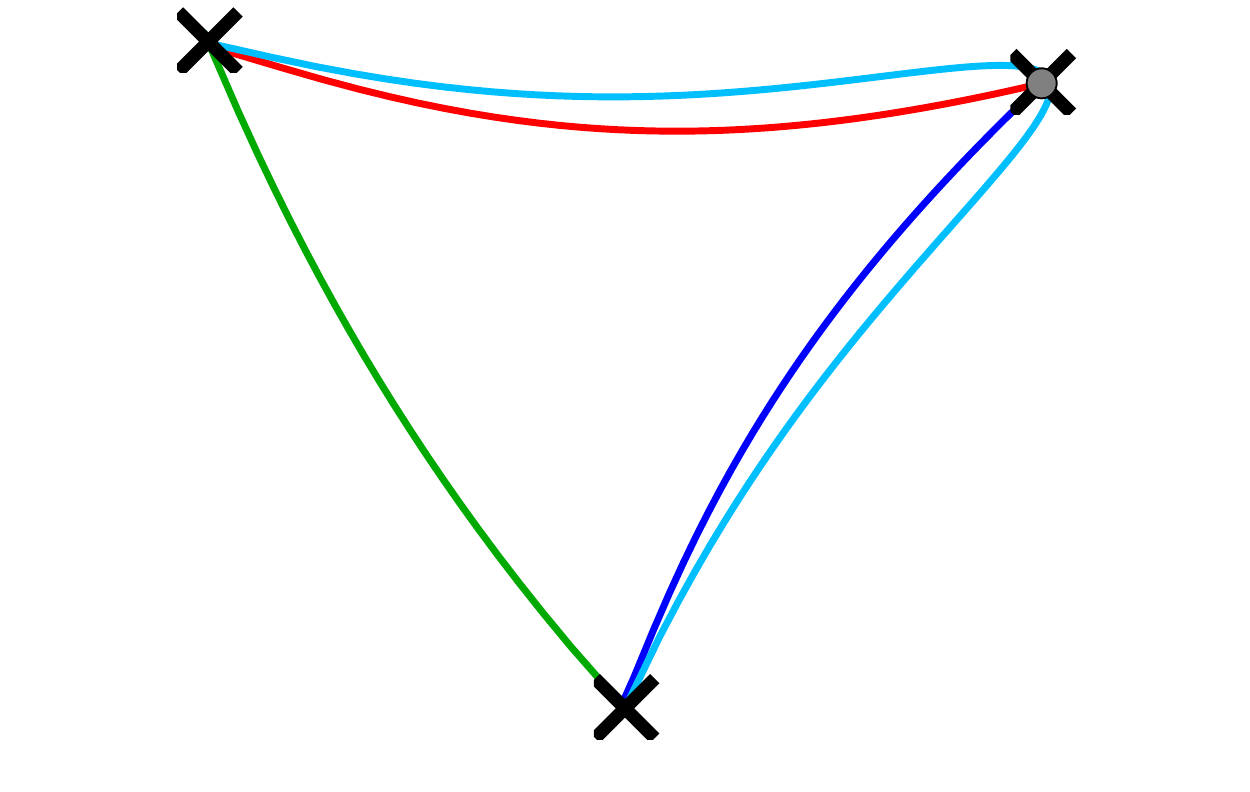}
	\end{subfigure}
	\begin{subfigure}{.22\textwidth}
		\centering
		\includegraphics[scale=.27]{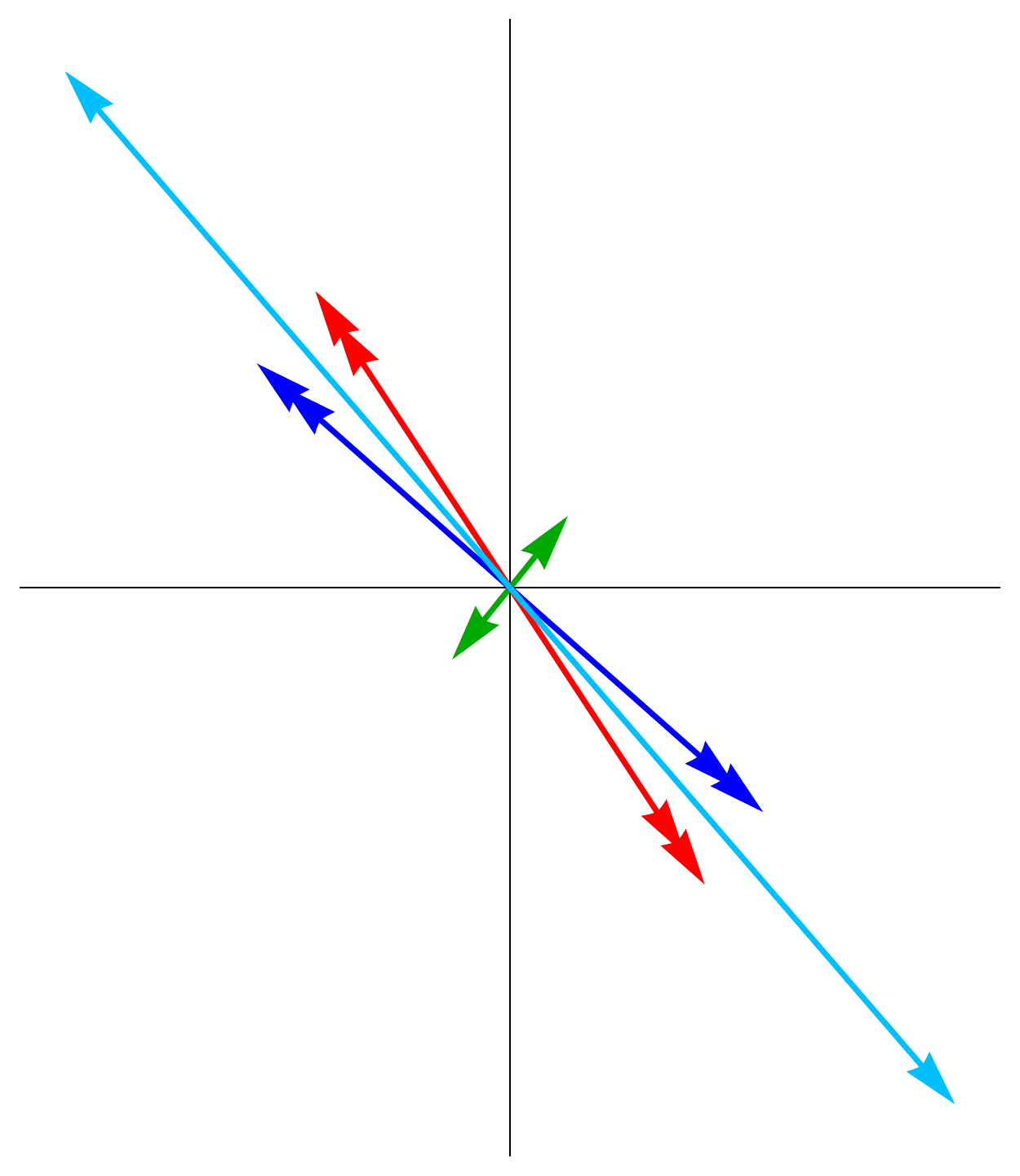}
	\end{subfigure}
		
	\caption{Wall-crossing of $\CS[A_{1}; \CD_\text{\rm reg}, \CD_{5}]$.}
	\label{fig:d3SN_wall_crossing}
\end{figure}

Next we consider the wall-crossing of the BPS spectra of $\CS[A_{1}; \CD_\text{\rm reg}, \CD_{5}]$. This wall-crossing mechanism, in combination with that of $\CS[A_{1}; \CD_{7}]$, will form building blocks for the wall-crossings of $\CS[A_{1}; \CD_\text{\rm reg}, \CD_{n+2}]$. Figure \ref{fig:d3SN_wall_crossing} shows how a wall-crossing happens in $\CS[A_{1}; \CD_\text{\rm reg}, \CD_{5}]$ with the $\SU(2)$ flavor symmetry as we move one of the branch points that is not on the puncture.

The second row of Figure \ref{fig:d3SN_wall_crossing} shows the finite $\CS$-walls and the central charges of the corresponding BPS states right after a wall-crossing happens. Again it happens when the central charges of two BPS states move across each other on the $Z$-plane, see the first two row of Figure \ref{fig:d3SN_wall_crossing}, which shows the central charge of a singlet BPS state going over that of the doublet. And similarly to the wall-crossing of $\CS[A_{1}; \CD_{n+5}]$ with the $\SU(2)$ flavor symmetry, which is shown in Figure \ref{fig:a3SN_SU_2_finite}, when the wall-crossing happens we have another doublet and an additional BPS state whose central charge is the sum of the central charges of two states from each doublet.

\begin{table}[h]
	\centering
	\begin{tabular}{c|c}
		state & $(e,m)$ \\ \hline
		$1$ & $(1,0)$ \\ \hline
		$2$ & $(0,1)$ \\ \hline
		$3$ & $(1,1)$ \\ \hline
		$4$ & $(2,1)$  
	\end{tabular}
	\caption{IR charges of the states in the maximal BPS spectrum of $\CS[A_{1}; \CD_\text{\rm reg}, \CD_{5}]$.}
	\label{tbl:d3SN_maximal_IR}
\end{table}

\begin{figure}[t]
	\centering
	
	\begin{subfigure}{.22\textwidth}
		\centering
		\includegraphics[scale=.3]{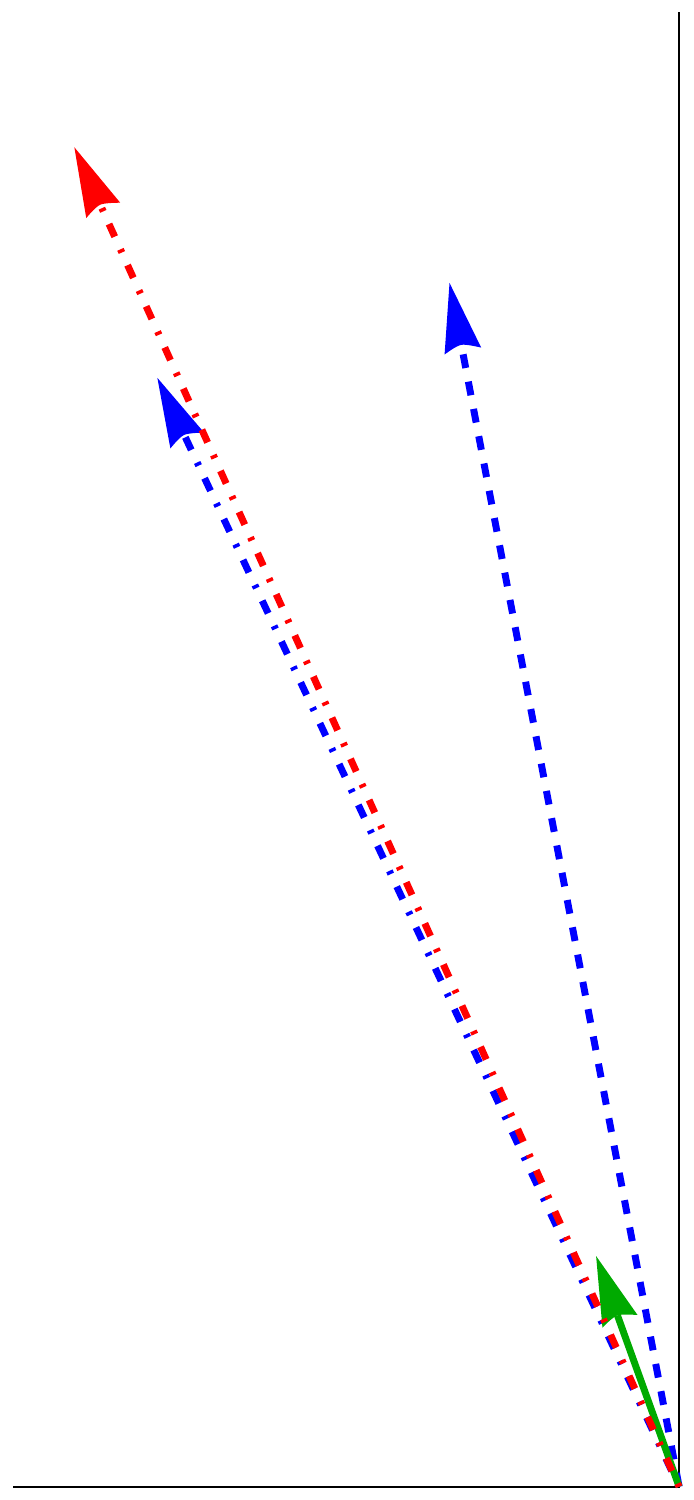}
	\end{subfigure}
	\begin{subfigure}{.22\textwidth}
		\centering
		\includegraphics[scale=.3]{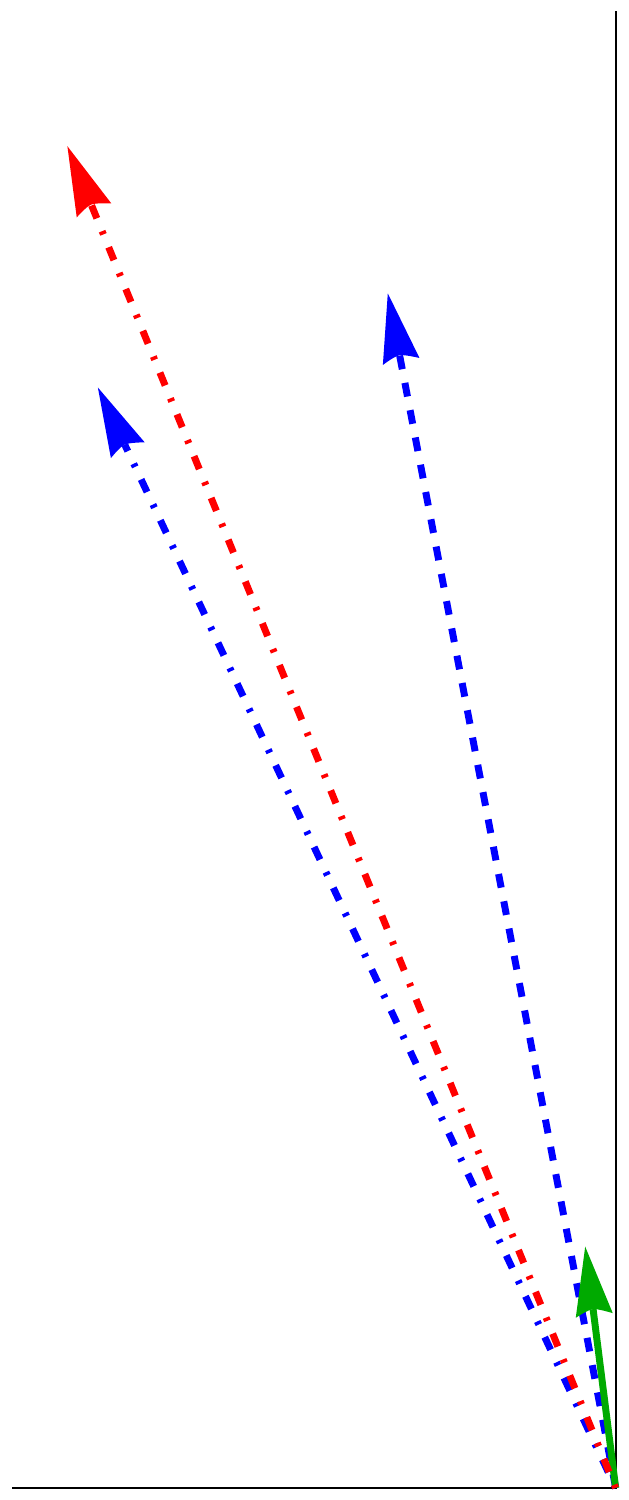}
	\end{subfigure}	
	\begin{subfigure}{.22\textwidth}
		\centering
		\includegraphics[scale=.3]{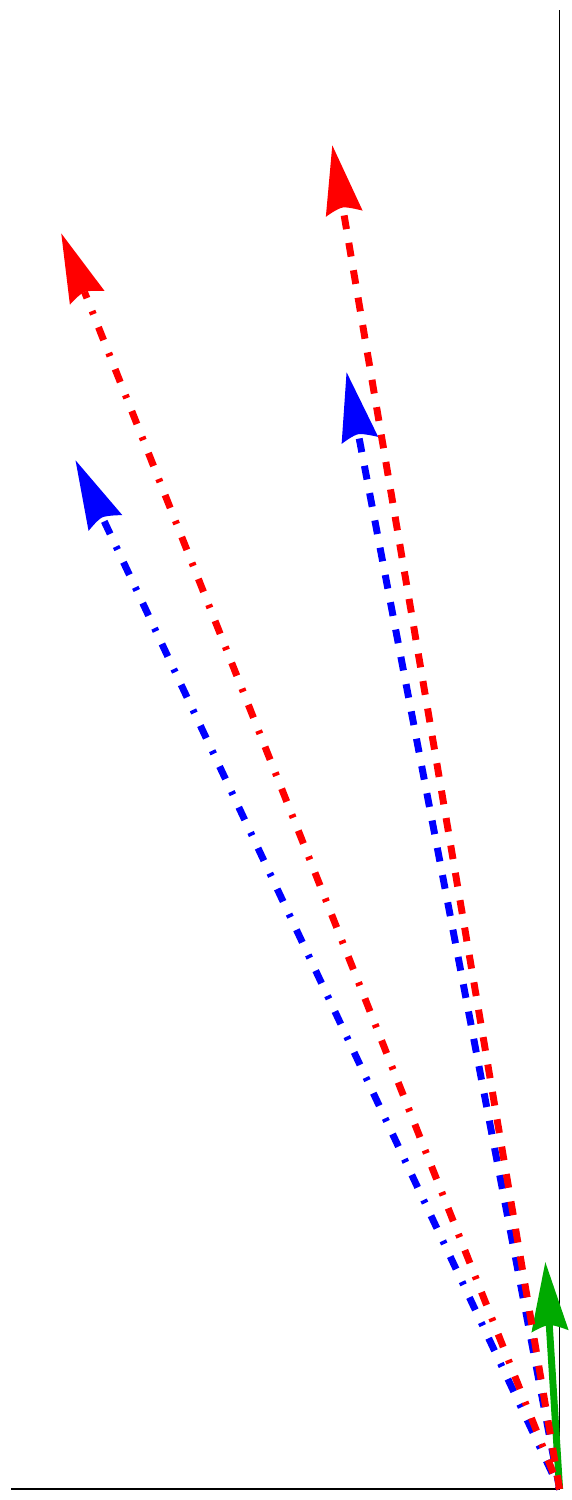}
	\end{subfigure}
	\begin{subfigure}{.22\textwidth}
		\centering
		\includegraphics[scale=.3]{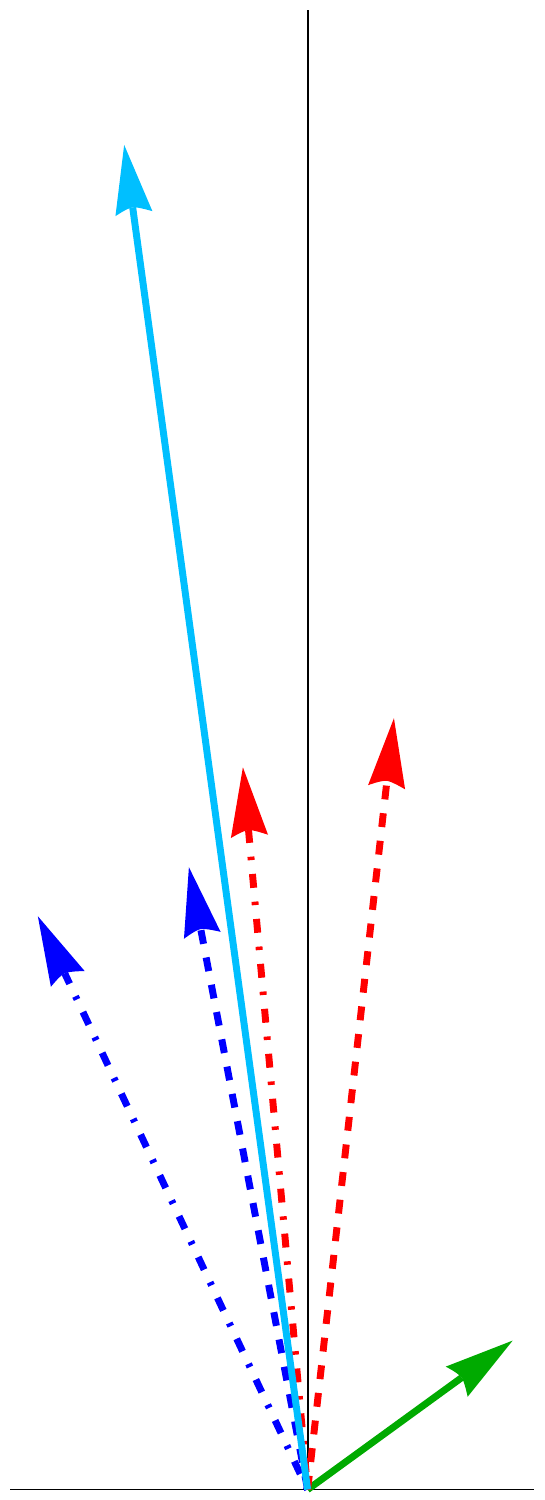}
	\end{subfigure}
				
	\caption{Wall-crossing of a doublet and a singlet.}
	\label{fig:d3SN_wall_crossing_resolved}
\end{figure}

To understand the wall-crossing involving a doublet, it is helpful to introduce an infinitesimal value of the mass parameter $m$ to resolve each doublet into two states with infinitesimal difference in their central charges, as shown in Figure \ref{fig:d3SN_wall_crossing_resolved}, where a series of three usual wall-crossings are illustrated. When $m \to 0$, the three wall-crossings happen at the same time, leading to the new wall-crossing phenomena shown in Figure \ref{fig:d3SN_wall_crossing}.

The last row of Figure \ref{fig:d3SN_wall_crossing} shows the maximal BPS spectrum of $\CS[A_{1}; \CD_\text{\rm reg}, \CD_{5}]$ with the $\SU(2)$ flavor symmetry, which consists of six BPS states with two doublets. This the same as that of $\CS[A_{1}; \CD_{8}]$ with vanishing residue at infinity, see the last row of Figure \ref{fig:a3SN_SU_2_finite}.

\paragraph{Equivalence of $\CS[A_{1}; \CD_\text{\rm reg}, \CD_{5}]$ and $\CS[A_{1}; \CD_{8}]$}
We have seen that the analysis of BPS spectra of $\CS[A_{1}; \CD_\text{\rm reg}, \CD_{5}]$ and $\CS[A_{1}; \CD_{8}]$ via spectral network provides good evidence for the equivalence of the two SCFTs. Another piece of evidence comes from comparing the central charges of the SCFTs, which are $a = \frac{11}{24}$ and $c = \frac{1}{2}$.  The central charge of $\CS[A_{1}; \CD_{8}]$ was computed in \cite{Shapere:1999xr}, see eq. (4.31) with $r=1$. The central charge of the $\CS[A_{1}; \CD_\text{\rm reg}, \CD_{5}]$ theory was computed in some papers, e.g. \cite{Xie:2013jc}, see $I_{2,1, F}$ in Table 3. 

Actually, we can show the SCFTs have the same Seiberg-Witten curves when both of them have the $\SU(2)$ flavor symmetry. Let us start with the curve of $\CS[A_{1}; \CD_{8}]$ with $c_3 = 0$,
\begin{align}
	x^2 = t^4 + c_2 t^2 + v_2,
\end{align}
where $\lambda  = x\, dt$. Now we change variables: first we take $t \to \sqrt{\tilde{t}}$,
\begin{align}
	\lambda = x\, dt = \frac{1}{2}\sqrt{\tilde{t} + c_2 + \frac{v_2}{\tilde{t}}} d\tilde{t} = \tilde{x}\,dt,
\end{align}
and then define $v = \tilde{t} \tilde{x}$, after which we have $\lambda = \left(v/\tilde{t}\right) d\tilde{t}$ and
\begin{align}
	v^2 = {\tilde{t}}^3 + c_2 {\tilde{t}}^2 + v_2 \tilde{t}.
\end{align}
These are equivalent to the Seiberg-Witten differential and curve of $\CS[A_{1}; \CD_\text{\rm reg}, \CD_{5}]$. 
Therefore we expect the two theories to be fully equivalent.

\subsubsection{$\CS[A_{1}; \CD_\text{\rm reg}, \CD_{6}]$ in $D_{4}$-class}
\label{sec:S[A_1,C;D_reg,D_6]}
The Seiberg-Witten curve of $\CS[A_{1}; \CD_\text{\rm reg}, \CD_{6}]$ is
\begin{align}
	v^2 = t^{4} + c_{1} t^{3} + c_2 t^2 + v_1 t + m^2.
\end{align}
The residue of the Seiberg-Witten differental $\lambda = \frac{v}{t} dt$ at $t = \infty$ is $\frac{1}{8} \left(c_1^2-4 c_2\right)$, which is associated with a $\UU(1)$ flavor symmetry. As in the previous case, the parameter $m$, which is the residue of the regular puncture at $x=0$, is associated with an $\SU(2)$ flavor symmetry. When both mass parameters vanish, we expect to have an enhanced $\SU(3)$ flavor symmetry \cite{Argyres:2012fu}, which we will confirm here from the analysis of spectral networks. When only $m=0$ but the residue of $\lambda$ at $t = \infty$ is nonzero, it serves as a good stepping stone to understand the cases of general $n$, as we will see later.

\begin{figure}[t]
	\centering
	\begin{subfigure}[b]{.27\textwidth}	
		\centering
		\includegraphics[width=\textwidth]{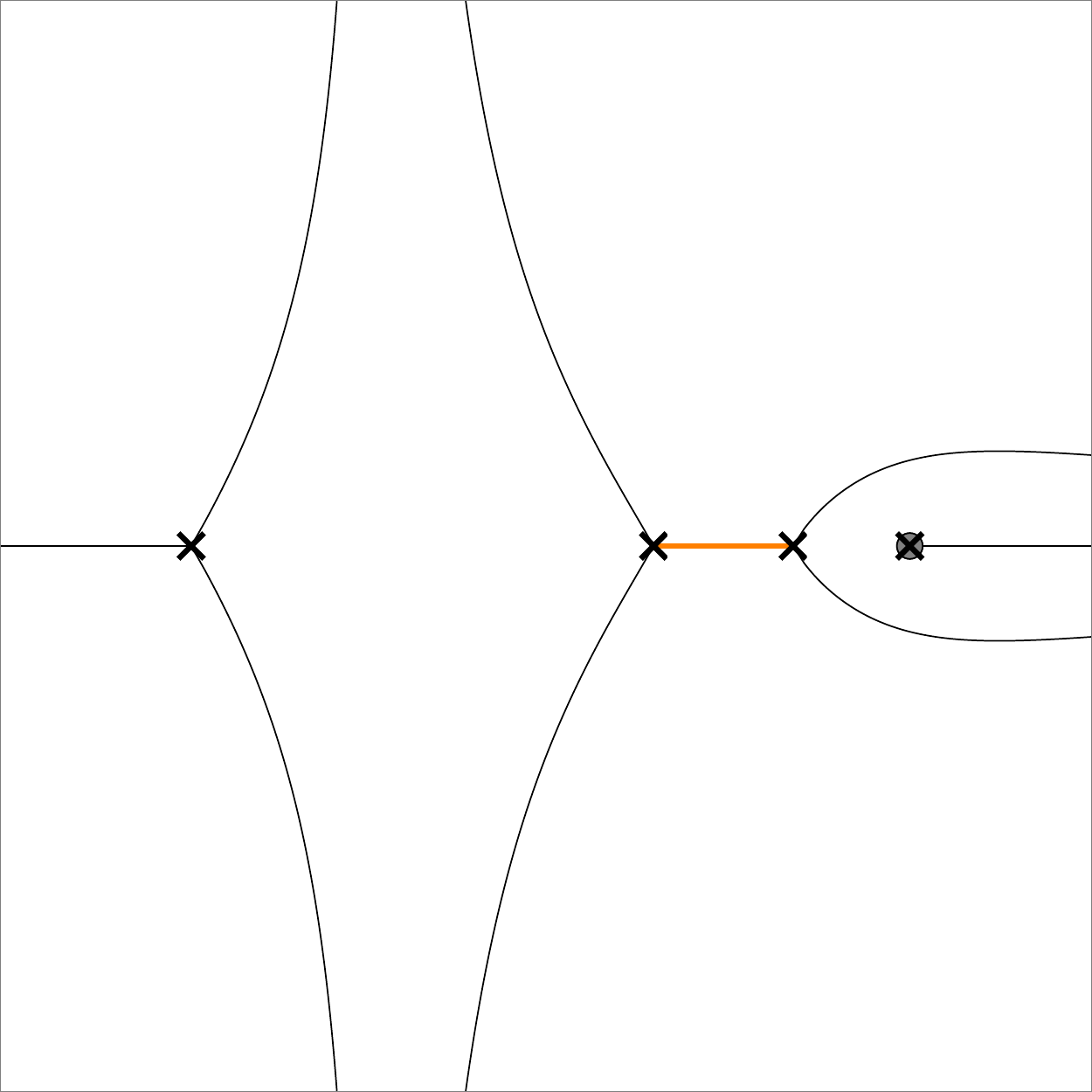}
		\caption{$\theta = \arg(Z_1)$}
		\label{fig:d4SN_SU_3_f_minimal_singlet}
	\end{subfigure}	
	\begin{subfigure}[b]{.27\textwidth}	
		\centering
		\includegraphics[width=\textwidth]{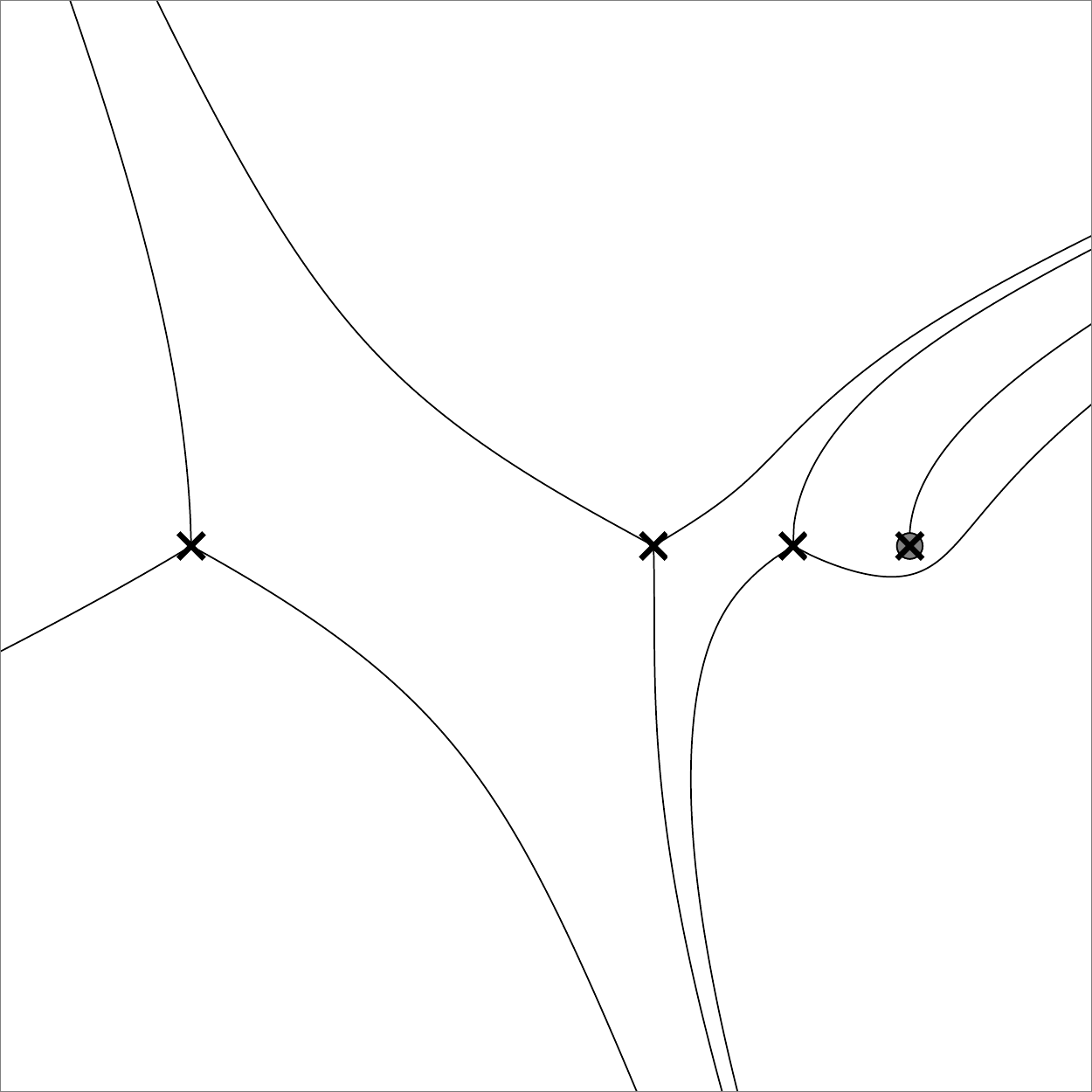}
		\caption{$\arg(Z_1) < \theta < \arg(Z_2^\mathbf{3})$}
		\label{fig:d4SN_SU_3_f_minimal_26}
	\end{subfigure}	
	\begin{subfigure}[b]{.27\textwidth}	
		\includegraphics[width=\textwidth]{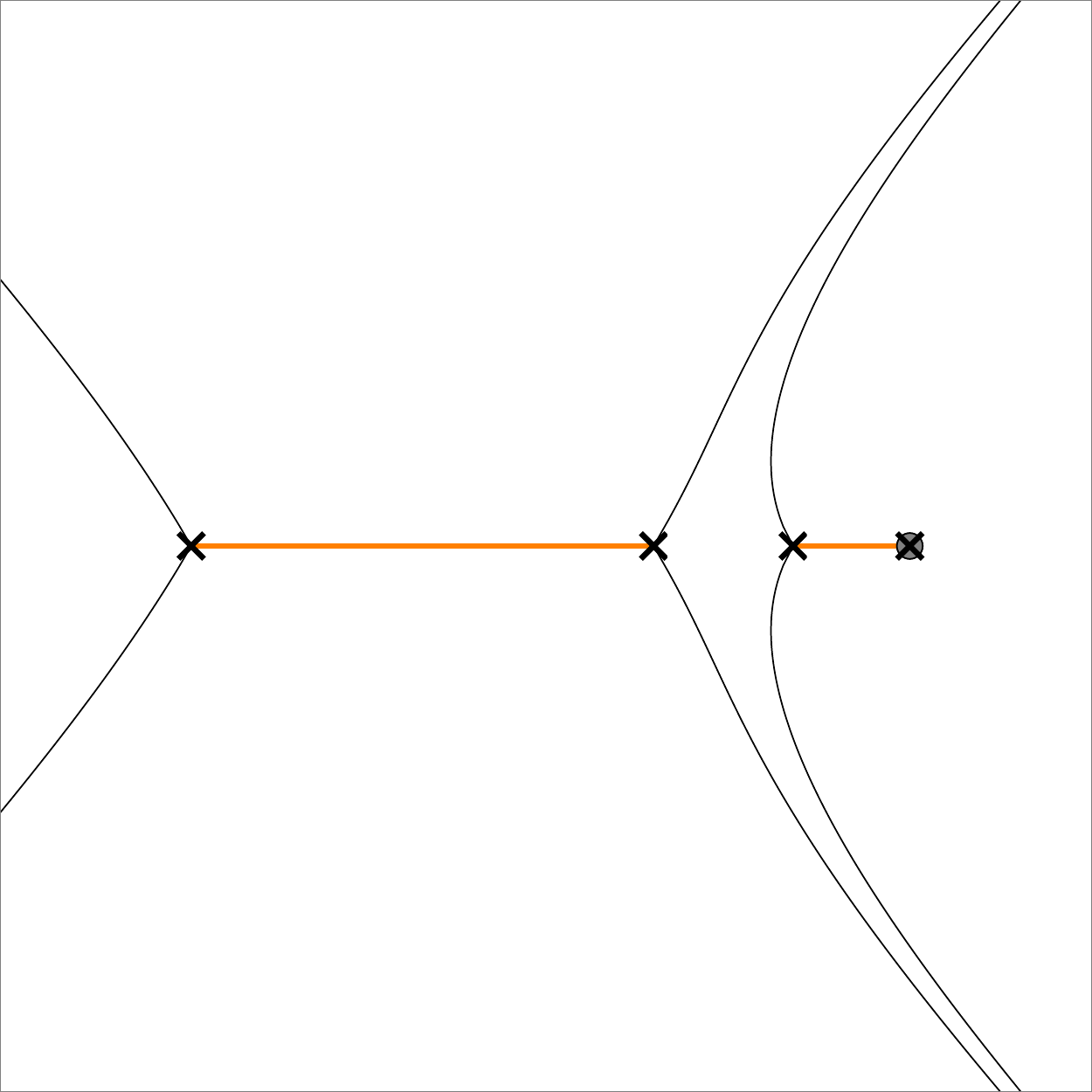}
		\caption{$\theta = \arg(Z_2^\mathbf{3})$}
		\label{fig:d4SN_SU_3_f_minimal_triplet}
	\end{subfigure}	

	\caption{Spectral networks of $\CS[A_{1}; \CD_\text{\rm reg}, \CD_{6}]$ with an $\SU(3)$ flavor symmetry and minimal BPS spectrum.}
	\label{fig:d4SN_SU_3_f_minimal}
\end{figure}

\begin{figure}[t]
	\centering
	\begin{subfigure}{.33\textwidth}	
		\includegraphics[width=\textwidth]{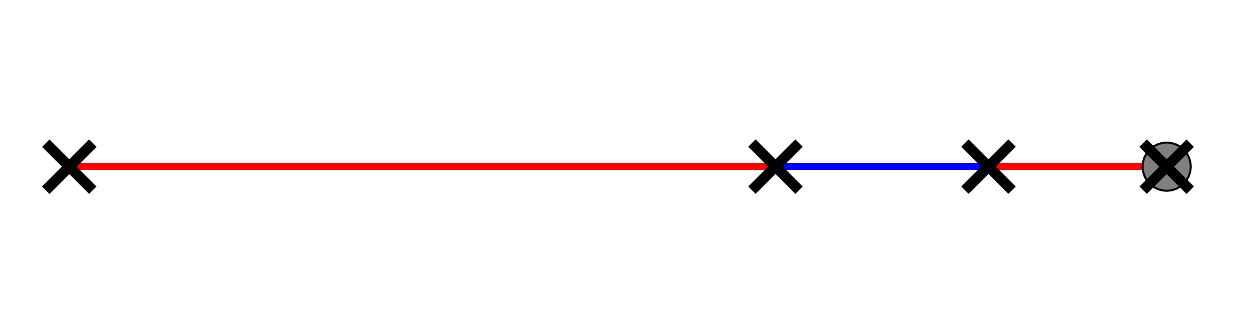}
	\end{subfigure}
	\begin{subfigure}{.25\textwidth}	
		\centering
		\includegraphics[scale=.3]{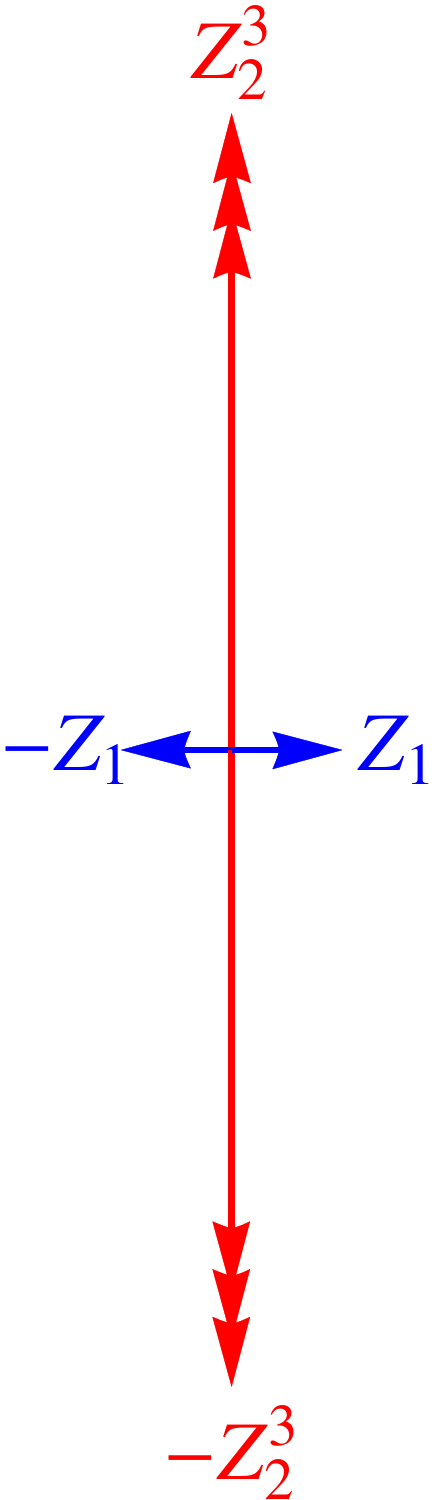}
	\end{subfigure}
	\begin{subfigure}{.33\textwidth}	
	\end{subfigure}

	\begin{subfigure}{.35\textwidth}	
		\includegraphics[width=\textwidth]{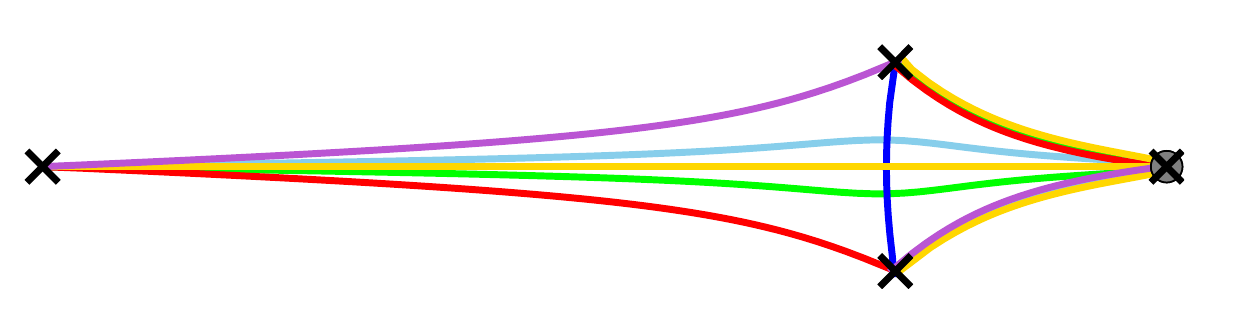}
	\end{subfigure}
	\begin{subfigure}{.23\textwidth}	
		\centering
		\includegraphics[scale=.3]{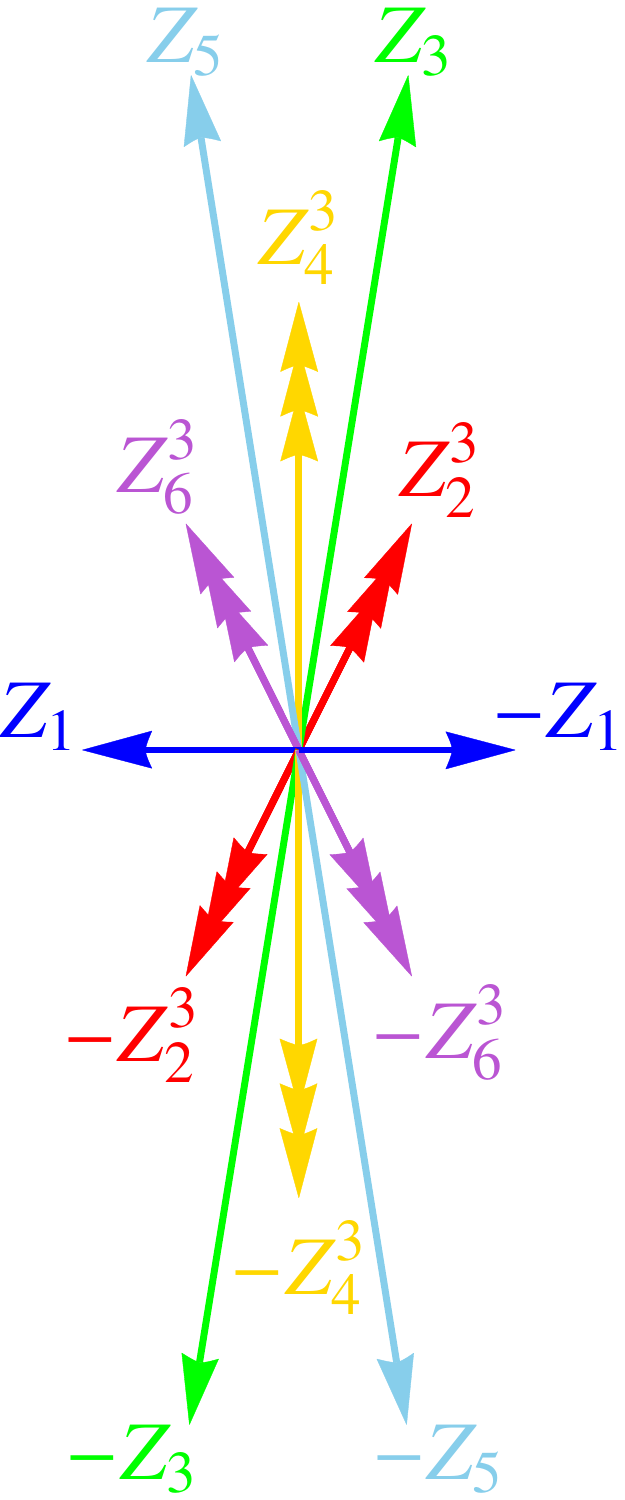}
	\end{subfigure}
		
	\begin{subfigure}{.33\textwidth}	
		\centering
		\includegraphics[width=\textwidth]{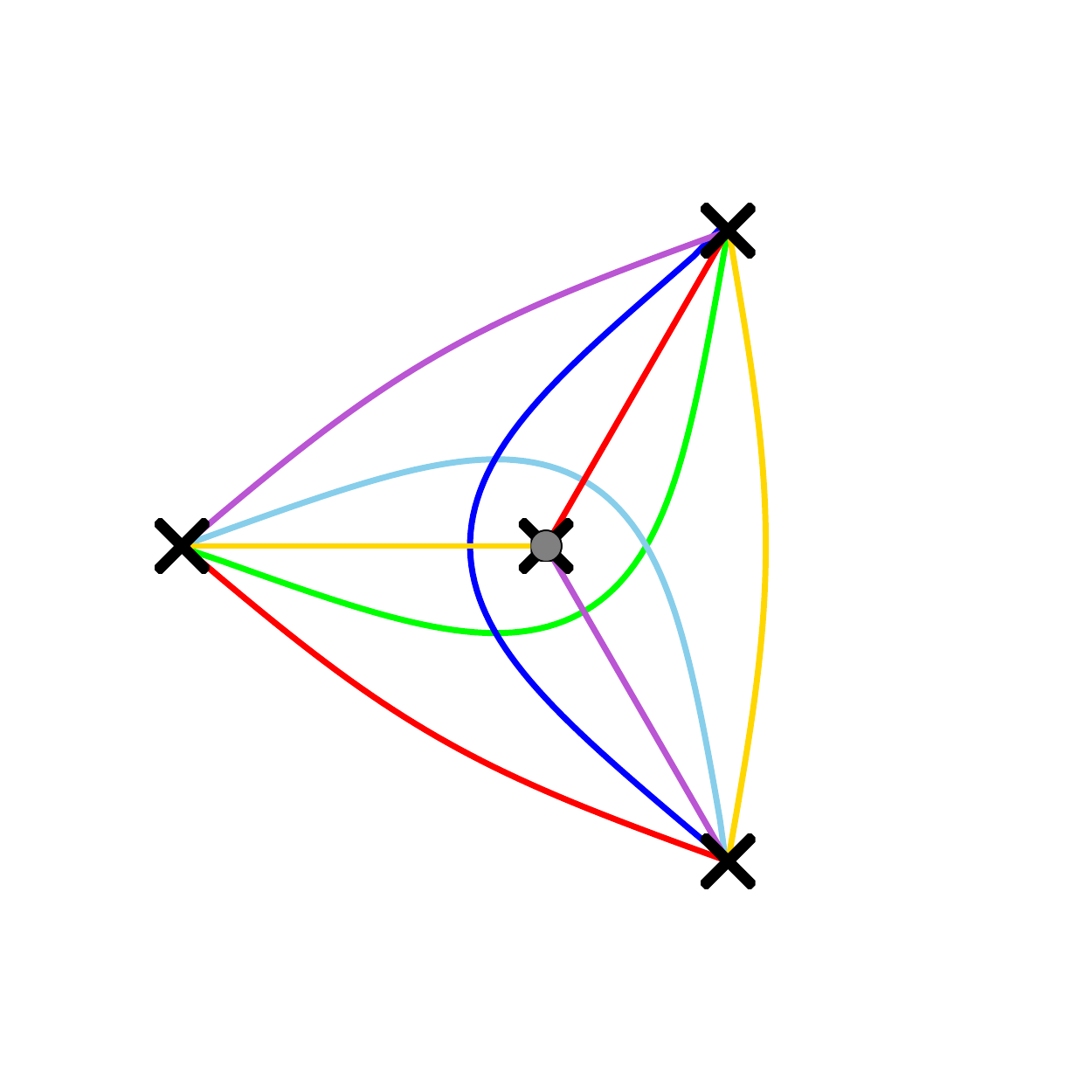}
	\end{subfigure}
	\begin{subfigure}{.25\textwidth}
		\centering
		\includegraphics[width=\textwidth]{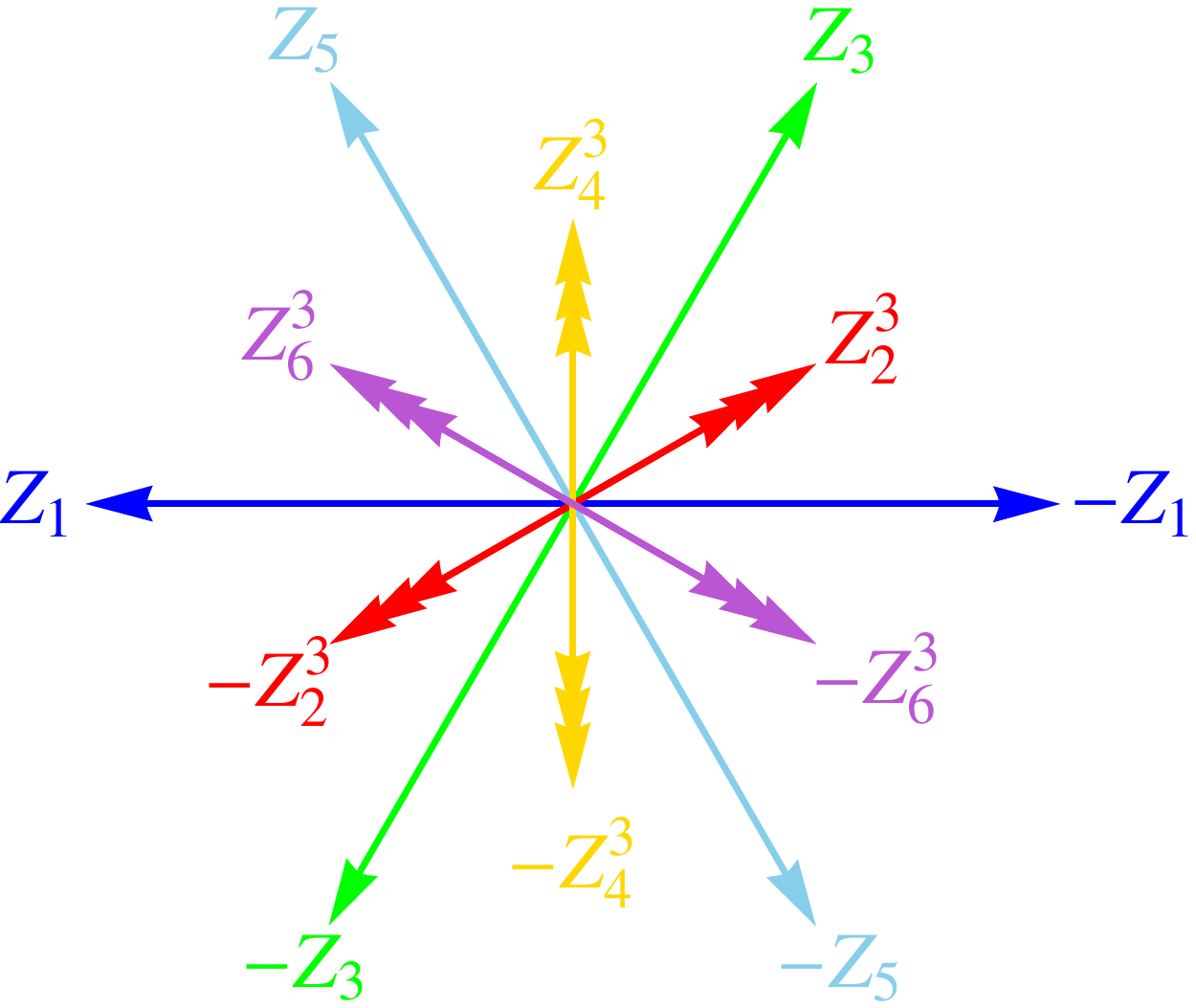}
	\end{subfigure}
	
	\caption{Wall-crossing of $\CS[A_{1}; \CD_\text{\rm reg}, \CD_{6}]$ with an $\SU(3)$ flavor symmetry.}
	\label{fig:d4SN_wall_crossing_SU_3_f}
\end{figure}

\paragraph{Minimal BPS spectrum of $\CS[A_{1}; \CD_\text{\rm reg}, \CD_{6}]$ with an $\SU(3)$ flavor symmetry}
When we set both $m = 0$ and ${c_1}^2 = 4 c_2$ and set $v_1 = {c_1}^3/54 - \delta$, where $\delta$ is a small number, $\CS[A_{1}; \CD_\text{\rm reg}, \CD_{6}]$ has a spectral network shown in Figure \ref{fig:d4SN_SU_3_f_minimal}. There are three BPS states of the same central charge $Z_2^\mathbf{3}$, which is represented as a three-headed arrow in the first row of Figure \ref{fig:d4SN_wall_crossing_SU_3_f}. This shows that when both the mass parameters vanish 
we indeed have an $\SU(3)$ flavor symmetry, and that there is a triplet of the $\SU(3)$. 

\begin{figure}[h]
	\centering
	\begin{subfigure}[b]{.3\textwidth}
		\centering
		\begin{tabular}{c|c}
			state  & $(e,m)$ \\ \hline
			1 & $(1,0)$ \\ \hline
			2 & $(0,1)$
		\end{tabular}
		\vspace{1em}
		\caption{IR charges}
		\label{table:d4A1SU3minimal}
		\label{tbl:d4SN_SU_3_f_minimal_IR}
	\end{subfigure}
	\begin{subfigure}[b]{.3\textwidth}
		\centering
		\begin{tikzpicture}
			\node[W,red] (21) at (0,0) {2};
			\node[W,blue] (1) at (1,0) {1};
			\node[W,red] (22) at (1.5,1.732/2) {2};
			\node[W,red] (23) at (1.5,-1.732/2) {2};
			
			\path (1) edge[->] (21);
			\path (1) edge[->] (22);
			\path (1) edge[->] (23);	
		\end{tikzpicture}
		\caption{BPS quiver}
		\label{fig:d4SN_SU_3_f_minimal_quiver}
	\end{subfigure}
	\caption{Minimal BPS spectrum of $\CS[A_{1}; \CD_\text{\rm reg}, \CD_{6}]$ with an $\SU(3)$ flavor symmetry.}
	\label{fig:d4SN_SU_3_f_minimal_BPS}
\end{figure}
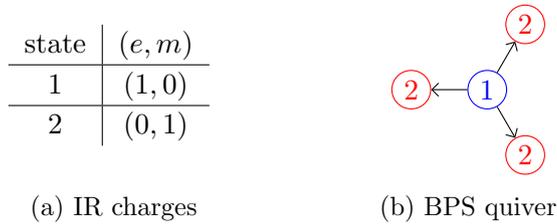

We can determine the IR charges of BPS states from Figure \ref{fig:d4SN_SU_3_f_minimal} after picking up a suitable basis. If one chooses the two cuts along the triplet $\CS$-walls at $\theta = \arg(Z_2^\mathbf{3})$ of the minimal spectrum in Figure \ref{fig:d4SN_SU_3_f_minimal_triplet}, we define the cycle corresponding to the singlet finite $\CS$-wall at $\theta = \arg(Z_1)$, say $\CS_1$, as an A-cycle and one of the triplet finite $\CS$-walls at $\theta = \arg(Z_2^\mathbf{3})$, say $\CS_2$, as a B-cycle. Their intersection number is $\intsec{\CS_{1}}{\CS_{2}}=1$ with a proper choice of orientations of the cycles, and from this we get the IR charges as described in Table \ref{tbl:d4SN_SU_3_f_minimal_IR}. This BPS spectrum can be represented by a $D_4$ quiver as shown in Figure \ref{fig:d4SN_SU_3_f_minimal_quiver}.

\paragraph{Wall-crossing of $\CS[A_{1}; \CD_\text{\rm reg}, \CD_{6}]$ with an $\SU(3)$ flavor symmetry}
\label{sec:Wall-crossing of d_4 with SU(3) flavor symmetry maintained}
When we maintain the maximal flavor symmetry of $\CS[A_{1}; \CD_\text{\rm reg}, \CD_{6}]$ during the wall-crossing, we observe that the BPS spectrum jumps from the minimal to the maximal one at once. Figure \ref{fig:d4SN_wall_crossing_SU_3_f} illustrates such a wall-crossing.

When $\delta \to 0$, the singlet becomes massless and the BPS spectrum jump from the minimal one to the maximal one at once, and the result is as shown in the second row of \ref{fig:d4SN_wall_crossing_SU_3_f}, where now $v_1 = {c_1}^3/54 + \delta$. The last row of Figure \ref{fig:d4SN_wall_crossing_SU_3_f} shows the case of $c_1 = c_2 = m =0$ and $v_1 \neq 0$, which has the $\SU(3)$ flavor symmetry and also symmetric arrangement of branch points and therefore a symmetric BPS spectrum.
Under the same basis the charges of BPS states as we found to get Table \ref{tbl:d4SN_SU_3_f_minimal_IR}, states in the maximal BPS spectrum has IR charges as described in Table \ref{tbl:d4SN_SU_3_maximal_IR}.

\begin{table}[h]
	\begin{center}
		\begin{tabular}{c|c}
			state & $(e,m)$ \\ \hline
			$1$ & $(1,0)$ \\ \hline
			$2$ & $(0,1)$ \\ \hline
			$3$ & $(1,3)$ \\ \hline			
			$4$ & $(1,2)$ \\ \hline
			$5$ & $(2,3)$ \\ \hline
			$6$ & $(1,1)$ 
		\end{tabular}
	\end{center}
	\caption{IR charges of the states in the maximal BPS spectrum of $\CS[A_{1}; \CD_\text{\rm reg}, \CD_{6}]$ with an $\SU(3)$ flavor symmetry.}
	\label{tbl:d4SN_SU_3_maximal_IR}
\end{table}

\begin{figure}[t]
	\centering
	
	\begin{subfigure}{.19\textwidth}
		\centering
		\includegraphics[scale=.28]{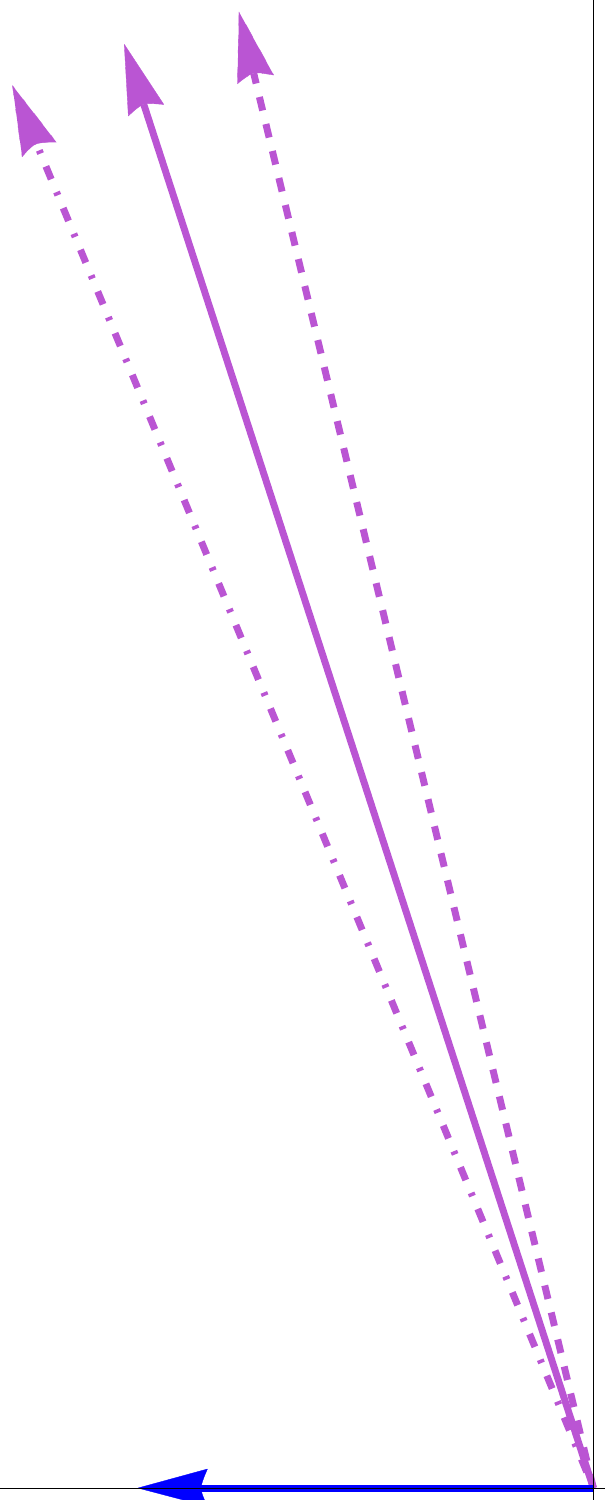}
	\end{subfigure}
	\begin{subfigure}{.19\textwidth}
		\centering
		\includegraphics[scale=.28]{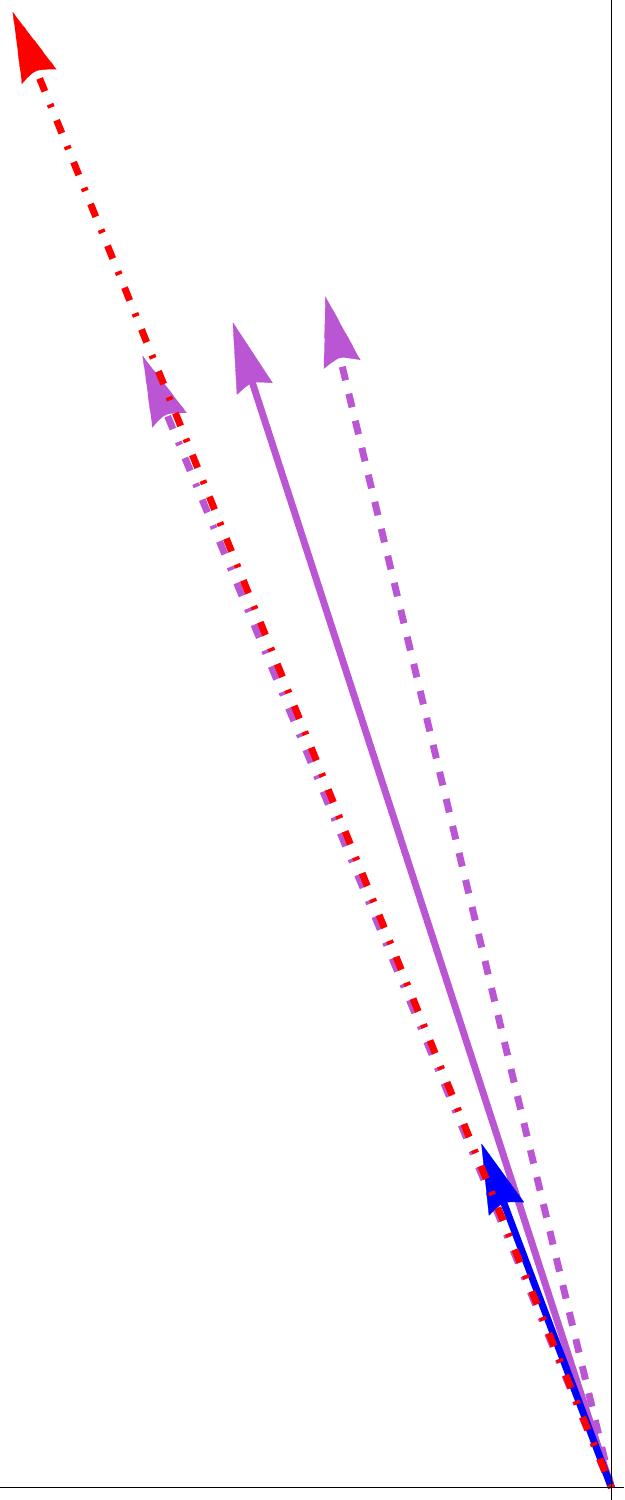}
	\end{subfigure}	
	\begin{subfigure}{.19\textwidth}
		\centering
		\includegraphics[scale=.28]{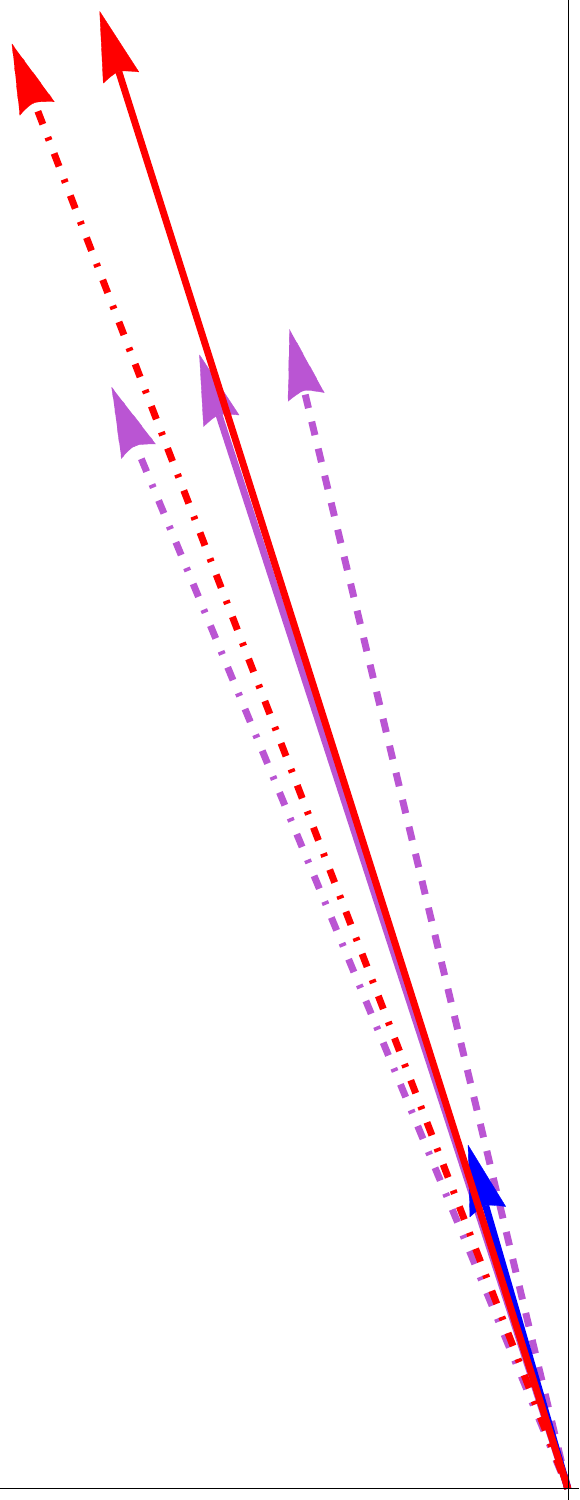}
	\end{subfigure}
	\begin{subfigure}{.19\textwidth}
		\centering
		\includegraphics[scale=.28]{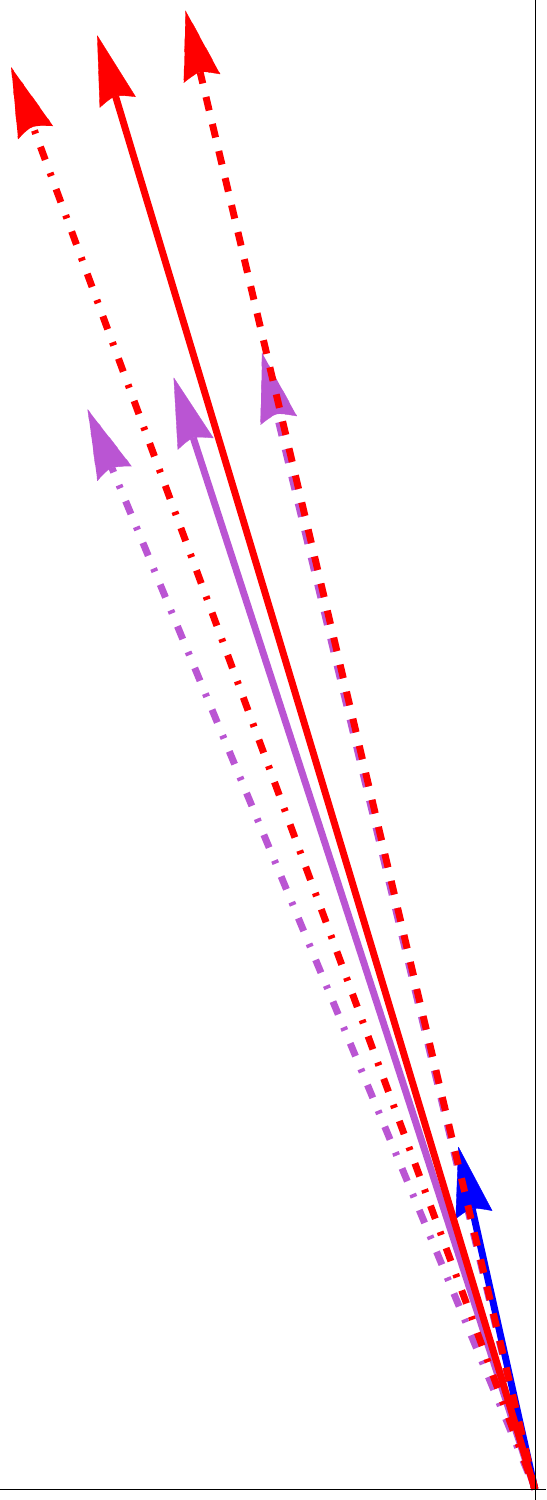}
	\end{subfigure}
	\begin{subfigure}{.19\textwidth}
		\centering
		\includegraphics[scale=.28]{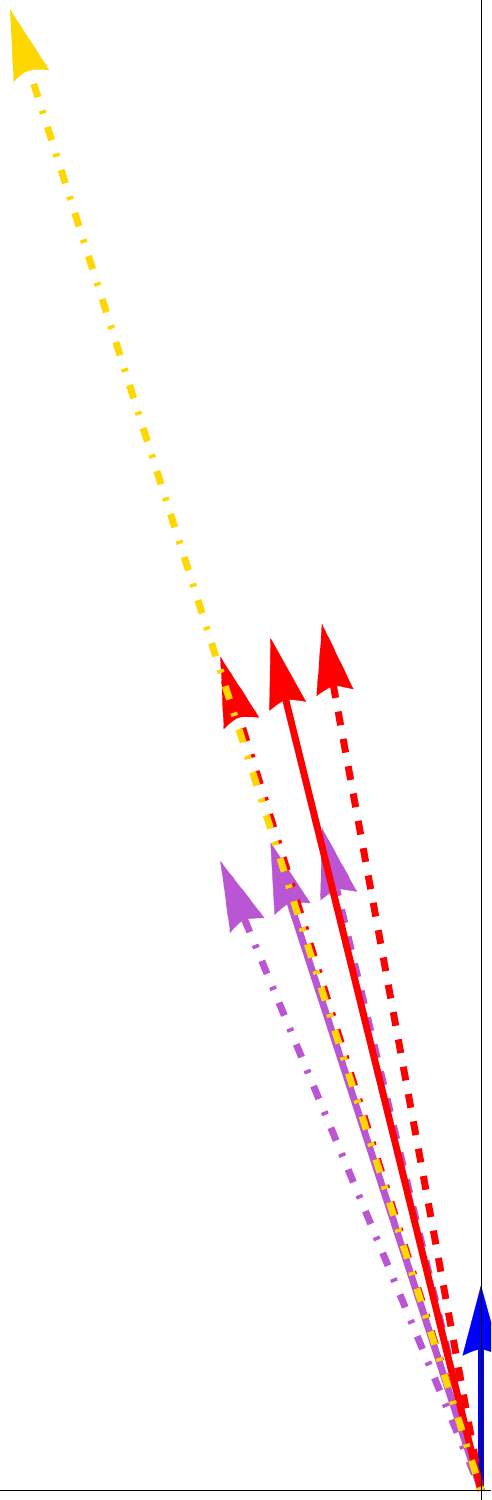}
	\end{subfigure}

\vspace{1em}	

	\begin{subfigure}{.2\textwidth}
		\centering
		\includegraphics[scale=.37]{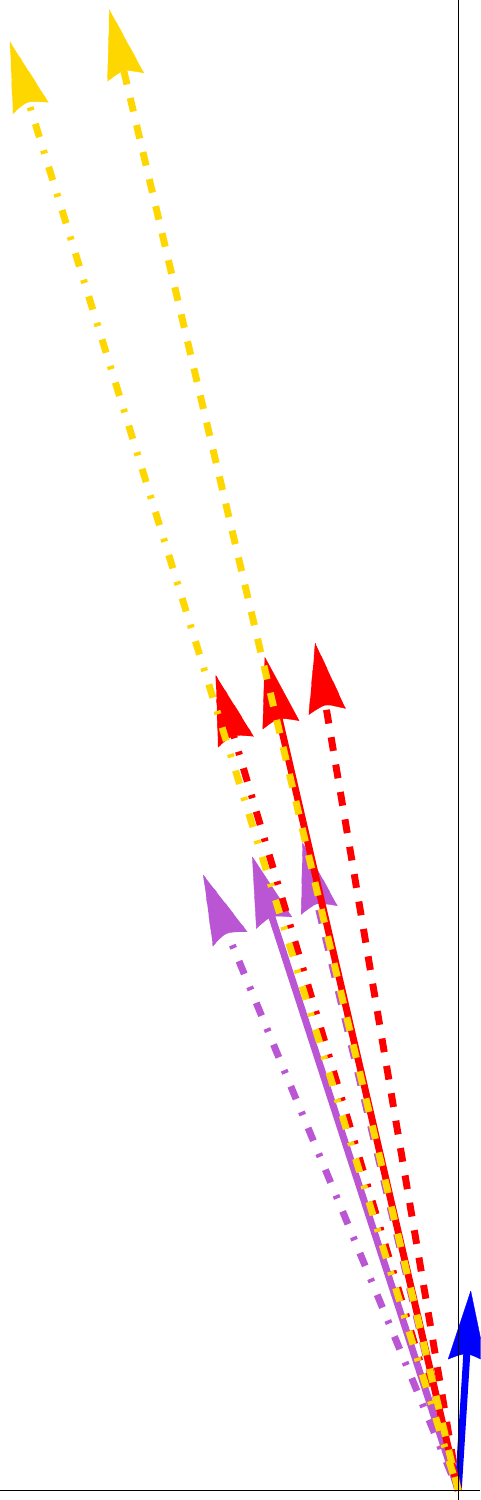}
	\end{subfigure}	
	\begin{subfigure}{.2\textwidth}
		\centering
		\includegraphics[scale=.37]{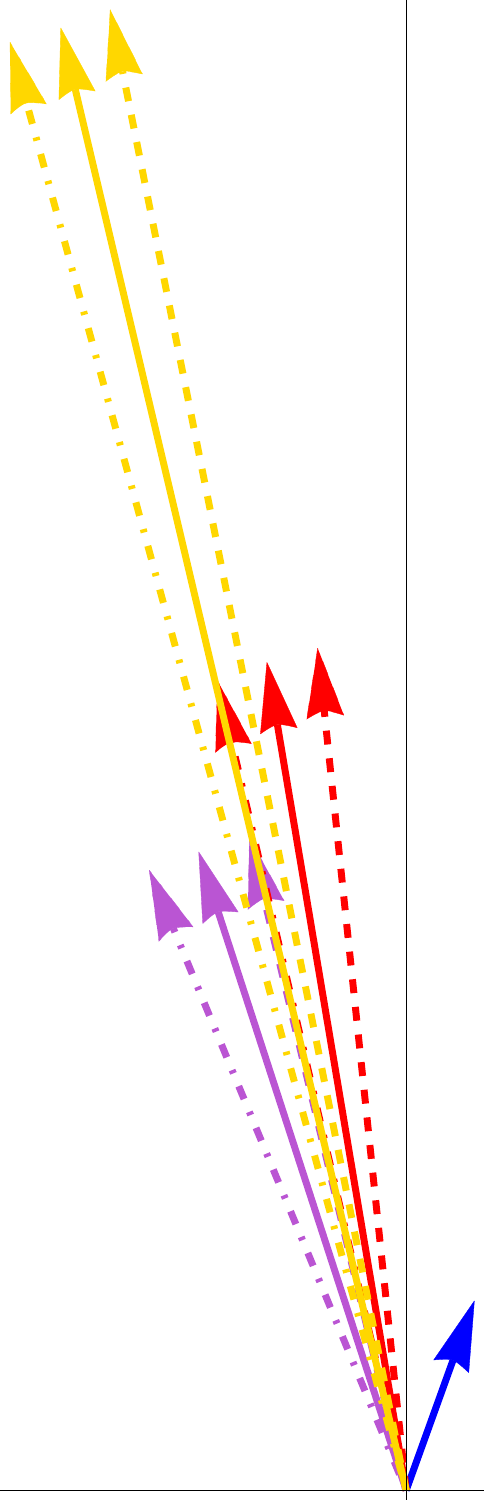}
	\end{subfigure}
	\begin{subfigure}{.2\textwidth}
		\centering
		\includegraphics[scale=.37]{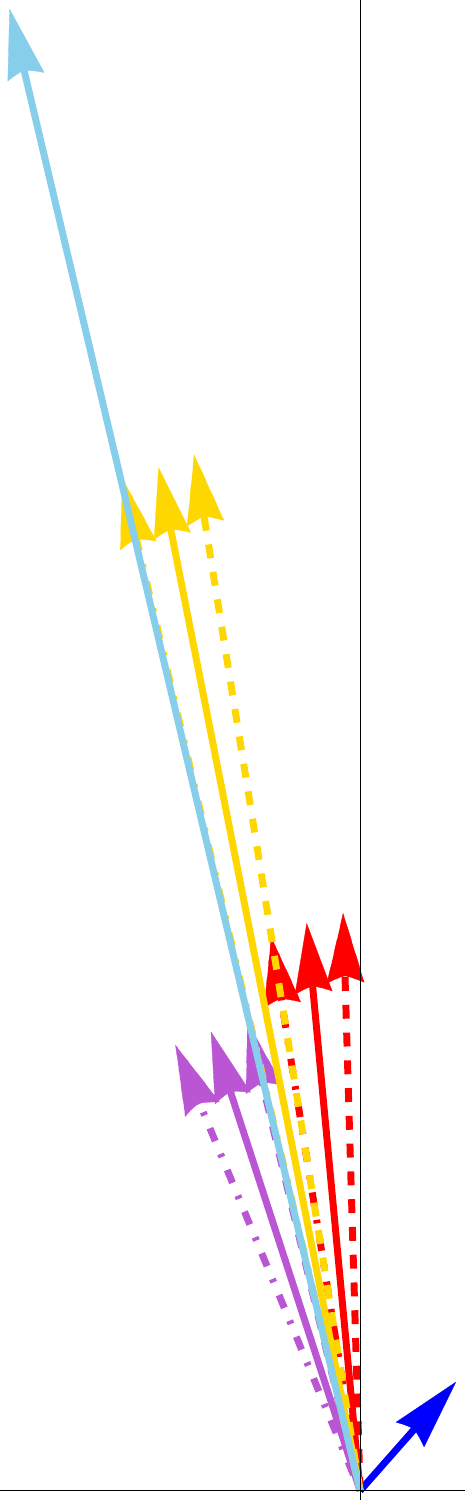}
	\end{subfigure}	
	\begin{subfigure}{.2\textwidth}
		\centering
		\includegraphics[scale=.37]{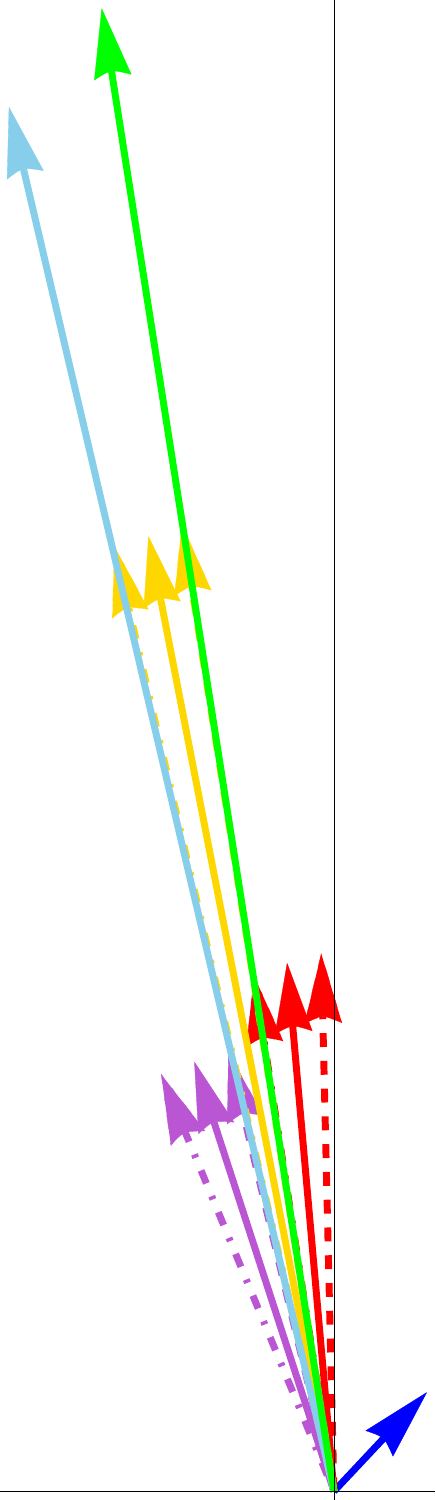}
	\end{subfigure}	
				
	\caption{Wall-crossing of a singlet and a triplet.}
	\label{fig:d4SN_wall_crossing_resolved}
\end{figure}

Figure \ref{fig:d4SN_wall_crossing_resolved} provides an explanation of such a wall-crossing by resolving the triplet into three BPS states and consider a series of usual wall-crossings between a singlet and a triplet, as we have resolved the doublet of  $\CS[A_{1}; \CD_\text{\rm reg}, \CD_{5}]$ to understand the wall-crossing of its BPS spectrum, see Figure \ref{fig:d3SN_wall_crossing_resolved}. In the limit of the three resolved BPS states becoming a triplet, these eight wall-crossings happen at the same time, resulting in two additional triplets and two additional singlets. We can also see that the structures of the triplets and the singlets suit well with what we have in Figure \ref{fig:d4SN_wall_crossing_SU_3_f}.

\subsubsection{$\CS[A_{1}; \CD_\text{\rm reg}, \CD_{n+2}]$ in $D_n$-class, $n \geq 4$}

Now we consider $\CS[A_{1}; \CD_\text{\rm reg}, \CD_{n+2}]$ for a general $n$. When $n = 2k+1$, the Seiberg-witten curve is
\begin{align}
	v^2 = t^{2k+1} + c_{1} t^{2k} + \cdots + c_k t^{k+1} + v_k t^k + \cdots + v_1 t + m^2,
\end{align}
and when $n=2k$ the Seiberg-Witten curve is
\begin{align}
	v^2 = t^{2k} + c_{1} t^{2k-1} + \cdots + c_{k-1} t^{k+1} + c_k t^k + v_{k-1} t^{k-1} + \cdots + v_1 t + m^2.
\end{align}
We will focus on the case of $m = 0$, where the BPS spectrum has an $\SU(2)$ flavor symmetry.

\begin{figure}[t]
	\centering
	\begin{subfigure}[b]{.26\textwidth}	
		\centering
		\includegraphics[width=\textwidth]{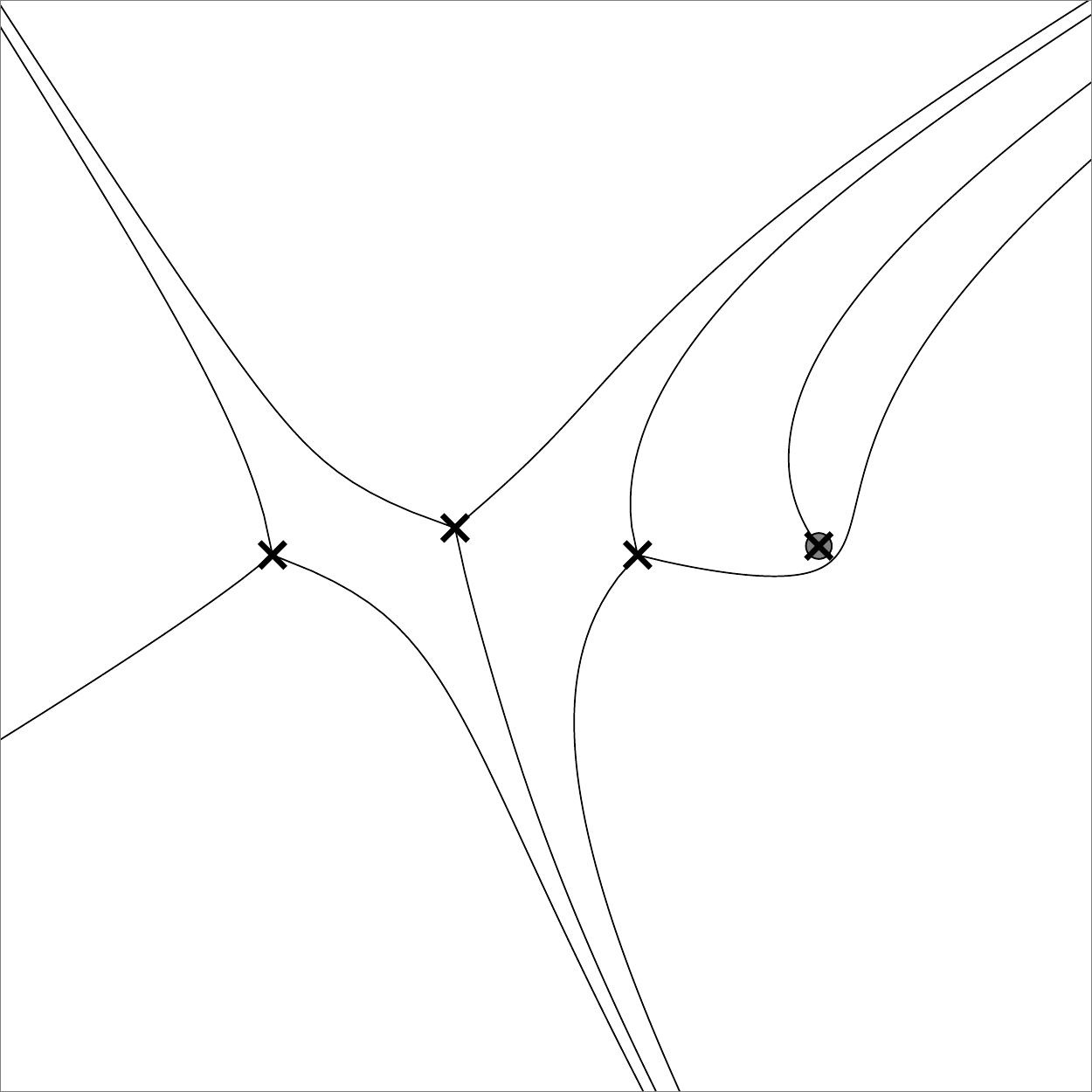}
		\caption{at general $\theta$}
		\label{fig:d4SN_SU_2_minimal_general}
	\end{subfigure}	
	\begin{subfigure}[b]{.4\textwidth}	
		\includegraphics[width=\textwidth]{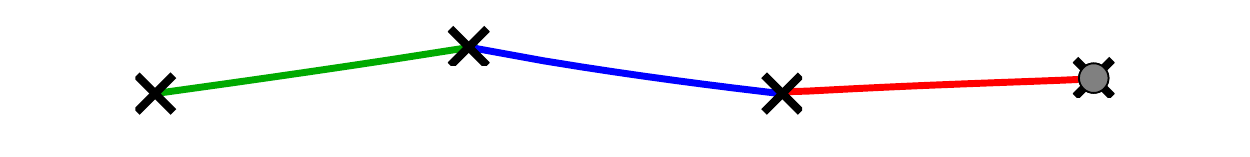}
		\vspace{3em}
		\caption{finite $\CS$-walls}
		\label{fig:d4SN_SU_2_minimal_finite}
	\end{subfigure}

	\caption{Spectral networks of $\CS[A_{1}; \CD_\text{\rm reg}, \CD_{6}]$ with an $\UU(1) \times \SU(2)$ flavor symmetry and minimal BPS spectrum.}
	\label{fig:d4SN_SU_2_minimal}
\end{figure}

\paragraph{$\CS[A_{1}; \CD_\text{\rm reg}, \CD_{n+2}]$ with a minimal BPS spectrum \& an $\SU(2)$ flavor symmetry}
To consider the generalization, let us go back to the previous example
and focus on the case with $m=0$ and general values of $c_{1}$, $c_{2}$ and $v_{1}$
where the flavor symmetry is $\UU(1) \times \SU(2)$.
Figure \ref{fig:d4SN_SU_2_minimal} shows the spectral network of $\CS[A_{1}; \CD_\text{\rm reg}, \CD_{6}]$
of the choice of the parameters. 
In Figure \ref{fig:d4SN_SU_2_minimal_finite} we have an $\SU(2)$ doublet of the finite $\CS$-walls connecting the puncture and one of the other branch point, as we have seen from the spectral network of $\CS[A_{1}; \CD_\text{\rm reg}, \CD_{5}]$.

We can see that the spectral network can be considered as a combination of an $\CS[A_{1}; \CD_{7}]$ spectral network and an $\CS[A_{1}; \CD_\text{\rm reg}, \CD_{5}]$ spectral network. This also applies for every $\CS[A_{1}; \CD_\text{\rm reg}, \CD_{n+2}]$ with $n>4$, whose spectral network can be considered as a combination of an $\CS[A_{1}; \CD_\text{\rm reg}, D_{5}]$ spectral network and an $\CS[A_{1}; \CD_{n+3}]$ spectral network, which in turn consists of spectral networks of $\CS[A_{1}; \CD_{7}]$.

\begin{figure}[h]
	\centering
	\begin{subfigure}[b]{.3\textwidth}
		\centering
		\includegraphics[width=\textwidth]{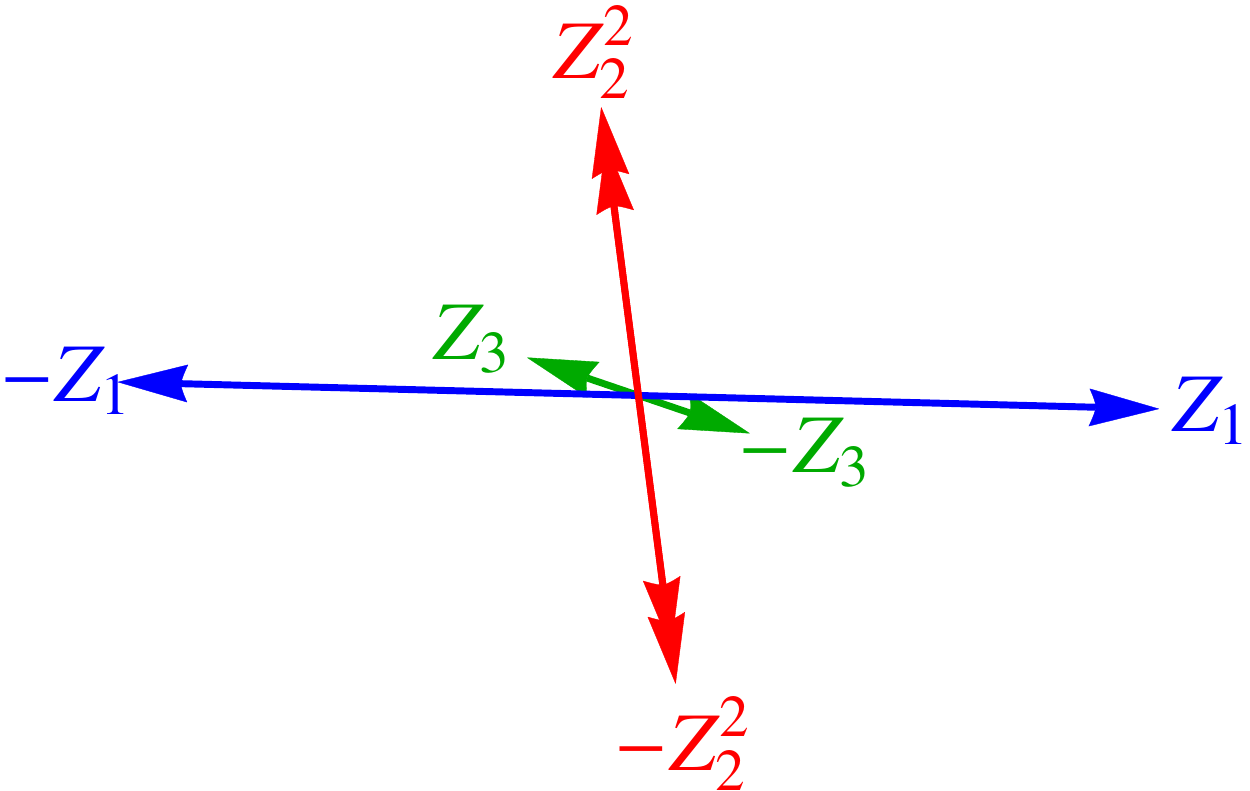}
		\caption{central charges}
		\label{fig:d4SN_SU_2_minimal_Z}
	\end{subfigure}
	\begin{subfigure}[b]{.3\textwidth}	
		\centering
		\begin{tabular}{c|c}
		   state  & $(e,m)$ \\ \hline
		  1 & $(1,0)$ \\ \hline
		  2 & $(0,1)$ \\ \hline
		  3 & $(0,1)$
		\end{tabular}
		\vspace{1em}
		\caption{IR charges}
		\label{tbl:d4SN_SU_2_minimal_IR}
	\end{subfigure}	
	\begin{subfigure}[b]{.25\textwidth}	
		\centering
		\begin{tikzpicture}[scale=.5]
			\node[W,darkergreen] (1) at (0,0) {3};
			\node[W,blue] (2) at (2,0) {1};
			\node[W,red] (31) at (3,1.732) {2};
			\node[W,red] (32) at (3,-1.732) {2};
			\path (2) edge[<-] (31);
			\path (2) edge[<-] (32);
			\path (2) edge[<-] (1);	
		\end{tikzpicture}
		\caption{BPS quiver}
		\label{fig:d4SN_SU_2_minimal_quiver}
	\end{subfigure}
	\renewcommand{\figurename}{Figure \& Table}
	\caption{Minimal BPS spectrum of $\CS[A_{1}; \CD_\text{\rm reg}, \CD_{6}]$ with an $\UU(1) \times \SU(2)$ flavor symmetry.}
	\label{figntbl:d4SN_SU_2_minimal_BPS}
\end{figure}

Figure \ref{fig:d4SN_SU_2_minimal_Z} and Table \ref{tbl:d4SN_SU_2_minimal_IR} describe the minimal BPS spectrum of $\CS[A_{1}; \CD_\text{\rm reg}, \CD_{6}]$ with only the $\UU(1) \times \SU(2)$ flavor symmetry. For general values of $c_1$ and $c_2$, we have a nonzero residue of $\lambda$ at $t = \infty$. Then the central charge of the doublet is differ by the residue from the central charge of one of the other two BPS states, whose corresponding $\CS$-wall is the same cycle of the elliptic curve from the Seiberg-Witten curve as the $\CS$-wall for the doublet.

\begin{figure}[t]
	\centering
	
	\begin{subfigure}{.3\textwidth}	
		\includegraphics[width=\textwidth]{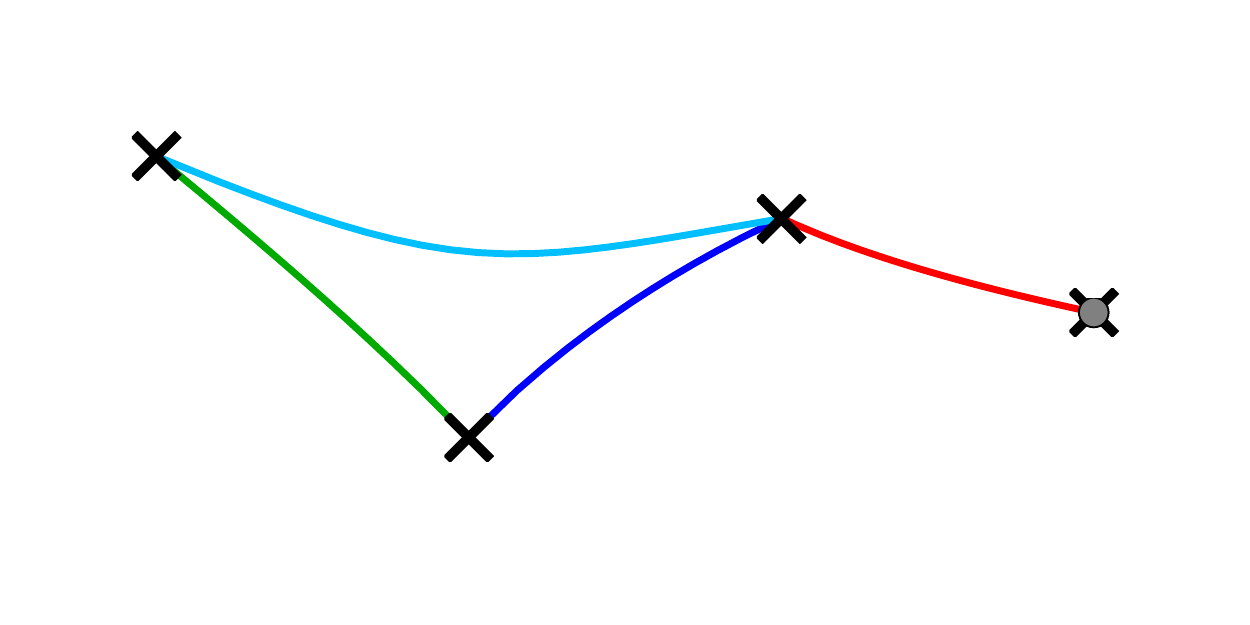}
	\end{subfigure}
	\begin{subfigure}{.25\textwidth}
		\centering
		\includegraphics[scale=.25]{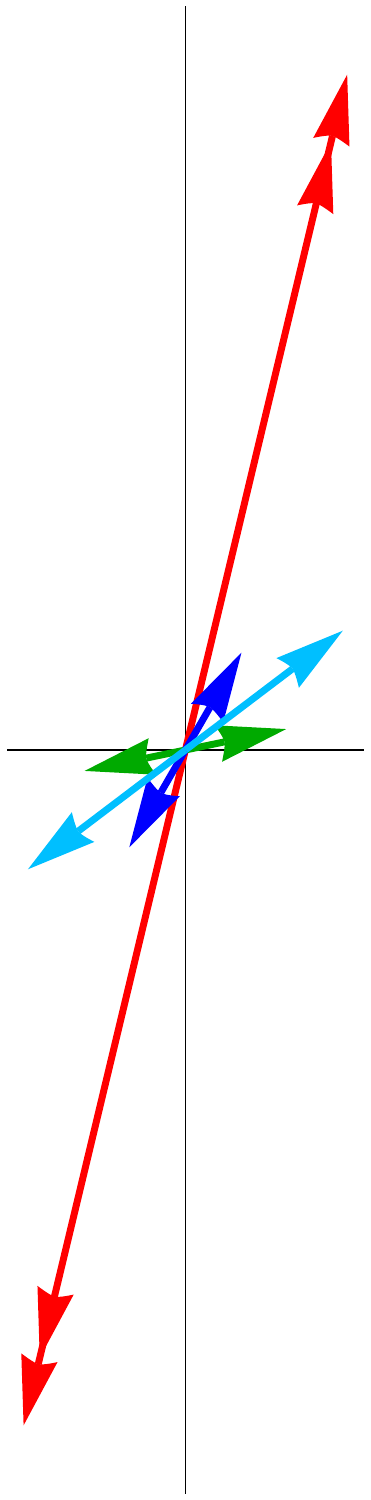}
	\end{subfigure}
	
	\begin{subfigure}{.3\textwidth}	
		\includegraphics[width=\textwidth]{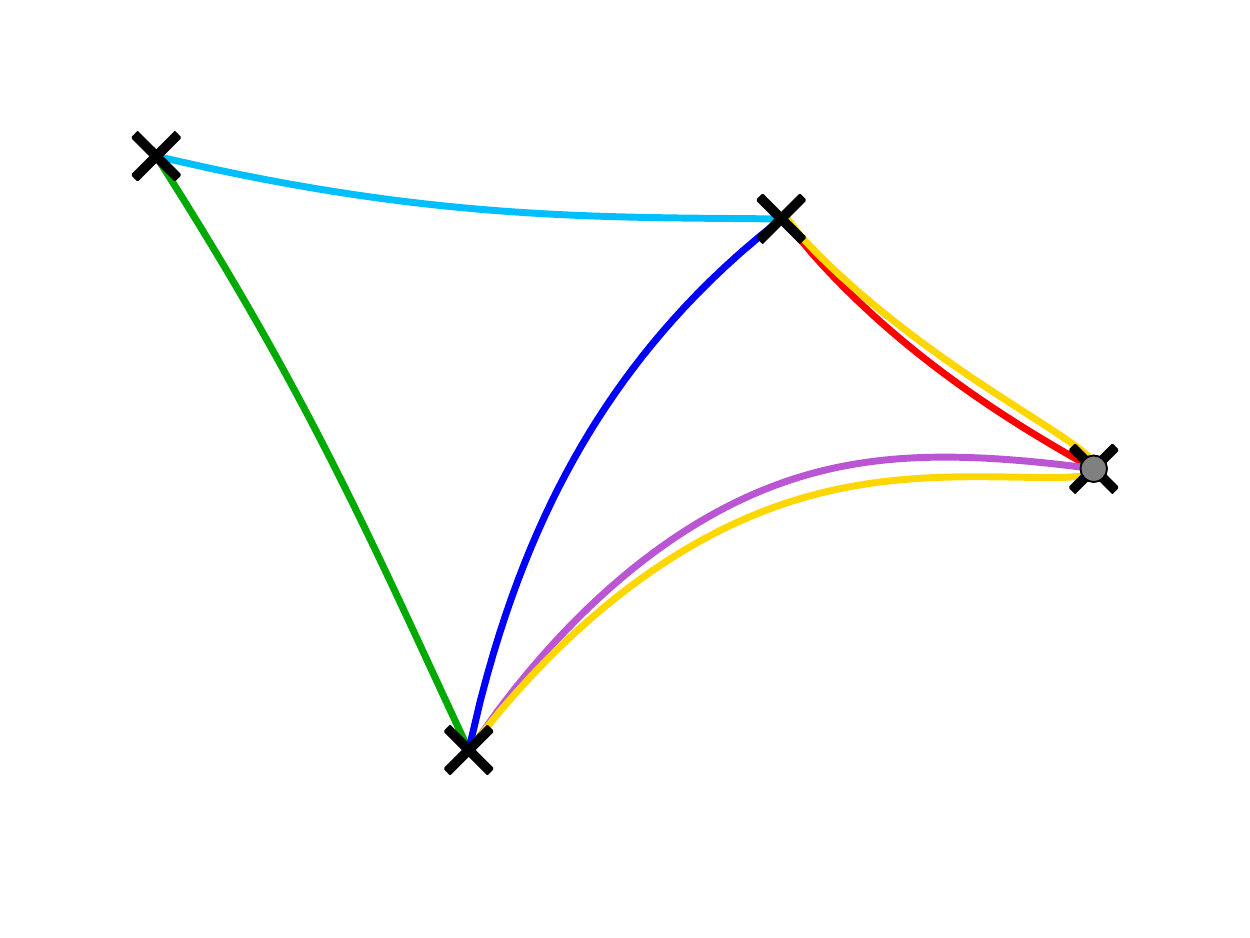}
	\end{subfigure}
	\begin{subfigure}{.25\textwidth}
		\centering
		\includegraphics[scale=.25]{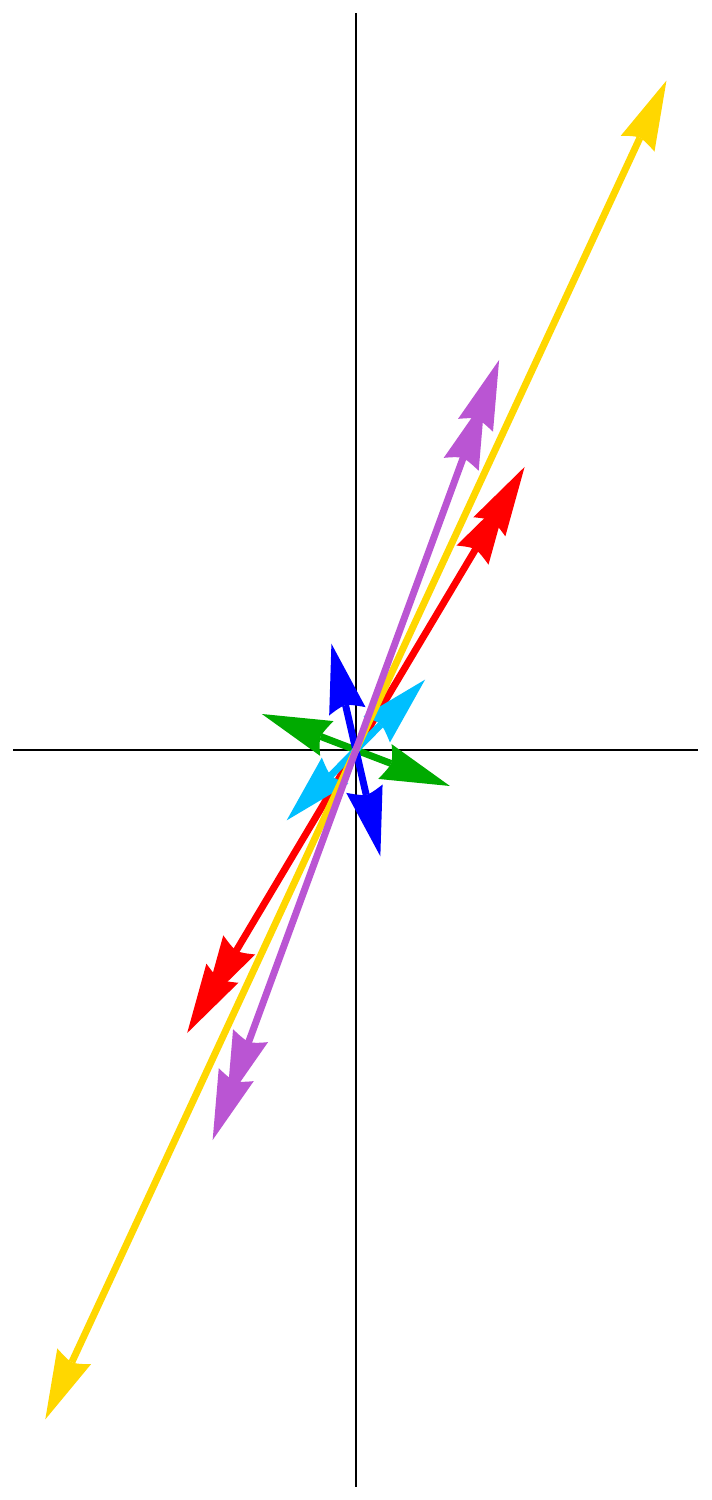}
	\end{subfigure}
		
	\begin{subfigure}{.3\textwidth}	
		\includegraphics[width=\textwidth]{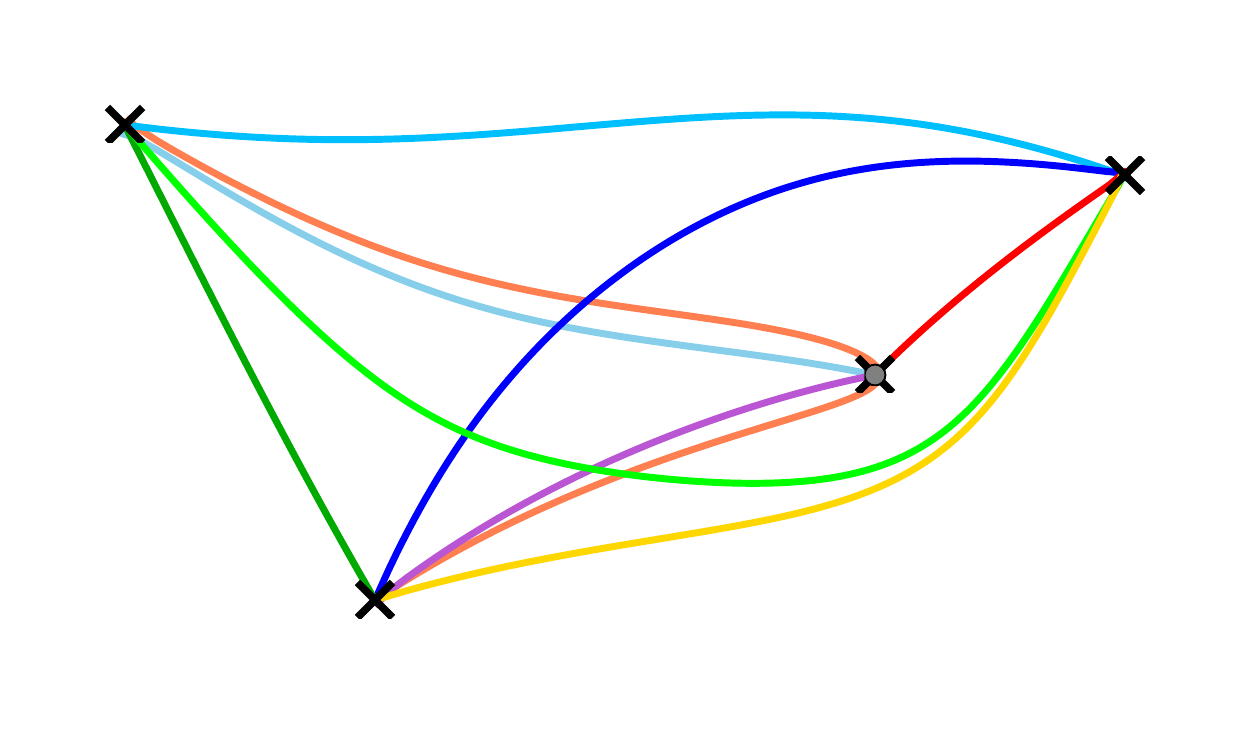}
	\end{subfigure}
	\begin{subfigure}{.25\textwidth}
		\centering
		\includegraphics[width=\textwidth]{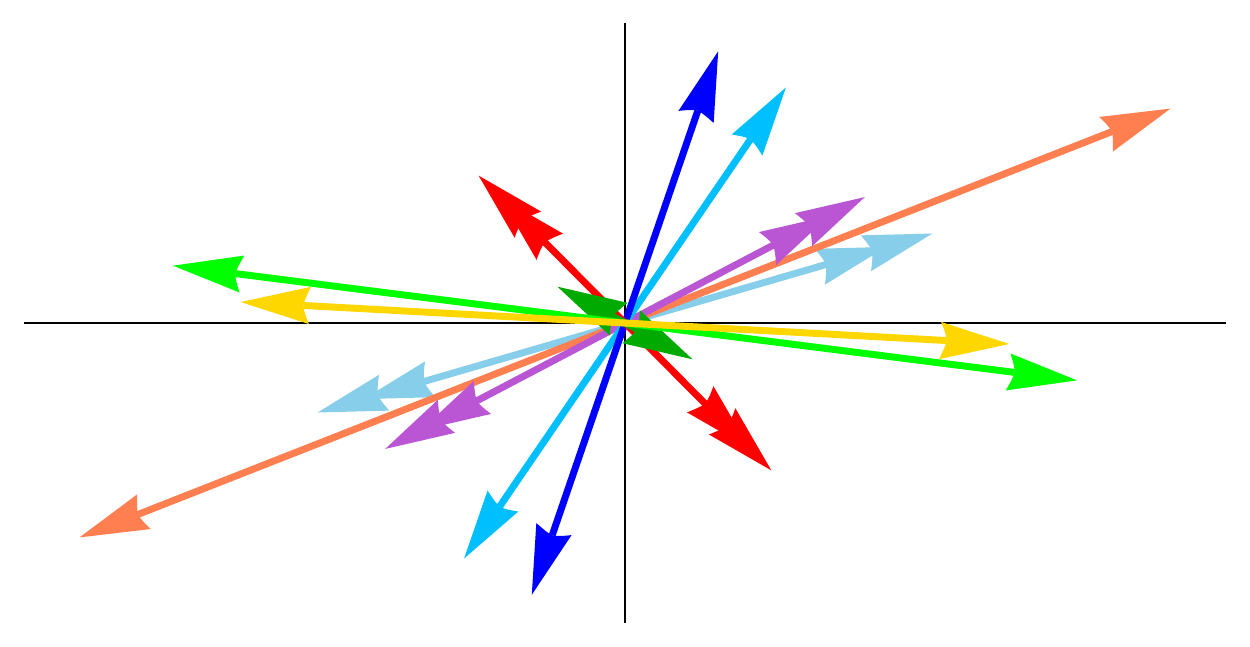}
	\end{subfigure}

	\caption{Wall-crossing of $\CS[A_{1}; \CD_\text{\rm reg}, \CD_{6}]$ with a $\UU(1) \times \SU(2)$ flavor symmetry.}
	\label{fig:d4SN_SU_2_wall_crossing}
\end{figure}

\paragraph{Wall-crossing of $\CS[A_{1}; \CD_\text{\rm reg}, \CD_{n+2}]$ with a $\UU(1) \times \SU(2)$ flavor symmetry}
The wall-crossings of $\CS[A_{1}; \CD_\text{\rm reg}, \CD_{6}]$ consists of the wall-crossings from $\CS[A_{1}; \CD_{7}]$ and $\CS[A_{1}; \CD_\text{\rm reg}, \CD_{5}]$, as illustrated in Figure \ref{fig:d4SN_SU_2_wall_crossing}. Starting from the minimal BPS spectrum of Figures \ref{fig:d4SN_SU_2_minimal_finite} and \ref{fig:d4SN_SU_2_minimal_Z}, after an $\CS[A_{1}; \CD_{7}]$ wall-crossing we get the BPS spectrum at the first row of Figure \ref{fig:d4SN_SU_2_wall_crossing}. Between the first row and the second row is another $\CS[A_{1}; \CD_\text{\rm reg}, D_{5}]$ wall-crossing, and after two additional $\CS[A_{1}; \CD_\text{\rm reg}, \CD_{5}]$ wall-crossings we arrive at the last row of Figure \ref{fig:d4SN_SU_2_wall_crossing}, where we have the maximal number of BPS state, six BPS states and six anti-states.

\paragraph{Maximal BPS spectrum of $\CS[A_{1}; \CD_\text{\rm reg}, \CD_{n+2}]$ with a $\UU(1) \times \SU(2)$ flavor symmetry}
The configuration of the spectral network of $\CS[A_{1}; \CD_\text{\rm reg}, \CD_{n+2}]$ is a straightforward generalization of the previous discussions and the resulting minimal BPS spectrum has the BPS quiver of $D_{n}$. However its maximal BPS spectrum, which has $\binom{n}{2} \times 2$ states and their anti-states, is more complicated, so here we describe an example of the maximal BPS spectrum, having in mind that this will be used to analyze the equivalence of $\CS[A_{1}; \CD_\text{\rm reg}, \CD_{n+2}]$ and $\CS[A_{n-2}; \CD_{\rm II}]$.

\begin{figure}[t]
	\centering
	\begin{subfigure}[b]{.27\textwidth}	
		\includegraphics[width=\textwidth]{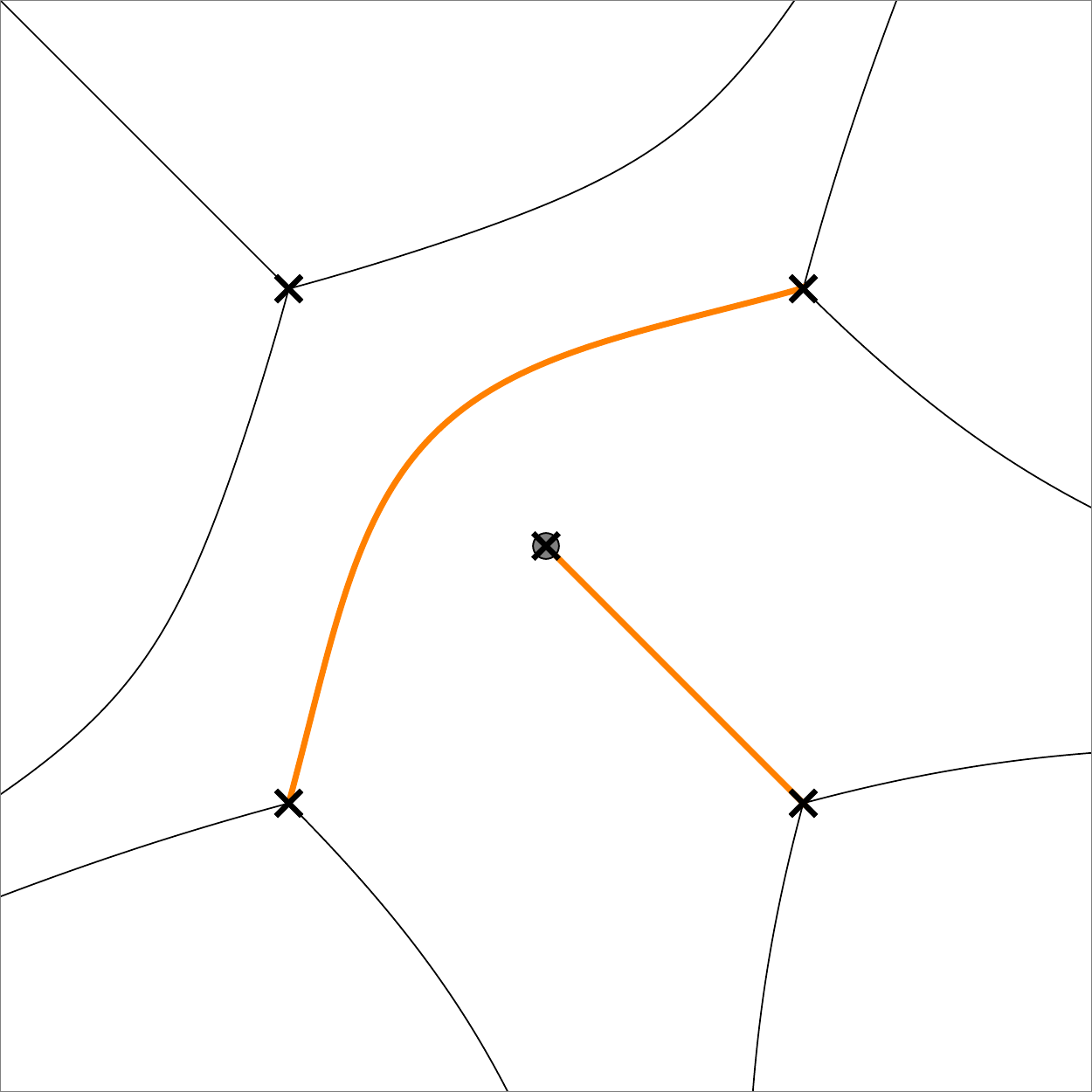}
		\caption{$\theta \approx \arg(Z_{2}) = \arg(Z_{4}^\mathbf{2})$}
		\label{fig:d5_from_A_1_maximal_symmetric_SN2}
	\end{subfigure}
	\begin{subfigure}[b]{.27\textwidth}	
		\includegraphics[width=\textwidth]{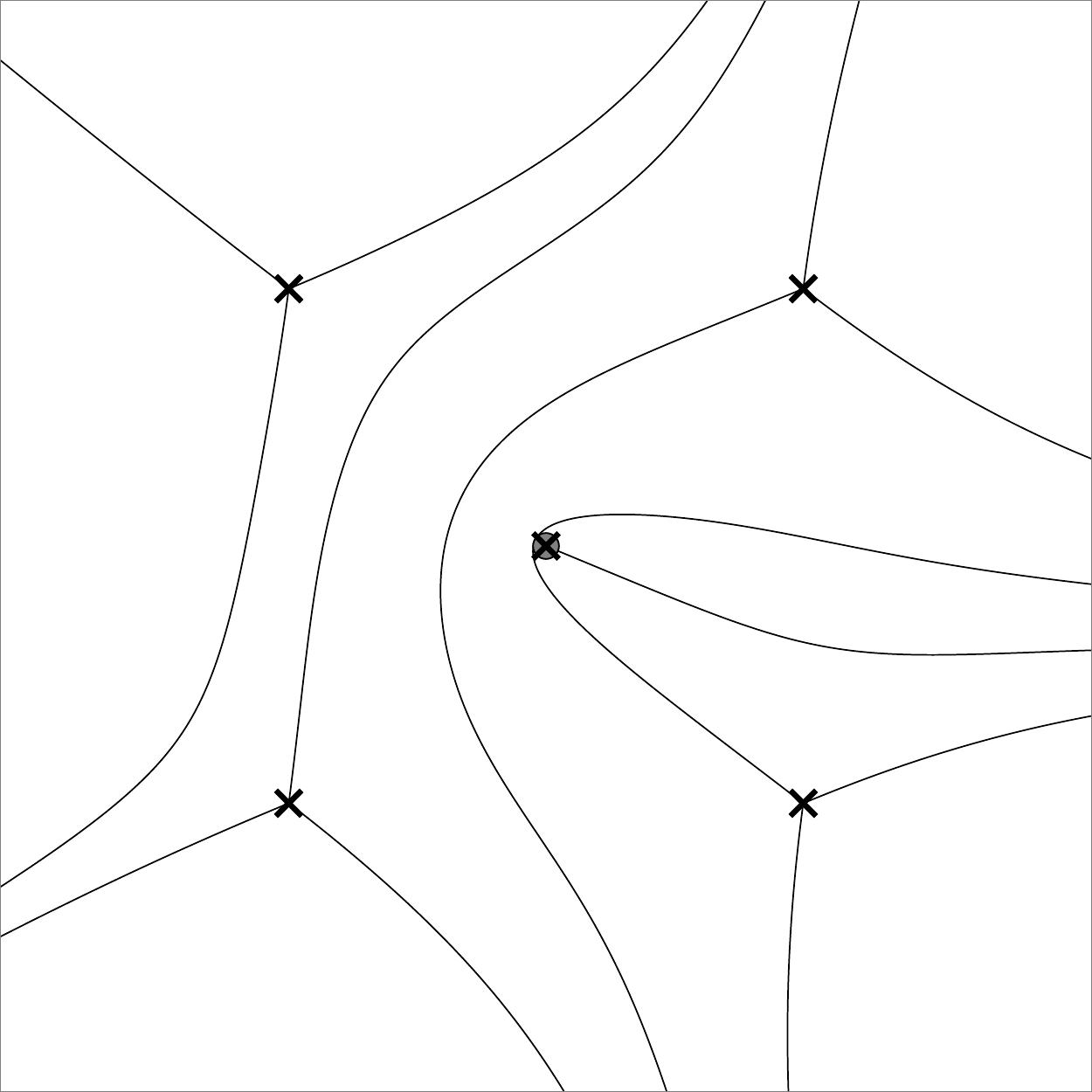}
		\caption{$\arg(Z_{2}) < \theta < \arg(Z_{1})$}
		\label{fig:d5_from_A_1_maximal_symmetric_SN3}
	\end{subfigure}

	\begin{subfigure}[b]{.27\textwidth}	
		\includegraphics[width=\textwidth]{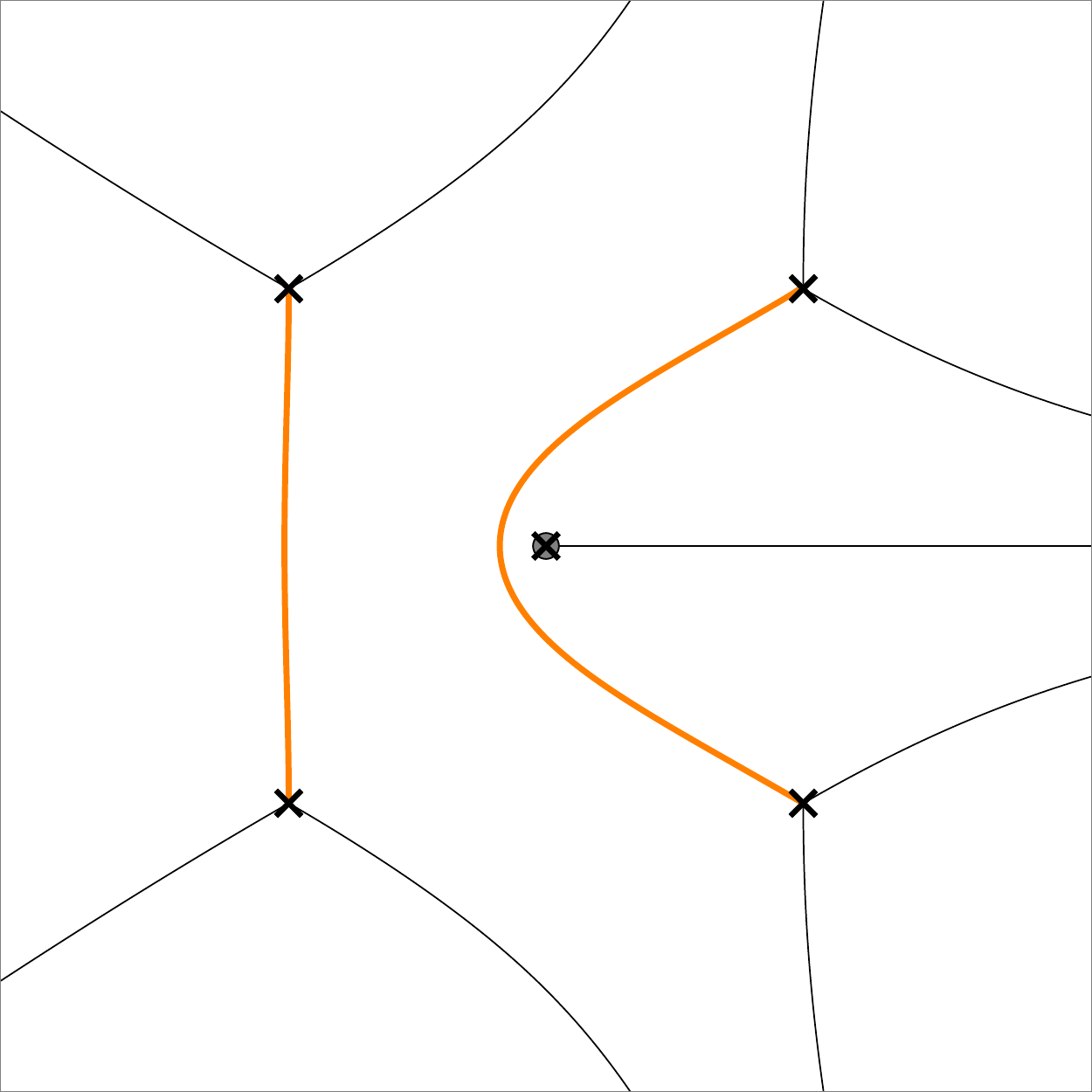}
		\caption{$\theta = \arg(Z_{1}) = \arg(Z_{3})$}
		\label{fig:d5_from_A_1_maximal_symmetric_SN1}
	\end{subfigure}
	\begin{subfigure}[b]{.27\textwidth}	
		\centering
		\includegraphics[width=\textwidth]{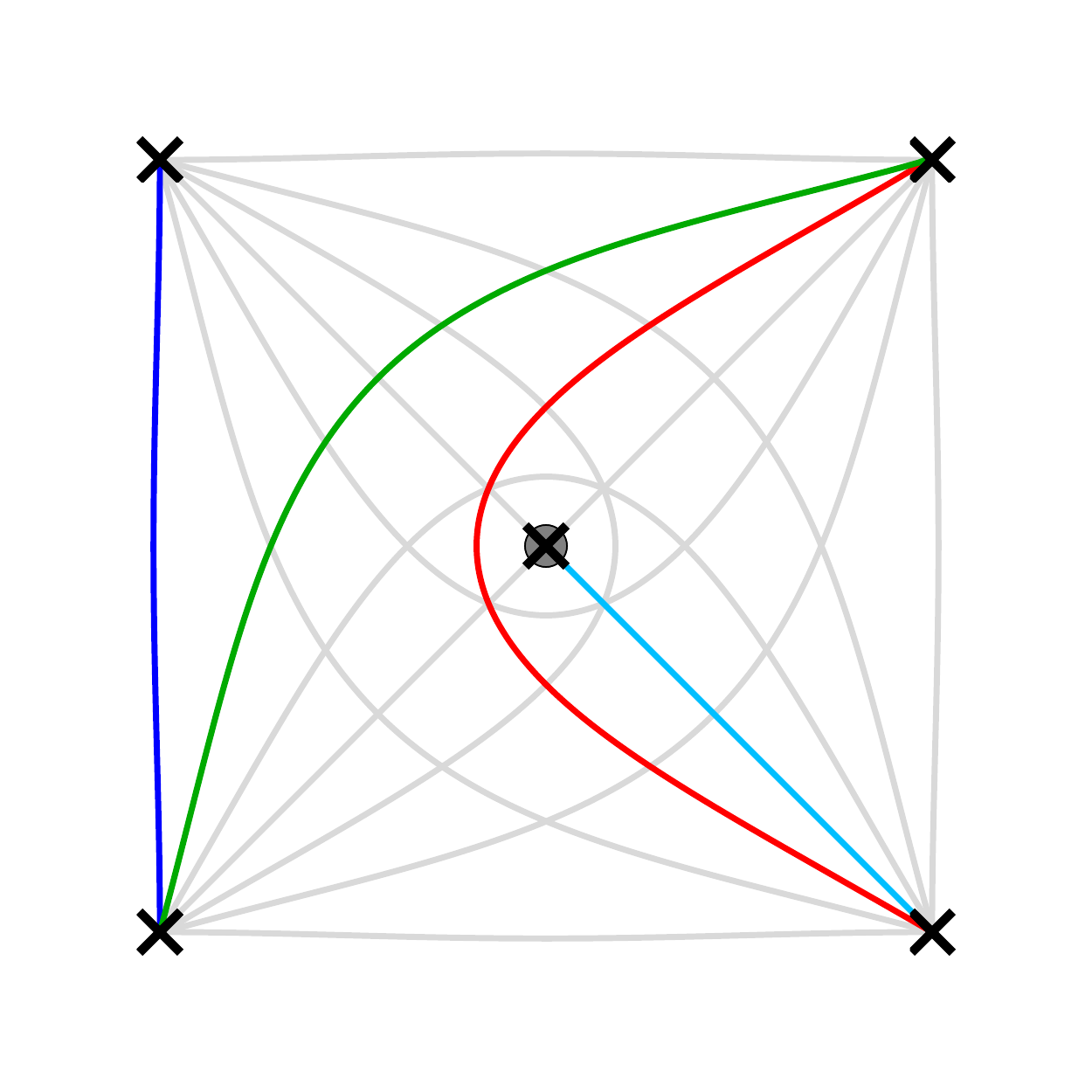}
		\caption{finite $\CS$-walls}
		\label{fig:d5_from_A_1_maximal_symmetric_finite}
	\end{subfigure}	
	\caption{spectral network of $\CS[A_{1}; \CD_\text{\rm reg}, \CD_{7}]$ with maximal, symmetric BPS spectrum.}
	\label{fig:d5_from_A_1_maximal_symmetric_SN}
\end{figure}

\begin{figure}[h]
	\centering

	\begin{subfigure}[b]{.4\textwidth}	
		\centering
		\includegraphics[width=\textwidth]{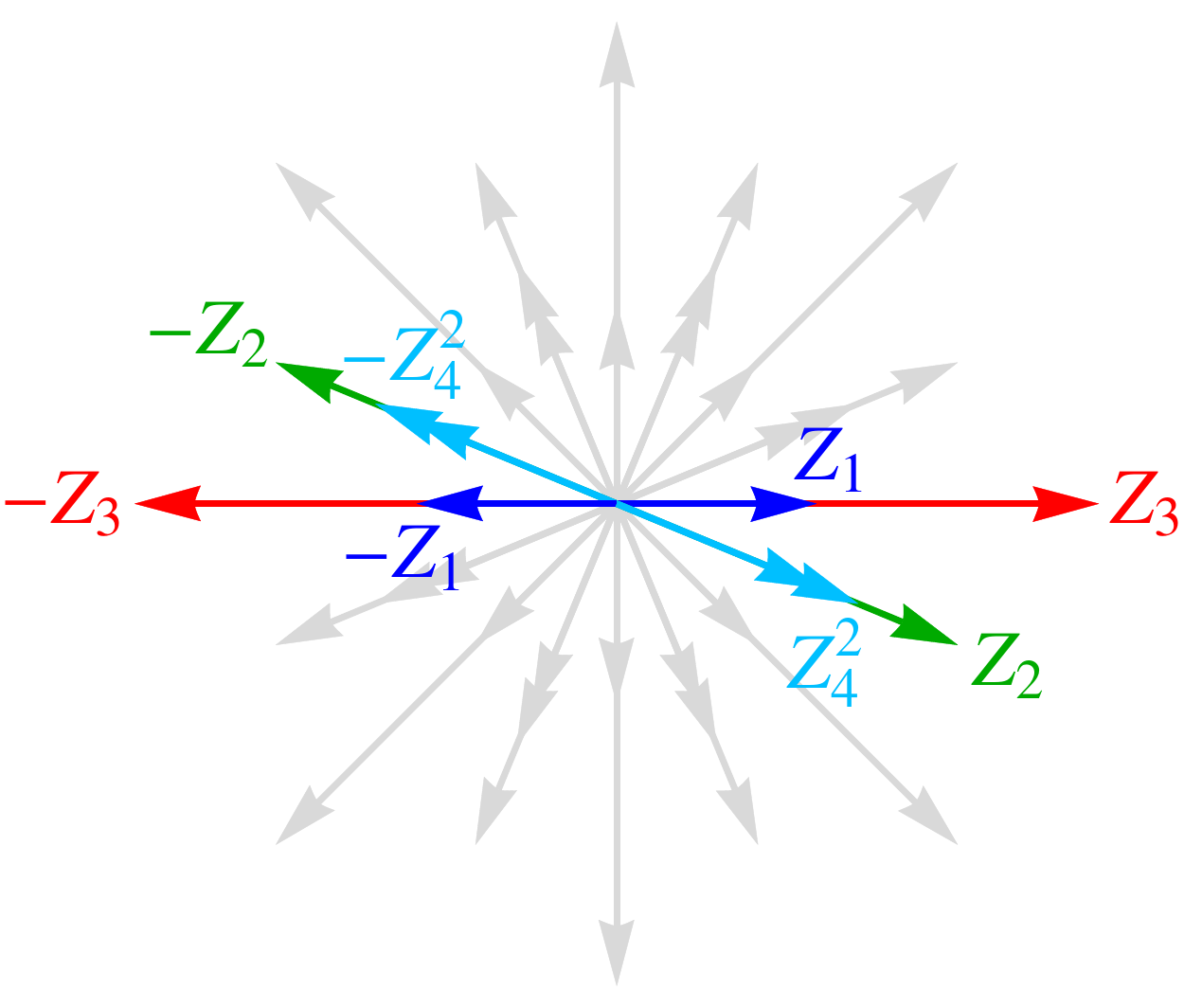}
		\vspace{5em}
		\caption{central charges}
		\label{fig:d5_from_A_1_maximal_symmetric_Z}
	\end{subfigure}
	\begin{subfigure}[b]{.4\textwidth}	
		\centering
		\begin{tabular}{c|c|c}
			state & $\UU(1)_{1}$ & $\UU(1)_{2}$\\ \hline \hline
			1 & $(1,0)$ & $(0,0)$ \\ \hline
			2 & $(0,1)$ & $(0,0)$ \\ \hline
			3 & $(1,0)$ & $(1,0)$ \\ \hline
			4 & $(0,0)$ & $(0,1)$ \\ \hline \hline
			5 & $(1,1)$ & $(1,0)$ \\ \hline
			6 & $(2,1)$ & $(1,0)$ \\ \hline
			7 & $(3,1)$ & $(2,2)$ \\ \hline
			8 & $(1,0)$ & $(1,1)$ \\ \hline \hline
			9 & $(1,0)$ & $(1,2)$ \\ \hline
			10 & $(2,1)$ & $(2,2)$ \\ \hline
			11 & $(3,2)$ & $(2,2)$ \\ \hline
			12 & $(2,1)$ & $(1,1)$ \\ \hline \hline
			13 & $(1,1)$ & $(0,0)$ \\ \hline
			14 & $(2,1)$ & $(1,2)$ \\ \hline
			15 & $(1,1)$ & $(1,2)$ \\ \hline
			16 & $(1,1)$ & $(1,1)$
		\end{tabular}
		\caption{IR charges}
		\label{tbl:d5_from_A_1_maximal_symmetric_IR}
	\end{subfigure}
	\renewcommand{\figurename}{Figure \& Table}
	\caption{Maximal BPS spectrum of $\CS[A_{1}; \CD_\text{\rm reg}, \CD_{7}]$.}
	\label{figntbl:d5_from_A_1_maximal_symmetric_BPS}
\end{figure}

Consider $\CS[A_{1}; \CD_\text{\rm reg}, \CD_{7}]$. When only $v_1 \neq 0$ and all the other parameters vanish, we have a symmetric arrangement of branch points around the massless puncture, which results in the spectral network shown in Figure \ref{fig:d5_from_A_1_maximal_symmetric_SN}. From the finite $\CS$-walls shown in Figure \ref{fig:d5_from_A_1_maximal_symmetric_finite}, we can find the maximal, symmetric BPS spectrum of $\CS[A_{1}; \CD_\text{\rm reg}, \CD_{7}]$, described in Figure 
\& Table \ref{figntbl:d5_from_A_1_maximal_symmetric_BPS}, where each state is labeled such that $Z_{i+4} = Z_i e^{i\pi/4}$. Between the minimal and the maximal BPS spectra there is a series of wall-crossings relating the two spectra.

\subsection{$\CS[A_{N-1}; \CD_{{\rm II}}]$ Theories}
\label{sec:S[A_N-1,C;D_II]}
  Here we study the BPS spectrum of the SCFT associated with the sphere with one irregular puncture
  of degree \eqref{IIpole}.\footnote{When $N=3$, the degree of $\phi_2$ at the irregular puncture is $6$, so we have $\CC_{(6,8)}$, which is an exception compared to $N>3$.}
  We claim that this class of SCFTs is the same as $\CS[A_{1}; \CD_\text{\rm reg}, \CD_{N+3}]$,
  namely the maximal conformal point of $\CN=2$ $\SU(N)$ gauge theory with two flavors.
  Indeed, as we will see in appendix \ref{sec:II2flavors},
  staring from the $N$-sheeted cover form of the Seiberg-Witten curve of the latter theory
  we can obtain the irregular singularity as described above.

  The Seiberg-Witten curve of this theory is
  again of canonical form 
\begin{align}  
  x^{N} + \sum_{i=2}^{N} \phi_{i} x^{N-i} = 0,
\end{align}
where
    \bea
    \phi_{i}
    &=&    c_{i},~~(i=2, \ldots, [\frac{N+1}{2}]),
           \nonumber \\
    \phi_{i}
    &=&    v_{N-i+1},~~(i=[\frac{N+1}{2}]+1, \ldots, N-2),
           \nonumber \\
    \phi_{N-1}
    &=&    t^{2} + v_{2},~~~
    \phi_{N}
     =     c_{1} t^{2} + C_{2} t + v_{1}.
    \eea
  The dimensions of the parameters are easily obtained as
    \bea
    \Delta(v_{i})
     =     2 - \frac{2i}{N+1},~~~
    \Delta(c_{i})
     =     \frac{2i}{N+1},~~~
    \Delta(C_{1})
     =     \Delta(C_{2})
     =     1,
    \eea
  for $i=1, \ldots, [N/2]$,
  where $C_{1} := c_{[(N+1)/2]}$ with dimension-one exists only when $N$ is odd.
  Note that the sum of dimensions of $v_{k}$ and $c_{k}$ is $2$.
  This is the same set of operators as that of $\CS[A_{1}; \CD_\text{\rm reg}, \CD_{N+3}]$.

\subsubsection{$\CS[A_{2}; \CD_{{\rm II}}]$ in $D_4$-class}
\label{sec:S[A_2,C;D_II]}
We first study $\CS[A_{2}; \CD_{{\rm II}}]$, which has a couple of interesting features 
because of its $\SU(3)$ flavor symmetry. The curve is $x^{3} + \phi_{2} x + \phi_{3} = 0$ with
    \bea
    \phi_{2} = t^{2} + C_{1}, ~~~
    \phi_{3} = c_1 t^{2} + C_{2} t + v_1,
    \eea
 where the scaling dimensions of the parameters are
 $\Delta(v_1) = \frac{3}{2}$, $\Delta(c_1) = \frac{1}{2}$ and $\Delta(C_{1})= \Delta(C_{2})= 1$.
This is the same spectrum of operators as that of $\CS[A_{1};\CD_\text{\rm reg},\CD_6]$.
Note that we can only see a $\U(1)^{2}$ flavor symmetry whose mass parameters are combinations of $C_{1}$, 
$C_{2}$ and $c_{1}^{~2}$.
We propose that this is enhanced to $\SU(3)$, which is the case of $\CS[A_{1};\CD_\text{\rm reg},\CD_6]$.

Each $\lambda_{ij}$ has a residue at the irregular puncture $t = \infty$, which are of the form
\begin{align}
	\text{Res} \left( \lambda_{ij}(t), \infty \right) = \alpha_{ij} C_1 + \beta_{ij} C_2 + \gamma_{ij} {c_1}^2,
\end{align}
where $\alpha_{ij}$, $\beta_{ij}$, and $\gamma_{ij}$ are numerical coefficients. The residues are mass parameters for the $\SU(3)$ flavor symmetry. By requiring the mass parameters to vanish, we find the relations between the three parameters
\begin{align}
	C_1 = - \frac{3}{4}{c_1}^2,\ 	C_2 = 0,
\end{align}
which ensures the theory to have the maximal flavor symmetry. From now on we will fix $C_1$ and $C_2$ to satisfy the above relations.

Now we have two complex parameters $c_1$ and $v_1$ that can be changed. 
The discriminant of the equation that describes the branch points is
\begin{align}
	\Delta_t g (c_1,v_1) \propto \left( v_1 - \frac{{c_1}^3}{4}  \right)^3 \left( v_1 + \frac{{c_1}^3}{4}  \right) \left( v_1 + \frac{3}{4} {c_1}^3 \right)^6.
\end{align}
The choice of $\frac{v_1}{{c_1}^3} = a_1 = \frac{1}{4}$ corresponds to the singularity where we have a massless triplet of the flavor $\SU(3)$, whereas the choice of $\frac{v_1}{{c_1}^3} = a_2 = -\frac{1}{4}$ results in a massless singlet. The third choice, $\frac{v_1}{{c_1}^3} = a_3 = -\frac{3}{4}$ does not correspond to any singularity but gives us a branch point of index 3.

\begin{figure}[t]
	\centering
	\begin{subfigure}[b]{.3\textwidth}	
		\centering
		\includegraphics[width=\textwidth]{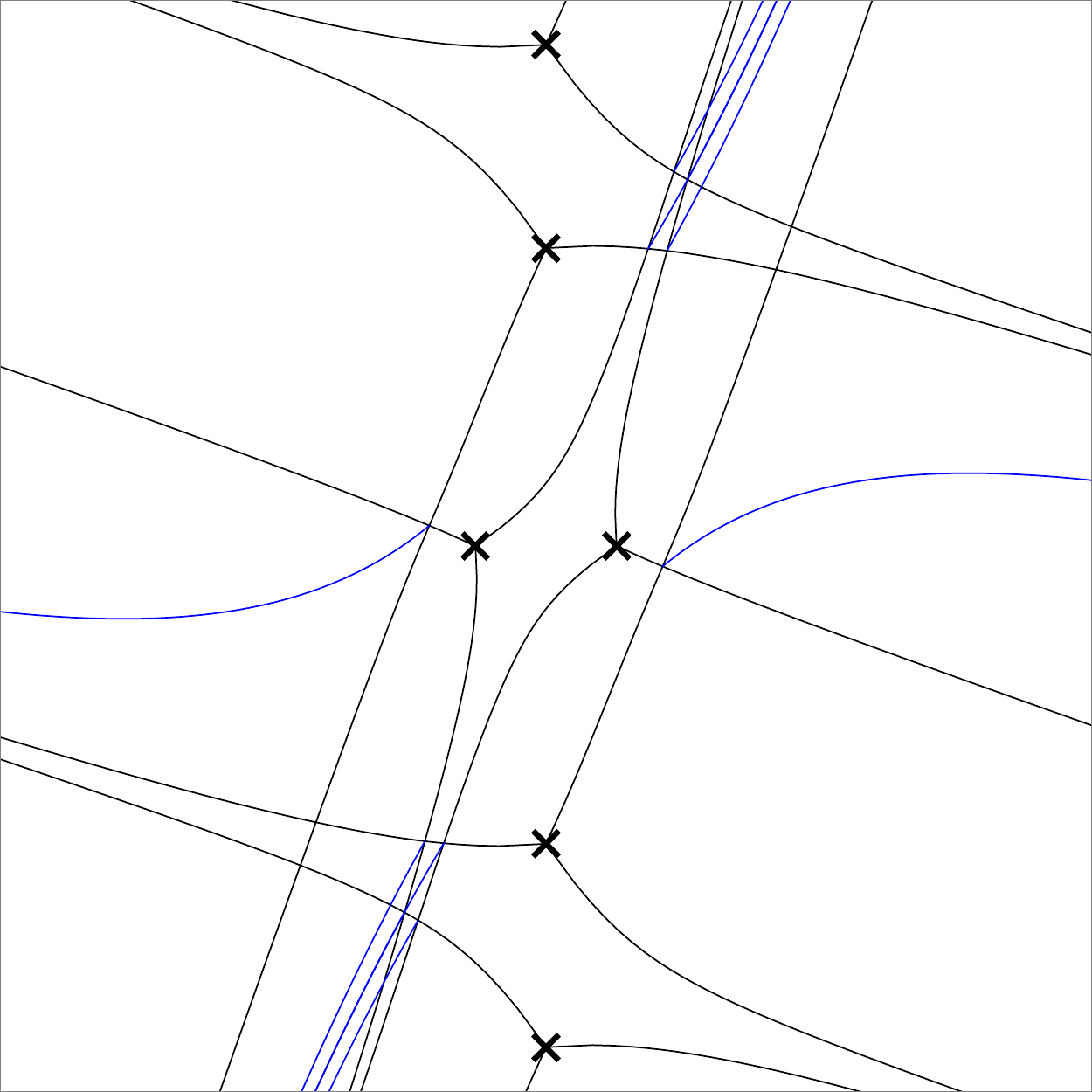}
		\vspace{1em}
		\caption{spectral network}
		\label{AD_of_SU_3_N_f_2_SU_3_f_minimal_SN_30}
	\end{subfigure}
	\begin{subfigure}[b]{.18\textwidth}	
		\centering
		\includegraphics[width=\textwidth]{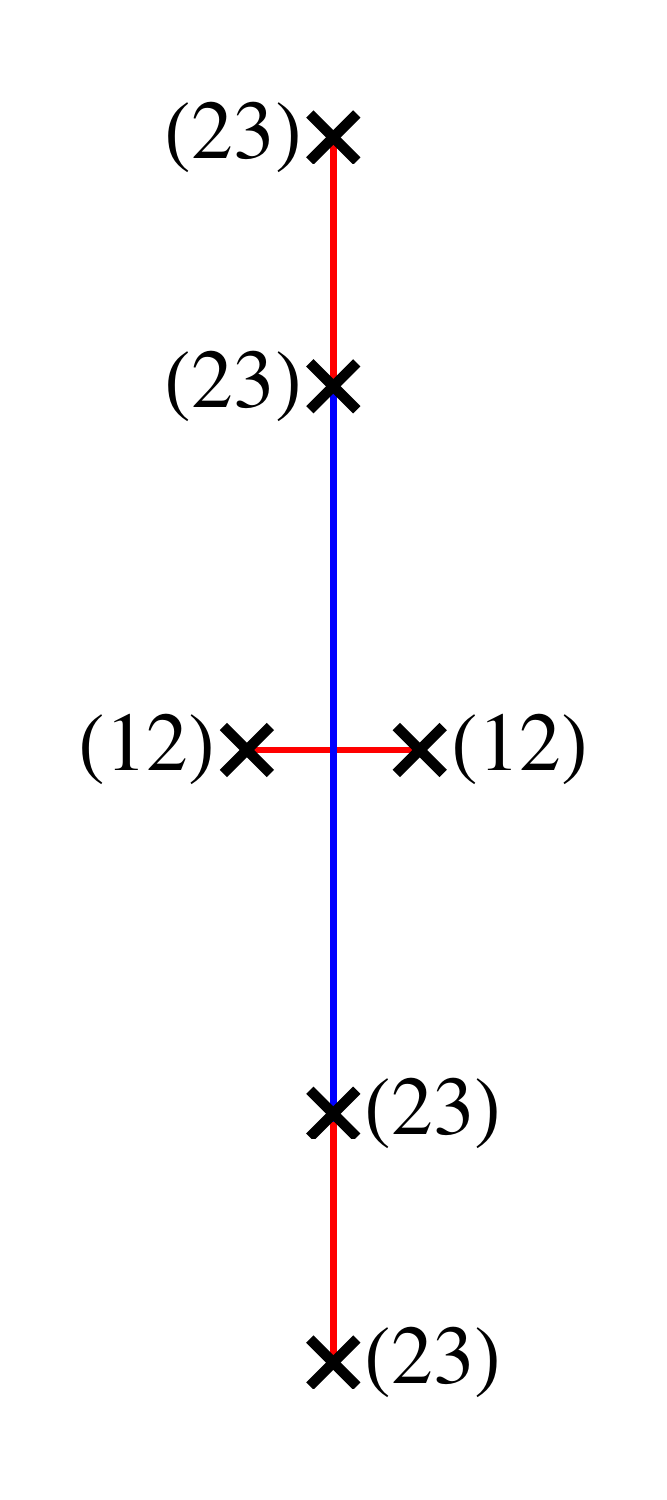}
		\caption{finite $\CS$-walls}
		\label{AD_of_SU_3_N_f_2_SU_3_f_minimal_finite}
	\end{subfigure}
	\caption{Spectral network of $\CS[A_{2}; \CD_{{\rm II}}]$ with a minimal BPS spectrum.}
	\label{fig:AD_of_SU_3_N_f_2_SU_3_f_minimal_SN}
\end{figure}

\paragraph{Minimal BPS spectrum}
Let us first consider the case of $a_2 < \frac{v_1}{{c_1}^3} < a_1$, when the theory has its minimal BPS spectrum. Its spectral network at a general value of $\theta$ and the resulting finite $\CS$-walls are shown in Figure \ref{fig:AD_of_SU_3_N_f_2_SU_3_f_minimal_SN}.

\begin{figure}[h]
	\centering	
	\begin{subfigure}[b]{.2\textwidth}	
		\centering
		\includegraphics[width=\textwidth]{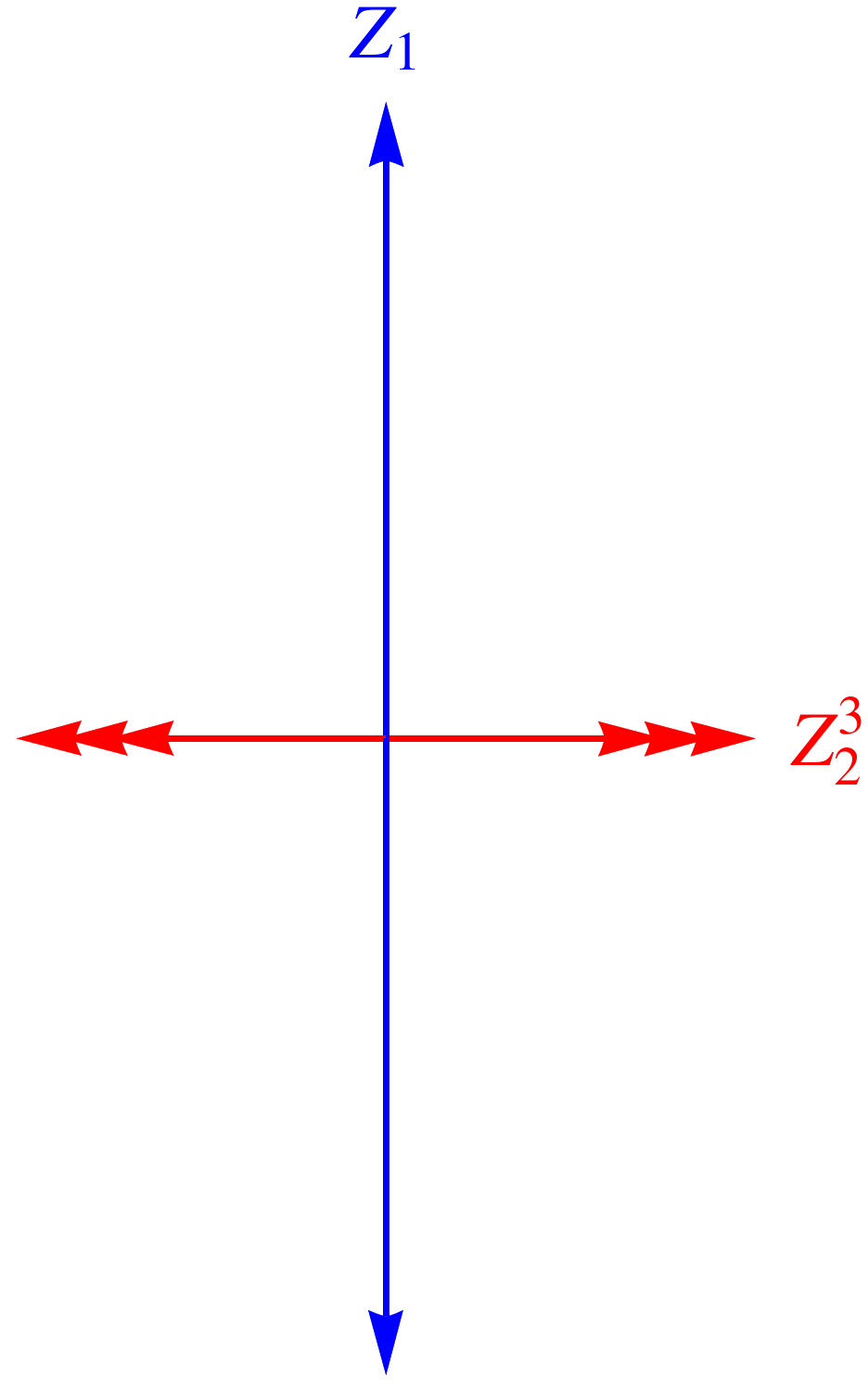}
		\caption{central charges}
		\label{fig:AD_of_SU_3_N_f_2_SU_3_f_minimal_Z}
	\end{subfigure}
	\begin{subfigure}[b]{.4\textwidth}	
		\centering
		\begin{tabular}{c|c}
		   state  & $(e,m)$ \\ \hline
		  1 & $(1,0)$ \\ \hline
		  2 & $(0,1)$
		\end{tabular}
		\vspace{4em}
		\caption{IR charges}
		\label{tbl:AD_of_SU_3_N_f_2_SU_3_f_minimal_IR}
	\end{subfigure}	
	\begin{subfigure}[b]{.15\textwidth}	
		\centering	
		\begin{tikzpicture}
			\node[W,red] (21) at (0,0) {2};
			\node[W,blue] (1) at (1,0) {1};
			\node[W,red] (22) at (1.5,1.732/2) {2};
			\node[W,red] (23) at (1.5,-1.732/2) {2};
			
			\path (1) edge[->] (21);
			\path (1) edge[->] (22);
			\path (1) edge[->] (23);	
		\end{tikzpicture}
		\vspace{2.5em}
		\caption{BPS quiver}
		\label{fig:AD_of_SU_3_N_f_2_SU_3_f_minimal_quiver}
	\end{subfigure}	
	\renewcommand{\figurename}{Figure \& Table}
	\caption{Minimal BPS spectrum of $\CS[A_{2}; \CD_{{\rm II}}]$.}
	\label{figntbl:AD_of_SU_3_N_f_2_SU_3_f_minimal_BPS}
\end{figure}

There are six branch points of index 2. Because the irregular puncture at $t = \infty$ is not a branch point for this case, the $(12)$-branch cut should terminate at two $(12)$-branch points without intersecting two $(23)$-branch cuts connecting each pair of $(23)$-branch points.
Thus we can figure out the intersections of the cycles corresponding to the finite $\CS$-walls. The finite $\CS$-walls corresponding to the BPS states of the triplet correspond to the same cycle, say A-cycle, of the Seiberg-Witten curve, which is a genus-1 curve in this case. The other finite $\CS$-wall corresponding to the singlet is a B-cycle that has intersection number 1 with the A-cycle. This intersection corresponds to the IR $\UU(1)$ charges described in Table \ref{tbl:AD_of_SU_3_N_f_2_SU_3_f_minimal_IR}, and we can represent this BPS spectrum with a BPS quiver of $D_4$ as shown in Figure \ref{fig:AD_of_SU_3_N_f_2_SU_3_f_minimal_quiver}, which illustrates the $\SU(3)$ flavor symmetry. This is the same BPS spectrum as that of the minimal BPS spectrum of $\CS[A_{1}; \CD_\text{\rm reg},\CD_{6}]$, see Figure \ref{fig:d4SN_SU_3_f_minimal_BPS} and the first row of Figure \ref{fig:d4SN_wall_crossing_SU_3_f}.
	
\paragraph{Wall-crossing to the maximal BPS spectrum}

When $\frac{v_1}{{c_1}^3}$ approaches the value of $a_1$, each pair of two branch points of the same indices collides, thereby giving a massless triplet. From the consideration of the wall-crossing of $\CS[A_{1}; \CD_\text{\rm reg},\CD_{6}]$ with the $\SU(3)$ flavor symmetry that we studied previously in Section \ref{sec:S[A_1,C;D_reg,D_6]}, we expect $\CS[A_{2}; \CD_{{\rm II}}]$ to have a maximal BPS spectrum as we go across the BPS wall. When the value of $\frac{v_1}{{c_1}^3}$ approaches $a_2$, now it's the singlet that becomes massless, and again a similar wall-crossing will give us a maximal BPS spectrum after we go over the BPS wall. Therefore when we consider the plane of the value of $\frac{v_1}{{c_1}^3}$, there is one chamber of the minimal BPS spectrum and the rest is another chamber of the maximal BPS spectrum, and the BPS wall goes through $\frac{v_1}{{c_1}^3} = a_1$ and $\frac{v_1}{{c_1}^3} = a_2$. 

\begin{figure}[t]
	\centering
	\begin{subfigure}{.27\textwidth}	
		\includegraphics[width=\textwidth]{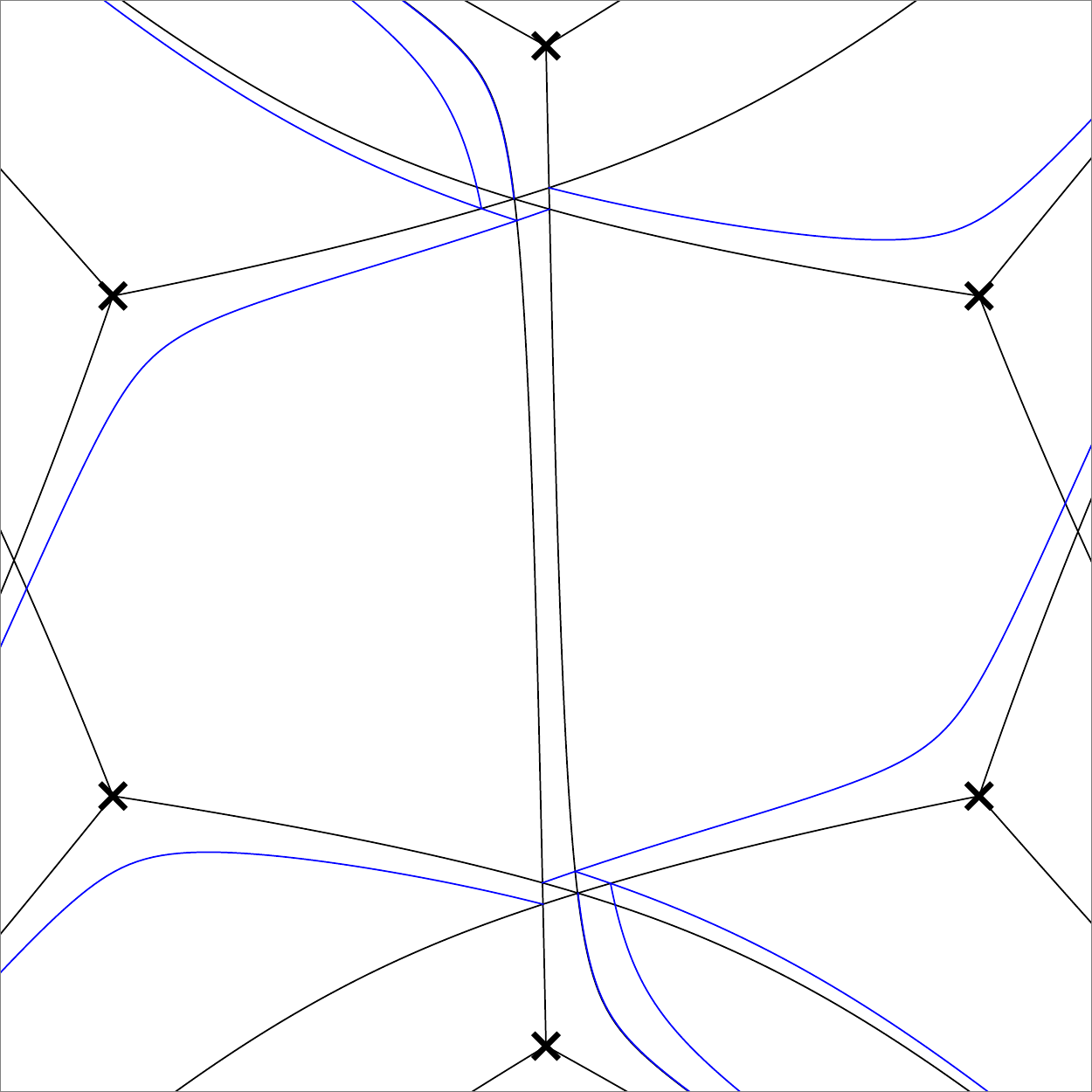}
		\caption{near $\theta = \arg(Z_{1}^{\mathbf{3}})$}
		\label{fig:AD_of_SU_3_N_f_2_SU_3_f_maximal_symmetric_SN_near_triplet}
	\end{subfigure}
	\begin{subfigure}{.27\textwidth}	
		\includegraphics[width=\textwidth]{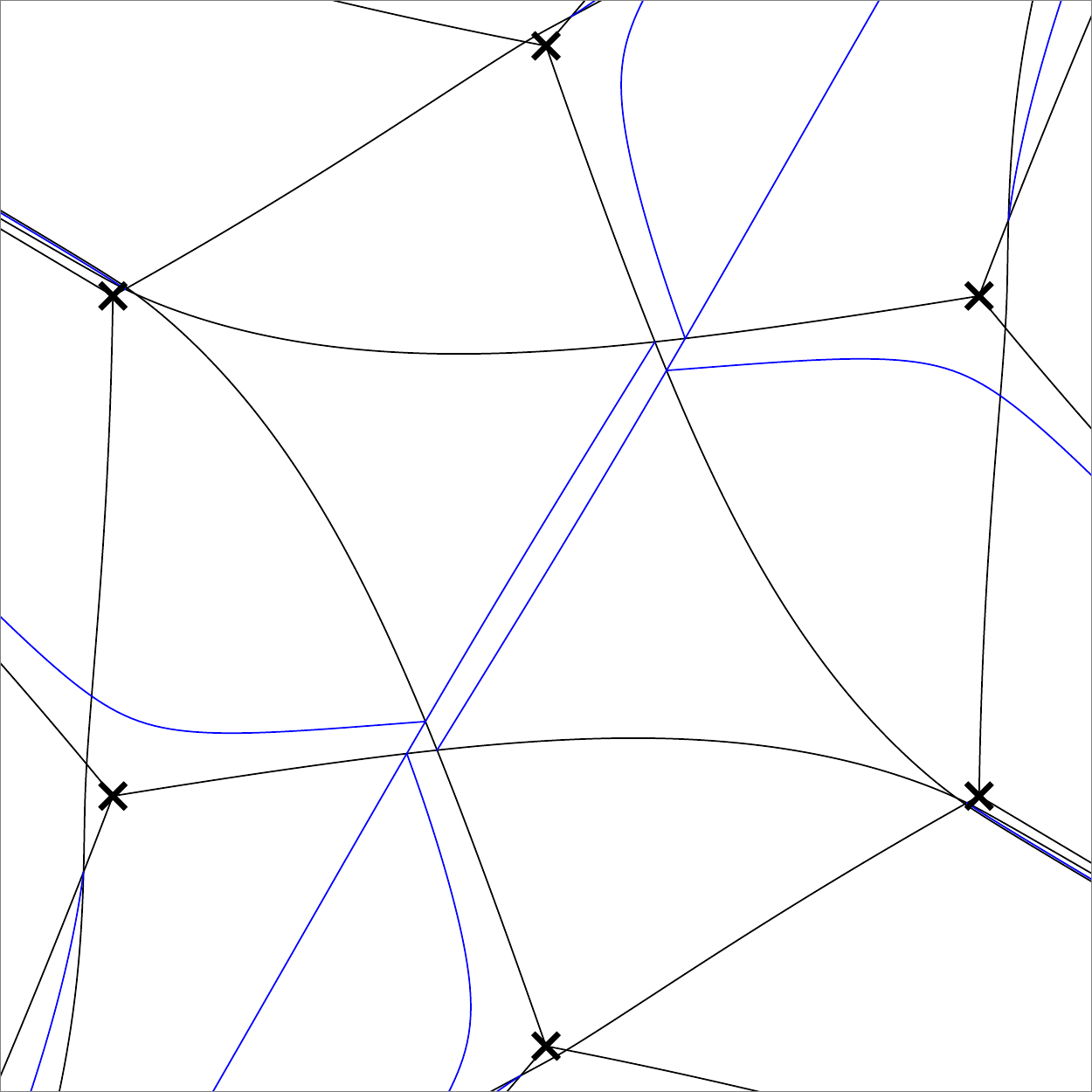}
		\caption{near $\theta = \arg(Z_{2})$}
		\label{fig:AD_of_SU_3_N_f_2_SU_3_f_maximal_symmetric_SN_near_singlet}
	\end{subfigure}
	\begin{subfigure}{.27\textwidth}	
		\includegraphics[width=\textwidth]{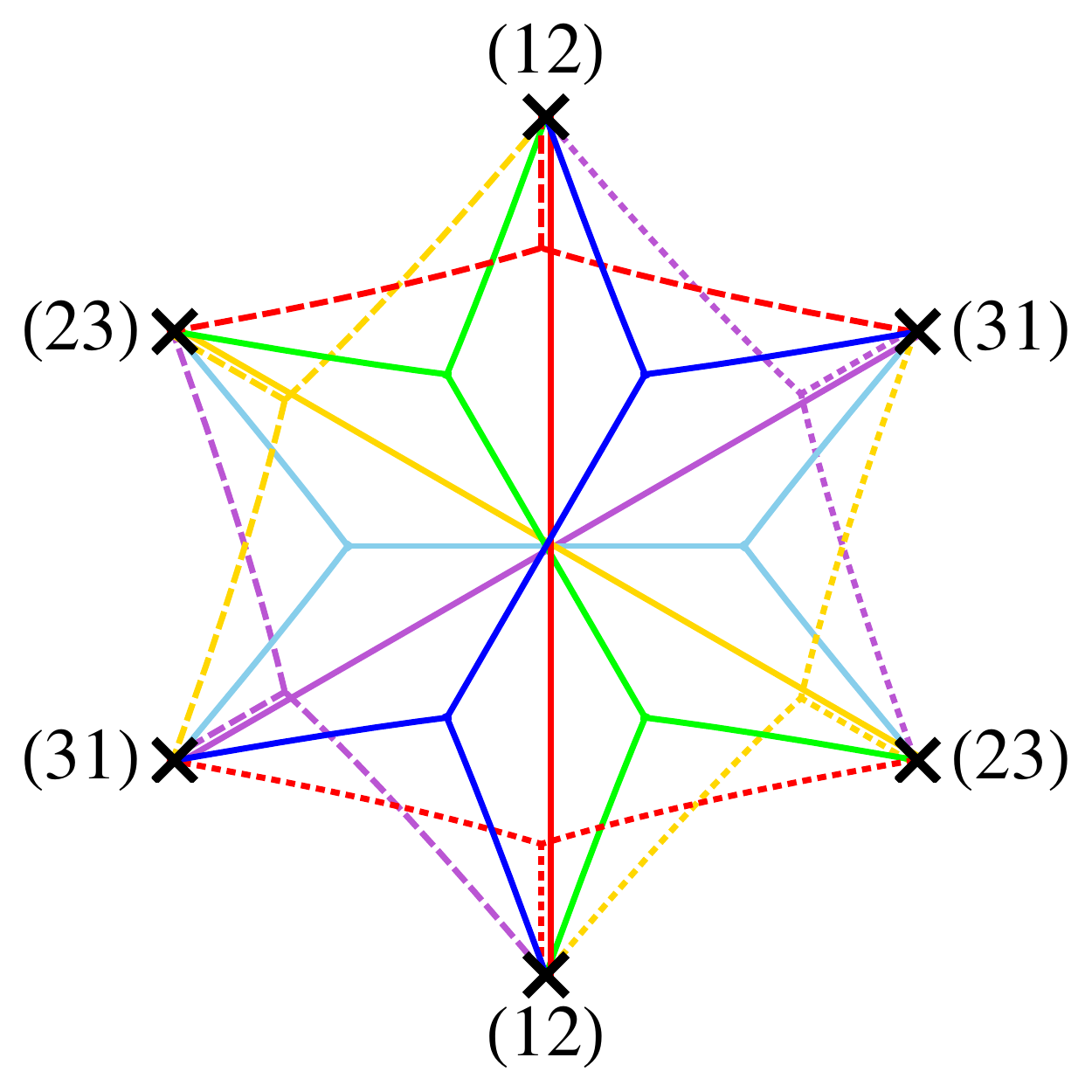}
		\caption{finite $\CS$-walls}
		\label{fig:AD_of_SU_3_N_f_2_SU_3_f_maximal_symmetric_finite}
	\end{subfigure}

	\caption{spectral network of $\CS[A_{2}; \CD_{{\rm II}}]$ with a maximal \& symmetric BPS spectrum.}
	\label{fig:AD_of_SU_3_N_f_2_SU_3_f_maximal_symmetric_SN}	
\end{figure}

Deep in the chamber of the maximal BPS spectrum is the point $c = C_1 = C_2 = 0$, $v \neq 0$, where we have a symmetric arrangement of branch points that leads to a symmetric BPS spectrum. Figure \ref{fig:AD_of_SU_3_N_f_2_SU_3_f_maximal_symmetric_SN} shows two examples of its spectral network, one near the value of $\theta$ for an $\SU(3)$ triplet and the other for a singlet. 

\begin{figure}[h]
	\centering
	\begin{subfigure}[b]{.3\textwidth}	
		\includegraphics[width=\textwidth]{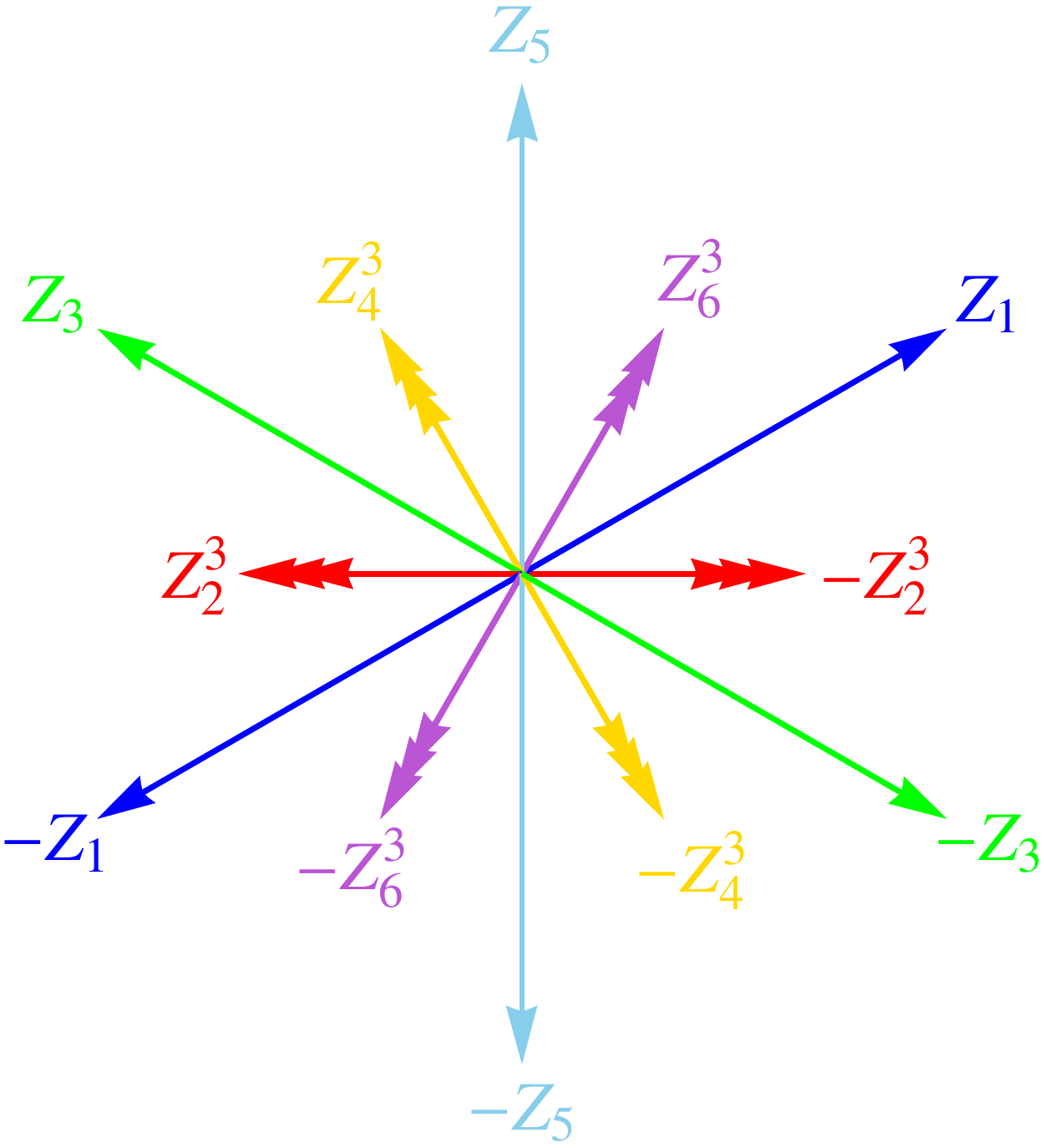}
		\caption{central charges}
		\label{fig:AD_of_SU_3_N_f_2_SU_3_f_maximal_symmetric_Z}
	\end{subfigure}
	\begin{subfigure}[b]{.35\textwidth}	
		\centering
		\begin{tabular}{c|c}
			state & $(e,m)$ \\ \hline
			$1$ & $(1,0)$ \\ \hline
			$2$ & $(0,1)$ \\ \hline
			$3$ & $(1,3)$ \\ \hline			
			$4$ & $(1,2)$ \\ \hline
			$5$ & $(2,3)$ \\ \hline
			$6$ & $(1,1)$  		
		\end{tabular}
		\vspace{2em}
		\caption{IR charges}
		\label{tbl:AD_of_SU_3_N_f_2_SU_3_f_maximal_symmetric_IR}
	\end{subfigure}
	\renewcommand{\figurename}{Figure \& Table}
	\caption{Maximal \& symmetric BPS spectrum of $\CS[A_{2}; \CD_{{\rm II}}]$.}
	\label{figntbl:AD_of_SU_3_N_f_2_SU_3_f_maximal_symmetric_BPS}
\end{figure}

Figure \ref{fig:AD_of_SU_3_N_f_2_SU_3_f_maximal_symmetric_Z} and Table \ref{tbl:AD_of_SU_3_N_f_2_SU_3_f_maximal_symmetric_IR} describe the maximal symmetric BPS spectrum that has three triplets and three singlets, which can be identified with the maximal, symmetric BPS spectrum of $\CS[A_{1}; \CD_\text{\rm reg},\CD_{6}]$, see the last row of Figure \ref{fig:d4SN_wall_crossing_SU_3_f}. Upon suitable choice of A and B-cyles, the BPS states in the spectrum have the same electric and magnetic charges as those in the maximal BPS spectrum of $\CS[A_{1}; \CD_\text{\rm reg},\CD_{6}]$, compare Table \ref{tbl:AD_of_SU_3_N_f_2_SU_3_f_maximal_symmetric_IR} and Table \ref{tbl:d4SN_SU_3_maximal_IR}.

\subsubsection{Exactly Marginal Deformation of \texorpdfstring{$\CS[A_{2}; \CD_{\rm II}]$}{S[A_2, C;D_II]}}
When and only when $N=3$, we can add a marginal deformation to the theory at the fixed point. 
The corresponding curve is 
    \bea
    \phi_{2}
     =     t^{2} + C_{1},~~~
    \phi_{3}
     =     t^{3} + c_1 t^{2} + C_{2} t + v_1.
    \eea
Compared to the previous curve of $\CS[A_{2}, \CC; D_{\rm II}]$ there is an additional $t^3$ term
whose coefficient is dimensionless which we have fixed to $1$.
We claim this corresponds to an exactly marginal deformation 
by showing that the BPS spectrum of the SCFT and its wall-crossing is the same as those of the theory 
without the deformation.

By requiring the residue of $\lambda_{ij}$ to vanish, we find the relations between the three parameters such that the theory has the maximal $\SU(3)$ flavor symmetry are $C_1 = - \frac{3}{31}{c_1}^2$ and $C_2 = \frac{9}{31}{c_1}^2$.
From now on we will fix $C_1$ and $C_2$ to satisfy the above relations.
The discriminant for the given Seiberg-Witten curve is
\begin{align}
	\Delta_t g (c_1,v_1) \propto  \left( v_1 - \frac{23{c_1}^3}{31^2}  \right) \left( v_1 - \frac{{c_1}^3}{31} \right)  \left( v_1 - \frac{3{c_1}^3}{31^2}(4\sqrt{93} - 31) \right) \left( v_1 + \frac{3{c_1}^3}{31^2}(4\sqrt{93} + 31) \right).
\end{align}
The choice of $\frac{v_1}{{c_1}^3} = a_1 = \frac{23}{31^2}$ corresponds to the singularity where we have a massless triplet of the flavor $\SU(3)$, whereas the choice of $\frac{v_1}{{c_1}^3} = a_2 = \frac{1}{31}$ results in a massless singlet. The other two choices, $\frac{v_1}{{c_1}^3} = a_\pm = \pm \frac{3}{31^2}(4\sqrt{93} \mp 31)$ do not correspond to any singularity but give us a branch point of index 3. Therefore the singularity structure is the same as that of $\CS[A_{2}, \CC; D_{\rm II}]$, and we observe the same wall-crossing.

\begin{figure}[t]
	\centering
	\begin{subfigure}[b]{.27\textwidth}	
		\centering
		\includegraphics[width=\textwidth]{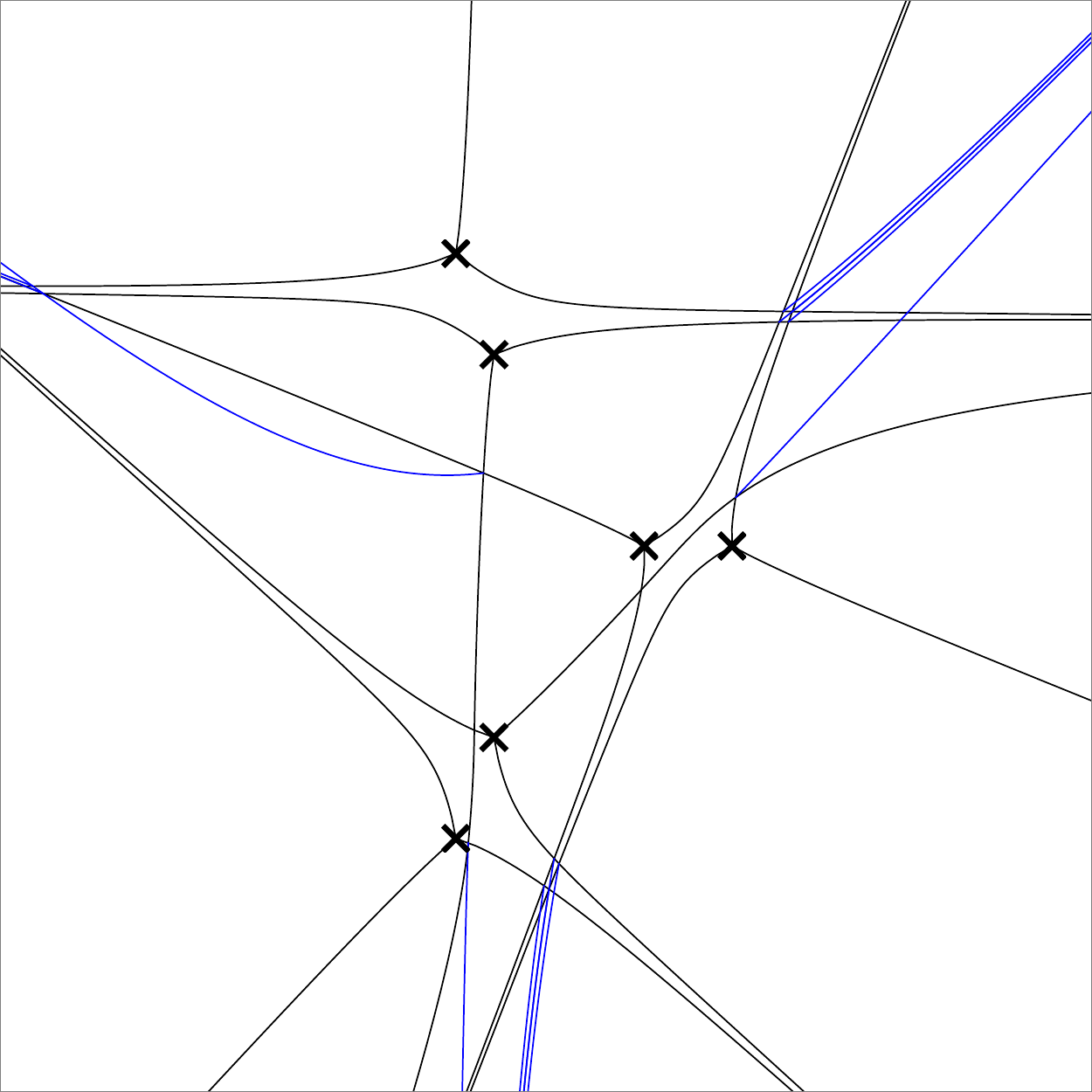}
		\vspace{1em}
		\caption{at a general $\theta$}
		\label{fig:AD_of_SU_3_N_f_2_w_3_SU_3_f_SN_26}
	\end{subfigure}
	\begin{subfigure}[b]{.27\textwidth}	
		\centering
		\includegraphics[width=\textwidth]{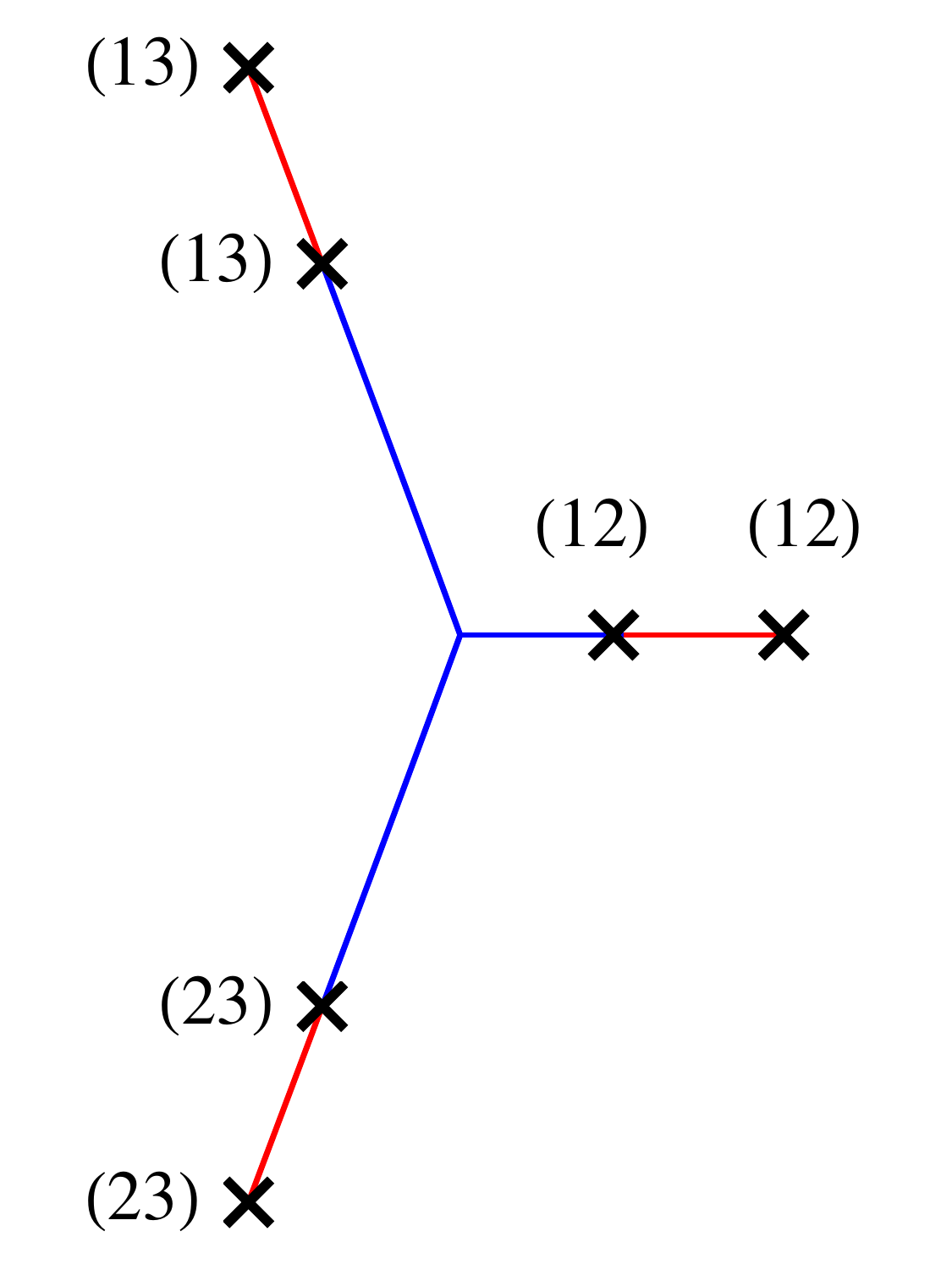}
		\caption{finite $\CS$-walls}
		\label{fig:AD_of_SU_3_N_f_2_w_3_SU_3_f_minimal_finite}
	\end{subfigure}
	\caption{Spectral network of the deformation of $\CS[A_{2}; \CD_{\rm II}]$ by a $t^3$-term, with minimal BPS spectrum.}
	\label{fig:AD_of_SU_3_N_f_2_w_3_SU_3_f_minimal_SN}
\end{figure}

\begin{figure}[h]
	\centering	
	\begin{subfigure}[b]{.23\textwidth}	
		\centering
		\includegraphics[width=\textwidth]{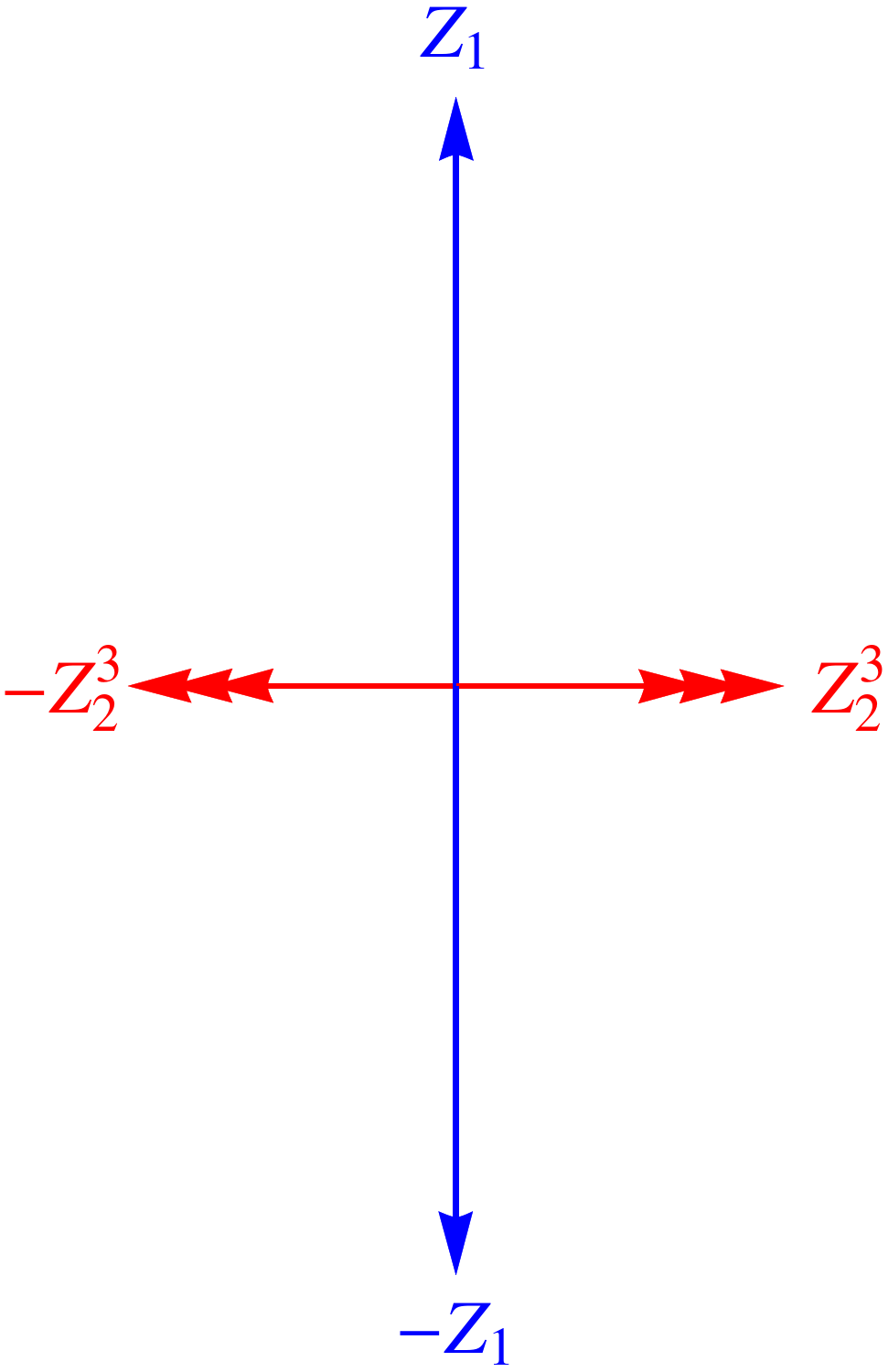}
		\caption{central charges}
		\label{fig:AD_of_SU_3_N_f_2_w_3_SU_3_f_minimal_Z}
	\end{subfigure}
	\begin{subfigure}[b]{.4\textwidth}	
		\centering
		\begin{tabular}{c|c}
		   state  & $(e,m)$ \\ \hline
		  1 & $(1,0)$ \\ \hline
		  2 & $(0,1)$
		\end{tabular}
		\vspace{6em}
		\caption{IR charges}
		\label{tbl:AD_of_SU_3_N_f_2_w_3_SU_3_f_minimal_IR}
	\end{subfigure}	
	\begin{subfigure}[b]{.15\textwidth}	
		\centering	
		\begin{tikzpicture}
			\node[W,red] (21) at (0,0) {2};
			\node[W,blue] (1) at (1,0) {1};
			\node[W,red] (22) at (1.5,1.732/2) {2};
			\node[W,red] (23) at (1.5,-1.732/2) {2};
			
			\path (1) edge[->] (21);
			\path (1) edge[->] (22);
			\path (1) edge[->] (23);	
		\end{tikzpicture}
		\vspace{5em}
		\caption{BPS quiver}
		\label{fig:AD_of_SU_3_N_f_2_w_3_SU_3_f_minimal_quiver}
	\end{subfigure}	
	\caption{Minimal BPS spectrum of the deformation of $\CS[A_{2}; \CD_{\rm II}]$ by a $t^3$-term.}
	\label{figntbl:AD_of_SU_3_N_f_2_w_3_SU_3_f_minimal_BPS}
\end{figure}

\paragraph{Minimal BPS spectrum}
Let us first consider the case of $a_2 < \frac{v_1}{{c_1}^3} < a_1$. An example of its spectral network and the resulting finite $\CS$-walls are shown in Figure \ref{fig:AD_of_SU_3_N_f_2_w_3_SU_3_f_minimal_SN}. Indeed the configuration of the spectral network and the $\CS$-walls are different from what we had for $\CS[A_{2}; \CD_{\rm II}]$ without the $t^3$-term, but the BPS spectrum is the same, see Figure \ref{fig:AD_of_SU_3_N_f_2_w_3_SU_3_f_minimal_Z} and Table \ref{tbl:AD_of_SU_3_N_f_2_w_3_SU_3_f_minimal_IR}. Its BPS quiver is again a $D_4$ quiver as shown in Figure \ref{fig:AD_of_SU_3_N_f_2_w_3_SU_3_f_minimal_quiver}, which exhibits the $\SU(3)$ flavor symmetry.

\paragraph{Wall-crossing to a maximal BPS spectrum}
When we change the value of $v_1$ from $(a_1 - \delta) {c_1}^3$ to $(a_1 + \delta) {c_1}^3$, where $\delta$ is a small positive real number, we are on the other side of the BPS wall in the Coulomb branch moduli space, where the triplet went through the phase of becoming massless. 

\begin{figure}[t]
	\centering
	\begin{subfigure}{.27\textwidth}	
		\includegraphics[width=\textwidth]{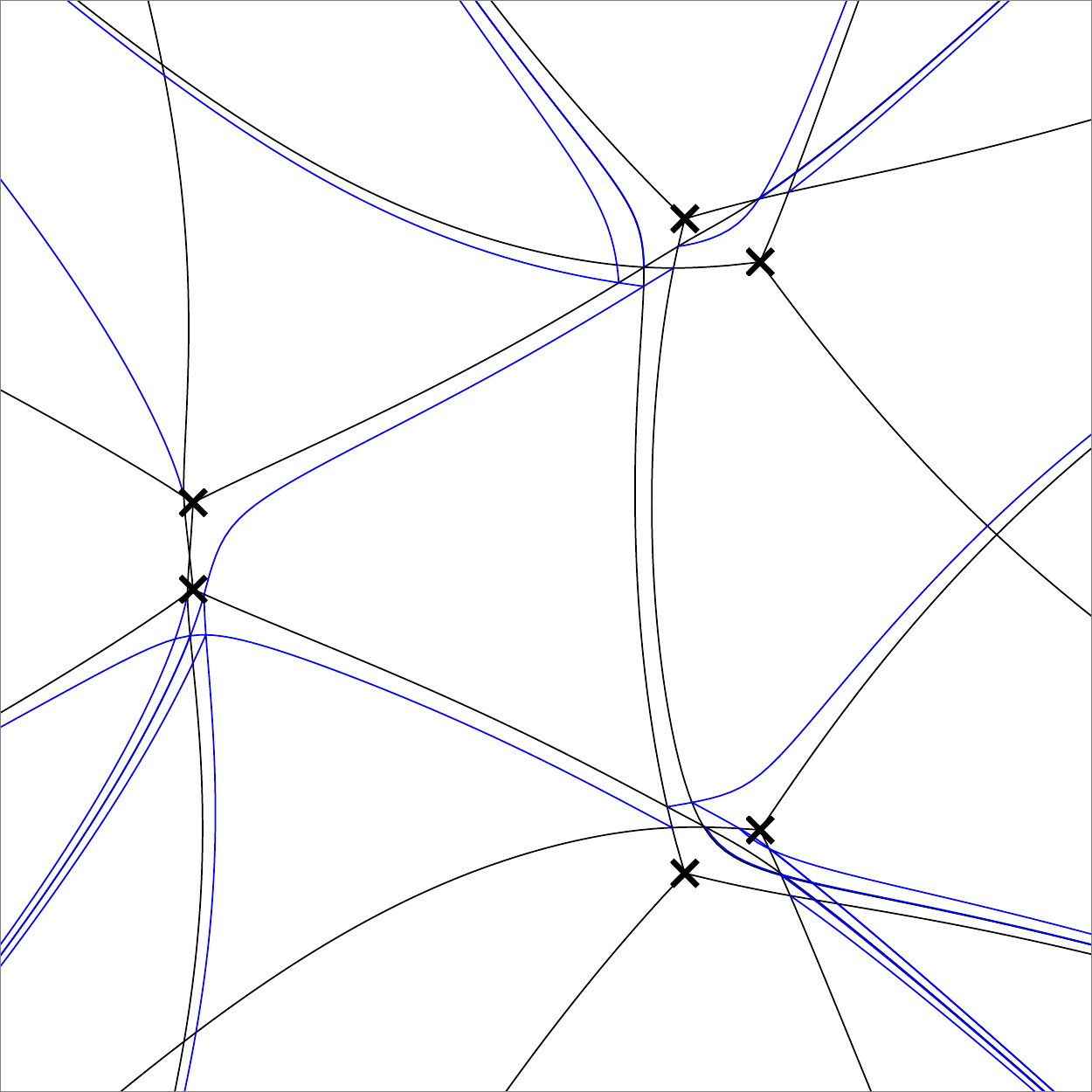}
		\caption{near $\theta = \arg(Z_{1}^{\mathbf{3}})$}
		\label{fig:AD_of_SU_3_N_f_2_w_3_SU_3_f_maximal_symmetric_SN_near_triplet}
	\end{subfigure}
	\begin{subfigure}{.27\textwidth}	
		\includegraphics[width=\textwidth]{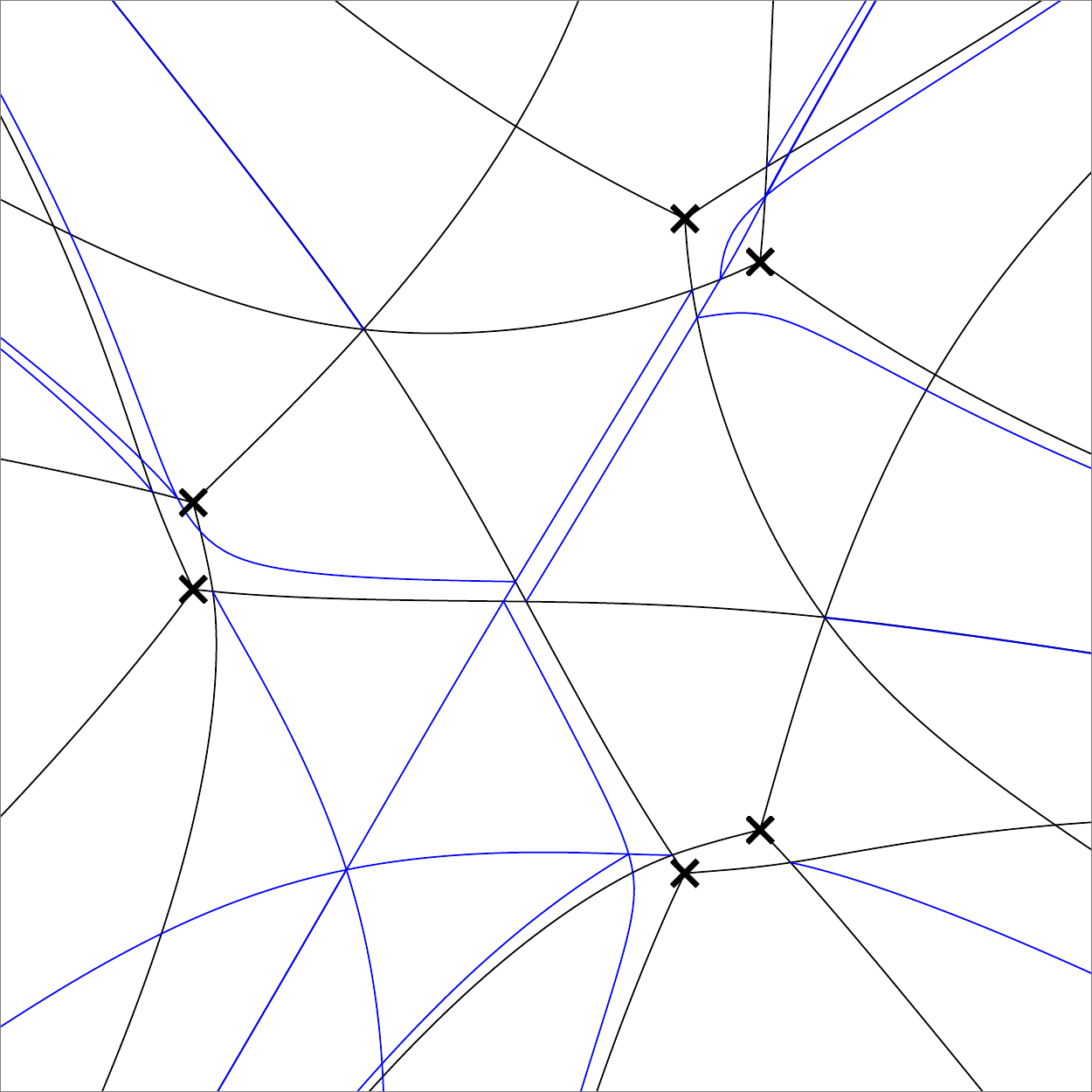}
		\caption{near $\theta = \arg(Z_{2})$}
		\label{fig:AD_of_SU_3_N_f_2_w_3_SU_3_f_maximal_symmetric_SN_near_singlet}
	\end{subfigure}
	\begin{subfigure}{.27\textwidth}	
		\includegraphics[width=\textwidth]{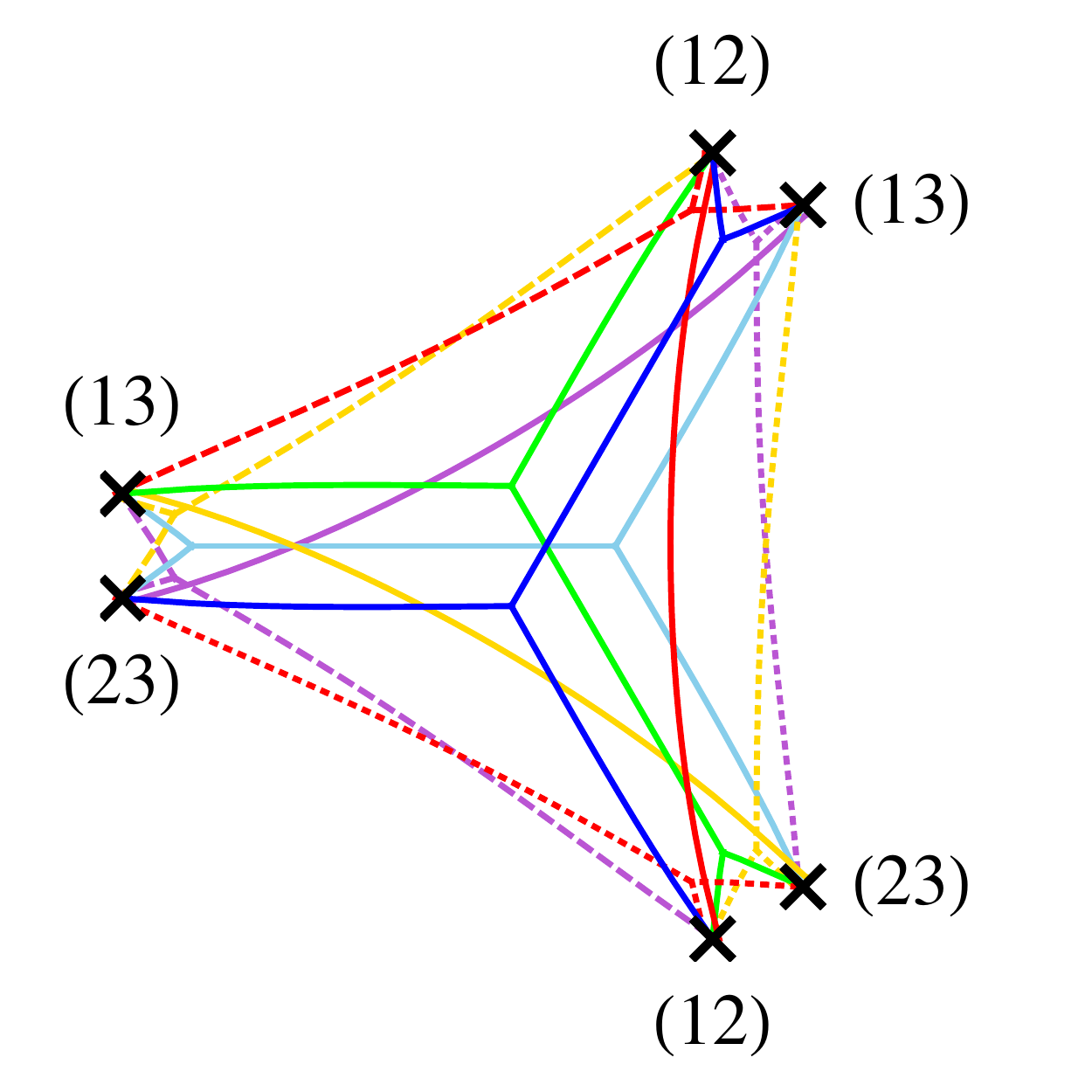}
		\caption{finite $\CS$-walls}
		\label{fig:AD_of_SU_3_N_f_2_w_3_SU_3_f_maximal_symmetric_finite}
	\end{subfigure}
	\caption{spectral network of the deformation of $\CS[A_{2}; \CD_{{\rm II}}]$ by $t^3$, with a maximal \& symmetric BPS spectrum.}
	\label{fig:AD_of_SU_3_N_f_2_w_3_SU_3_f_maximal_symmetric_SN}	
\end{figure}

When $c_1 = C_1 = C_2 = 0$, $v_1 \neq 0$ the deformed theory has the maximal, symmetric BPS spectrum, as $\CS[A_{2}; \CD_{\rm II}]$ did. Figures \ref{fig:AD_of_SU_3_N_f_2_w_3_SU_3_f_maximal_symmetric_SN_near_triplet} and \ref{fig:AD_of_SU_3_N_f_2_w_3_SU_3_f_maximal_symmetric_SN_near_singlet} show its spectral network when $\theta$ is close to having a triplet and a singlet, respectively. Again, spectral networks are different from those of the undeformed theory.

\begin{figure}[h]
	\centering
	\begin{subfigure}[b]{.3\textwidth}	
		\includegraphics[width=\textwidth]{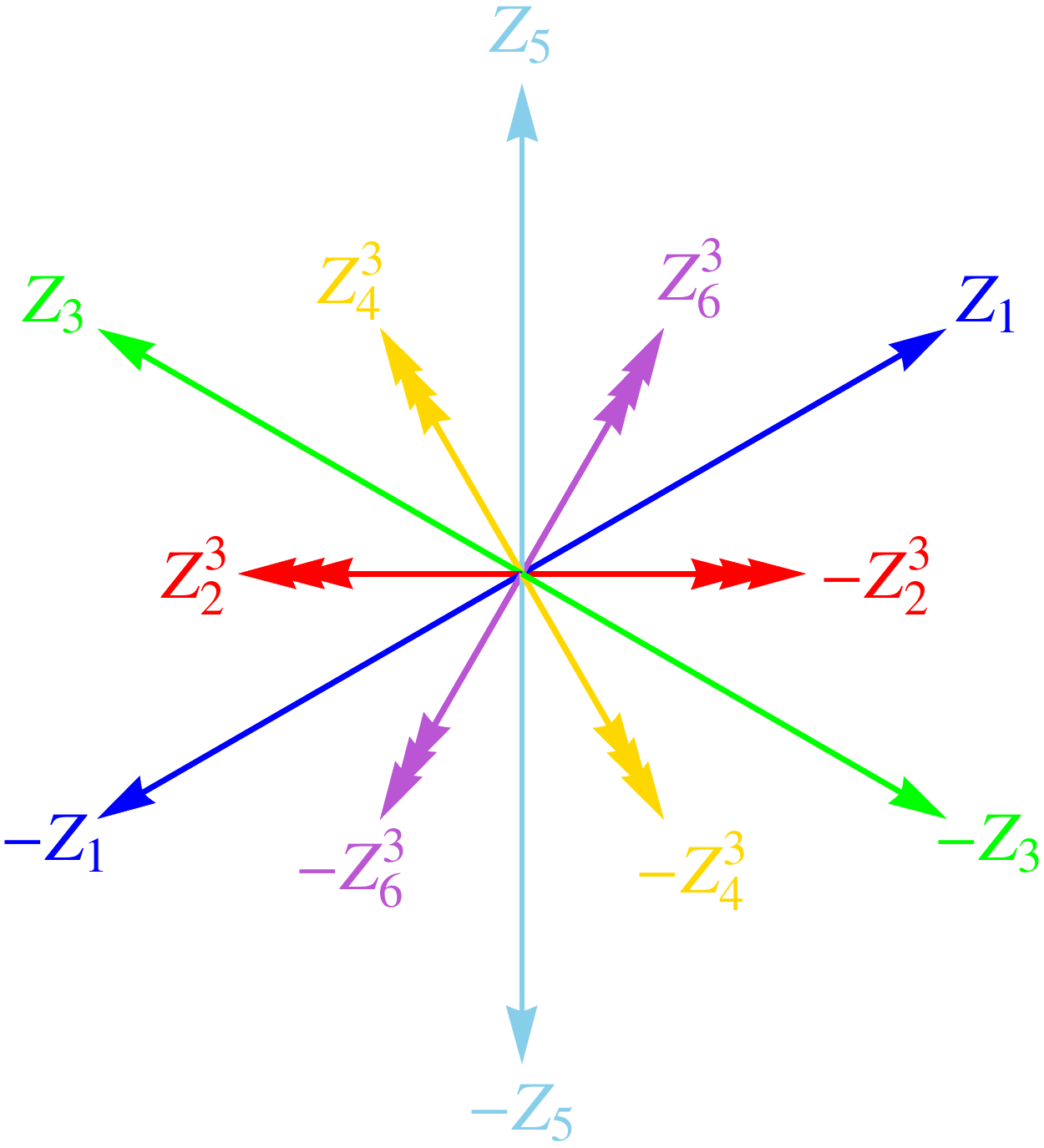}
		\caption{central charges}
		\label{fig:AD_of_SU_3_N_f_2_w_3_SU_3_f_maximal_symmetric_Z}
	\end{subfigure}
	\begin{subfigure}[b]{.35\textwidth}	
		\centering
		\begin{tabular}{c|c}
			state & $(e,m)$ \\ \hline
			$1$ & $(1,0)$ \\ \hline
			$2$ & $(0,1)$ \\ \hline
			$3$ & $(1,3)$ \\ \hline			
			$4$ & $(1,2)$ \\ \hline
			$5$ & $(2,3)$ \\ \hline
			$6$ & $(1,1)$ 
		\end{tabular}
		\vspace{2em}
		\caption{IR charges}
		\label{tbl:AD_of_SU_3_N_f_2_w_3_SU_3_f_maximal_symmetric_IR}
	\end{subfigure}
	\renewcommand{\figurename}{Figure \& Table}
	\caption{Maximal \& symmetric BPS spectrum of the deformed $\CS[A_{2};\CD_{\rm II}]$.}
	\label{figntbl:AD_of_SU_3_N_f_2_w_3_SU_3_f_maximal_symmetric_BPS}
\end{figure}

Figure \ref{fig:AD_of_SU_3_N_f_2_w_3_SU_3_f_maximal_symmetric_Z} and Table \ref{tbl:AD_of_SU_3_N_f_2_w_3_SU_3_f_maximal_symmetric_IR} describe the BPS spectrum of the deformed theory, which can be identified with the maximal, symmetric BPS spectrum of $\CS[A_{2}; \CD_{{\rm II}}]$, see Figure \ref{fig:AD_of_SU_3_N_f_2_SU_3_f_maximal_symmetric_Z} and Table \ref{tbl:AD_of_SU_3_N_f_2_SU_3_f_maximal_symmetric_IR}. Although the two theories have different spectral networks, their BPS spectra agree and this is good evidence for the claim that $t^3$-term corresponds to an exactly marginal deformation for the 4d SCFT.

\subsubsection{$\CS[A_{N-1};\CD_{\rm II}]$ in $D_{N+1}$-class}
Now we want to consider $\CS[A_{N-1};\CD_{\rm II}]$ with general $N$ and show its BPS spectra and their wall-crossings are the same as those of $\CS[A_1;\CD_\text{\rm reg},\CD_{N+3}]$. One difference from $N=3$ case is that when $N>3$ the maximal flavor symmetry is $\SU(2)$ (or $\SU(2) \times \UU(1)$ when $N$ is odd), 
and we only have doublets rather than triplets of $\SU(3)$ that exist in the BPS spectrum of $\CS[A_{2};\CD_{\rm II}]$.

\begin{figure}[t]
	\centering
	\begin{subfigure}[b]{.27\textwidth}	
		\centering
		\includegraphics[width=\textwidth]{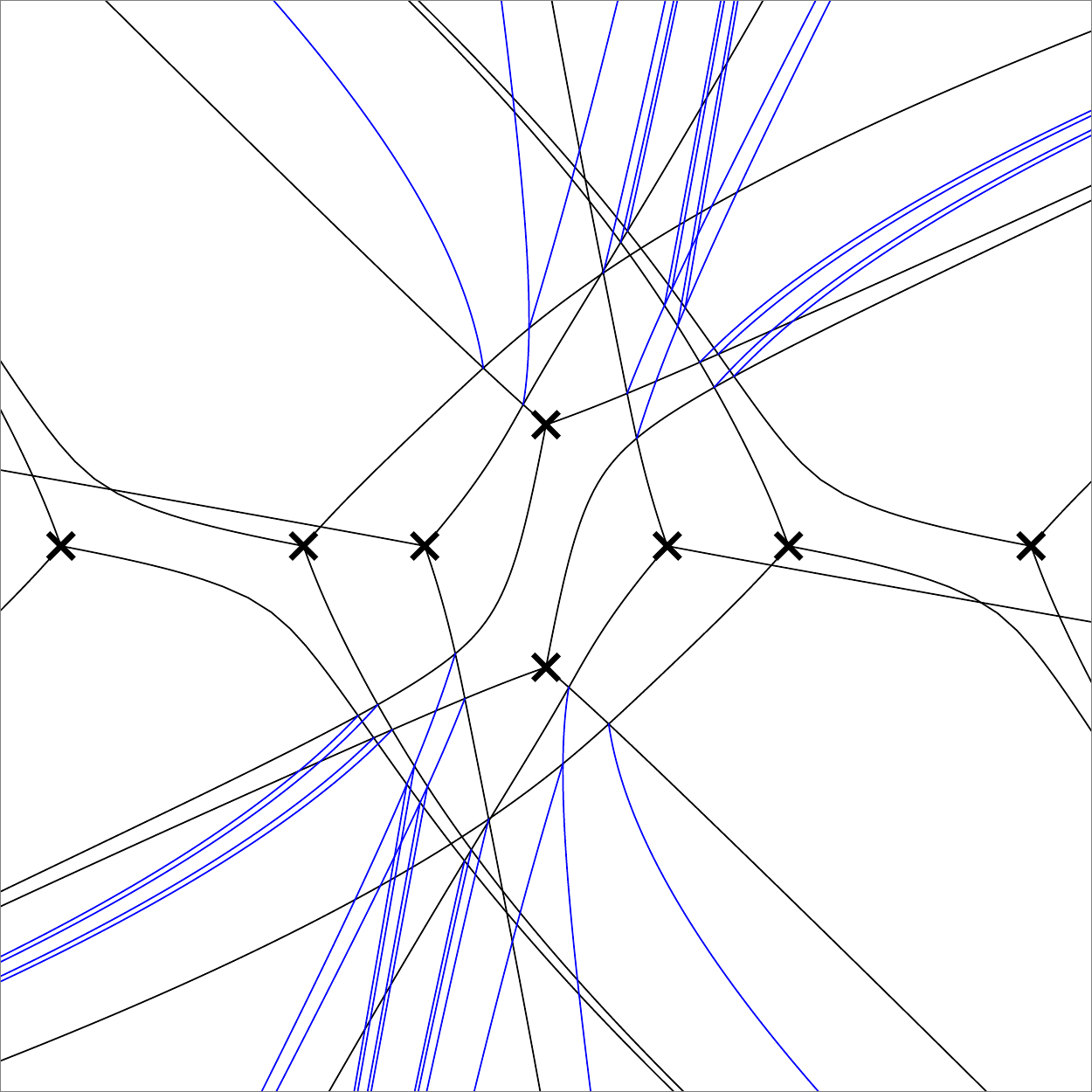}
		\caption{general $\theta$}
		\label{fig:AD_of_SU_4_N_f_2_minimal_SN_42}
	\end{subfigure}
	\begin{subfigure}[b]{.5\textwidth}	
		\centering
		\includegraphics[width=\textwidth]{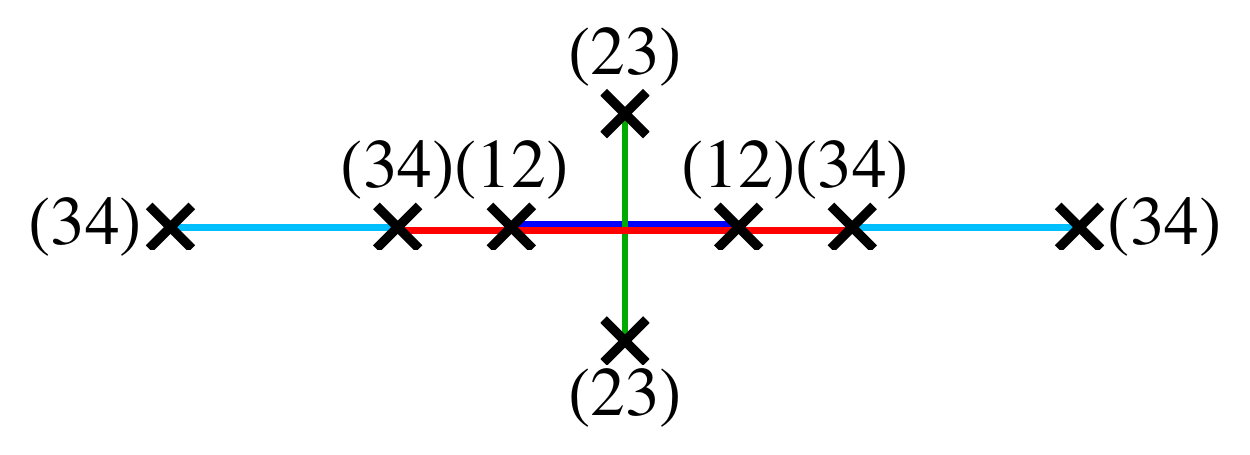}
		\vspace{1em}
		\caption{finite $\CS$-walls}
		\label{fig:AD_of_SU_4_N_f_2_minimal_finite}
	\end{subfigure}
	\caption{Spectral network of $\CS[A_{3};\CD_{\rm II}]$ with minimal BPS spectrum.}
	\label{fig:AD_of_SU_4_N_f_2_minimal_SN}
\end{figure}

Let us start with the minimal BPS spectrum. The configuration of a spectral network that provides the minimal BPS spectra of $\CS[A_{N-1};\CD_{\rm II}]$ has branch points of index 2 aligned along two perpendicular lines on the $t$-plane. Because the BPS spectrum is represented by a $D_{N+1}$ quiver, which contains $A_{N-1}$ quiver in it, we expect the spectral network of $\CS[A_{N-1};\CD_{\rm I}]$ to be a part of that of $\CS[A_{N-1};\CD_{\rm II}]$ and it is indeed the case, see Figures \ref{fig:AD_of_SU_3_N_f_2_SU_3_f_minimal_SN} and \ref{fig:AD_of_SU_4_N_f_2_minimal_SN}, which represent $N=3$ and $N=4$ cases, respectively, and contain the spectral networks of $\CS[A_{2};\CD_{\rm I}]$ and $\CS[A_{3};\CD_{\rm I}]$, respectively, see Figures \ref{fig:a2SNFromA2_rho_inf_SN} and \ref{fig:a3SNFromA3_05_SN}.

\begin{figure}[h]
	\centering
	\begin{subfigure}[b]{.37\textwidth}	
		\centering
		\includegraphics[width=\textwidth]{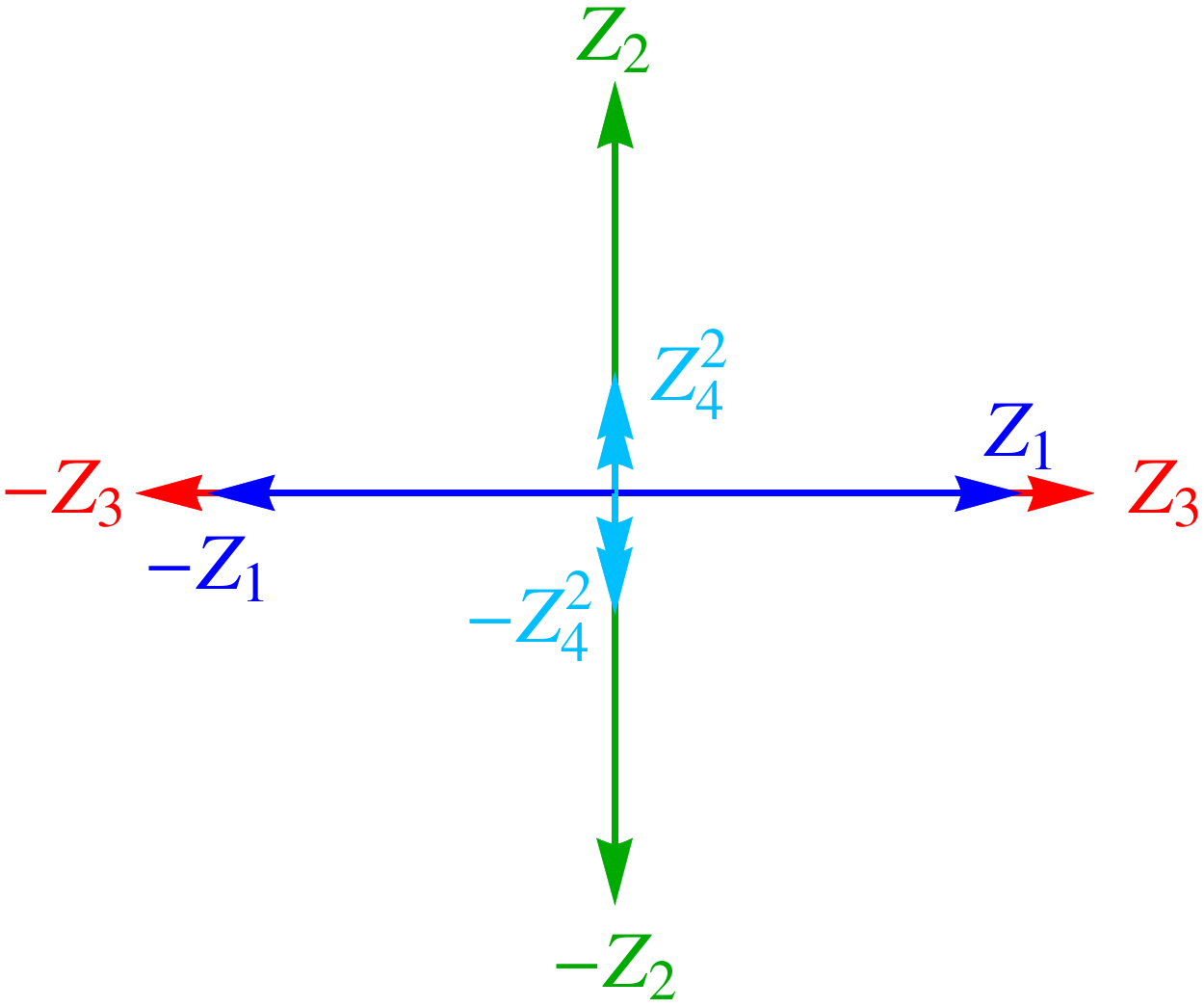}
		\caption{central charges}
		\label{fig:AD_of_SU_4_Nf_2_minimal_Z}
	\end{subfigure}
	\hspace{1em}
	\begin{subfigure}[b]{.5\textwidth}	
		\centering
		\begin{tabular}{c|c|c}
			state & $\UU(1)_{1}$ & $\UU(1)_{2}$\\ \hline
			1 & $(1,0)$ & $(0,0)$ \\ \hline
			2 & $(0,1)$ & $(0,0)$ \\ \hline
			3 & $(1,0)$ & $(1,0)$ \\ \hline
			4 & $(0,0)$ & $(0,1)$
		\end{tabular}
		\vspace{3.5em}
		\caption{IR charges}
		\label{tbl:AD_of_SU_4_Nf_2_minimal_IR}
	\end{subfigure}

	\begin{subfigure}{.3\textwidth}	
		\centering
		\begin{tikzpicture}
			\node[W,blue] (1) at (-1,0) {1};
			\node[W,darkergreen] (2) at (0,0) {2};
			\node[W,red] (3) at (1,0) {3};
			\node[W,deepskyblue] (41) at (1.5,1.732/2) {4};
			\node[W,deepskyblue] (42) at (1.5,-1.732/2) {4};
			
			\path (1) edge[->] (2);
			\path (3) edge[->] (2);
			\path (3) edge[->] (41);
			\path (3) edge[->] (42);	
		\end{tikzpicture}
		\caption{BPS quiver}
		\label{fig:AD_of_SU_4_Nf_2_minimal_quiver}
	\end{subfigure}
	\renewcommand{\figurename}{Figure \& Table}
	\caption{Minimal BPS spectrum of $\CS[A_{3};\CD_{\rm II}]$.}
	\label{fig:AD_of_SU_4_Nf_2_minimal_BPS}
\end{figure}

The Seiberg-Witten curve of $\CS[A_{N-1};\CD_{\rm II}]$ has two more branch points of index 2 in addition to those of the curve of $\CS[A_{N-1};\CD_{\rm I}]$. Remember that a trivialization of $\CS[A_{N-1};\CD_{\rm I}]$ is achieved by putting a branch cut between every branch point at finite $t$ to $t = \infty$, where either there is a branch point of index $N$ when $N$ is odd, or there are two branch points of index $N/2$ when $N$ is even. 
The two additional branch points are connecting the $(N-1)$-th and $N$-th sheets. These branch points intercept two branch cuts from the other two branch points of the same kind, therefore at $t = \infty$ the curve of $\CS[A_{N-1};\CD_{\rm II}]$ has either a branch points of index $N-1$ when $N$ is even, or two branch points of index $(N-1)/2$ when $N$ is odd. Note that by applying Riemann-Hurwitz formula we find the genus of the Seiberg-Witten curve to be $g = N/2$ when $N$ is even and $g = (N-1)/2$ when $N$ is odd. With this trivialization of the curve of $\CS[A_{N-1};\CD_{\rm II}]$, and that we have two more finite $\CS$-walls compared to $\CS[A_{N-1};\CD_{\rm I}]$ with minimal BPS spectrum, it is straightforward that the minimal BPS spectrum of $\CS[A_{N-1};\CD_{\rm II}]$ is equivalent to that of $\CS[A_1;\CD_\text{\rm reg},\CD_{N+3}]$ and can be represented by a $D_{N+1}$ quiver. Figure \ref{fig:AD_of_SU_4_Nf_2_minimal_Z} and Table \ref{tbl:AD_of_SU_4_Nf_2_minimal_IR} describe the minimal BPS spectrum of $\CS[A_{3};\CD_{\rm II}]$, which can be represented with a $D_5$ quiver as shown in Figure \ref{fig:AD_of_SU_4_Nf_2_minimal_quiver}.

To show that $\CS[A_{N-1};\CD_{\rm II}]$ is equivalent to $\CS[A_1;\CD_\text{\rm reg},\CD_{N+3}]$, we also compare the maximal, symmetric BPS spectrum of the two. When only $v_1 \neq 0$ and all the other parameters vanish, $\CS[A_{N-1};\CD_{\rm II}]$ has a symmetric arrangement of branch points around $t = 0$. There are $2N$ branch points of index 2, each pair of them having the same indices and being located oppositely from $t=0$. The indices of the first $N$ branch points are $(12),\ (23),\ \ldots,\ (N-1,N),\ (N,1)$, and each branch point is an end point of a branch cut that goes to $t = \infty$. When $N=3$ this leads to no branch point at $t = \infty$ as we have seen previously. When $N>3$, if $N$ is even there is a branch point of index $N-1$ at $t = \infty$, and if $N$ is odd there are two branch points of index $(N-1)/2$ at $t = \infty$. 

\begin{figure}[t]
	\centering
	\begin{subfigure}[b]{.27\textwidth}	
		\includegraphics[width=\textwidth]{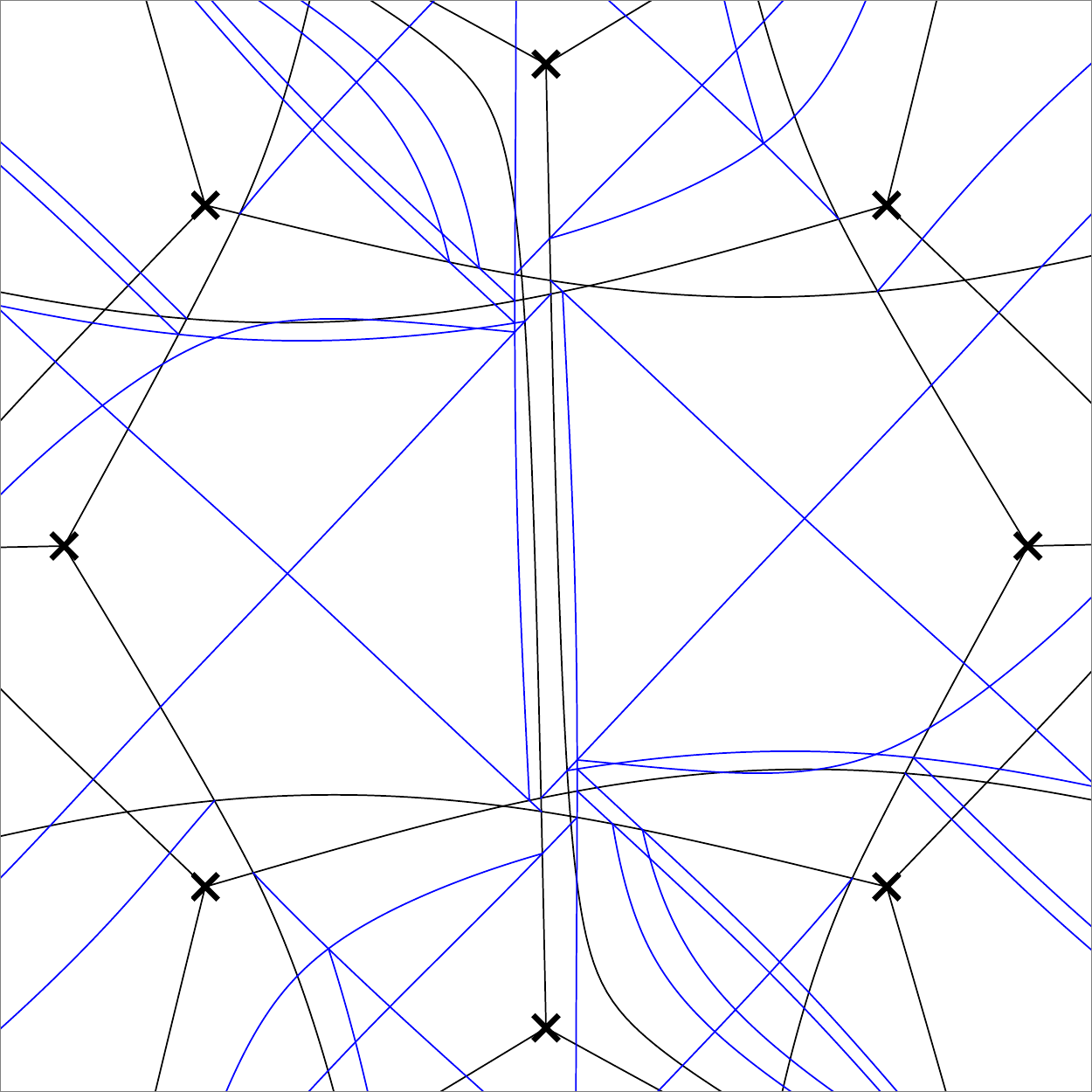}
		\vspace{0pt}
		\caption{$\theta \approx \arg(Z_{1}) = \arg(Z_{3})$}
		\label{fig:AD_of_SU_4_N_f_2_maximal_SN_02}
	\end{subfigure}
	\begin{subfigure}[b]{.27\textwidth}	
		\includegraphics[width=\textwidth]{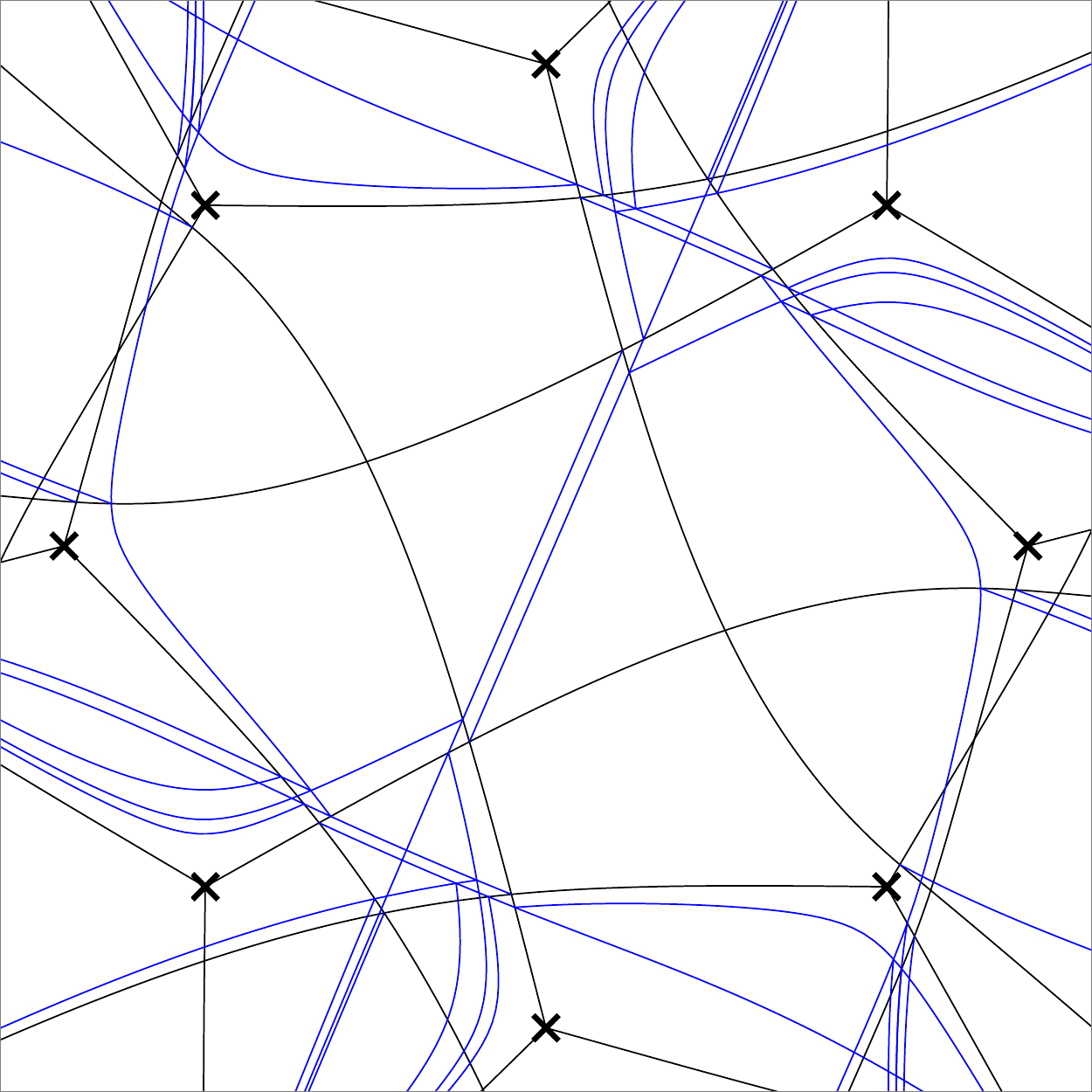}
		\vspace{0pt}
		\caption{$\theta \approx \arg(Z_{2}) = \arg(Z_{4}^\mathbf{2})$}
		\label{fig:AD_of_SU_4_N_f_2_maximal_SN_13}
	\end{subfigure}
	\begin{subfigure}[b]{.33\textwidth}	
		\centering
		\includegraphics[width=\textwidth]{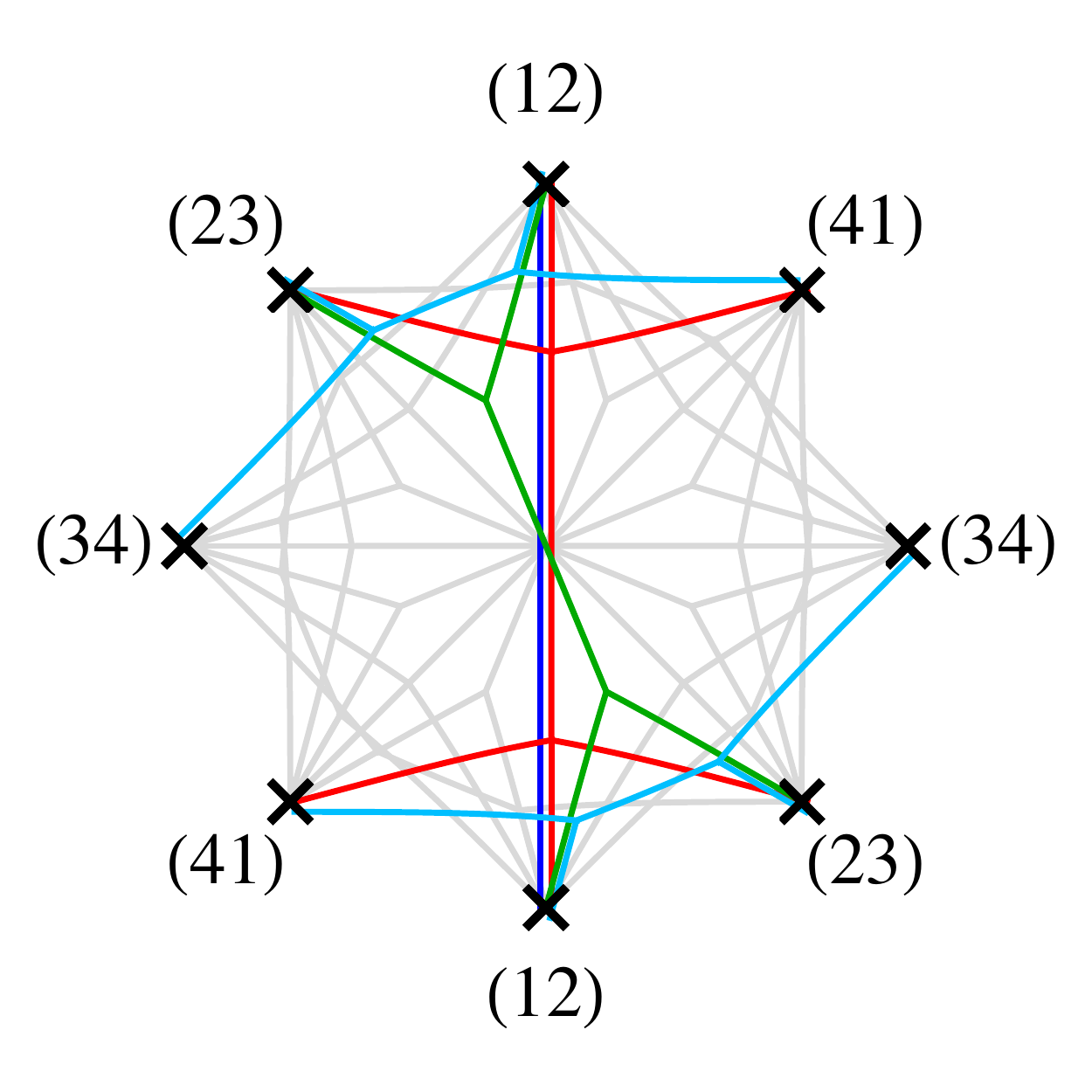}
		\caption{finite $\CS$-walls}
		\label{fig:AD_of_SU_4_N_f_2_maximal_symmetric_finite}
	\end{subfigure}	
	\caption{Spectral network of $\CS[A_{3};\CD_{\rm II}]$ with maximal, symmetric BPS spectrum.}
	\label{fig:AD_of_SU_4_N_f_2_maximal_SN}
\end{figure}

\begin{figure}[h]
	\centering
	\begin{subfigure}[b]{.45\textwidth}	
		\centering
		\includegraphics[width=\textwidth]{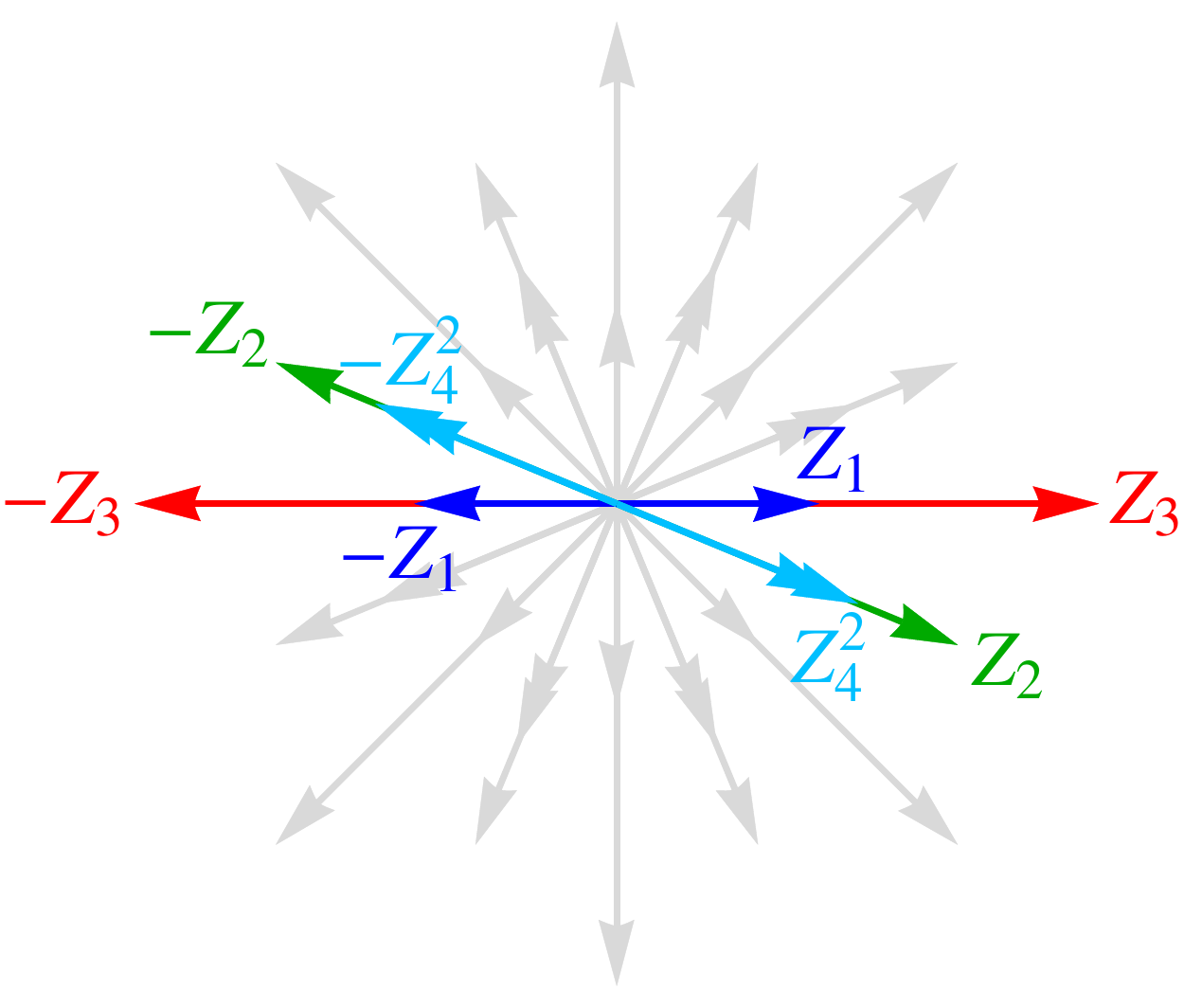}
		\vspace{3em}
		\caption{central charges}
		\label{fig:AD_of_SU_4_N_f_2_maximal_symmetric_Z}
	\end{subfigure}
	\begin{subfigure}[b]{.5\textwidth}	
		\centering
		\begin{tabular}{c|c|c}
			state & $\UU(1)_{1}$ & $\UU(1)_{2}$\\ \hline \hline
			1 & $(1,0)$ & $(0,0)$ \\ \hline
			2 & $(0,1)$ & $(0,0)$ \\ \hline
			3 & $(1,0)$ & $(1,0)$ \\ \hline
			4 & $(0,0)$ & $(0,1)$ \\ \hline \hline
			5 & $(1,1)$ & $(1,0)$ \\ \hline
			6 & $(2,1)$ & $(1,0)$ \\ \hline
			7 & $(3,1)$ & $(2,2)$ \\ \hline
			8 & $(1,0)$ & $(1,1)$ \\ \hline \hline
			9 & $(1,0)$ & $(1,2)$ \\ \hline
			10 & $(2,1)$ & $(2,2)$ \\ \hline
			11 & $(3,2)$ & $(2,2)$ \\ \hline
			12 & $(2,1)$ & $(1,1)$ \\ \hline \hline
			13 & $(1,1)$ & $(0,0)$ \\ \hline
			14 & $(2,1)$ & $(1,2)$ \\ \hline
			15 & $(1,1)$ & $(1,2)$ \\ \hline
			16 & $(1,1)$ & $(1,1)$
		\end{tabular}
		\caption{IR charges}
		\label{tbl:AD_of_SU_4_N_f_2_maximal_symmetric_IR}
	\end{subfigure}
	\renewcommand{\figurename}{Figure \& Table}
	\caption{Maximal, symmetric BPS spectrum of $\CS[A_{3};\CD_{\rm II}]$.}
	\label{fig:AD_of_SU_4_N_f_2_maximal_symmetric_BPS}
\end{figure}

This configuration leads to a symmetric, maximal BPS spectrum, where there are $N(N-1)$ singlets and $N$ doublets, including anti-states, which results in $N(N-1) + 2 \times N = N(N+1) = 2 \binom{N+1}{2}$ states in the BPS spectrum, which is the same number of the states in the maximal BPS spectrum of $\CS[A_1;\CD_\text{\rm reg},\CD_{N+3}]$.
An example of spectral networks for the $N=4$ case is shown in Figure \ref{fig:AD_of_SU_4_N_f_2_maximal_SN}. The resulting BPS spectrum is described by Figure \ref{fig:AD_of_SU_4_N_f_2_maximal_symmetric_Z} and Table \ref{tbl:AD_of_SU_4_N_f_2_maximal_symmetric_IR}, which is the same as the maximal symmetric BPS spectrum of $\CS[A_1;\CD_\text{\rm reg},\CD_{7}]$, see Figure \ref{fig:d5_from_A_1_maximal_symmetric_Z} and Table \ref{tbl:d5_from_A_1_maximal_symmetric_IR}.

\subsection{$\CS[A_{2}; \CD_\text{\rm reg}, \CD_{{\rm III}}]$ Theories}
\label{sec:S[A_2,C;D_reg,D_III]}
Here we will see an example of 4d SCFT from the 6d $(2,0)$ $A_{N-1}$ theory with $N>2$ compactified on a Riemann surface with both a regular puncture and an irregular one.
The Seiberg-Witten curve is $v^{3} + \tilde{\phi}_{2} v + \tilde{\phi}_{3} = 0$ with
    \bea
    \tilde{\phi}_{2}
    &=&    c\, {t} + \left({C_{2} - m_{+}^{2}/3}\right),
           \nonumber \\
    \tilde{\phi}_{3}
    &=&   t^2 - v t - \left({C_{3} - \frac{C_{2}m_{+}}{3} +\frac{2m_{+}^{3}}{27}}\right),
    \eea
 where the regular puncture is at $t=0$ and the irregular puncture is at $t=\infty$. The Seiberg-Witten differential is $\lambda = \frac{v}{t} dt$.
The dimensions of the parameters are
    \bea
    \Delta(C_{3})
     =     3, ~~~
    \Delta(C_{2})
     =     2, ~~~
    \Delta(m_{+})
     =     1, ~~~
    \Delta(c)
     =     \frac{1}{2},~~~
    \Delta(v)
     =     \frac{3}{2}.
    \eea
These are the same as those of the class 2 SCFT of $\SU(3)$ with $N_{f} = 3$ in \cite{Eguchi:1996vu}. We will show that the BPS spectra of this theory and their wall-crossings are the same as those of $\CS[A_{1}; \CD_\text{\rm reg}, \CD_{6}]$ and $\CS[A_2;\CD_{\rm II}]$, all three of which are in the same $D_4$-class..

The irregular singularity at $t = \infty$ is a branch point of index 3 and has no residue. For general values of parameters, we have four branch points of index 2. When we set the values of $C_2$ and $C_3$ as
\begin{align}
	C_2 = \frac{m_+^3}{3},\ C_3 = \frac{m_+^3}{27},
\end{align}
two among the four branch points collide with the regular puncture at $t = 0$, forming a branch point of index 3. This choice corresponds to enhancing the flavor symmetry to $\SU(3)$, and the puncture has two triplets of $\CS$-walls coming out of it. 

With values of $C_i$ fixed as above, the discriminant of the equation of branch points is
\begin{align}
	\Delta_w g (c,v) \propto c^3\left(v - \frac{c^3}{27}\right).
\end{align}
$v = c^3/27$ corresponds to the singularity where a singlet becomes massless. $c = 0$ does not correspond to a singularity but a collision of two branch points of index 2, forming a single branch point of index 3.

\begin{figure}[t]
	\centering
	\begin{subfigure}[b]{.27\textwidth}	
		\includegraphics[width=\textwidth]{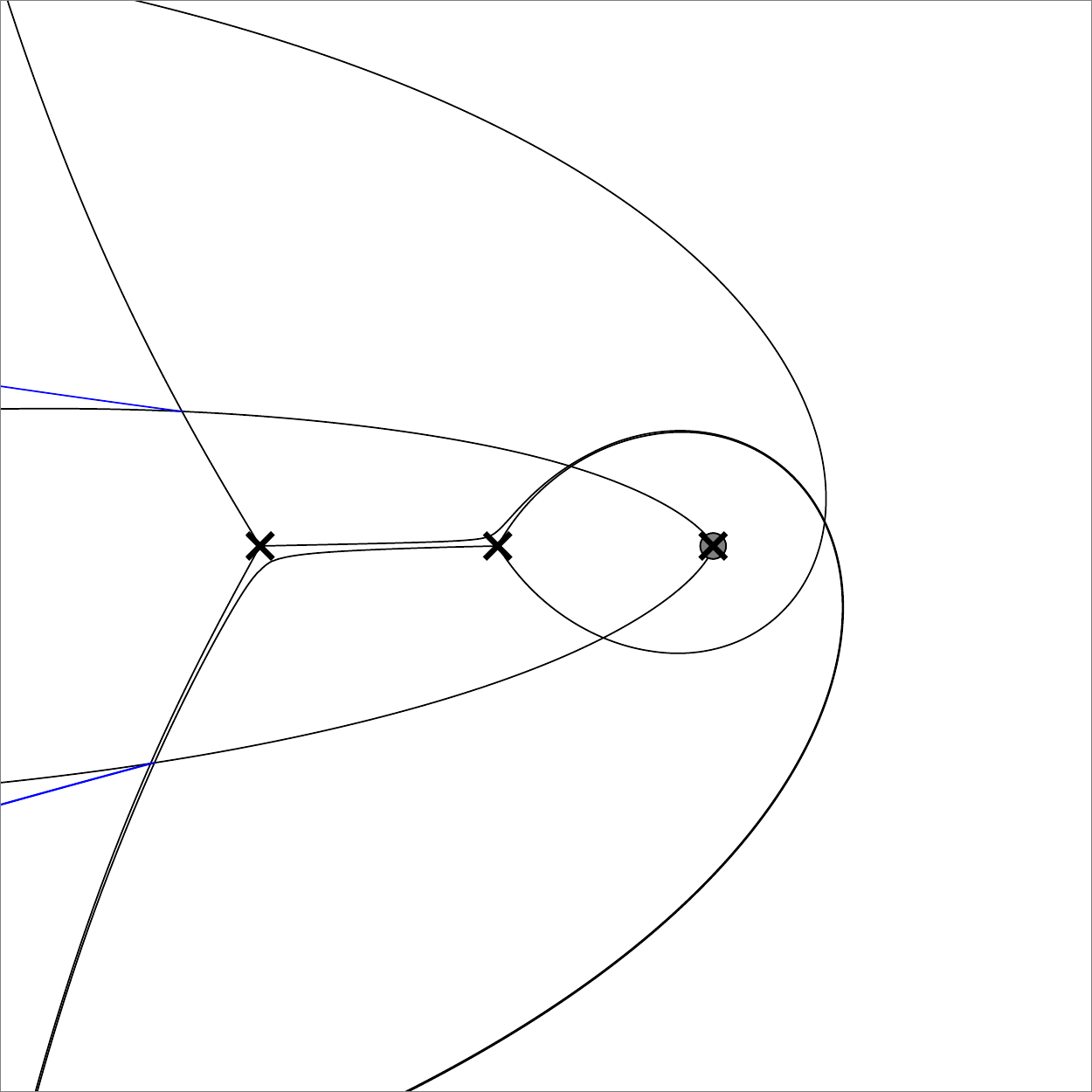}
		\caption{near $\theta = \arg(Z_{1})$}
		\label{fig:AD_of_SU_3_N_f_3_SU_3_f_minimal_SN_near_singlet}
	\end{subfigure}
	\begin{subfigure}[b]{.27\textwidth}	
		\includegraphics[width=\textwidth]{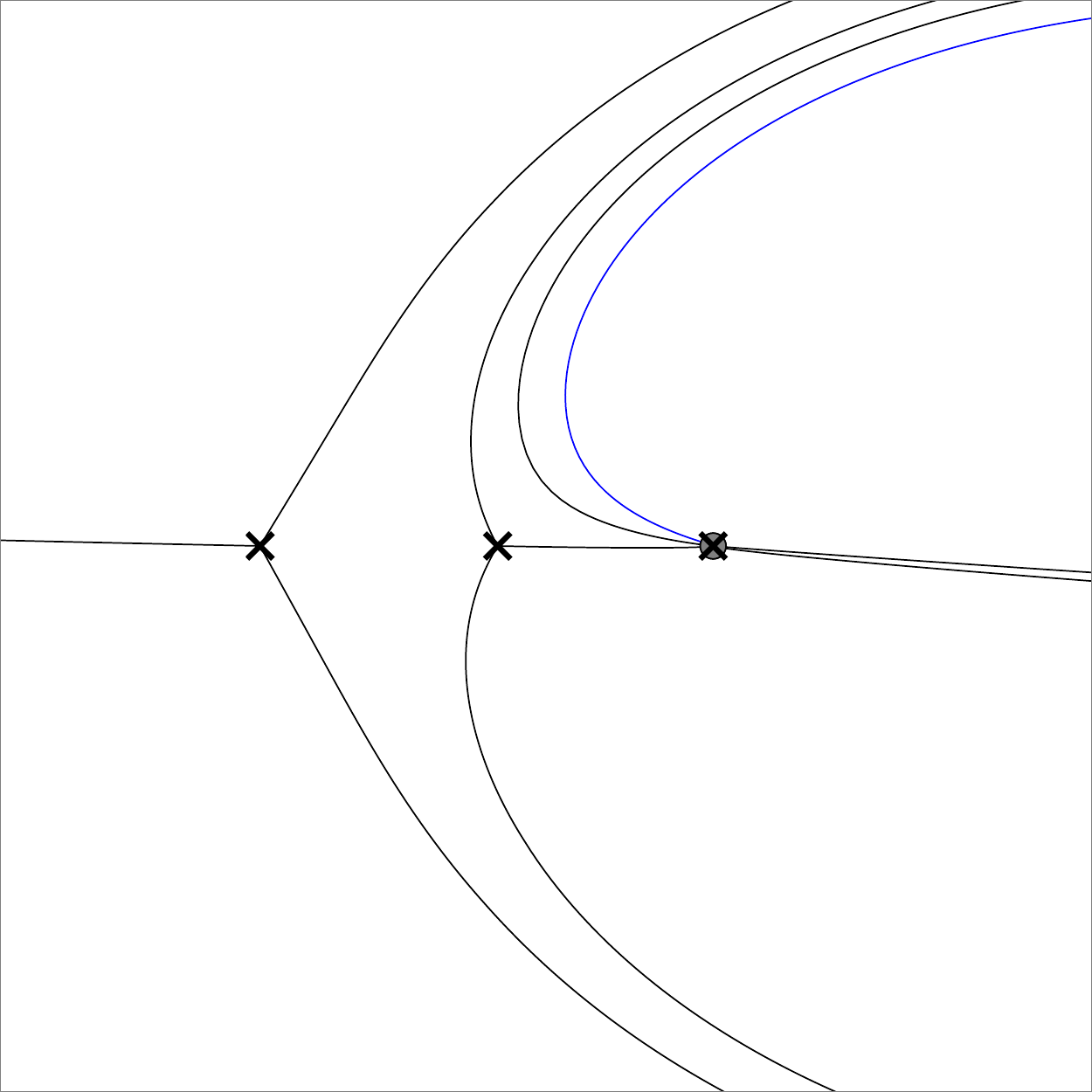}
		\caption{near $\theta = \arg(Z_{2}^\mathbf{3})$}
		\label{fig:AD_of_SU_3_N_f_3_SU_3_f_minimal_SN_near_triplet}
	\end{subfigure}
	\begin{subfigure}[b]{.35\textwidth}	
		\centering
 		\includegraphics[width=\textwidth]{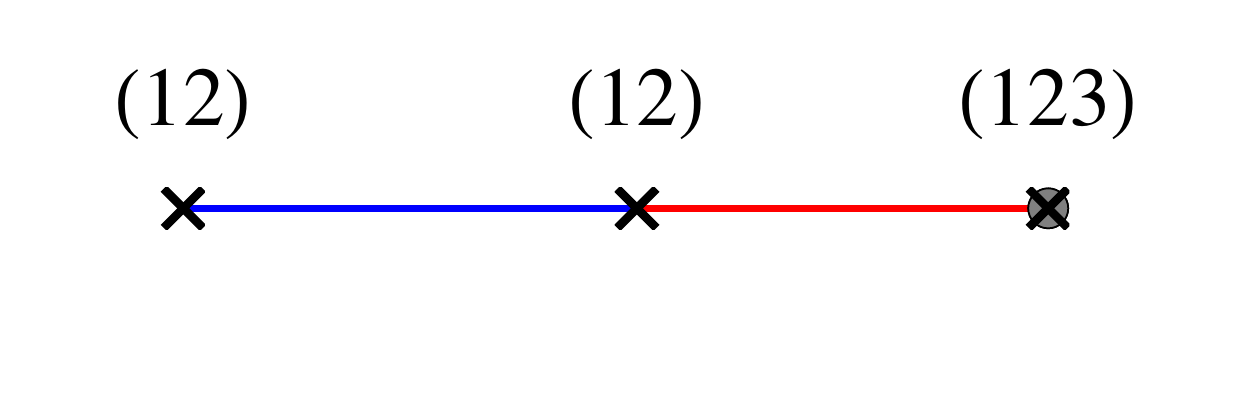}
		\vspace{2em}
		\caption{finite $\CS$-walls}
		\label{fig:AD_of_SU_3_N_f_3_SU_3_f_minimal_finite}
	\end{subfigure}

	\caption{Spectral network of $\CS[A_{2};\CD_{\rm III}]$ with minimal BPS spectrum.}
	\label{fig:AD_of_SU_3_N_f_3_SU_3_f_minimal_SN}	
\end{figure}

\paragraph{Minimal BPS spectrum}
When $c$ is fixed as a real number and $v = c^3/27 - \delta$, where $\delta$ is a small real number, we have two branch points of ramification index 2 in addition to the puncture of index 3, as shown in Figure \ref{fig:AD_of_SU_3_N_f_3_SU_3_f_minimal_SN}, where finite $\CS$-walls corresponding to BPS states are also depicted.

Note that, in addition to a finite $\CS$-wall connecting the two branch points of index 2 that corresponds to a singlet, there is a triplet of $\CS$-walls from the puncture, i.e.\
 there are three coincident finite $\CS$-walls connecting the puncture and one of the branch points of index 2, which gives us an $\SU(3)$ triplet. 

\begin{figure}[ht]
	\centering
	\begin{subfigure}[b]{.4\textwidth}	
		\centering
		\includegraphics[width=\textwidth]{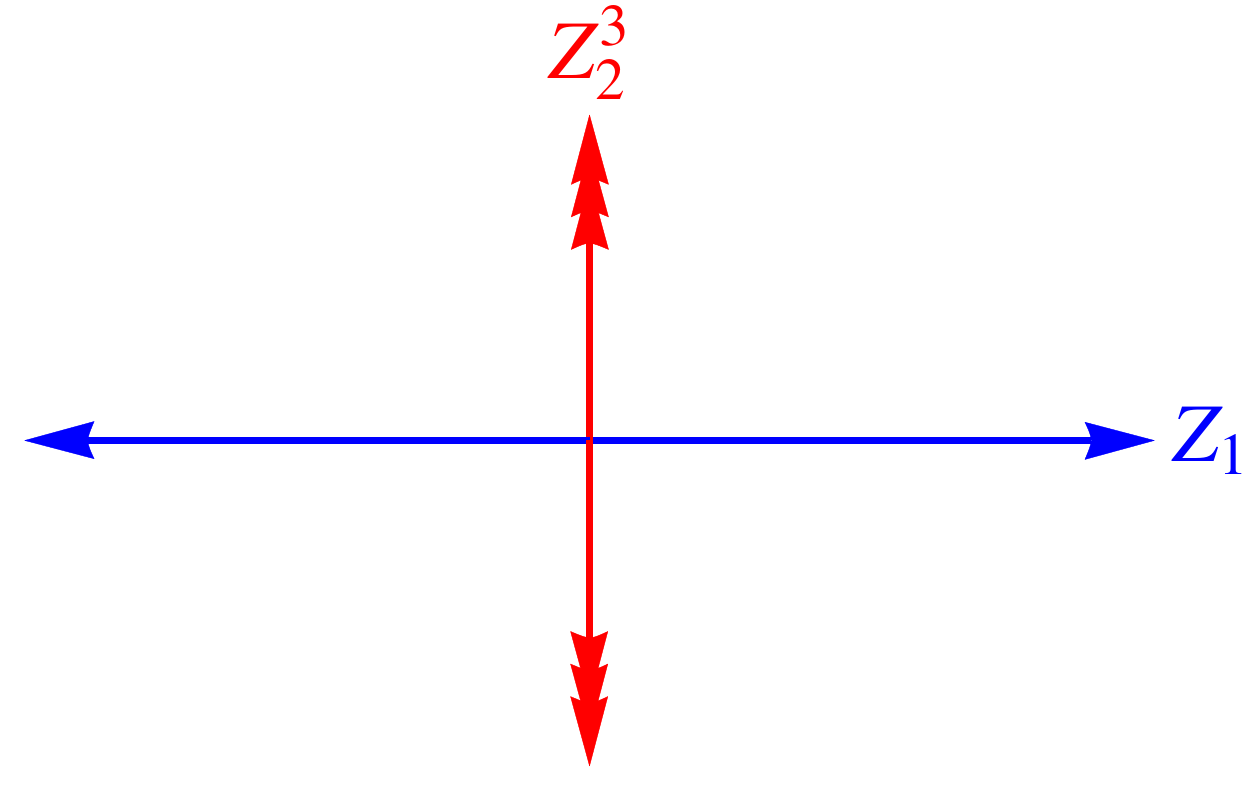}
		\caption{central charges}
		\label{fig:AD_of_SU_3_N_f_3_SU_3_f_minimal_Z}
	\end{subfigure}
	\begin{subfigure}[b]{.35\textwidth}	
		\centering
		\begin{tabular}{c|c}
		   state  & $(e,m)$ \\ \hline
		  1 & $(1,0)$ \\ \hline
		  2 & $(0,1)$
		\end{tabular}
		\vspace{2em}
		\caption{IR charges}
		\label{tbl:AD_of_SU_3_N_f_3_SU_3_f_minimal_IR}
	\end{subfigure}		
	\begin{subfigure}[b]{.15\textwidth}	
		\centering	
		\begin{tikzpicture}
			\node[W,red] (21) at (0,0) {2};
			\node[W,blue] (1) at (1,0) {1};
			\node[W,red] (22) at (1.5,1.732/2) {2};
			\node[W,red] (23) at (1.5,-1.732/2) {2};
			
			\path (1) edge[->] (21);
			\path (1) edge[->] (22);
			\path (1) edge[->] (23);	
		\end{tikzpicture}
		\vspace{1em}
		\caption{BPS quiver}
		\label{fig:AD_of_SU_3_N_f_3_SU_3_f_minimal_quiver}
	\end{subfigure}	
	\renewcommand{\figurename}{Figure \& Table}
	\caption{minimal BPS spectrum of $\CS[A_{2};\CD_{\rm III}]$.}
	\label{fig:AD_of_SU_3_N_f_3_SU_3_f_minimal}
\end{figure}

The intersections of the cycles corresponding to the $\CS$-walls can be easily read out if we consider the trivialization of the Seiberg-Witten curve by introducing a branch cut between the two branch points of index 2 and another branch cut connecting the puncture and the branch point of index 3 at infinity. The resulting BPS spectrum, described in Figure \ref{fig:AD_of_SU_3_N_f_3_SU_3_f_minimal_Z} and Table \ref{tbl:AD_of_SU_3_N_f_3_SU_3_f_minimal_IR}, is the same as the minimal BPS spectrum of $\CS[A_{1}; \CD_\text{\rm reg}, D_{6}]$ (Figure \ref{fig:d4SN_SU_3_f_minimal_BPS}) and $\CS[A_{2};\CD_{\rm II}]$ (Figure \& Table \ref{figntbl:AD_of_SU_3_N_f_2_SU_3_f_minimal_BPS}), all of which can be represented with a $D_4$ quiver as shown in Figure \ref{fig:AD_of_SU_3_N_f_3_SU_3_f_minimal_quiver}.

\paragraph{Wall-crossing to the maximal BPS spectrum}

As $\delta \to 0$, the two branch points of index 2 approach each other, corresponding to the BPS state from the $\CS$-wall connecting the two becoming massless, and as we go across the wall at $\delta = 0$ the BPS spectrum undergoes a wall-crossing to the maximum BPS spectrum, which is similar to what we have observed for the other theories with a $D_4$ BPS spectrum and an $\SU(3)$ flavor symmetry. Now we fix the value of $v$ and take $c \to 0$, then the the two branch points of index 2 move to the other side of the puncture, and one of the two branch points goes through the branch cut connecting the puncture and infinity, resulting in a branch point connecting different sheets. 

\begin{figure}[t]
	\centering
	\begin{subfigure}[b]{.27\textwidth}	
		\includegraphics[width=\textwidth]{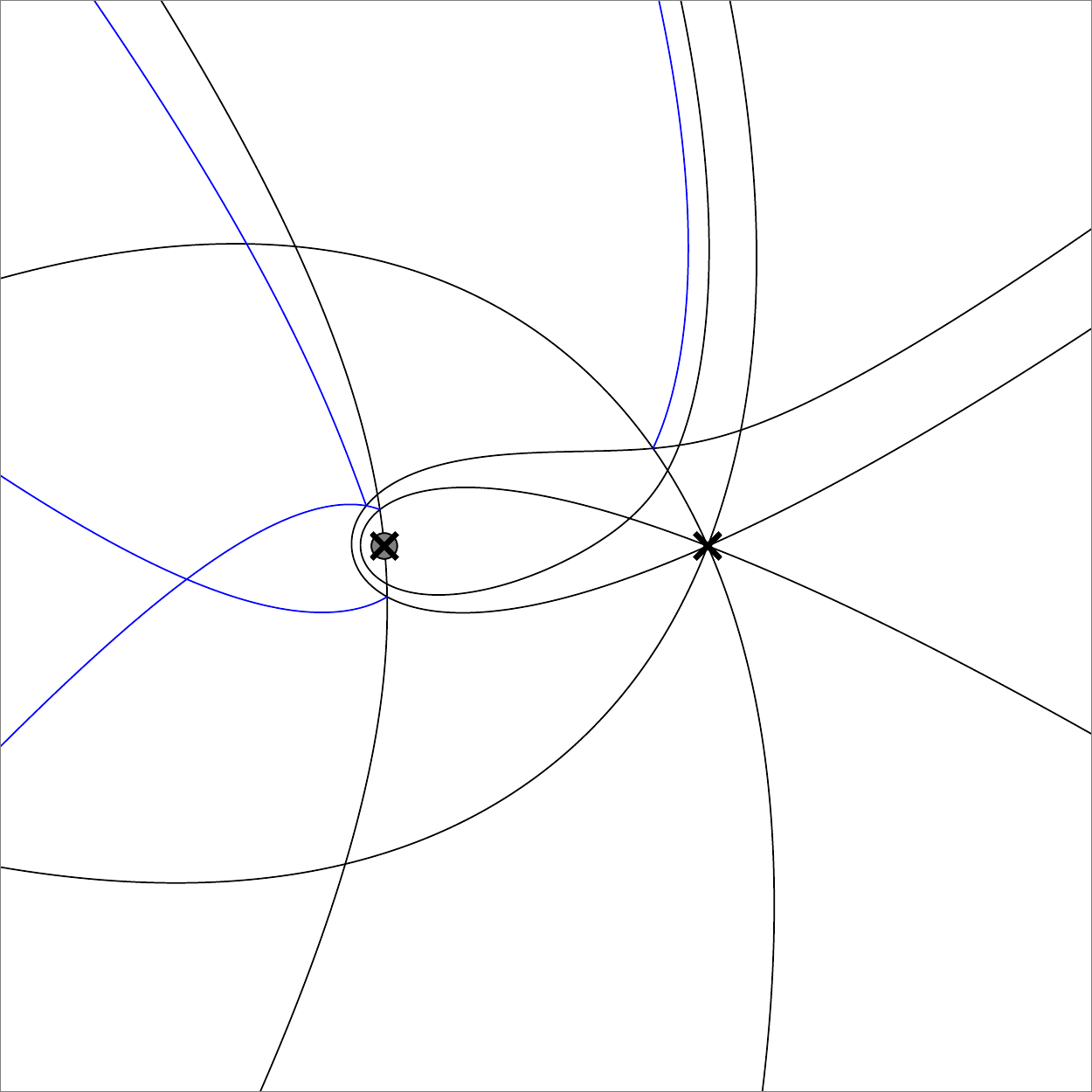}
		\caption{near $\theta = \arg(Z_{1})$}
		\label{fig:AD_of_SU_3_N_f_3_SU_3_f_maximal_SN_near_singlet}
	\end{subfigure}
	\begin{subfigure}[b]{.27\textwidth}	
		\includegraphics[width=\textwidth]{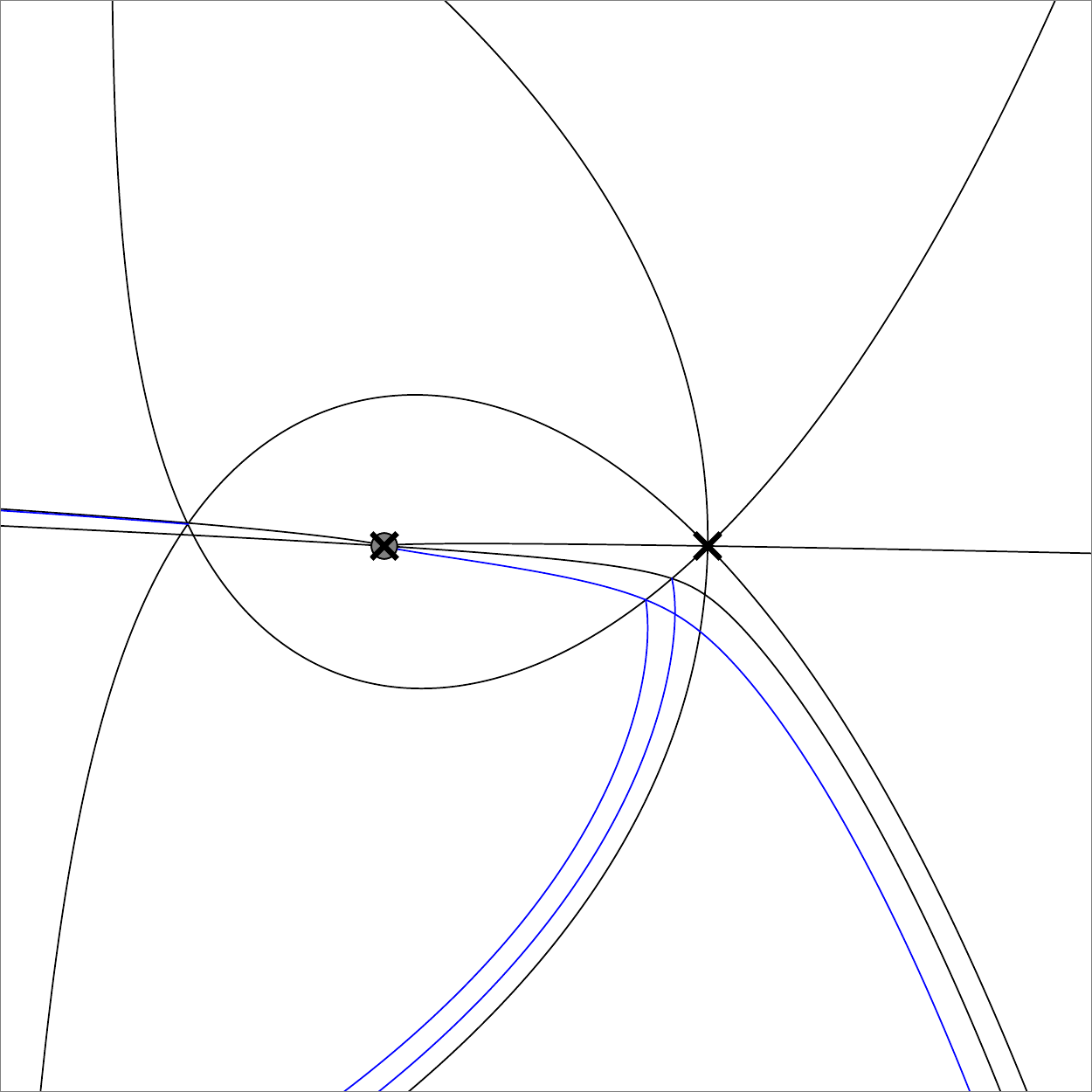}
		\caption{near $\theta = \arg(Z_{2}^\mathbf{3})$}
		\label{fig:AD_of_SU_3_N_f_3_SU_3_f_maximal_SN_near_triplet}
	\end{subfigure}
	\begin{subfigure}[b]{.3\textwidth}	
		\centering
		\includegraphics[width=\textwidth]{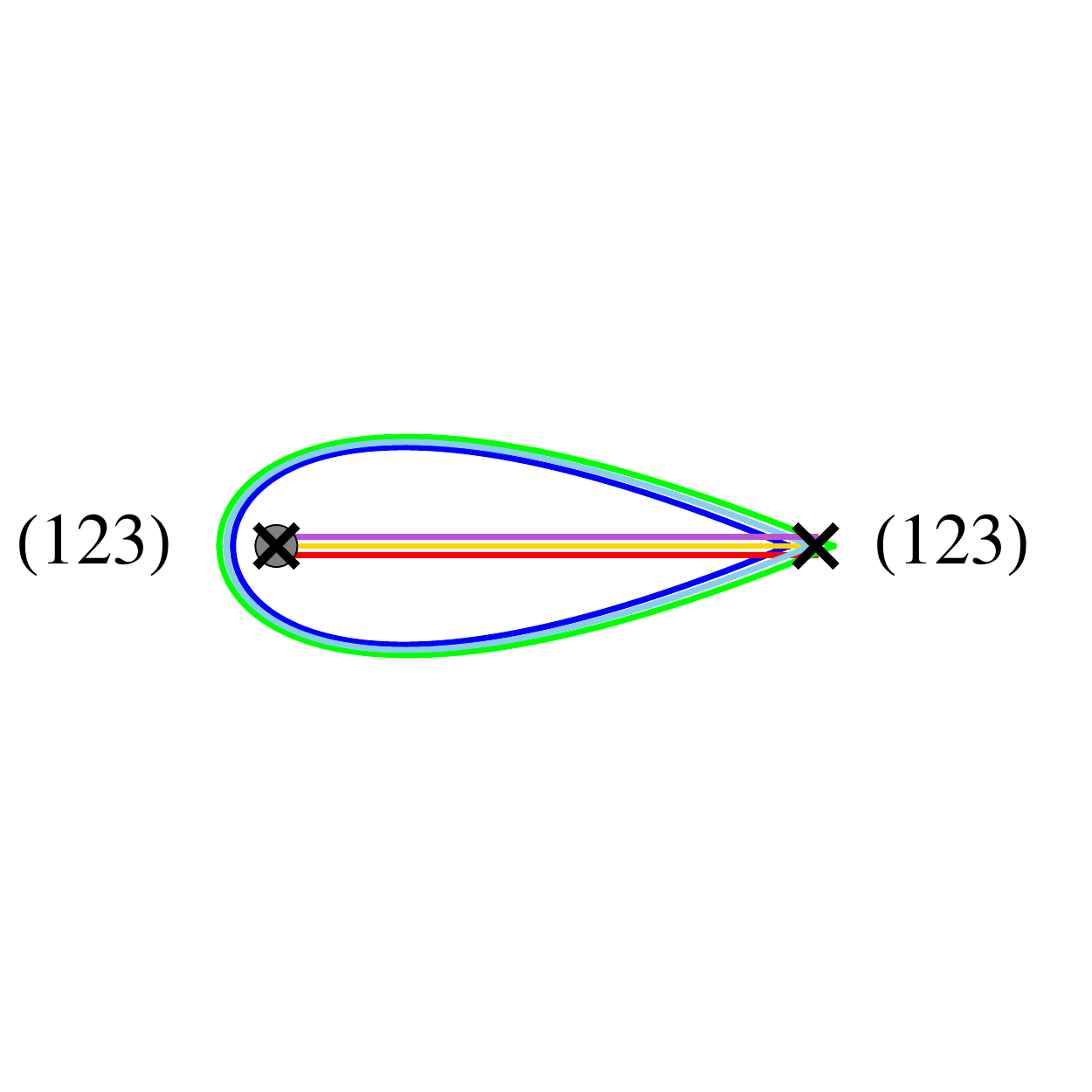}
		\caption{finite $\CS$-walls}
		\label{fig:AD_of_SU_3_N_f_3_SU_3_f_maximal_symmetric_finite}
	\end{subfigure}
	
	\caption{spectral network of $\CS[A_{2},\CC;D_{\rm III}]$ with maximal BPS spectrum}
	\label{fig:AD_of_SU_3_N_f_3_SU_3_f_maximal_SN}	
\end{figure}

\begin{figure}[h]
	\centering
	\begin{subfigure}[b]{.36\textwidth}	
		\centering
		\includegraphics[width=\textwidth]{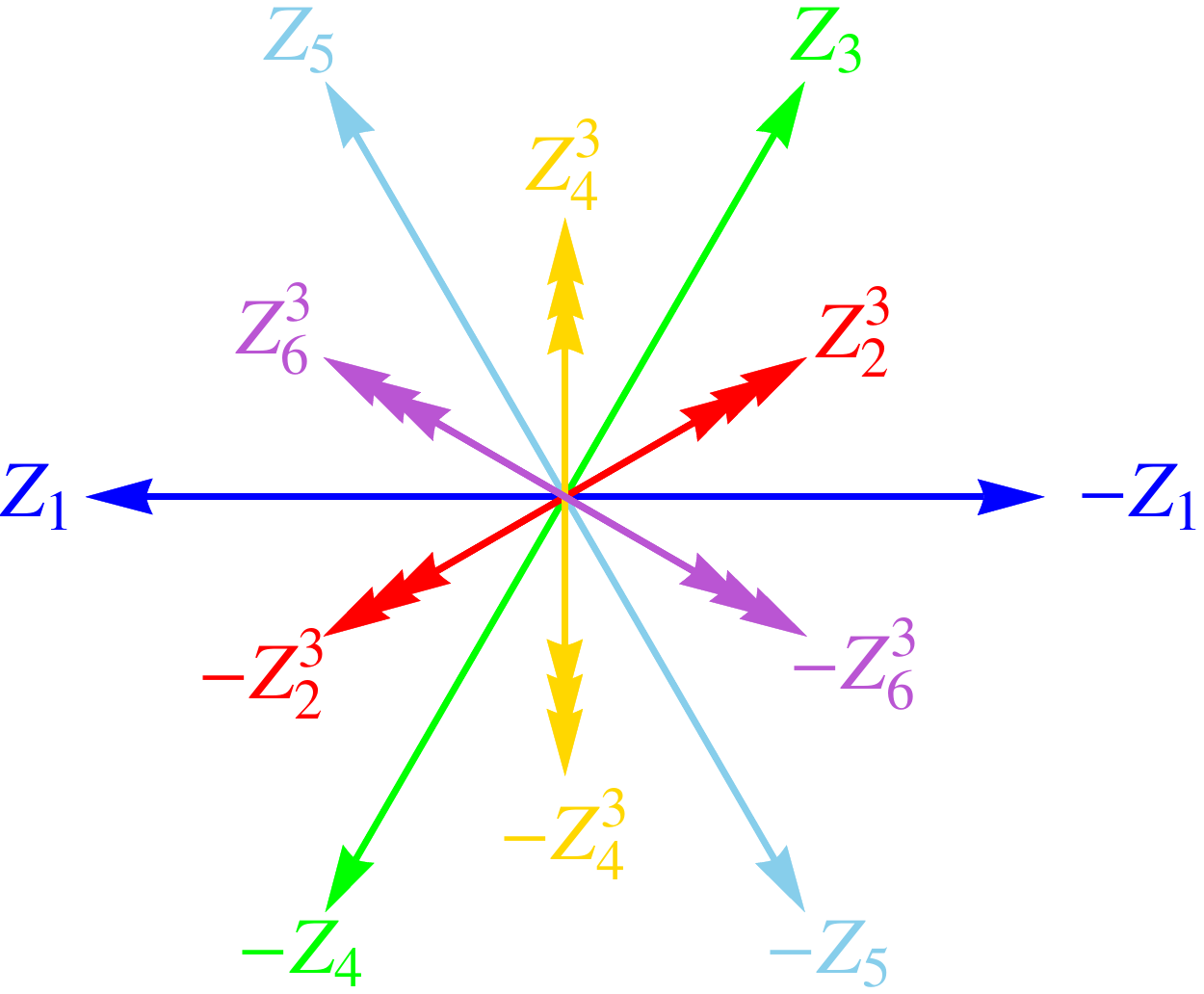}
		\caption{central charges}
		\label{fig:AD_of_SU_3_N_f_3_SU_3_f_maximal_symmetric_Z}
	\end{subfigure}
	\begin{subfigure}[b]{.35\textwidth}	
		\centering
		\begin{tabular}{c|c}
			state  & $(e,m)$ \\ \hline
			$1$ & $(1,0)$ \\ \hline
			$2$ & $(0,1)$ \\ \hline
			$3$ & $(1,3)$ \\ \hline			
			$4$ & $(1,2)$ \\ \hline
			$5$ & $(2,3)$ \\ \hline
			$6$ & $(1,1)$ 
		\end{tabular}
		\vspace{2em}
		\caption{IR charges}
		\label{tbl:AD_of_SU_3_N_f_3_w_3_SU_3_f_maximal_symmetric_IR}
	\end{subfigure}
	\renewcommand{\figurename}{Figure \& Table}
	\caption{Maximal BPS spectrum of $\CS[A_{2};\CD_{\rm III}]$.}
	\label{fig:AD_of_SU_3_N_f_3_SU_3_f_maximal_symmetric}
\end{figure}

When we eventually set $c = 0$, the two branch points collide to form a single branch point of index 3. 
This is a symmetric configuration of three branch points (including one at infinity), considering the locations of three points on a complex plane does not introduce any modulus. Figure \ref{fig:AD_of_SU_3_N_f_3_SU_3_f_maximal_SN} shows its spectral networks and finite $\CS$-walls, from which we get the maximal, symmetric BPS spectrum of $\CS[A_{2};\CD_{\rm III}]$ described in Figure \ref{fig:AD_of_SU_3_N_f_3_SU_3_f_maximal_symmetric_Z} and Table \ref{tbl:AD_of_SU_3_N_f_3_w_3_SU_3_f_maximal_symmetric_IR}. This spectrum can be identified with those of $\CS[A_{1}; \CD_\text{\rm reg}, \CD_{6}]$ (Figure \ref{fig:d4SN_wall_crossing_SU_3_f} and Table \ref{tbl:d4SN_SU_3_maximal_IR}) and $\CS[A_{2};\CD_{\rm II}]$ (Figure \& Table \ref{figntbl:AD_of_SU_3_N_f_2_SU_3_f_maximal_symmetric_BPS}), thereby providing good evidence for the equivalence of the three theories.

\section{Discussion and Outlook}
\label{sec:Discussion and outlook}
  In this paper, we have studied the spectral networks of four-dimensional SCFT of Argyres-Douglas type
  focusing on the minimal and maximal chambers on its Coulomb branch.
  For the SCFTs obtained from the $A_{N-1}$ theory we have studied a particular class of SCFTs
  in sections \ref{sec:S[A_N-1,C;D_I]}, \ref{sec:S[A_N-1,C;D_II]} and \ref{sec:S[A_2,C;D_reg,D_III]}: 
  a compactification on a sphere with the irregular puncture $D_{{\rm I}}$, $D_{{\rm II}}$
  and $D_{{\rm III}}$.
  It would be interesting thus to analyze the spectral networks of more general SCFTs
  associated with other types of irregular punctures studied in \cite{Xie:2012hs, Kanno:2013vi}.
  In the presence of the regular puncture, the SCFT would be related with the one studied 
  in \cite{Cecotti:2012jx,Cecotti:2013lda}.

  The BPS quiver (and quiver mutation) method developed in \cite{Cecotti:2010fi,Cecotti:2011rv,Alim:2011ae}
  is another useful way to study the BPS spectrum of four-dimensional $\CN=2$ theory. The BPS quiver method is useful when figuring out which chambers are in the Coulomb branch moduli space of a theory. In comparison, the method using spectral network provides a straightforward way to get the BPS spectrum at a particular location in the moduli space.
  It would be interesting to understand how to use spectral networks and quiver mutations 
  in a complementary way to each other to find BPS spectra and study their wall-crossings.
  
  The spectral network has been introduced in \cite{Gaiotto:2012rg} in connection with the 2d-4d BPS states
  associated with the four-dimensional theory in presence of a surface defect.
  It is interesting to study these BPS states and the corresponding two-dimensional BPS spectrum 
  of the SCFTs studied in this paper.

We can find the BPS spectrum of any theory of class $\CS$ using spectral network, which can be very useful when we lack of perturbative understanding it. For example, in \cite{Gaiotto:2010jf} it is shown than the low-energy physcis of a 4d $\CN=2$ $\SU(N)$ theory with $2n$ flavors is described by two strongly-coupled SCFTs coupled by an IR-free $\SU(2)$ gauge multiplet, and it will be interesting to study such a theory using spectral network. Also, it will be interesting to study BPS spectra of SCFTs in \cite{Chacaltana:2010ks} that are obtained from compactifying 6d $A_{N-1}$ theory on a three-punctured sphere.


\section*{Acknowledgments}
It is a pleasure to thank Yuji Tachikawa for discussions during the early stage of this work, 
and to Kimyeong Lee, Yu Nakayama, Andy Neitzke, John H. Schwarz, Dan Xie, and Piljin Yi for discussions during the final stage of this work.
We thank the organizers of the 2013 Summer Simons Workshop at the SCGP for hospitality while this work was in progress.
K.M.~and C.Y.P.~would also like to thank the organizers of the second ``N=2 JAAZ'' workshop at McGill University, C.Y.P.~would like to thank the organizers of GAP 2013 at the CRM, KIAS, and Kavli IPMU, and K.M.~and W.Y. would like to thank the organizers of Kavli IPMU workshop on Gauge and String Theory, where part of this work has been done, for hospitality and support.
The work of K.M. is supported by a JSPS postdoctoral fellowship for research abroad.
The work of W.Y. is supported in part by the Sherman Fairchild scholarship and by DOE grant DE-FG02-92-ER40701.

\appendix


\section{SCFTs of Argyres-Douglas Type}
\label{sec:AD points of SU(3) SQCD}
\label{sec:SU(3)}

\subsection{$\CS[A_{N-1}; \CD_{{\rm I}}]$ Theories from $\SU(N)$ Pure SYM}
\label{sec:ISYM}
  The M-theory curve of $\CN=2$ $\SU(N)$ pure SYM theory is
    \bea
    \Lambda^{N} t^{2} + P_{N}(v) t + \Lambda^{N}
     =     0,
    \eea
  where $P_{N} = v^{N} + \sum_{i=2}^{N} u_{i} v^{N-i}$ and $u_{i}$ are the Coulomb moduli parameters.
  By setting $v = xt$, we get the following form of the curve:
    \bea
    x^{N} + \sum_{i=2}^{N} \phi_{i}(t) x^{N-i}
     =     0,
    \eea
  where
    \bea
    \phi_{i}
    &=&    \frac{u_{i}}{t^{i}},~~(i=2,\ldots,N-1), ~~{\rm and}~~
    \phi_{N}
     =     \frac{\Lambda^{N}}{t^{N+1}} + \frac{u_{N}}{t^{N}} + \frac{\Lambda^{N}}{t^{N-1}}.
    \eea
  The Seiberg-Witten differential is $\lambda = x dt$.
  This denotes that there are two irregular singularities at $t=0$ and $t= \infty$.

  The maximal conformal point is at $u_{i} = \pm 2 \Lambda^{2N} \delta^{iN}$
  (we choose the minus sign here), at which the curve becomes
     \bea
     x^{N}
      + \frac{\Lambda^{N}}{t^{N+1}} (t-1)^{2}
      =     0.
     \eea
   To consider the small deformation from this point, let us define the parameters as
   $u_{i} = \hat{u}_{i} - 2 \Lambda^{2N} \delta^{iN}$ by which the curve is
     \bea
     x^{N}
      + \frac{\Lambda^{N}}{t^{N+1}} (t-1)^{2}
      + \sum_{i=2}^{N} \frac{\hat{u}_{i}}{t^{i}} x^{N-i}
      =     0
     \eea

   Let us look at the region close to $t=1$. To do this, we introduce the new coordinate
   $\tilde{t}= (t-1)\Lambda^{a}$ with $a>0$ and take the limit $\Lambda \rightarrow \infty$.
   This is written as $t = 1 + \frac{\tilde{t}}{\Lambda^{a}}$.
   The SW differential is now written as
   $\lambda = \frac{x}{\Lambda^{a}} d \tilde{t}$, so we define $\tilde{x} = \frac{x}{\Lambda^{a}}$
   such that the differential is of canonical form.
   The curve in terms of these coordinates is written as
     \bea
     \tilde{x}^{N} + \Lambda^{N-(2+N)a} \tilde{t}^{2}
      + \sum_{i=2}^{N} \frac{\hat{u}_{i}}{\Lambda^{ia} (1+\CO(\Lambda^{-a}))^{i}} \tilde{x}^{N-i}
      =     0.
     \eea
   We want to keep the second term finite which means $a=\frac{N}{N+2}$.
   In order for the deformation terms to be finite, we also have
     \bea
     \hat{u}_{i}
      =     \Lambda^{\frac{N}{N+2}i} c_{i},
     \eea
   where $i=2,\ldots,N$.
   We can easily see that the scaling dimension of $c_{i}$ is $\frac{2i}{N+2}$.
   Let us define $v_{i} = c_{N-i+2}$ for $i=2, \ldots, [\frac{N+1}{2}]$ such that the dimensions are
     \bea
     \Delta(v_{i})
      =     2 - \frac{2i}{N+2},~~~
     \Delta(c_{i})
      =     \frac{2i}{N+2}.
     \eea
   for $i=2,\ldots,[\frac{N+1}{2}]$.
   Note that they satisfy $\Delta(v_{i}) + \Delta(v_{i}) = 2$.
   When $N=2k$, there is a mass parameter $c_{k+1}$ with dimension $1$.
   The final form of the curve is
     \bea
     \tilde{x}^{N} + \sum_{i=2}^{N} \phi_{i}(t) \tilde{x}^{N-i}
      =     0,
            \label{SWcanonical}
     \eea
   where
     \bea
     \phi_{i}
     &=&    c_{i},~~(i=2,\ldots,[\frac{N+2}{2}]),
            \nonumber \\
     \phi_{i}
     &=&    v_{N-i+2},~~(i=[\frac{N+2}{2}]+1, \ldots, N-1),~~{\rm and}~~
     \phi_{N}
      =     \tilde{t}^{2} + v_{2}.
     \eea

\subsection{$\CS[A_{N-1}; \CD_{{\rm II}}]$ from $\SU(N)$ with $N_{f}=2$}
\label{sec:II2flavors}
  The Seiberg-Witten curve of $\SU(N)$ gauge theory with two flavors with masses $m_{1,2}$ is
    \bea
    \Lambda^{N-1}(v-m_{1}) t^{2} + P_{N}(v) t + \Lambda^{N-1}(v-m_{2})
     =     0.
    \eea
  The most singular point of this curve is at
    \bea
    u_{k} = \pm 2 \Lambda^{N-1} \delta_{k,N-1},~~~
    m_{1} = m_{2} = 0.
    \eea
  Indeed the curve is factorized at this point into
    \bea
    x^{N} + x \Lambda^{N-1}\frac{(t\pm 1)^{2}}{t^{N}}
     =     0,
           \label{ADSU(N)2flavors}
    \eea
  where $v=xt$.

  We parameterized the parameters as $u_{k} = - 2\Lambda^{2} + \hat{u}_{k}$
  and the coordinate as $\Lambda^{a} (t-1)  = \tilde{t}$.
  The SW differential is $\lambda = x dt = \tilde{x} d \tilde{t}$ where $\tilde{x} = \frac{x}{\Lambda^{a}}$.
  By substituting these into the curve we obtain $a = \frac{N-1}{N+1}$
  in order to keep the second term in \eqref{ADSU(N)2flavors} finite.
  Therefore we obtain the curve of the form \eqref{SWcanonical} where
    \bea
    \phi_{i}
    &=&    c_{i},~~(i=2, \ldots, [\frac{N+1}{2}])
           \nonumber \\
    \phi_{i}
    &=&    v_{N-i+1}, ~~(i=[\frac{N+1}{2}]+1, \ldots, N-2)
           \nonumber \\
    \phi_{N-1}
    &=&    \tilde{t}^{2} + v_{2},~~~
    \phi_{N}
     =     c_{1} \tilde{t}^{2} + C_{2} \tilde{t} + v_{1},
    \eea
  where the last terms have been obtained from the expansion of $x^{0}$ terms
    \bea
    \frac{1}{\Lambda^{Na}}
    \left(- \frac{m_{2}\Lambda^{N-1}}{(1+\tilde{t}/\Lambda^{a})^{N+1}} + \frac{\hat{u}_{N}}{(1+\tilde{t}/\Lambda^{a})^{N}}
     - \frac{m_{1} \Lambda^{N-1}}{(1+\tilde{t}/\Lambda^{a})^{N-1}} \right).
    \eea
  The dimensions of the parameters are easily obtained as
    \bea
    \Delta(v_{i})
     =     2 - \frac{2i}{N+1},~~~
    \Delta(c_{i})
     =     \frac{2i}{N+1},~~~
    \Delta(C_{1})
     =     \Delta(C_{2})
     =     1,
    \eea
  for $i=1, \ldots, [N/2]$,
  where $C_{1} := c_{[(N+1)/2]}$ with dimension-one exists only when $N$ is odd.

\subsection{$\CS[A_{2}; \CD_{\rm reg}, \CD_{{\rm III}}]$ Theories from $\SU(3)$ with $N_{f}=3$}
\label{sec:III3flavors}
  Let us next consider the AD point of $\SU(3)$ with $N_{f}=3$.
  The Seiberg-Witten curve is given by
    \bea
    \phi_{2}
    &=&  - \frac{m_{+}^{2}}{3(t-1)^{2}} + \frac{C_{2}}{t^{2}} + \frac{u_{2}}{t^{2}(t-1)},
           \nonumber \\
    \phi_{3}
    &=&    \frac{2 m_{+}^{3}}{27 (t-1)^{3}} + \frac{\Lambda^{3}}{t^{4}} + \frac{C_{3} - \frac{2C_{2}m_{+}}{3}}{t^{3}}
         + \frac{u_{3}}{t^{3}(t-1)} - \frac{u_{2} m_{+}}{3t^{2}(t-1)^{2}}.
           \label{Nf3}
    \eea
  There are a simple regular puncture at $t=1$ and a full regular puncture at $t = \infty$.
  The puncture at $t=0$ is irregular of $\{ 2, 4 \}$,
  which corresponds to no hypermultiplet.

  We first consider $\phi_{2}$ whose expansion is, by setting $t = \Lambda^{a} w$,
    \bea
    \Lambda^{2a} \left(  \frac{C_{2} - m_{+}^{2}/3}{\Lambda^{2a} w^{2}}
     + \frac{u_{2}-2m_{+}^{2}/3}{\Lambda^{3a} w^{3}}
     + \frac{u_{2}-m_{+}^{2}}{\Lambda^{4a} w^{4}} + \ldots \right).
     \label{phi2phi2}
    \eea
  Let $m_{+}$ be finite parameter here.
  It follows that the second term can be kept finite by $u_{2} = \Lambda^{a} v$
  and the higher order terms are suppressed.
  So, we get
    \bea
    \phi_{2}
     =     \frac{C_{2} - m_{+}^{2}/3}{w^{2}} + \frac{v}{w^{3}}.
    \eea
  We next consider $\phi_{3}$ whose expansion is
    \bea
    \Lambda^{3a} \left(  \frac{C_{3} - \frac{C_{2}m_{+}}{3} +\frac{2m_{+}^{3}}{27}}{\Lambda^{3a} w^{3}}
     + \frac{\frac{2m_{+}^{3}}{9} + \Lambda^{3} + u_{3}- \frac{u_{2}m_{+}}{3}}{\Lambda^{4a} w^{4}}
     + \frac{\frac{4m_{+}^{3}}{9} + u_{3}- \frac{2u_{2} m_{+}}{3}}{\Lambda^{5a} w^{5}} + \ldots \right).
     \label{phi3phi3}
    \eea
  Since $m_{+}$ is finite and $u_{2} \sim \Lambda^{a}$, in order to have $1/w^{5}$ term we need to set
  $u_{3} = - \Lambda^{3} + \Lambda^{3/2} v_{2}$ and $a = 3/2$.
  By this, we get
    \bea
    \phi_{3}
     =     \frac{C_{3} - \frac{C_{2}m_{+}}{3} +\frac{2m_{+}^{3}}{27}}{w^{3}} + \frac{v_{2}}{w^{4}} - \frac{1}{w^{5}}.
    \eea

  The dimensions of the parameters are
    \bea
    \Delta(C_{3})
     =     3, ~~~
    \Delta(C_{2})
     =     2, ~~~
    \Delta(m_{+})
     =     1, ~~~
    \Delta(v_{1})
     =     \frac{3}{2},~~~
    \Delta(v_{2})
     =     \frac{1}{2}.
    \eea
  These are the same as those of the class 2 SCFT of $\SU(3)$ with $N_{f} = 3$ in \cite{Eguchi:1996vu}.

  In terms of a Riemann surface,
  this AD point corresponds to a sphere with one regular full puncture at $t = \infty$
  and one irregular puncture of $\{ 3, 5 \}$.
  Note that these degrees are lower than those of the two hypermultiplets, which is $\{ 4, 6 \}$.

  By the transformation $w \rightarrow 1/w$, the Seiberg-Witten curve is
    \bea
    \phi_{2}
    &=&    \frac{v}{w} + \frac{C_{2} - m_{+}^{2}/3}{w^{2}},
           \nonumber \\
    \phi_{3}
    &=&   \frac{1}{w} - \frac{v_{2}}{w^{2}} - \frac{C_{3} - \frac{C_{2}m_{+}}{3} +\frac{2m_{+}^{3}}{27}}{w^{3}}
    \eea
  where the regular puncture is at $t=0$ and the irregular puncture is at $t=\infty$.
  The Seiberg-Witten differential is $\lambda = x dw$,
  where $x^{3} + \phi_{2} x + \phi_{3} = 0$.

\section{Finite $\CS$-walls, 1-cycles and intersection numbers}
\label{sec:IntNumberSWall}

In section \ref{sec:4dBPSFromSwall} we briefly mentioned that cycles correspond to BPS states can be identified with finite $\CS$-walls. In this section we review how to relate finite $\CS$-walls to 1-cycles on the Seiberg-Witten curve, and how to calculate intersection numbers between the cycles from the finite $\CS$-walls, which is crucial in determining the $\UU(1)$ IR charges of the corresponding BPS states.

\begin{figure}[ht]
\begin{center}
  \begin{tabular}{cc}
	  \begin{tabular}{c}
		  \begin{subfigure}{.25\textwidth}
		    \includegraphics[width=\textwidth]{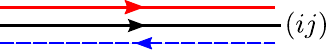}
		    \vspace{.75em}
		    \caption{}
		    \label{fig:swall2cycle}
		  \end{subfigure}
		  \\
		  \\
		  \begin{subfigure}{.4\textwidth}
		    \includegraphics[width=\textwidth]{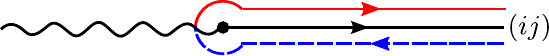}
		    \vspace{.75em}
		    \caption{}
		    \label{fig:swall2cyclewBP}
		  \end{subfigure}
	  \end{tabular} 
	  &
	  \begin{subfigure}{.3\textwidth}
	    \includegraphics[width=1\textwidth]{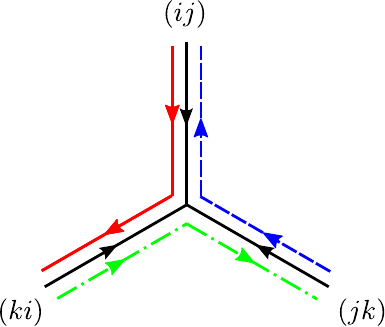}
	    \caption{\label{fig:swall2cyclemulti}}
	  \end{subfigure}
  \end{tabular}
\end{center}
\caption{\label{fig:swall2cycleall} $\CS$-walls and the corresponding 1-cycles.  Black (solid) line: $\CS$-walls. Red (solid) line: cycles on $i$-th sheet. Blue (dash) line: cycles on $j$-th sheet. Green (dash-dot) line: cycles on $k$-th sheet. Black dot: the branch point $(ij)$. Wiggled line: the branch cut separating $i$-th sheet and $j$-th sheet.}
\end{figure}

The direction of a finite $\CS_{ij}$-wall determines the orientation of the corresponding 1-cycle on the Seiberg-Witten curve. Figure \ref{fig:swall2cycle} illustrates the case when one part of the cycle in the $i$-th sheet goes along the direction of the $\CS_{ij}$-wall, while the other part in the $j$-th sheet goes along the opposite direction of the $\CS$-wall. Figure \ref{fig:swall2cyclewBP} shows that when an $\CS$-wall is connected to a branch point, the corresponding 1-cycle goes across the branch cut and moves from the $i$-th sheet into the $j$-th sheet. The condition on a joint of multiple $\CS$-walls guarantees that there is a consistent definition of the corresponding 1-cycle, See Figure \ref{fig:swall2cyclemulti}, where we have a joint of three $\CS$-walls. In this way the direction of a 1-cycle from a finite $\CS$-wall is completely determined. One can reverse the directions of the finite $\CS$-wall to obtain a 1-cycle of the opposite orientation. 

\begin{figure}[ht]
	\centering
	\begin{subfigure}[b]{.4\textwidth}
		\includegraphics[width=\textwidth]{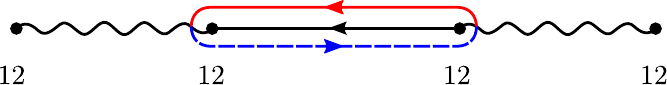}
		\vspace{4em}
		\caption{}
		\label{fig:swall2cycle2cut}
	\end{subfigure}
	\hspace{1em}
	\begin{subfigure}[b]{.4\textwidth}
		\includegraphics[width=\textwidth]{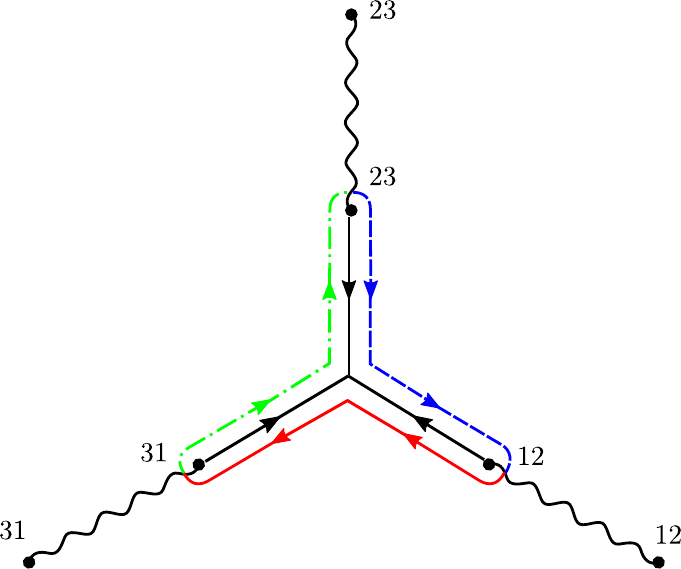}
		\caption{}
		\label{fig:swall2cycle3cut}
	\end{subfigure}
	\caption{Finite $\CS$-walls and corresponding 1-cycles}
	\label{fig:swall2cyclesfullall}
\end{figure}

Examples of 1-cycles from finite $\CS$-walls are shown in Figure \ref{fig:swall2cyclesfullall}, where the same colors and line shapes as those of Figure \ref{fig:swall2cycleall} are used to represent $\CS$-walls, 1-cycles, and branch points/cuts. Figure \ref{fig:swall2cycle2cut} shows a finite $\CS$-wall connecting two branch points, which gives a 1-cycle going from one sheet to the other. Figure \ref{fig:swall2cycle3cut} shows a finite $\CS$-wall connecting three different branch points. 

\begin{figure}[ht]
	\centering
	\begin{subfigure}{.2\textwidth}
	    \includegraphics[width=\textwidth]{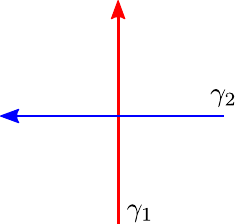}
	    \caption{$\intsec{\gamma_1}{\gamma_2}=+1$}
	    \label{fig:intpositive}
	\end{subfigure}\hspace{1cm}
	\begin{subfigure}{.2\textwidth}
	    \includegraphics[width=1\textwidth]{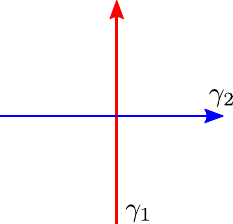}
	    \caption{\label{fig:intnegative}$\intsec{\gamma_1}{\gamma_2}=-1$}
	\end{subfigure}
	\caption{The convention for intersection number.}
	\label{fig:intnumconvention}
\end{figure}

From the intersection number of two 1-cycles we can determine the $\UU(1)$ IR charges of the corresponding BPS states. The convention for an intersection number is summarized in Figure \ref{fig:intnumconvention}. The intersection number is $+1$ if the first cycle goes across the second cycle from its left to its right, while the intersection number is $-1$ in the opposite case.

\begin{figure}[ht]
\begin{center}
  \begin{subfigure}[b]{.3\textwidth}
    \includegraphics[width=\textwidth]{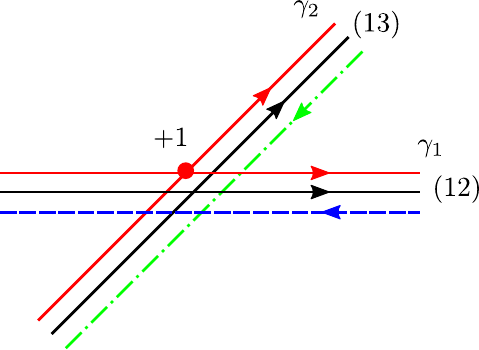}
    \caption{$\intsec{\gamma_1}{\gamma_2}=+1$}
    \label{fig:intreg}
  \end{subfigure}\hspace{1cm}
  \begin{subfigure}[b]{.4\textwidth}
    \includegraphics[width=1\textwidth]{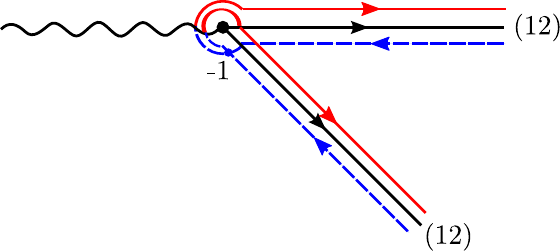}
    \caption{\label{fig:intBP}$\intsec{\gamma_1}{\gamma_2}=-1$}
  \end{subfigure}
\end{center}
\caption{Local intersection number of 1-cycles}
\end{figure}

Since each finite $\CS$-wall completely determines the corresponding 1-cycle, one can read off the intersection number between 1-cycles from the corresponding $\CS$-walls. For example, if a finite $\CS_{ij}$-wall crosses over another finite $\CS_{ik}$-wall, the corresponding 1-cycles will have an intersection on the $i$-th sheet, and the direction of each cycle comes from the direction of each finite $\CS$-wall. Figure \ref{fig:intreg} shows two finite $\CS$-walls crossing over each other. Since one is an $\CS_{12}$ and the other is $\CS_{13}$, the corresponding 1-cycles intersect only on the first sheet. If two $\CS$-walls meet at the same branch point, they will again produce an intersection of the corresponding 1-cycles. Figure \ref{fig:intBP} shows two finite $\CS$-walls meeting at the same branch point. The corresponding 1-cycles have intersection number $\pm1$. 

\begin{figure}[h]
	\centering
    \includegraphics[width=.45\textwidth]{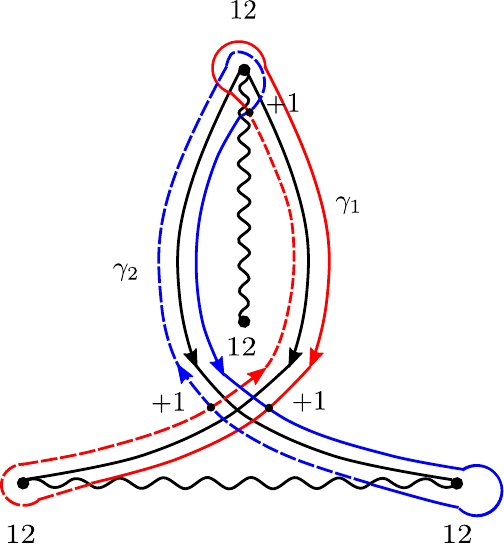}
	\caption{A 1-cycle $\gamma_1$ (red) intersects another 1-cycle $\gamma_2$ (blue) at three points (small black dots), and the intersection number is $\intsec{\gamma_1}{\gamma_2}=+3$. Solid line: cycles on the first sheet. Dash line: cycles on the second sheet.}
	\label{fig:intersecex}
\end{figure}

The intersection number of two 1-cycles is given by summing all the local intersection numbers of them. An example of calculating an intersection number of 1-cycles from finite $\CS$-walls is shown in Figure \ref{fig:intersecex}.


\bibliographystyle{ytphys}
\bibliography{ref}

\end{document}